%% file: mainfile-physrept.tex
\documentclass[a4paper,fleqn]{cas-sc}

\usepackage[numbers,square,sort&compress]{natbib}
\usepackage{braket}
\usepackage{physics}
\usepackage{float}
\usepackage{graphicx}
\usepackage{caption}
\usepackage{slashed}
\def\be{\begin{eqnarray}}
\def\en{\end{eqnarray}}
\usepackage{upgreek}

\usepackage{tikz-feynman}
\tikzfeynmanset{compat=1.1.0}
\tikzfeynmanset{warn luatex=false}
\newcommand{\LbtopKpipi}{\Lambda_{b}^{0}\to pK^-\pi^+\pi^-}

\newcommand{\CP}{C\!P}
\newcommand{\Lb}{\Lambda_{b}^{0}}
\newcommand{\Lbr}{\overline{\Lambda}_{b}^{0}}

\def\tsc#1{\csdef{#1}{\textsc{\lowercase{#1}}\xspace}}
\tsc{WGM}
\tsc{QE}
\tsc{EP}
\tsc{PMS}
\tsc{BEC}
\tsc{DE}

\begin{document}
\let\WriteBookmarks\relax
\def\floatpagepagefraction{1}
\def\textpagefraction{.001}
\shorttitle{Baryonic $C\!P$ violation}
\shortauthors{J.-P. Wang et~al.}

\title [mode = title]{\texorpdfstring{$\CP$ Violation in Beauty Baryon Decays}{CP Violation in Beauty Baryon Decays}}

\author[1,2]{Jian-Peng Wang}[
orcid=0000-0002-3157-3827]
\cormark[1]
\ead{wangjianpeng@lzu.edu.cn}
\credit{Writing (theory)}

\author[3]{Qin Qin}[
orcid=0000-0003-1438-9873]
\cormark[1]
\ead{qqin@hust.edu.cn}
\credit{Writing (theory)}

\author[4]{Stephane Monteil}[
                        orcid=0000-0001-5015-3353]
\cormark[1]
\ead{monteil@in2p3.fr}
\credit{Writing (experiment)}

\author[1]{Fu-Sheng Yu}[orcid=0000-0001-6249-0568]
\cormark[1]
\ead{yufsh@lzu.edu}
\credit{Writing (theory)}

\author[5]{Yan-Xi Zhang}[orcid=0000-0002-0157-188X]
\cormark[1]
\ead{yanxi.zhang@pku.edu.cn}
\credit{Writing (experiment)}

\affiliation[1]{organization={Lanzhou University},
            city={Lanzhou},
            postcode={730000}, 
            country={China}} 
\affiliation[2]{organization={University of Siegen},
            city={Siegen},
            postcode={57072}, 
            country={Germany}}
\affiliation[3]{organization={Huazhong University of Science and Technology},
            city={Wuhan},
            postcode={430074}, 
            country={China}}
\affiliation[4]{organization={Universit\'{e} Clermont Auvergne},
            city={Clermont-Ferrand},
            postcode={63001}, 
            country={France}}
\affiliation[5]{organization={Peking University},
            city={Beijing},
            postcode={‌100871‌}, 
            country={China}}

\cortext[cor1]{Corresponding author}


\begin{abstract}
The 2025 discovery of $\CP$ violation in the baryon sector marks a tectonic shift in flavor physics, bridging a sixty-year gap since the initial discovery of $C\!P$ asymmetry in mesons. This review provides a comprehensive synthesis of the theoretical and experimental journey to this milestone. We integrate the development of dynamical calculation frameworks, such as perturbative QCD and the rescattering mechanism, with the experimental breakthrough achieved through rate asymmetry measurements. Furthermore, we establish a roadmap for the future of the field by detailing the construction of multi-dimensional observables, which are instrumental in advancing the next era of precision physics and refining the search for new physics beyond the standard model.

\end{abstract}




\maketitle
\tableofcontents

\section{Introduction} \label{sec:intro}
\input{sec1-intro}

\section{Basics of Baryonic $\CP$ Violation} \label{sec:theory}
\input{sec2-theory}

\section{Computational Frameworks for Nonleptonic Decays of Baryons} \label{sec:compute}
\input{sec3-computation}

\section{Experimental Landscape} \label{sec:experiment}
\input{sec4-experiment}

\section{The Landmark Discovery and Theoretical Interpretation} 
\label{sec:discovery}
\input{sec5-discovery}

\section{Multi-Dimensional Probes and New Observables}\label{sec:observable}
\input{sec6-observable}

\section{Summary and Prospects} \label{sec:summary}
\input{sec7-summary}

\section*{Acknowledgements}

This work is supported in part by the National Natural Science Foundation of China under Grant Nos. 12522506, 1252500590, 12247101, 12375086, 12335003, W2543009, the China Postdoctoral Science Foundation under Grant Number GZC20261850, the National Key Research and Development Program of China under Grant No.~2022YFA1601904, the Fundamental Research Funds for the Central Universities under No. lzujbky-2023-stlt01, lzujbky-2024-oy02, lzujbky-2025-jdzx07 and lzujbky-2025-eyt01, the Scientific Research Innovation Capability Support Project for Young Faculty under Grant No. ZYGXQNJSKYCXNLZCXMP2, the Natural Science Foundation of Gansu Province (No.25JRRA799), and the ‘111 Center’ under Grant No. B20063. We also thank Dong-Hao Li for providing some of the figures used in this review. One of us
(Y. Zhang) would like to thank Luciano Maiani for valuable discussions.



\printcredits

\bibliographystyle{elsarticle-num}
\bibliography{ref}


\end{document}

%% file: sec1-intro.tex
In 2025, the LHCb collaboration reported the first observation of $\CP$ violation in a baryon decay~\cite{LHCb:2025ray}, more than six decades after $\CP$ symmetry breaking was first established in meson decays in 1964~\cite{Christenson:1964fg}. Here, $\CP$ symmetry denotes invariance under the combined action of charge conjugation ($C$), which interchanges particles and antiparticles, and parity ($P$), which reverses spatial coordinates. Baryons form the class of particles that includes protons and neutrons, which account for most of the mass of ordinary matter, although the observed effect itself occurred in the decay of a short-lived beauty baryon. Like many major advances in particle physics, this milestone emerged
from a sustained interplay between theory and experiment. This review traces the principal theoretical and experimental developments that prepared the ground for this discovery and now shape its interpretation.

The significance of this discovery extends beyond the first observation of $\CP$ violation in a new class of particles. Owing to their three-quark structure and spin, beauty baryons provide information complementary to that available from meson decays, offering an independent arena in which to test our understanding of $\CP$ violation and to search for potential new dynamics. This discovery therefore marks the beginning of a new frontier in flavor physics rather than the completion of a long-standing search. To place this development in its broadest physical context, we first recall the role of $\CP$ violation in the origin of the cosmic matter-antimatter asymmetry.

\subsection{$\CP$ Violation and Cosmological Matter-Antimatter Asymmetry}

The overwhelming predominance of matter over antimatter is one of the central unresolved questions in modern physics. It is quantitatively characterized by the net baryon-to-photon ratio, inferred from observations of the cosmic microwave background and the primordial abundances of light elements~\cite{Planck:2018vyg,ParticleDataGroup:2024cfk}, 
\begin{equation}
    \eta_B \equiv \frac{n_B-n_{\bar B}}{n_\gamma}
    \simeq 6\times 10^{-10},
\end{equation}
where $n_B$, $n_{\bar B}$, and $n_\gamma$ denote the number densities of baryons, antibaryons, and photons, respectively. Although this number is small, it accounts for the ordinary matter observed today, after the annihilation of almost all baryon-antibaryon pairs in the early Universe. The hot Big Bang framework can in principle, although unnaturally, accommodate a nonzero value of $\eta_B$ as an initial condition, but does not explain its microscopic origin. Moreover, in conventional inflationary cosmology, any baryon asymmetry generated before inflation would be exponentially diluted~\cite{Dolgov:1991fr,Dine:2003ax}, so the observed asymmetry must have been produced at the end of inflation or during the subsequent evolution of the Universe. This dynamical generation of a net baryon number is referred to as baryogenesis~\cite{Canetti:2012zc}.

In 1967, Sakharov identified three conditions required for baryogenesis~\cite{Sakharov:1967dj}. First, baryon number violation is needed to evolve from a state with zero net baryon number to one with $n_B\neq n_{\bar B}$. Second, both $C$ symmetry and $\CP$ symmetry must be violated; otherwise, processes producing baryons and antibaryons would occur with equal probabilities and no net preference for matter could develop. Third, baryogenesis must take place out of thermal equilibrium; otherwise, inverse reactions would erase any asymmetry that might be produced. These three requirements concern distinct aspects of the problem: changing the total baryon number, distinguishing matter from antimatter, and preventing the resulting excess of matter from being washed out.

At first sight, the Standard Model (SM) of particle physics appears to contain all three ingredients required for baryogenesis. Electroweak sphaleron processes violate baryon plus lepton number~\cite{Kuzmin:1985mm}, weak interactions violate $C$ and $\CP$ symmetries, and the electroweak phase transition in the early Universe may provide the required departure from thermal equilibrium. Quantitatively, however, the last two ingredients are insufficient. For the observed Higgs-boson mass, the electroweak transition is a smooth crossover rather than a strongly first-order phase transition and therefore does not generate a sufficiently strong departure from equilibrium~\cite{Kajantie:1996mn,DOnofrio:2014rug}. Moreover, the $\CP$ violation associated with quark mixing in weak interactions is strongly suppressed in baryogenesis and produces an asymmetry many orders of magnitude below the observed value~\cite{Gavela:1993ts,Huet:1994jb}. These shortcomings motivate new dynamics capable of modifying the electroweak transition and providing additional sources of $\CP$ violation beyond the SM.

Electroweak baryogenesis provides one possible realization through weak-scale extensions of the SM~\cite{Cohen:1990py,Cohen:1990it,Riotto:1999yt,Morrissey:2012db} and can therefore leave experimentally accessible signatures. Additional degrees of freedom can render the electroweak transition strongly first order, proceeding through expanding bubbles whose walls provide the required departure from thermal equilibrium. New $\CP$-violating interactions with the walls generate particle-antiparticle asymmetries ahead of them, biasing sphaleron transitions and thereby producing a net baryon number. Other possibilities include grand-unified baryogenesis~\cite{Yoshimura:1978ex,Weinberg:1979bt}, leptogenesis~\cite{Fukugita:1986hr}, the Affleck--Dine mechanism~\cite{Affleck:1984fy}, $B$-mesogenesis~\cite{Elor:2018twp,Nelson:2019fln}, and fourth generation model~\cite{Li:2025bon}. These mechanisms operate at different energy scales and can place the relevant $\CP$-violating dynamics in different particle sectors.

Against this broad landscape, hadronic systems occupy a special position: they remain the only systems in which $\CP$ violation has been experimentally established~\cite{ParticleDataGroup:2024cfk}, making hadron decays and mixing a central arena in the search for additional $\CP$-violating dynamics that could account for the cosmic baryon asymmetry. The first observation of $\CP$ violation in a baryon~\cite{LHCb:2025ray} has now made beauty baryon decays a new laboratory for this program. A concrete illustration is provided by the general two-Higgs-doublet model, in which the enlarged scalar sector can strengthen the electroweak transition, while additional complex Yukawa couplings can provide the $\CP$ violation required for electroweak baryogenesis and contribute to low-energy flavor amplitudes~\cite{Branco:2011iw}. In such scenarios, $\CP$-violating observables in beauty-hadron decays can probe the new Yukawa sector relevant to electroweak baryogenesis~\cite{Cline:2011mm,Hou:2023ran}. Not all baryogenesis mechanisms, however, predict observable effects in hadron decays. In leptogenesis, for example, the relevant $\CP$ violation may reside in the heavy-neutrino and lepton sectors and need not leave an observable imprint in hadronic observables~\cite{Davidson:2008bu}. Nevertheless, hadronic measurements remain an essential component of a comprehensive experimental effort to uncover the origin of the cosmic matter-antimatter asymmetry, alongside electric-dipole-moment searches, neutrino experiments, Higgs measurements and direct searches at colliders.

\subsection{Historical Milestones of $C\!P$ violation}

\begin{figure}
    \centering
    \includegraphics[width=0.82\linewidth]{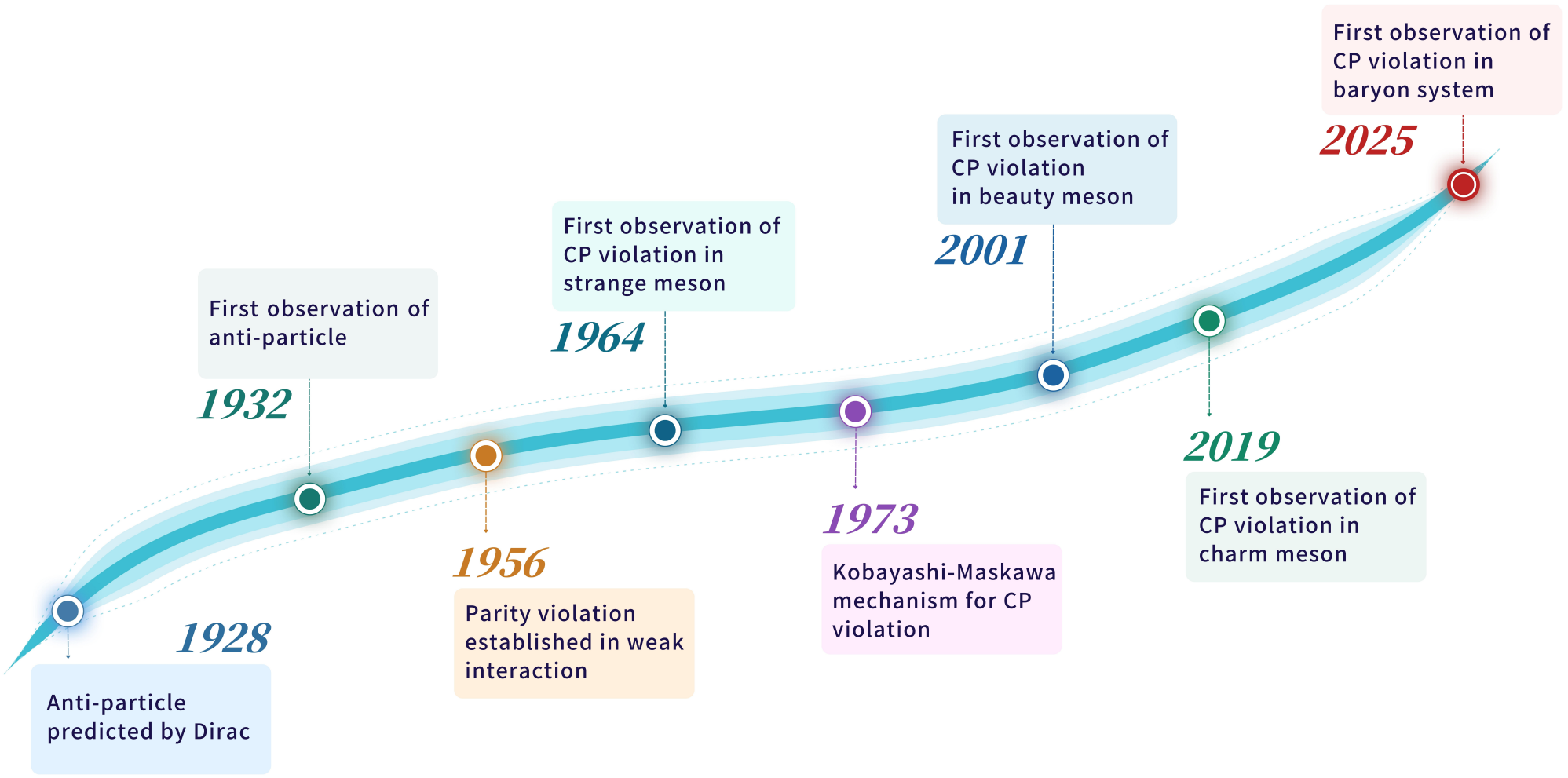}
    \caption{History and milestones of $\CP$ violation.}
    \label{fig:history}
\end{figure}

Having established that $\CP$ violation is an essential ingredient in any dynamical explanation of the cosmic matter-antimatter asymmetry, we now review the principal theoretical and experimental milestones that have shaped our understanding of $\CP$ symmetry and its violation.

The path to this understanding begins with the formulation and experimental testing of discrete symmetries that were once regarded as exact. Parity was among the first such symmetries to be formalized in quantum mechanics. In 1927, Wigner introduced the parity operator and used parity conservation in electromagnetic interactions to explain selection rules in atomic spectra~\cite{Wigner:1927ggq}. 
This was followed in 1928 by Dirac's equation in relativistic quantum mechanics, which admits solutions corresponding to particles with the same mass as known particles but opposite electric charge, {\it i.e.}, the antiparticles~\cite{Dirac:1928hu}. The existence of such antiparticles was spectacularly confirmed by Anderson's cloud-chamber observation of the positron in 1932~\cite{Anderson:1932zz}. Thus, the operation of charge conjugation ($C$), which interchanges particles and antiparticles, became a physically meaningful symmetry to investigate.

Parity and charge-conjugation invariance were subsequently assumed to hold in all fundamental interactions.
The first challenge to this assumption came in 1956. Prompted by the $\tau$--$\theta$ puzzle posed by two apparently identical particles decaying into final states of opposite parity, Lee and Yang pointed out that parity conservation had never been tested in weak interactions and proposed a series of experiments to examine it~\cite{Lee:1956qn}. Parity violation was soon demonstrated in the $\beta$ decay of polarized ${}^{60}\mathrm{Co}$ nuclei~\cite{Wu:1957my}, and in weak decays of hyperons~\cite{Crawford:1957zzb,Eisler:1957ih}, pions and muons~\cite{Garwin:1957hc}. They provided a key experimental foundation for the later establishment of the left-handed chiral structure of the weak interaction~\cite{Feynman:1958ty}. The experiment on pion and muon decays also established that weak interactions violate charge conjugation ($C$), which transforms particles into antiparticles~\cite{Garwin:1957hc}. These discoveries overturned the expectation that $P$ and $C$ are separately exact, while leaving open the possibility that their combination, $\CP$, might remain conserved~\cite{Landau:1957tp}.

The neutral-kaon system provided a particularly sensitive test of the hypothesis that $\CP$ symmetry is conserved. Gell-Mann and Pais had shown that the $K^0$ and $\bar K^0$ states can mix to form two states with markedly different lifetimes~\cite{Gell-Mann:1955ipe}. In the $\CP$-conserving limit, the short-lived state $K_S$ is $\CP$ even and can decay into the $\CP$-even two-pion final state, whereas the long-lived state $K_L$ is $\CP$ odd and cannot. Ioffe, Okun and Rudik, and independently Lee, Oehme and Yang, had presented some of the earliest analyses of possible $\CP$ violation and examined how it could manifest itself in neutral-kaon decays~\cite{Ioffe:1957oga,Lee:1957qq}. In 1964, Christenson, Cronin, Fitch and Turlay observed the $\CP$-forbidden decay $K_L\to\pi^+\pi^-$ with a branching fraction of only about $2\times10^{-3}$, thereby establishing for the first time that $\CP$ symmetry is violated~\cite{Christenson:1964fg}.

The discovery established the existence of $\CP$ violation but did not identify its origin. Wolfenstein proposed that it arose from a new superweak interaction confined to neutral-kaon mixing~\cite{Wolfenstein:1964ks}. Building on Cabibbo's two-generation framework for quark mixing in the weak interaction~\cite{Cabibbo:1963yz}, Kobayashi and Maskawa showed in 1973 that such mixing can generate $\CP$ violation if nature contains at least three generations of quarks~\cite{Kobayashi:1973fv}. This description became known as the Cabibbo--Kobayashi--Maskawa (CKM) mechanism. Subsequent experiments progressively distinguished between these possibilities. In 1999, KTeV and NA48 reported measurements of direct $\CP$ violation in neutral-kaon decays, establishing that $\CP$ violation also occurs in decay amplitudes and ruling out the original superweak scenario~\cite{KTeV:1999kad,NA48:1999szy}. In 2001, BaBar and Belle observed large $\CP$ asymmetries in neutral-$B$-meson decays, $B^0\to J/\psi K_{S(L)}^0$, a manifestation of $\CP$ violation in the third generation, providing strong support for the CKM mechanism~\cite{BaBar:2001pki,Belle:2001zzw}. 

Over the following two decades, the study of beauty mesons developed into a broad program of precision tests of the CKM mechanism. Direct $\CP$ violation was established in both $B^0$ and $B_s^0$ decays~\cite{BaBar:2004gyj,Belle:2004nch,LHCb:2013syl}, while time-dependent measurements in neutral-$B$ systems probed $\CP$ violation arising from the interplay between decay and meson-antimeson mixing~\cite{BaBar:2001pki,Belle:2001zzw,LHCb:2011aa}. Measurements of flavor-specific asymmetries also placed stringent constraints on $\CP$ violation in $B^0$--$\bar B^0$ mixing~\cite{CLEO:2001exw,BaBar:2013xng,LHCb:2014dcj} and $B_s^0$--$\bar B_s^0$ mixing~\cite{D0:2012hft,LHCb:2016ssr}. In parallel, multibody beauty-meson decays revealed large regional $\CP$ asymmetries across phase space, exposing structures that are hidden in phase-space-integrated observables~\cite{LHCb:2013ptu,LHCb:2014mir}. Together, these measurements have tested the CKM description across a wide range of decay topologies and dynamical environments, with no conclusive evidence for contributions beyond it~\cite{ParticleDataGroup:2024cfk,HeavyFlavorAveragingGroupHFLAV:2024ctg}.

For many years, established $\CP$ violation was confined to the strange and beauty meson sectors, both associated with down-type quarks. This changed in 2019, when the LHCb collaboration observed a nonzero difference between the time-integrated $\CP$ asymmetries of the decays $D^0\to K^+K^-$ and $D^0\to\pi^+\pi^-$ with a significance exceeding five standard deviations~\cite{LHCb:2019hro}. This result constituted the first observation of $\CP$ violation in the charm sector and extended the phenomenon to decays of an up-type quark.

Despite these achievements, every established instance of $\CP$ violation still involved only mesons. 
Searches for $\CP$ violation in baryon decays, spanning hyperons, charm baryons and beauty baryons, had been pursued over several decades without reaching the threshold for observation~\cite{HyperCP:2004zvh,BESIII:2021ypr,BESIII:2023jhj,BESIII:2018ciw,LHCb:2017hwf,LHCb:2024yzj}. 
The absence of established $\CP$ violation in baryons has thus constituted a long-standing puzzle, making the pursuit of $\CP$ violation in this sector a critical and highly anticipated frontier in flavor physics. 
In 2025, LHCb measured the phase-space-integrated $\CP$ asymmetry in $\Lb\to pK^-\pi^+\pi^-$ decays,
$$
A_{\CP} \equiv \frac{\Gamma(\Lb\to pK^-\pi^+\pi^-) - \Gamma(\Lbr\to \bar pK^+\pi^-\pi^+)}{ \Gamma(\Lb\to pK^-\pi^+\pi^-) + \Gamma(\Lbr\to \bar pK^+\pi^-\pi^+)} = (2.45 \pm 0.46 \pm 0.10) \% .
$$
This result differs from zero with a significance of $5.2$ standard deviations, establishing $\CP$ violation in a baryon decay for the first time~\cite{LHCb:2025ray}. 
This discovery extended experimentally established $\CP$ violation from mesons to a new class of particles, marking a major milestone in the history of $\CP$-violation studies.

\subsection{Distinctive Features of Baryonic $C\!P$ Violation}

Baryonic $\CP$ violation is of particular importance because the visible universe is overwhelmingly composed of baryons, the very building blocks of ordinary matter. Any successful baryogenesis mechanism must ultimately account for the generation of this baryon excess, making the baryon sector a crucial laboratory for probing the $\CP$-violating dynamics that operated in the early Universe. 
Moreover, owing to their three-quark structure and non-zero spin, baryons provide a complementary window to meson decays, offering an independent arena in which to test our understanding of $\CP$ violation and to search for potential new dynamics. 

\begin{figure}
    \centering
    \includegraphics[width=0.80\linewidth]{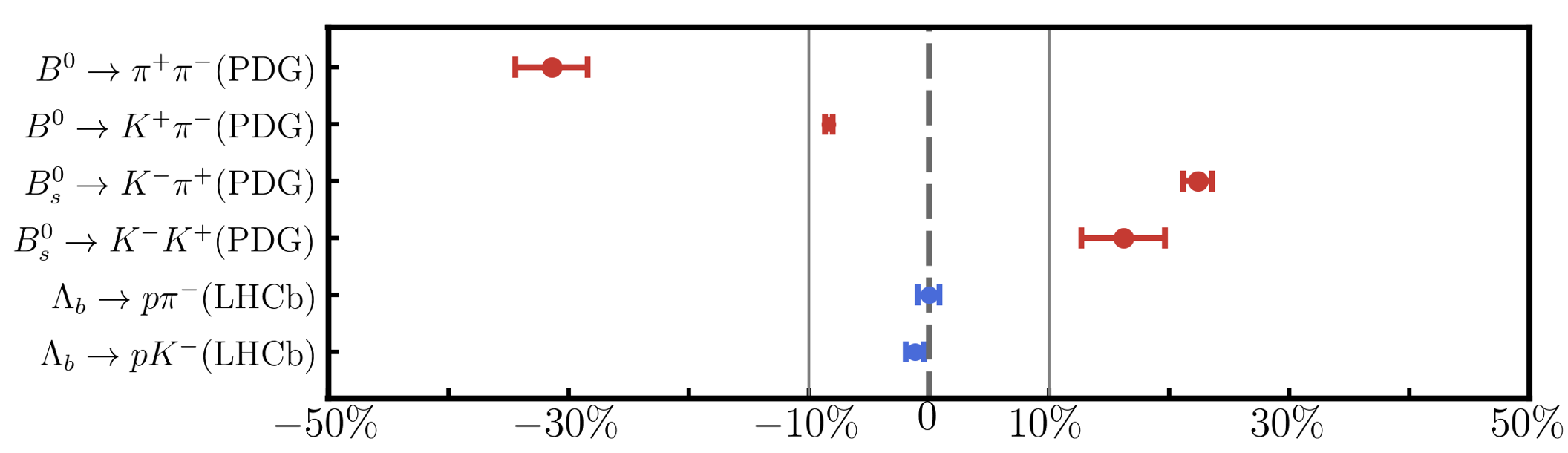}
    \caption{Comparison of $\CP$ violation in $B$-meson and $\Lb$ decays. Typical $\CP$ asymmetries are at the $\mathcal{O}(10\%)$ level in $B$ mesons, but only at the $\mathcal{O}(1\%)$ level in $\Lb$ baryons .}
    \label{fig:magnitude_CPV}
\end{figure}

Baryonic $\CP$ violation exhibits a distinctive pattern that sets it apart from the meson case. While $\CP$ asymmetries in $B$-meson decays typically reach the $\mathcal{O}(10\%)$ level, those in baryon decays are generically an order of magnitude smaller, at the $\mathcal{O}(1\%)$ level, seen in Fig.~\ref{fig:magnitude_CPV}. 
This is why baryonic $\CP$ violation was difficult to observe experimentally until recently.
A plausible explanation for this suppression is the non-zero spin of baryons, which can entail the existence of at least two amplitudes with different orbital angular momenta (partial waves) contributing to the same decay. Individual partial-wave $\CP$ asymmetries might themselves be as large as $\mathcal{O}(10\%)$, comparable to their mesonic counterparts. However, the tree-level and penguin operators enter with opposite chirality structures, namely $(V-A)$ and $(V+A)$, respectively, where $V$ and $A$ denote the vector and axial-vector currents. Consequently, the interference terms responsible for $\CP$ violation in different partial waves may acquire opposite signs, which could lead to a substantial cancellation and a small net $\CP$ asymmetry~\cite{Han:2024kgz}. 

A further distinctive feature concerns the discovery channels. In every meson system where $\CP$ violation was first established, the observation was made in a two-body decay, as discussed earlier. In baryons, by contrast, the first observation was achieved in a four-body decay, $\Lb \to pK^-\pi^+\pi^-$. The reason is that among baryon two-body modes, the only channels with sufficient statistics are the charged decays $\Lb \to p\pi^-$ and $\Lb \to pK^-$, whose measured $\CP$ asymmetries are consistent with zero with a precision at the sub-percent level. The search for baryonic $\CP$ violation has therefore necessarily shifted to multibody final states. Such processes involve a rich spectrum of resonances dominating different regions of phase space, thereby generating diverse $\CP$-violating dynamics and offering the opportunity for relatively enhanced local $\CP$ asymmetries.

Indeed, the discovery of baryonic $\CP$ violation in~\cite{LHCb:2025ray} is dominated by one of the four examined phase-space regions, specifically the low-$p\pi^+\pi^-$-mass region. Here, the $\CP$ asymmetry is measured to be 
\begin{equation}
{A}_{\CP} \bigl(\Lb\to R(p\pi^+\pi^-)K^-\bigr)_ {m_{p\pi^+\pi^-}<2.7~\mathrm{GeV}} = (5.4\pm0.9\pm0.1)\% ,\end{equation}
which reaches a significance of $6\sigma$. 
Nevertheless, many-body decays introduce the theoretical challenge of describing multibody hadronic dynamics and the associated strong phases.
A final-state rescattering mechanism based on nucleon-pion scattering data yields a $\CP$ violation of approximately $5.8\%$, which is consistent with the experimental measurement~\cite{Wang:2024oyi}.

With the existence of $\CP$ violation in baryon decays now established, the central question shifts from whether the effect occurs to what new information it can reveal. Baryons combine a three-quark internal structure with nonzero spin, creating dynamical features and experimental observables that have no direct counterpart in spinless meson decays. We next examine these distinctive features and explain why baryonic decays provide a qualitatively new laboratory for studies of $\CP$ violation.

The most immediate distinction between beauty baryons and the spin-zero mesons discussed above lies in the additional spin degree of freedom. A spin-$1/2$ beauty baryon can be produced in a polarized state, while its decay is generally described by multiple amplitudes corresponding to different helicity configurations, or equivalently to different partial waves. The magnitudes and interference of these amplitudes retain information about spin orientations and correlations that are lost when spins and decay angles are summed over. Beauty baryon decays, therefore, provide access to a broader set of observables than total decay rates alone.

Although the spin states of individual particles are generally not measured directly in collider experiments, information about them is encoded in the angular distributions of their decay products. For a simple weak decay of a spin-$1/2$ baryon into a spin-$1/2$ baryon and a pseudoscalar meson, the angular distribution takes the form ${\dd\Gamma}/{\dd\cos\theta}\propto 1+\alpha \mathcal{P}\cos\theta$~\cite{Lee:1957qs}, where $\mathcal{P}$ denotes the polarization of the parent baryon, $\theta$ is the angle of the daughter baryon relative to the polarization axis, and the parameter $\alpha$ characterizes the parity asymmetry of the decay. Consequently, in a baryon decay where the daughter baryon subsequently undergoes a parity-violating weak decay, the angular distribution of its decay products can in turn serve as an analyzer of the daughter-baryon polarization. More generally, multibody and cascade decays involve several decay angles and correlations between decay planes, whose joint distribution encodes additional spin correlations and interference terms~\cite{Durieux:2016nqr}. Which polarization components and spin correlations can be reconstructed therefore depends on the decay topology and on the available spin analyzers.

This additional information is particularly valuable in searches for $\CP$ violation. A phase-space-integrated rate asymmetry is constructed from rates summed over different helicity configurations, partial waves, intermediate resonances and kinematic regions, so distinct $\CP$-violating contributions may be diluted or cancel one another. Angular coefficients, which can be extracted through suitably weighted moments, can instead isolate selected amplitude contributions and interference terms, revealing differences between baryon and antibaryon decays that may be hidden in the total rates~\cite{Durieux:2016nqr,Zhang:2021fdd,Wang:2022fih}. Their systematic discussion will be shown in Sec.~\ref{sec:observable}. 

Beyond increasing the sensitivity to $\CP$ violation, spin-dependent observables can help characterize the Lorentz and Dirac structures of possible new-physics contributions to baryon decay amplitudes. Possible new interactions with left- or right-handed currents, or with scalar, pseudoscalar or tensor structures, contribute differently to helicity amplitudes and can therefore produce distinctive polarization and angular patterns~\cite{Hiller:2001zj,Das:2018sms}. When such interactions violate $\CP$, the resulting spin-dependent asymmetries can help discriminate among new-physics operators~\cite{Bensalem:2002ys}. Connecting an observed pattern to a specific new-physics interaction requires a combined treatment of short-distance interactions, strong dynamics and multidimensional observables, which motivates the organization of this review.

\subsection{Scope and Organization of the Review}

The expected $\CP$ violation scales vary considerably: $10^{-5}$--$10^{-4}$ for strange, $10^{-4}$--$10^{-3}$ for charmed, and $\sim 10^{-2}$ for beauty baryons, as inferred from direct $\CP$ asymmetries measured in the corresponding meson decays. Consequently, this review focuses primarily on beauty baryon decays, which provide the central laboratory for the established signal and for future precision studies. Results from other baryon sectors are included where they provide historical context, methodological guidance or complementary tests. In this respect, the present article serves as a companion to the review of studies of $\CP$ violation at the proposed Super Tau-Charm Facility (STCF)~\cite{Cheng:2025kpp}, which, among other topics, covers $\CP$ violation in hyperon production and decays as well as in charmed baryon decays. 

To capture the full scope of this emerging field, this review connects the theoretical description of weak decay amplitudes, whose quantitative calculation requires a reliable treatment of strong-interaction dynamics, with the evolution of the experimental methods that led to the first observation. Experimental measurements constrain theoretical descriptions by providing empirical information on the relevant strong-interaction effects, while theoretical calculations identify decay modes and observables expected to offer the greatest sensitivity to $\CP$ violation and thereby help set experimental priorities. Beyond explaining how the discovery was achieved, we aim to show how it opens a new window on $\CP$ violation through distinctive spin-dependent and multidimensional observables in beauty baryon decays, and how these observables can probe new $\CP$-violating dynamics. 

The remainder of this review is organized as follows. Section~\ref{sec:theory} introduces the theoretical foundations of $\CP$ violation in hadron decays and the particular challenges posed by strong dynamics in baryonic systems. Section~\ref{sec:compute} reviews the theoretical frameworks for calculating beauty baryon decay amplitudes, including factorization-based approaches, perturbative QCD and final-state rescattering. Section~\ref{sec:experiment} surveys the experimental landscape and the searches that preceded the first observation, while Section~\ref{sec:discovery} discusses the landmark measurement and its theoretical interpretation. Section~\ref{sec:observable} develops multidimensional probes beyond integrated rate asymmetries, with particular emphasis on spin-dependent and angular observables. Finally, Section~\ref{sec:summary} summarizes the present status and identifies high-priority directions for improving theoretical control, extending experimental measurements and exploiting future facilities.

%% file: sec2-theory.tex
This section introduces the theoretical foundations for describing $\CP$ violation in beauty-baryon decays. In the SM, $\CP$ violation in the quark sector arises from an irreducible complex phase in the CKM matrix, which describes quark mixing in charged-current weak interactions. The CKM matrix is predicted to be unitary in the SM. Precision measurements of its elements and tests of its unitarity therefore provide important probes for possible new flavor and $\CP$-violating interactions.

At energies well below the electroweak scale, hadronic weak decays are described by an effective Hamiltonian containing the effective operators and the corresponding coefficients. The CKM factors supply the weak phases, the Wilson coefficients encode the short-distance dynamics, and the matrix elements of the effective operators describe the strong-interaction effects. $\CP$-violating observables arise from interference between decay amplitudes carrying different weak phases. These amplitudes generally involve different operators or decay topologies and can therefore be shaped differently by strong interactions. A reliable treatment of these contributions is essential for predicting $\CP$-violating observables. The calculation of beauty-baryon decay amplitudes is particularly difficult because their three-quark structure leads to complicated strong dynamics, while their nonzero spin introduces distinct helicity amplitudes.

We first review the CKM mechanism and the current constraints on its parameters. We then introduce the effective weak Hamiltonian and the conditions for observable $\CP$ violation, before discussing the theoretical challenges posed by strong interactions in baryon decays.

\subsection{CKM Mechanism} \label{sec:ckm}

The CKM mechanism has been extensively discussed in reviews and textbooks; see, e.g., Refs.~\cite{Bigi:2000yz,Beyer:2002tj}. For completeness and to fix the notation used in the following, we briefly summarize it here. In the SM, quark masses arise from the Yukawa interactions of the left-handed quark doublets $Q'_{L}=(u'_{L},d'_{L})^{T}$ and the right-handed quark singlets $u'_{R}$ and $d'_{R}$ with the Higgs doublet $\Phi$,
\begin{equation}
\mathcal{L}_{Y} = -\overline{Q}'_{L}Y_{d}\Phi d'_{R} -\overline{Q}'_{L}Y_{u}\widetilde{\Phi}u'_{R} +\mathrm{h.c.},
\end{equation}
where $\widetilde{\Phi}=i\sigma_{2}\Phi^{*}$, $Y_{u,d}$ are generally complex matrices in the quark generation space, and the primed fields denote fields in the weak interaction basis. After electroweak symmetry breaking, the Higgs field acquires the vacuum expectation value $\langle\Phi\rangle=(0,v/\sqrt{2})^{T}$, and the Yukawa matrices give the quark mass matrices $M_{u,d}=vY_{u,d}/\sqrt{2}$. Since these mass matrices are not diagonal in general, they are diagonalized by independent unitary transformations of the left- and right-handed quark fields,
\begin{equation}
U_{uL}^{\dagger}M_{u}U_{uR}=\widehat{M}_{u},
\qquad
U_{dL}^{\dagger}M_{d}U_{dR}=\widehat{M}_{d},
\end{equation}
where $\widehat{M}_{u,d}$ are diagonal matrices that contain the physical masses of the quarks.

The charged-current weak interaction couples the left-handed up- and down-type weak eigenstates through the $W^\pm$ gauge fields, with gauge coupling $g$. After transforming to the mass eigenstates, $u_L=U_{uL}^{\dagger}u'_L$ and $d_L=U_{dL}^{\dagger}d'_L$, it becomes
\begin{equation}
\begin{aligned}
\mathcal{L}_{\mathrm{cc}} &= -\frac{g}{\sqrt{2}}\, \overline{u}'_{L}\gamma^{\mu}d'_{L}W_{\mu}^{+} +\mathrm{h.c.}\\
&=-\frac{g}{\sqrt{2}}\, \overline{u}_{L}\gamma^{\mu}Vd_{L}W_{\mu}^{+} +\mathrm{h.c.}, \qquad
V=U_{uL}^{\dagger}U_{dL}.
\end{aligned}
\end{equation}
The quark mixing matrix $V$ is therefore unitary. It appears because the transformations that diagonalize the up- and down-type mass matrices are generally different.

For $n$ generations with nondegenerate quark masses, a general $n\times n$ unitary matrix contains $n(n-1)/2$ mixing angles and $n(n+1)/2$ phases. Not all these phases are physical. By rephasing the quark fields without changing their masses or kinetic terms, the rows and columns of the matrix are correspondingly rephased, and some of the phases can be hence removed. Of the $2n$ possible quark-field rephasings, one common rephasing leaves the mixing matrix unchanged. Therefore, $2n-1$ phases can be removed, leaving $(n-1)(n-2)/2$ physical phases. For two generations, the mixing matrix therefore contains only one physical parameter and can be written as a real rotation characterized by the Cabibbo angle $\theta_C$~\cite{Cabibbo:1963yz}. In this case, no physical complex phase is present. For three or more generations, by contrast, the mixing matrix can contain physical complex phases. Under a $\CP$ transformation, the charged-current couplings are complex conjugated, $V_{ij}\to V_{ij}^{*}$, and thus the $\CP$ symmetry can be violated. In the three-generation SM, $V$ contains three mixing angles and one physical phase. A nonzero value of this phase therefore makes the charged-current interaction $\CP$ violating~\cite{Kobayashi:1973fv}. The resulting three-generation mixing matrix is known as the CKM matrix.

After choosing a proper quark-field rephasing configuration, the three-generation CKM matrix can be written in terms of three mixing angles, $\theta_{12}$, $\theta_{23}$ and $\theta_{13}$, and one physical phase $\delta$. The standard parametrization is~\cite{ParticleDataGroup:2024cfk}
\begin{equation}\label{eq:ckm_standard}
V_{\rm CKM}= 
\begin{pmatrix} 
c_{12}c_{13} & s_{12}c_{13} & s_{13}e^{-i\delta} \\
-s_{12}c_{23}-c_{12}s_{23}s_{13}e^{i\delta} & c_{12}c_{23}-s_{12}s_{23}s_{13}e^{i\delta} & s_{23}c_{13} \\
s_{12}s_{23}-c_{12}c_{23}s_{13}e^{i\delta} & -c_{12}s_{23}-s_{12}c_{23}s_{13}e^{i\delta} & c_{23}c_{13} 
\end{pmatrix},
\end{equation}
where $s_{ij}\equiv\sin\theta_{ij}$ and $c_{ij}\equiv\cos\theta_{ij}$. This parametrization is exact and preserves the unitarity of the CKM matrix. The observed mixing angles exhibit a strong hierarchy. It is conveniently described by the Wolfenstein parametrization~\cite{Wolfenstein:1983yz}, defined through $s_{12}=\lambda$, $s_{23}=A\lambda^{2}$ and $s_{13}e^{-i\delta}=A\lambda^{3}(\rho-i\eta)$. Expanding the exact parametrization in the small parameter $\lambda\simeq |V_{us}|$ gives
\begin{equation}\label{eq:ckm_wolfenstein}
V_{\rm CKM} =
\begin{pmatrix}
1-\lambda^2/2 & \lambda & A\lambda^3(\rho-i\eta) \\
-\lambda & 1-\lambda^2/2 & A\lambda^2 \\
A\lambda^3(1-\rho-i\eta) & -A\lambda^2 & 1
\end{pmatrix}
+\mathcal{O}(\lambda^4).
\end{equation}
The four real parameters $\lambda$, $A$, $\rho$ and $\eta$ correspond to the three mixing angles and the phase in Eq.~\eqref{eq:ckm_standard}. The CKM matrix is nearly diagonal: mixing between the first two generations appears at order $\lambda$, that between the second and third generations at order $\lambda^2$, and that between the first and third generations at order $\lambda^3$. The parameters $\rho$ and $\eta$ determine the magnitude and phase of the first--third-generation mixing. In this convention, a nonzero value of $\eta$ corresponds to a nonzero complex phase, which first appear in $V_{ub}$ and $V_{td}$ at order $\lambda^3$. For precision studies, it is useful to introduce the rephasing-invariant parameters
\begin{equation}\label{eq:rhobar_etabar}
\bar\rho+i\bar\eta
\equiv
-\frac{V_{ud}V_{ub}^{*}}{V_{cd}V_{cb}^{*}}
=
(\rho+i\eta)
\left[
1-\frac{\lambda^2}{2}
+\mathcal{O}(\lambda^4)
\right].
\end{equation}
A phase-convention-independent measure of CKM $\CP$ violation is the rephasing invariant $J$, introduced in the works of Wu and Jarlskog~\cite{Wu:1985ea,Jarlskog:1985ht} and referred to here as the Wu--Jarlskog invariant,
\begin{equation}\label{eq:wu_jarlskog}
\begin{aligned}
J \equiv
\operatorname{Im}\left(V_{ud}V_{cs}V_{us}^{*}V_{cd}^{*}\right) 
= c_{12}c_{23}c_{13}^{2} s_{12}s_{23}s_{13}\sin\delta
= A^{2}\lambda^{6}\eta+\mathcal{O}(\lambda^{8}).
\end{aligned}
\end{equation}
A nonzero value of $J$ signals $\CP$ violation in the CKM matrix, and shows that $\CP$ violation requires a nonzero phase and nonzero mixing among all three generations.

An important property of the CKM matrix is its unitarity, which implies
\begin{equation}\label{eq:ckm_unitarity_general}
\sum_{\alpha=d,s,b}
V_{i\alpha}V_{j\alpha}^{*}
=
\delta_{ij},
\qquad
\sum_{i=u,c,t}
V_{i\alpha}^{*}V_{i\beta}
=
\delta_{\alpha\beta},
\end{equation}
where $i,j=u,c,t$ and $\alpha,\beta=d,s,b$. The off-diagonal relations, with $i\neq j$ or $\alpha\neq\beta$, define six triangles in the complex plane. The area of every unnormalized unitarity triangle is $|J|/2$, although their side lengths and shapes can differ substantially.

The most commonly studied relation follows from the orthogonality of the $d$ and $b$ columns,
\begin{equation}\label{eq:db_unitarity}
V_{ud}V_{ub}^{*}
+
V_{cd}V_{cb}^{*}
+
V_{td}V_{tb}^{*}
=
0.
\end{equation}
All three terms are of order $\lambda^{3}$, so the corresponding $db$ triangle is not strongly flattened. This feature makes this unitarity triangle particularly useful for experimental tests of the CKM mechanism. In particular, precision measurements of its sides and angles provide stringent tests of CKM unitarity and therefore constitute an important probe of possible new physics associated with the flavor structure. 
Its three internal angles are conventionally defined as~\cite{ParticleDataGroup:2024cfk}
\begin{equation}\label{eq:ut_angles}
\begin{aligned}
\alpha = \phi_{2} &\equiv \arg\left( -\frac{V_{td}V_{tb}^{*}}{V_{ud}V_{ub}^{*}} \right),\\
\beta = \phi_{1} &\equiv \arg\left( -\frac{V_{cd}V_{cb}^{*}}{V_{td}V_{tb}^{*}} \right),\\
\gamma=\phi_{3} &\equiv \arg\left( -\frac{V_{ud}V_{ub}^{*}}{V_{cd}V_{cb}^{*}} \right),
\end{aligned}
\qquad \alpha+\beta+\gamma=\pi.
\end{equation}

From the definition of the angles \eqref{eq:ut_angles}, it can be seen that the standard unitarity triangle is determined mainly from weak decays and mixing of beauty hadrons. The first observations of mixing-induced $\CP$ violation in $B^0\to J/\psi K_S$ decays by BaBar and Belle established that the large effect predicted by the CKM mechanism is present in the beauty sector~\cite{Bigi:1981qs,Belle:2001zzw,BaBar:2001pki}. Subsequent measurements have turned this program from establishing $\CP$ violation into precision tests, in which the three angles and two independent sides of the triangle are constrained by processes with different theoretical and experimental uncertainties.

The angle $\beta$ is most precisely extracted from the time-dependent $\CP$ asymmetry in $B_d^0\to J/\psi K_S$, yielding a current world average of $\beta=(22.63^{+0.45}_{-0.44})^\circ$ as reported by the Heavy Flavor Averaging Group (HFLAV)~\cite{HFLAV:Summer2025}. For $\gamma$, the primary determination relies on tree-level $B^-\to DK^-$ processes, exploiting the interference between the $B^- \to D^0 K^-$ and $B^- \to \bar D^0 K^-$ amplitudes when the neutral $D$ meson decays into a common final state $f$. Assuming $\CP$ conservation in these $D$ decays, the HFLAV average is $\gamma=(66.4^{+2.7}_{-2.8})^\circ$~\cite{HFLAV:Summer2025}. Meanwhile, $\alpha$ is constrained primarily through charmless hadronic $B$-meson decays, especially $B\to\pi\pi$, $B\to\rho\pi$, and $B\to\rho\rho$, yielding a direct HFLAV value of $\alpha=(84.1^{+3.7}_{-3.0})^\circ$~\cite{HFLAV:Summer2025}.
The three measured angles are mutually consistent within their present uncertainties. Their HFLAV combination determines the triangle apex from angle information alone as $(\bar\rho,\bar\eta)=(0.139\pm0.014,0.360\pm0.008)$~\cite{HFLAV:Summer2025}.  This is a direct experimental construction and should be distinguished from an SM global fit, which also uses measurements of CKM magnitudes, neutral-meson mixing and kaon observables.

Combining these and other flavor constraints gives a stringent overconstrained test of the CKM mechanism. The CKMfitter Summer 2025 SM fit finds
\begin{equation}\label{eq:ckmfitter_summer2025}
    \bar\rho=0.156^{+0.010}_{-0.005},
    \qquad
    \bar\eta=0.356^{+0.006}_{-0.007},
\end{equation}
at $68\%$ confidence level (CL)~\cite{Charles:2004jd,CKMfitter:Summer2025}. The constraints from angles, sides, mixing and kaon $\CP$ violation select a common region. The present global fit therefore gives no conclusive indication of a breakdown of three-generation CKM unitarity. An independent global analysis by the UTfit collaboration obtains compatible constraints and likewise supports the conclusion~\cite{UTfit:2022hsi,UTfit:2026qxi}.


The other unitarity relations provide complementary but geometrically less favorable tests.  For the $bs$ relation $V_{us}V_{ub}^{*}+V_{cs}V_{cb}^{*}+V_{ts}V_{tb}^{*}=0$, the three sides scale as $\lambda^4$, $\lambda^2$ and $\lambda^2$.  Its small opening angle is closely related to the weak phase measured in $B_s$ mixing and decay.  For the $cu$ relation $V_{ud}V_{cd}^{*}+V_{us}V_{cs}^{*}+V_{ub}V_{cb}^{*}=0$, the sides scale as $\lambda$, $\lambda$ and $\lambda^5$.  This strong suppression helps explain why CKM $\CP$ violation in charm decays is small. Within current precision, the tests of these relations are also consistent with the three-generation CKM framework~\cite{CKMfitter:Summer2025,UTfit:2026qxi}.

Despite this global consistency, the CKM picture still faces several open questions. Long-standing tensions remain in a number of observables that enter the global flavor fit, motivating more precise measurements and improved theoretical calculations. One important issue is the determination of $V_{cb}$ and $V_{ub}$ from semileptonic $B$ decays. For both CKM elements, the values extracted from exclusive and inclusive analyses have shown persistent differences at the level of roughly 2--3$\sigma$~\cite{Fael:2024rys,HeavyFlavorAveragingGroupHFLAV:2024ctg,Gambino:2020jvv,
Belle-II:2018jsg,
DiCanto:2022icc,
USQCD:2022mmc}. Recent determinations of $V_{ud}$ and $V_{us}$, based on nuclear $\beta$ decays, kaon decays and $\tau$ decays, show mild inconsistencies in tests of first-row CKM unitarity, often referred to as the Cabibbo anomaly. The size of the tension is typically at the level of about $3\sigma$ \cite{Seng:2022ufm,Seng:2022wcw,
Seng:2021nar}, depending on the chosen inputs and theoretical corrections. This discrepancy may originate from experimental systematics or from hadronic and electroweak radiative corrections \cite{Seng:2018yzq,Seng:2018qru,
Czarnecki:2019mwq,Hayen:2020cxh,Shiells:2020fqp,Feng:2020zdc,Hardy:2020qwl,Gorchtein:2018fxl,FlavourLatticeAveragingGroupFLAG:2024oxs,
Ishikawa:2022ulx}, but it also provides
interesting hints for new physics \cite{Crivellin:2020lzu,
Crivellin:2021njn}. Another tension is associated with indirect $\CP$ violation in the neutral kaon system. When the CKM parameters are fixed by other observables, the SM prediction for $|\epsilon_K|$ differs from the experimental value. This comparison is particularly sensitive to $V_{cb}$, and the size of the tension depends strongly on whether the inclusive or exclusive determination of $V_{cb}$ is adopted \cite{Lee:2023lxz}. Finally, rare $B$-decay observables provide another set of complementary tests. Some exclusive $b\to s\ell\ell$ branching fractions and angular observables still differ from SM predictions by about $2$--$3\sigma$ \cite{LHCb:2020lmf,
LHCb:2023gpo,
LHCb:2024onj,
CMS:2024atz,
Bordone:2024hui,
LHCb:2025mqb}. These observables are not direct inputs to the standard unitarity triangle construction, but they are sensitive to loop-induced penguin
and box amplitudes and therefore provide important complementary information on the flavor structure, which may indirectly affect the interpretation of the global CKM picture.

Besides the above mesonic studies, it is worth noting that the rapid progress of LHCb in heavy baryon physics has opened new possibilities for using baryon decays as independent and complementary probes of the CKM matrix elements. For example, semi-leptonic decays such as $\Lb\to p\ell\nu$ and $\Lb\to \Lambda_c^+\ell\nu$ provide alternative determinations of $V_{ub}$ and $V_{cb}$ \cite{Detmold:2015aaa,
LHCb:2015eia,
Zhou:2026fui,Faustov:2016pal}, while decays $\Lambda^{0}_{b}\to \Lambda \ell^{+}\ell^{-}$ and $\Lambda^{0}_{b}\to \Lambda^{+}_{c}\tau\nu$ offer valuable tests of flavor anomalies and constrain possible new physics effects \cite{LHCb:2013uqx,
LHCb:2018jna,
Detmold:2016pkz,
LHCb:2022piu,
Datta:2017aue,
Blake:2017une,
Meinel:2020owd,
Mu:2019bin,
Das:2018sms,
Blake:2019guk}. Several classes of baryon based methods have also been proposed to determine $\gamma$ by exploiting even- and odd-partial-wave interference in three-body $\Xi_b$ decays, interference between tree amplitudes in modes such as $\Lb\to\Lambda D$, and spin-dependent information in the correlated $\Lambda D$ final state~\cite{Sinha:2021mmx,Giri:2001ju,Zhang:2021sit,Rui:2026ihu,Du:2026qsy}. These proposals exploit spin correlations and partial-wave interference in ways that distinguish them from conventional meson-based methods. The detailed observables are discussed in Sec.~\ref{sec:observable}. After reviewing the CKM matrix elements and their current constraints, we next turn to the formulas of $\CP$ violation in heavy baryon decays.

\subsection{Formalism for $C\!P$ Violation in Heavy-Baryons Decays}

To date, three classes of $\CP$ violation have been well established experimentally: direct $\CP$ violation in decays, indirect $\CP$ violation from neutral-meson mixing, and mixed $\CP$ violation arising from the interference between neutral-meson mixing and its decay~\cite{Bigi:2000yz,ParticleDataGroup:2024cfk}. 
Additional types of $\CP$ violation have been proposed based on double mixings~\cite{Shen:2023nuw,Song:2024jjn} or on the interplay between parent decay and daughter mixing~\cite{Yu:2017oky}. 
Since baryon number is conserved, baryons cannot undergo particle-antiparticle mixing. Consequently, they can exhibit only direct $\CP$ violation in baryon decays.

A description of baryonic $\CP$ asymmetries requires the calculation of the relevant decay amplitudes.
Weak decays of hadrons containing heavy quarks play an important role in the extraction of CKM phases, precision tests of the SM, and the search for possible new physics effects. To achieve these goals, experimental measurements and theoretical predictions are both indispensable. On the theoretical side, an accurate treatment of the weak interactions of quarks and reliable methods for describing the strong dynamics involved at different energy scales are required to understand and predict $\CP$ asymmetries in weak hadronic decays. There are three characteristic scales in heavy hadron decays: the electroweak scale $m_{W}$, the heavy quark mass scale $m_Q$, and the hadronic binding scale $\Lambda_{\rm QCD}$. These three scales exhibit a strong hierarchy $m_{W}\gg m_Q\gg \Lambda_{\rm QCD}$. In modern quantum field theory, a well-established and efficient way to separate dynamics associated with different energy scales is to construct an effective field theory. More precisely, from the viewpoint of path integral quantization, the high-energy quantum fluctuations are integrated out and absorbed into a set of effective couplings, known as Wilson coefficients. The remaining interactions at the heavy quark mass scale are then described by the hadronic matrix elements of a set of effective local operators~\cite{Buchalla:1995vs}. 

We illustrate the formalism here with the weak effective Hamiltonian for $b$-quark decays, since this review mainly focuses on beauty baryon decays and their $\CP$-violation. For charmless $\Delta B=1$ transitions, the effective Hamiltonian within the SM is given by~\cite{Buchalla:1995vs}
\begin{equation}\label{EFF HAM}
	\begin{aligned} 
    \mathcal{H}_{\rm eff} \supset  \frac{G_F}{\sqrt{2}}\Big\{ & \sum_{p=u,c}V_{pb} V_{pq}^*\left[C_1(\mu) \mathcal{O}_1^p(\mu)+C_2(\mu) \mathcal{O}_2^p(\mu)\right] \\
    &-V_{t b} V_{t q}^*\left[\sum_{i=3}^{10} C_i(\mu) \mathcal{O}_i(\mu) +C_{7\gamma}(\mu)\mathcal{O}_{7\gamma}(\mu)
+C_{8g}(\mu)\mathcal{O}_{8g}(\mu) \right]\Big\}+\text { H.c.} \;.
	\end{aligned}
\end{equation}
Here $G_F$ denotes the Fermi constant, $q=d,s$, and $C_i(\mu)$, $C_{7\gamma}(\mu)$ and $C_{8g}(\mu)$ are the Wilson coefficients. The CKM unitarity, $\sum_{p=u,c,t}V_{pb}V_{pq}^{*}=0$, has been used to collect the penguin contributions under $-V_{tb}V_{tq}^{*}$. The Wilson coefficients $C_{i}(\mu)$ are obtained through a matching procedure, in which the Green’s functions defined in the effective theory are required to reproduce those of the full theory. By computing appropriate Green’s functions in each theory and equating them at the matching scale, one can extract these Wilson coefficients. This procedure is fundamentally perturbative for both QCD and electroweak interactions and can therefore be carried out with high precision. Complete NNLO QCD matching and renormalization-group evolution are available for the current--current and QCD-penguin coefficients $C_{1-6}$~\cite{Gorbahn:2004my}, as well as for the dipole coefficients~\cite{Misiak:2004ew,Gorbahn:2005sa,Czakon:2006ss}. For the electroweak-penguin coefficients $C_{7-10}$, the dominant, but not complete, NNLO QCD corrections are known~\cite{Buras:1999st}.

The operators $\mathcal O_1,\ldots,\mathcal O_{10}$ are local four-quark operators, whereas $\mathcal O_{7\gamma}$ and $\mathcal O_{8g}$ are dipole operators. Their matrix elements encode the strong-interaction dynamics at the hadronic scale. By solving the renormalization-group equations for the Wilson coefficients, one can resum the large logarithms $\ln(m_W^2/\mu^2)$ appearing in fixed-order perturbative calculations and evolve the coefficients from the matching scale $\mu_W\sim m_W$ to the characteristic scale $\mu\sim m_Q$ of heavy hadron decays. When the Wilson coefficients and renormalized matrix elements are evaluated at the same scale and in the same renormalization scheme, their scale and scheme dependences cancel through the perturbative order retained. The residual scale dependence is therefore a higher-order truncation effect. This implies that higher-order perturbative calculations are essential for reducing the residual $\mu$-dependence and hence the associated perturbative uncertainties of the prediction.

The effective operators are listed explicitly below: 
\begin{itemize}
\item Current-Current tree operators
    \begin{equation}\label{tree operators}
	\begin{aligned}
 & \mathcal{O}_1^p=\left(\bar{p}_\alpha b_\alpha\right)_{V-A}\left(\bar{q}_\beta p_\beta\right)_{V-A},\\
& \mathcal{O}_2^p=\left(\bar{p}_\alpha b_\beta\right)_{V-A}\left(\bar{q}_\beta p_\alpha\right)_{V-A}, 
	\end{aligned}
\end{equation}
\item QCD penguin operators
    \begin{equation}\label{QCD penguin operators}
	\begin{aligned}
 & \mathcal{O}_3=\left(\bar{q}_\alpha b_\alpha\right)_{V-A} \sum_{q^{\prime}}\left(\bar{q}_\beta^{\prime} q_\beta^{\prime}\right)_{V-A}, \\
& \mathcal{O}_4=\left(\bar{q}_\beta b_\alpha\right)_{V-A} \sum_{q^{\prime}}\left(\bar{q}_\alpha^{\prime} q_\beta^{\prime}\right)_{V-A}, \\
& \mathcal{O}_5=\left(\bar{q}_\alpha b_\alpha\right)_{V-A} \sum_{q^{\prime}}\left(\bar{q}_\beta^{\prime} q_\beta^{\prime}\right)_{V+A}, \\
& \mathcal{O}_6=\left(\bar{q}_\beta b_\alpha\right)_{V-A} \sum_{q^{\prime}}\left(\bar{q}_\alpha^{\prime} q_\beta^{\prime}\right)_{V+A},
	\end{aligned}
\end{equation}
\item Electroweak penguin operators
    \begin{equation}\label{EW penguin operators}
	\begin{aligned}
 \mathcal{O}_7 & =\frac{3}{2}\left(\bar{q}_\alpha b_\alpha\right)_{V-A} \sum_{q^{\prime}} e_{q^{\prime}}\left(\bar{q}_\beta^{\prime} q_\beta^{\prime}\right)_{V+A}, \\
\mathcal{O}_8 & =\frac{3}{2}\left(\bar{q}_\beta b_\alpha\right)_{V-A} \sum_{q^{\prime}} e_{q^{\prime}}\left(\bar{q}_\alpha^{\prime} q_\beta^{\prime}\right)_{V+A}, \\
\mathcal{O}_9 & =\frac{3}{2}\left(\bar{q}_\alpha b_\alpha\right)_{V-A} \sum_{q^{\prime}} e_{q^{\prime}}\left(\bar{q}_\beta^{\prime} q_\beta^{\prime}\right)_{V-A}, \\
\mathcal{O}_{10} & =\frac{3}{2}\left(\bar{q}_\beta b_\alpha\right)_{V-A} \sum_{q^{\prime}} e_{q^{\prime}}\left(\bar{q}_\alpha^{\prime} q_\beta^{\prime}\right)_{V-A},
	\end{aligned}
\end{equation}
\item Electromagnetic and chromomagnetic dipole operators
\begin{align}\label{eq:dipoleoperators}
\mathcal{O}_{7\gamma} &= \frac{e}{8\pi^{2}}\,m_b(\mu)\, \bar q_{\alpha}\sigma^{\mu\nu}(1+\gamma_{5})b_{\alpha}\, F_{\mu\nu}, \\
\mathcal{O}_{8g} &= \frac{g_s}{8\pi^{2}}\,m_b(\mu)\, \bar q_{\alpha}\sigma^{\mu\nu}(1+\gamma_{5}) T^{a}_{\alpha\beta}b_{\beta}\, G^{a}_{\mu\nu}.
\end{align}
\end{itemize}
Here $\alpha,\beta$ are color indices, $T^a$ are the generators of color $SU(3)$ group in the fundamental representation, $q^{\prime}=(u,d,s,c,b)$ runs over the active quark flavors at the $m_{b}$ scale, $e_{q'}$ is the fractional electric charge of $q'$, and $F_{\mu\nu}$ and $G^{a}_{\mu\nu}$ are the field strength tensors of the photon and gluon fields, respectively. The left- and right-handed currents are defined as $(\bar{q}_{\alpha}q^\prime_{\beta})_{V\mp A}=\bar{q}_{\alpha}\gamma_{\mu}(1\mp \gamma_{5})q^\prime_{\beta}$. We do not display the evanescent operators, which vanish in four dimensions but must be included in dimensional regularization for a consistent calculation of higher-order corrections~\cite{Buchalla:1995vs}.

The four-quark operator basis defined above is usually referred to as the conventional basis. This basis, especially the penguin operators, is not very convenient for higher-loop calculations because of the difficulty associated with traces involving the $\gamma_{5}$ matrix in dimensional regularization. To overcome this problem, an alternative operator basis, known as the CMM basis, has been developed and widely used~\cite{Chetyrkin:1997gb,Gorbahn:2004my}. The two bases are physically equivalent provided that the Wilson coefficients, renormalized matrix elements and evanescent-operator conventions are treated consistently. To remain consistent with most existing studies of beauty baryon decays, we use the conventional basis in the following discussion. 

The theoretical description of non-leptonic hadron decays then reduces to the calculation of the hadronic matrix elements of these local effective operators between the relevant initial and final hadron states. It is clear that the decay amplitude $\mathcal{A}$ defined by the effective Hamiltonian \eqref{EFF HAM} can be decomposed into two parts, which carry different CKM factors, 
    \begin{equation}\label{eq:tp}
	\begin{aligned}
\mathcal{A}=V_{ub}V^{*}_{uq}\mathcal{A}_{1}+V_{tb}V_{tq}^{*}\mathcal{A}_{2} .
	\end{aligned}
\end{equation}
The contributions arising from the tree operators $\mathcal{O}_{1,2}^u(\mu)$ enter the amplitude $\mathcal{A}_{1}$, the contributions arising from the penguin and dipole operators enter the amplitude $\mathcal{A}_{2}$, and the $\mathcal{O}_{1,2}^c(\mu)$ contributions enter $\mathcal{A}_{1}$ and $\mathcal{A}_{2}$ simultaneously after using CKM unitarity. Even for charmless decays, the operators $\mathcal{O}_{1,2}^c(\mu)$ can also contribute by charm loops. Usually, the amplitudes $\mathcal{A}_{1}$ and $\mathcal{A}_{2}$ are referred to as the tree amplitude and the penguin amplitude, respectively. With the CKM factors factored out explicitly, the amplitudes $\mathcal{A}_{1}$ and $\mathcal{A}_{2}$  can still contain nonzero phases induced by the strong interaction, which are referred to as strong phases. Now, we parameterize $\mathcal{A}$ as 
    \begin{equation}\label{eq:tpCP}
	\begin{aligned}
\mathcal{A}=V_{ub}V^{*}_{uq}|\mathcal{A}_{1}|e^{i\delta_{1}}+V_{tb}V_{tq}^{*}|\mathcal{A}_{2}|e^{i\delta_{2}} .
	\end{aligned}
\end{equation}
The $\CP$-conjugate amplitude can then be obtained straightforwardly by taking the complex conjugate of the CKM factors while keeping the remaining hadronic parts unchanged $\bar{\mathcal{A}}=V^{*}_{ub}V_{uq}|\mathcal{A}_{1}|e^{i\delta_{1}}+V^{*}_{tb}V_{tq}|\mathcal{A}_{2}|e^{i\delta_{2}}$. An intuitive observable for quantifying $\CP$ violation is direct $\CP$ asymmetry, which is defined as the relative difference between the decay rates of two $\CP$-conjugate processes
    \begin{equation}
	\begin{aligned}
A^{\rm dir}_{\CP}=\frac{|\mathcal{A}|^{2}-|\bar{\mathcal{A}}|^{2}}{|\mathcal{A}|^{2}+|\bar{\mathcal{A}}|^{2}}.
	\end{aligned}
\end{equation}
Substituting the parametrized amplitudes \eqref{eq:tp} and \eqref{eq:tpCP} into this definition, one obtains straightforwardly
\begin{equation}\label{eq:acp}
A_{\CP}^{\rm dir}=\frac{-2\operatorname{Im}\left[V_{ub}V_{uq}^{*}V_{tb}^{*}V_{tq}\right]
|\mathcal{A}_{1}||\mathcal{A}_{2}|\sin(\delta_1-\delta_2)
}{|V_{ub}V_{uq}^{*}|^2 |\mathcal{A}_{1}|^2+|V_{tb}V_{tq}^{*}|^2 |\mathcal{A}_{2}|^2
+2\operatorname{Re}\left[V_{ub}V_{uq}^{*}V_{tb}^{*}V_{tq}\right]
|\mathcal{A}_{1}||\mathcal{A}_{2}|\cos(\delta_1-\delta_2)} .
\end{equation}
The absolute value of the imaginary part of the CKM combination appearing in the numerator is exactly $|J|$, where $J$ is the Wu--Jarlskog invariant defined in Eq.~\eqref{eq:wu_jarlskog}. It is easy to see that, for a given decay process, a nonzero direct $\CP$ asymmetry requires not only the presence of a nonzero relative weak phase, but also a nonzero strong-phase difference between the two amplitudes. 

It is often convenient to define the ratio of the moduli of the two interfering amplitudes, for example, $r=|V_{tb}V_{tq}^{*}|^2 |\mathcal{A}_{2}|^2/|V_{ub}V_{uq}^{*}|^2 |\mathcal{A}_{1}|^2$ or its inverse, with $r$ chosen as the ratio of the smaller to the larger amplitude, so that $0<r\leq 1$. Introducing $\Delta\delta=\delta_{1}-\delta_{2}$ and the weak phase $\Delta\phi$ through $J=|V_{ub}V_{uq}^{*}V_{tb}^{*}V_{tq}|\sin\Delta\phi$, the direct $\CP$ asymmetry $A_{\CP}^{\rm dir}$ can then be written as 
\begin{equation}\label{eq:acpn}
A_{\CP}^{\rm dir}=\frac{-2\,r\sin\Delta\phi\sin\Delta\delta}{1+r^{2}+2r\cos\Delta\phi\cos\Delta\delta}.
\end{equation}
For $r\ll1$,  i.e. when one of the tree and penguin amplitudes is much larger than the other, its denominator can be approximated by unity, and the direct $\CP$ asymmetry reduces to
\begin{equation}\label{eq:acp-rratio}
A_{\CP}^{\rm dir}\approx-2\,r\sin\Delta\phi\sin\Delta\delta.
\end{equation}
It provides a very direct intuition for estimating the characteristic direct $\CP$ asymmetries in different decays. The $A_{\CP}^{\rm dir}$ is suppressed when $r$ is very small, as well as when the strong phase difference $\Delta\delta$ or the weak phase $\Delta\phi$ is small. All three of these issues are obviously channel dependent. 
For example, in charm decays, penguin amplitudes are strongly suppressed by the CKM factor $V_{cb}V^{*}_{ub}$, and the associated $A^{\rm dir}_{CP}$ is expected to be very small. In contrast, in beauty decays, penguin contributions can sometimes be comparable to tree amplitudes, and hence are particularly attractive for experimental measurements. 
Equations~(\ref{eq:acp}) and~(\ref{eq:acp-rratio}) are general formulas for direct $\CP$ violation, valid for both mesons and baryons.

\begin{figure}
\begin{center}
\includegraphics[width=0.40\textwidth]{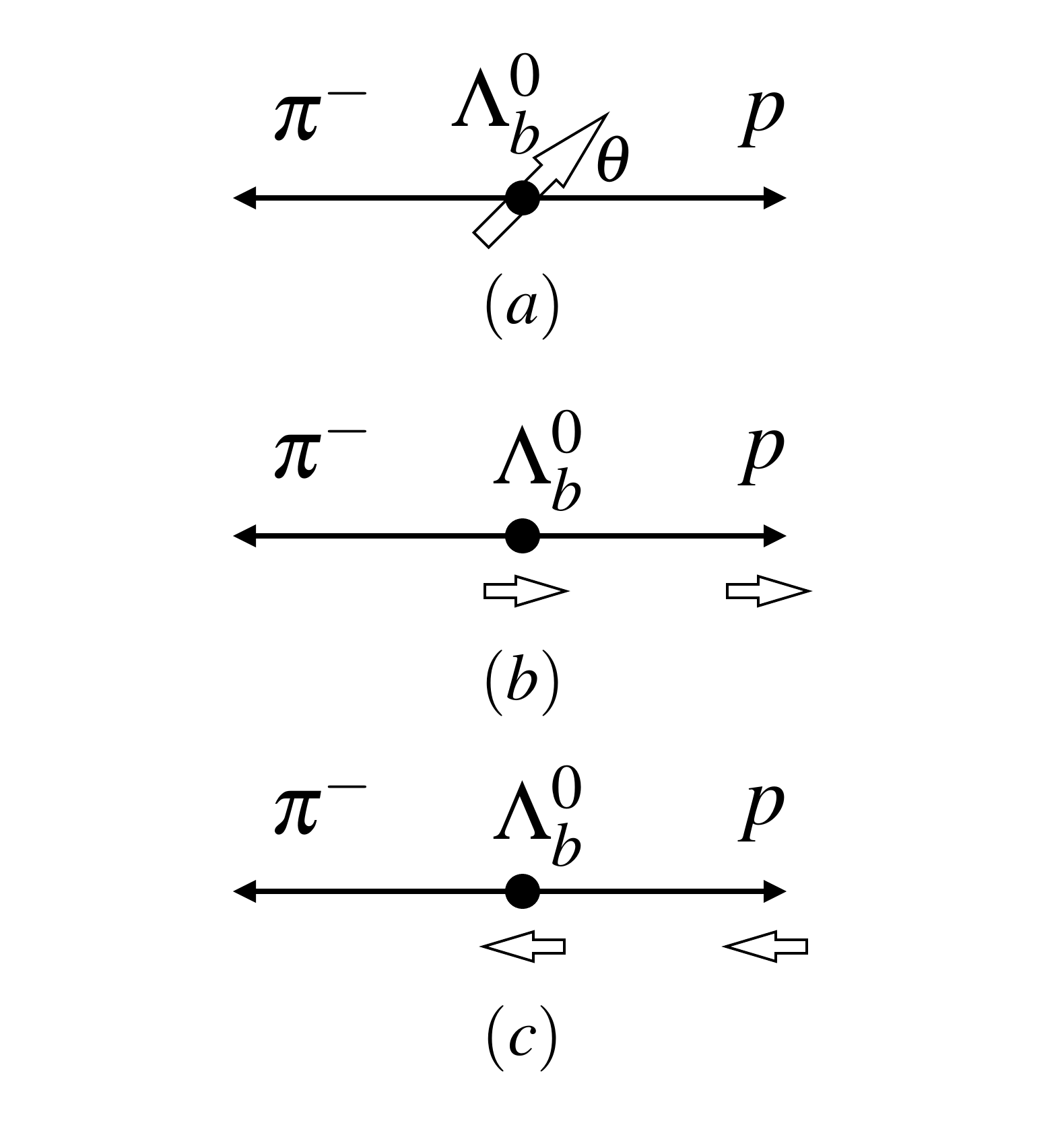}
\captionsetup{type=figure}
\captionof{figure}{(a) Decay of a polarized spin-$1/2$ baryon, where $\theta$ is the angle between the parent-baryon polarization and the proton momentum; (b) For a fully polarized parent baryon and $\theta=0$, only the positive-helicity amplitude $h_+$ contributes, which can be expressed in terms of the partial-wave amplitudes as $h_+=(S+P)/\sqrt{2}$; (c) For a fully polarized parent baryon and $\theta=\pi$, only the positive-helicity amplitude $h_-$ contributes, which can be expressed as $h_-=(S-P)/\sqrt{2}$. }
\label{fig-LambdaPolarization}
\end{center}
\end{figure}

In beauty baryon decays, the non-zero spin of the baryons implies that the decay amplitude consists of multiple components rather than a single one. A spin-$1/2$ baryon has two independent polarization (helicity) states, which can be recast in terms of partial-wave amplitudes with different orbital angular momenta. Taking the simplest process $\Lb \to p h^-$ ($h=\pi^-,K^-$) as an example, both the initial $\Lb$  and the final proton have spin $1/2$, while the final meson $h^-$ is spinless. Angular momentum conservation therefore allows both $S$-wave and $P$-wave contributions. The decay amplitude can be written as
\begin{equation}\label{eq:amplitude-SP}
\mathcal{A}
= i\bar{u}_p(p_p)\left[S + P\gamma_5\right]u_{\Lambda_b}(p_{\Lambda_b}),
\end{equation}
where $S$ and $P$ denote the parity-violating $S$-wave and parity-conserving $P$-wave amplitudes, respectively. 
The partial-wave amplitudes $S$ and $P$ can be related to the two independent helicity amplitudes $h_+$ and $h_-$ illustrated in Fig.~\ref{fig-LambdaPolarization} as
\begin{equation}
h_+=\frac{1}{\sqrt{2}}(S+P), \qquad
h_-=\frac{1}{\sqrt{2}}(S-P).
\end{equation}

Each partial-wave amplitude receives both tree and penguin contributions,
\begin{equation}\label{eq:SPWwave}
\begin{aligned}
S&=\lambda_{\mathcal T}|S_{\mathcal T}|e^{i\delta_{\mathcal T}^{S}} +\lambda_{\mathcal P}|S_{\mathcal P}|e^{i\delta_{\mathcal P}^{S}}, \qquad 
P=\lambda_{\mathcal T}|P_{\mathcal T}|e^{i\delta_{\mathcal T}^{P}} +\lambda_{\mathcal P}|P_{\mathcal P}|e^{i\delta_{\mathcal P}^{P}},
\end{aligned}
\end{equation}
with $\lambda_{\mathcal T}=V_{ub}V_{uq}^{*}$,
$\lambda_{\mathcal P}=V_{tb}V_{tq}^{*}$, and $q=d,s$. Here the calligraphic subscripts $\mathcal T$ and $\mathcal P$ on the amplitudes and strong phases denote the tree and penguin contributions, respectively, whereas the amplitude symbols $S$ and $P$ and the superscripts $S$ and $P$ on the strong phases denote the $S$-wave and $P$-wave partial waves. For example, $S_{\mathcal T}$ is the $S$-wave tree amplitude, $S_{\mathcal P}$ is the $S$-wave penguin amplitude, and $\delta_{\mathcal T}^{S}$ is the strong phase of the $S$-wave tree contribution. The $\CP$ conjugates of the amplitudes are given by~\cite{Donoghue:1986hh}
\begin{equation}\label{eq:SPbarWwave}
\begin{aligned}
\bar{S}=-\lambda^{*}_{\mathcal{T}}|S_{\mathcal{T}}|e^{i\delta^{S}_{\mathcal{T}}}-\lambda^{*}_{\mathcal{P}}|S_{\mathcal{P}}|e^{i\delta^{S}_{\mathcal{P}}}, \qquad 
\bar{P}=\lambda^{*}_{\mathcal{T}}|P_{\mathcal{T}}|e^{i\delta^{P}_{\mathcal{T}}}+\lambda^{*}_{\mathcal{P}}|P_{\mathcal{P}}|e^{i\delta^{P}_{\mathcal{P}}}, \\
\end{aligned}
\end{equation}
where the minus sign in $\bar{S}$ reflects the fact that vector and axial-vector currents transform differently under charge conjugation. The corresponding $\CP$ partial-wave asymmetries are
\begin{equation}\label{eq:acp-partial-wave}
A_{\CP}^{S}=\frac{|S|^2-|\bar S|^2}{|S|^2+|\bar S|^2},
\qquad
A_{\CP}^{P}=\frac{|P|^2-|\bar P|^2}{|P|^2+|\bar P|^2}.
\end{equation}

Substituting Eqs.~\eqref{eq:SPWwave} and~\eqref{eq:SPbarWwave} into Eq.~\eqref{eq:acp-partial-wave}, the partial-wave $\CP$ asymmetries are proportional to the interference terms between the tree and penguin amplitudes,
\begin{equation}\label{eq:acp-partial-wave-explicit}
\begin{aligned}
A_{\CP}^{S} &\propto \operatorname{Im}\left(\lambda_{\mathcal T}\lambda_{\mathcal P}^*\right) |S_{\mathcal T}||S_{\mathcal P}| \sin\left(\delta_{\mathcal T}^{S}-\delta_{\mathcal P}^{S}\right), \\
A_{\CP}^{P} &\propto \operatorname{Im}\left(\lambda_{\mathcal T}\lambda_{\mathcal P}^*\right) |P_{\mathcal T}||P_{\mathcal P}| \sin\left(\delta_{\mathcal T}^{P}-\delta_{\mathcal P}^{P}\right).
\end{aligned}
\end{equation}
These expressions show that $A_{\CP}^{S}$ and $A_{\CP}^{P}$ are generated by the interference between the tree and penguin contributions within each partial wave, and that they can have different signs and magnitudes because the corresponding strong-phase differences $\delta_{\mathcal T}^{S}-\delta_{\mathcal P}^{S}$ and $\delta_{\mathcal T}^{P}-\delta_{\mathcal P}^{P}$ are generally different.

Consequently, the integrated direct $\CP$ asymmetry is not an independent quantity in baryon decays, but a weighted sum of the partial-wave asymmetries,
\begin{equation}\label{eq:acp-total}
A_{\CP}^{\rm dir}
=\kappa_S A_{\CP}^{S}+\kappa_P A_{\CP}^{P},
\end{equation}
where the weight coefficients are
\begin{equation}\label{eq:acp-weights}
\kappa_S=\frac{|S|^2+|\bar S|^2}
{|S|^2+|\bar S|^2+\eta\left(|P|^2+|\bar P|^2\right)},
\qquad
\kappa_P=1-\kappa_S,
\end{equation}
with the kinematic factor
\begin{equation}\label{eq:eta-kinematic}
\eta=\frac{(m_{\Lambda_b}-m_p)^2-m_h^2}{(m_{\Lambda_b}+m_p)^2-m_h^2}.
\end{equation}
These weights reflect the relative $S$- and $P$-wave contributions to the $\CP$-averaged decay rate.

\begin{figure}
\begin{center}
\includegraphics[width=0.50\textwidth]{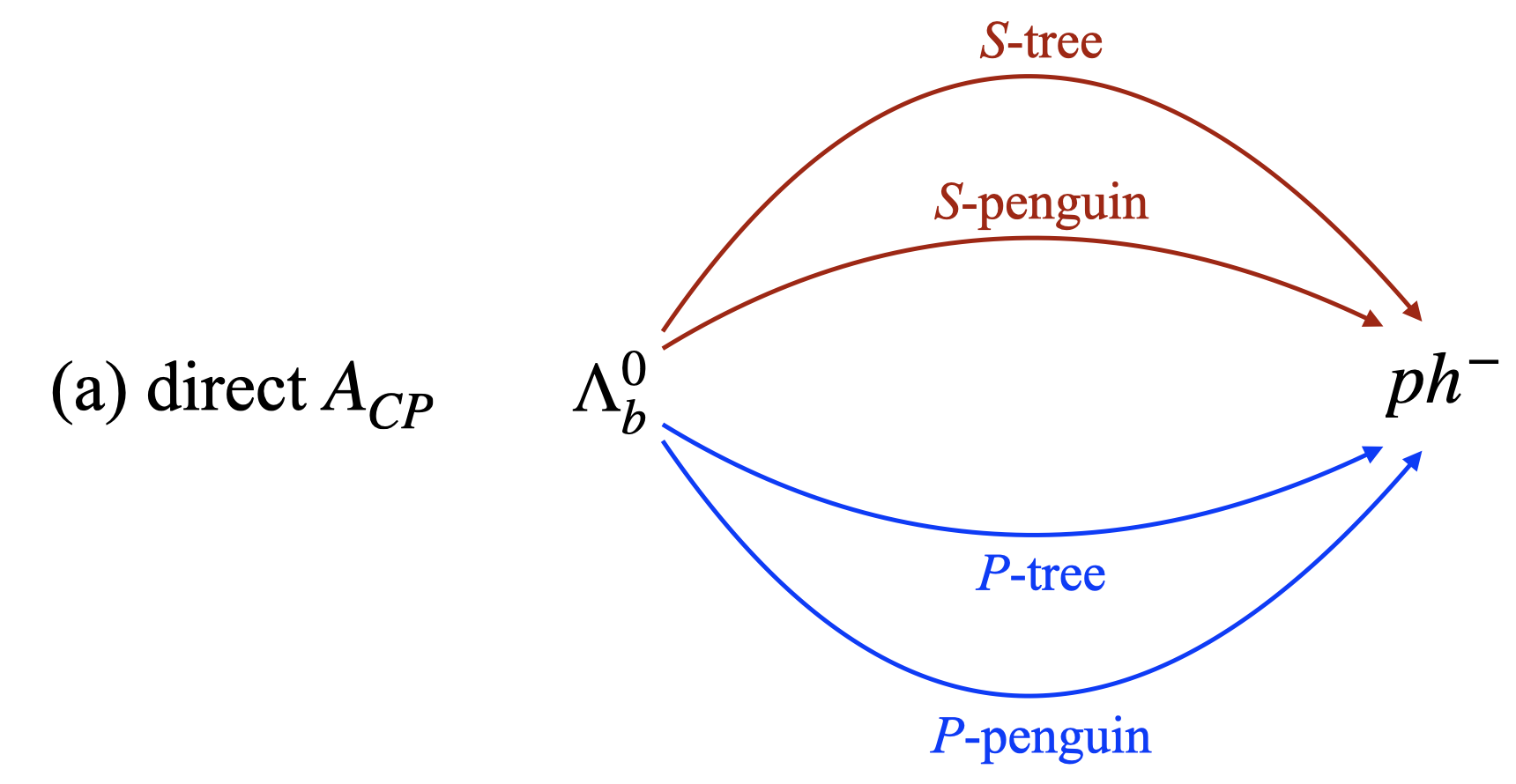}
\vspace{5mm}

\includegraphics[width=0.50\textwidth]{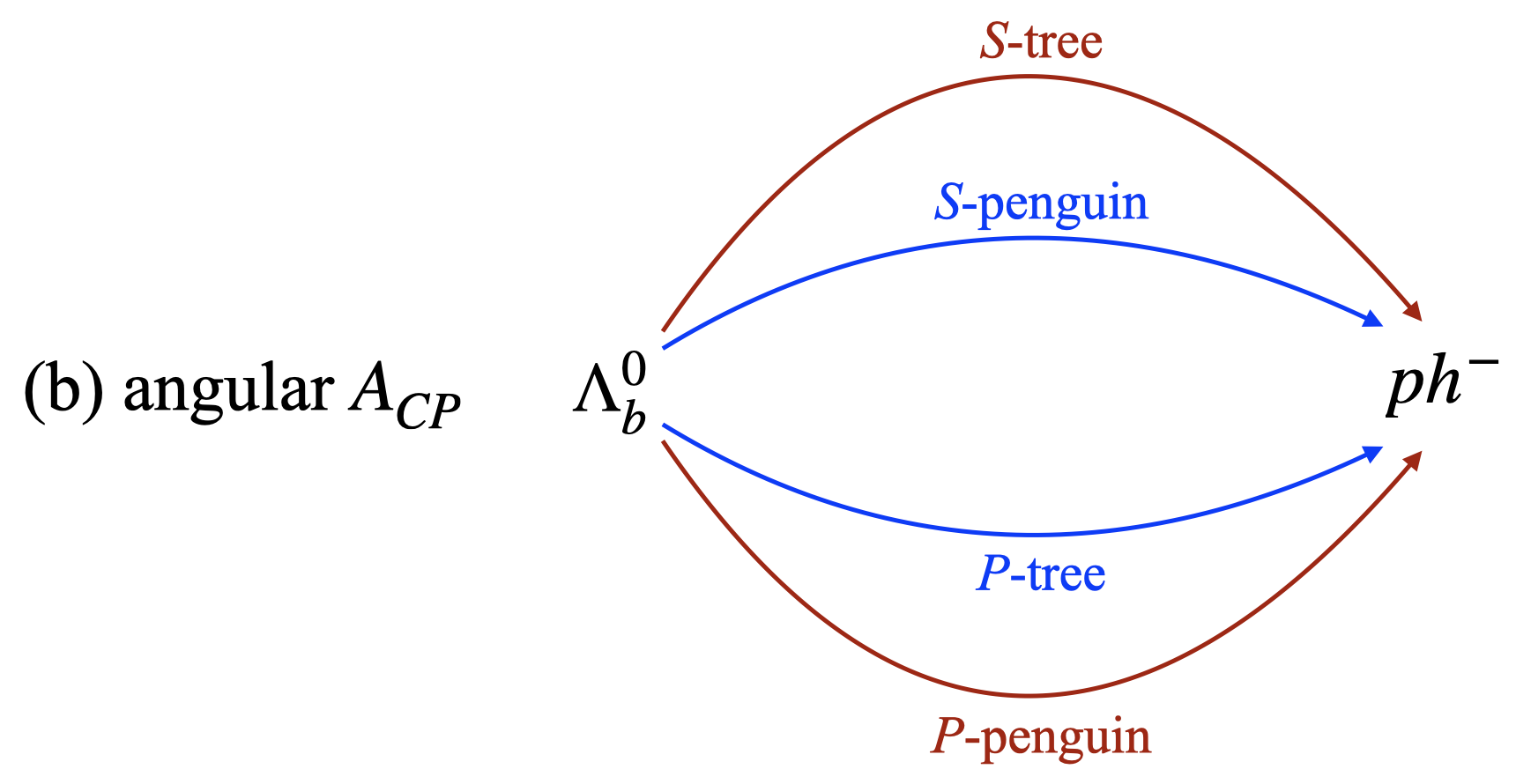}
\caption{Schematic illustration of the interference mechanisms contributing to different \(\CP\)-violating observables in \(\Lb \to p h^-\) decays. Curves of the same color denote the two interfering amplitudes. (a) The direct \(\CP\) asymmetry \(A_{CP}^{\rm dir}\) arises from the interference between tree and penguin amplitudes within the same partial wave (i.e., \(S\)-wave tree with \(S\)-wave penguin, and \(P\)-wave tree with \(P\)-wave penguin). (b) The \(\CP\) violation in angular distribution observables originates from the interference between tree and penguin amplitudes belonging to different partial waves (i.e., \(S\)-wave tree with \(P\)-wave penguin, and \(P\)-wave tree with \(S\)-wave penguin).}
\label{fig-Acp-direct-angular}
\end{center}
\end{figure}

Beyond the integrated rate asymmetry in Eq.~\eqref{eq:acp-total}, additional $\CP$-violating information is encoded in the angular distributions of baryon decays. For a fully polarized parent baryon, the angular distribution is given by
\begin{equation}\label{eq:angular-polarized}
\frac{1}{\Gamma}\frac{\mathrm{d}\Gamma}{\mathrm{d}\cos\theta}
= \frac{1}{2}\left(1+\alpha\cos\theta\right),
\end{equation}
where the Lee-Yang decay asymmetry parameter $\alpha$ is defined in terms of the helicity amplitudes as
\begin{equation}\label{eq:alpha-helicity}
\alpha
= \frac{|h_+|^2-|h_-|^2}{|h_+|^2+|h_-|^2}.
\end{equation}
In terms of partial-wave amplitudes, one obtains
\begin{equation}\label{eq:alpha-SP}
\alpha
= \frac{2\,\mathrm{Re}(S^*P)}{|S|^2+|P|^2}.
\end{equation}
Thus, $\alpha$ directly probes the interference between the $S$- and $P$-wave amplitudes. Its $\CP$ asymmetry is defined as
\begin{equation}\label{eq:acp-alpha}
a_{\CP}^{\alpha}
= \frac{\alpha+\bar{\alpha}}{\alpha-\bar{\alpha}},
\qquad
\bar{\alpha}
= \frac{2\,\mathrm{Re}(\bar S^*\bar P)}{|\bar S|^2+|\bar P|^2},
\end{equation}
where the plus sign in the numerator follows from the parity-odd nature of $\alpha$.

Analogously, the \(\CP\) asymmetry of the Lee-Yang parameter \(\alpha\), which arises from the interference between the \(S\)- and \(P\)-wave amplitudes, is proportional to the interference terms between the tree and penguin contributions in different partial waves,
\begin{equation}\label{eq:acp-alpha-explicit}
a_{\CP}^{\alpha} \propto \operatorname{Im}\left(\lambda_{\mathcal T}\lambda_{\mathcal P}^*\right) \left[ |S_{\mathcal T}||P_{\mathcal P}| \sin\left(\delta_{\mathcal P}^{P}-\delta_{\mathcal T}^{S}\right) - |S_{\mathcal P}||P_{\mathcal T}| \sin\left(\delta_{\mathcal T}^{P}-\delta_{\mathcal P}^{S}\right) \right].
\end{equation}
Its difference from the direct $\CP$ asymmetries in Eq.~(\ref{eq:acp-partial-wave-explicit}) is schematically illustrated in Fig.~\ref{fig-Acp-direct-angular}.

Therefore, besides the integrated direct $\CP$ asymmetry, which is a weighted sum of partial-wave asymmetries, baryon decays also exhibit $\CP$ violation in angular distributions generated by the interference between different partial-wave amplitudes and their charge-conjugate counterparts. 
This angular $\CP$ violation, exemplified by $a_{\CP}^{\alpha}$, provides a complementary probe to $A_{\CP}^{\rm dir}$, since it can remain nonzero even when the latter is suppressed or vanishes.

The discussion above is based on the effective Hamiltonian~\eqref{EFF HAM}, in which the tree and penguin operators contribute to the same charmless final state. A second class of $\Delta B=1$ transitions deserves separate attention, namely $b\to c\bar{u}q$ and $b\to u\bar{c}q$ with $q=d,s$. The four quark flavors involved are all different, so that no penguin operator can contribute and the effective Hamiltonian contains only current--current tree operators,
\begin{equation}
\mathcal{H}_{\rm eff} \supset \frac{G_{F}}{\sqrt{2}}
\Big\{V_{cb}V^{*}_{uq}\big[C_{1}(\mu)\mathcal{O}^{c\bar{u}}_{1}(\mu)
+C_{2}(\mu)\mathcal{O}^{c\bar{u}}_{2}(\mu)\big]
+V_{ub}V^{*}_{cq}\big[C_{1}(\mu)\mathcal{O}^{u\bar{c}}_{1}(\mu)
+C_{2}(\mu)\mathcal{O}^{u\bar{c}}_{2}(\mu)\big]\Big\}+{\rm H.c.} \; ,
\end{equation}
with
\begin{align}\label{b to c operator}
\mathcal{O}^{c\bar{u}}_{1}&=(\bar{c}_{\alpha}b_{\alpha})_{V-A}(\bar{q}_{\beta}u_{\beta})_{V-A},
&\mathcal{O}^{c\bar{u}}_{2}&=(\bar{c}_{\alpha}b_{\beta})_{V-A}(\bar{q}_{\beta}u_{\alpha})_{V-A},
\nonumber\\
\mathcal{O}^{u\bar{c}}_{1}&=(\bar{u}_{\alpha}b_{\alpha})_{V-A}(\bar{q}_{\beta}c_{\beta})_{V-A},
&\mathcal{O}^{u\bar{c}}_{2}&=(\bar{u}_{\alpha}b_{\beta})_{V-A}(\bar{q}_{\beta}c_{\alpha})_{V-A}.
\end{align}
Since QCD is flavor blind, these operators mix under renormalization in exactly the way characterized by the anomalous-dimension submatrix of $\mathcal{O}^{p}_{1,2}$ in Eq.~\eqref{tree operators}, and their Wilson coefficients are given by the same functions $C_{1,2}(\mu)$.

In decay modes containing a neutral $D$ meson such as $B^{-}\to DK^{-}$ discussed in Sec.~\ref{sec:ckm}, the two terms above produce $D^0$ and $\bar D^0$ states. Interference arises when the neutral $D$ mesons are reconstructed in a final state $f$ accessible to both. Neglecting neutral-$D$ mixing and $\CP$ violation in $D$-meson decays, the full cascade amplitude, including the $D\to f$ decay amplitudes, has the same two-term structure as Eq.~\eqref{eq:tp}, but with different CKM factors. Its direct $\CP$ asymmetry is therefore of the form in Eq.~\eqref{eq:acp}, with the relevant strong-phase difference arising from the two complete decay routes.

The two cases discussed above show that a direct partial-rate asymmetry generated by two interfering contributions requires both a nonzero relative weak-phase difference and a nonzero strong-phase difference. This strong-phase requirement is specific to this observable rather than a condition for $\CP$ violation in the underlying interaction, which in the CKM mechanism is characterized by the rephasing-invariant condition $J\neq0$. In fact, there exist other types of $\CP$-violating observables that have different reliance on strong-phase difference and need not vanish when it is zero. Such observables, together with the direct $\CP$ asymmetries, are expected to play a central role in experimental studies and phenomenological analyses. We will return to the issue of $\CP$-violating observables in baryon decays and discuss it in more detail in Sec.~\ref{sec:observable}.

\subsection{Complexity of Strong Dynamics in Baryon Decays}

In theoretical predictions of $\CP$ asymmetries in heavy-hadron decays, the main challenge lies in treating the QCD dynamics contained in the hadronic matrix elements. As discussed in the preceding subsection, reliable predictions require control over the relative magnitudes and strong phases of the interfering amplitudes. Although mesonic and baryonic decays are governed by the same underlying weak interactions, their hadronic matrix elements involve different strong dynamics. Even without going into the technical details, it can be seen that the three-quark structure of baryons introduces additional difficulties both in factorization-based calculations and in phenomenological descriptions of long-distance effects.

In factorization based approaches, beauty baryon decays involve two spectator quarks, whereas beauty meson decays involve only one. Consequently, for a heavy-to-light baryon transition at large recoil region, the hard scattering contribution generally requires two hard-collinear gluon exchanges to transfer momentum to two spectator quarks, so that the corresponding kernel starts at $\mathcal{O}(\alpha_s^2)$~\cite{Wang:2011uv,Huang:2024ugd,Anikin:2013aka}. The additional quark and gluon connections increase the number of Feynman diagrams and complicate the power counting of the contributing amplitudes. Calculating these contributions and determining which could be negligible or not therefore require more extensive analyses than that in the meson decays. The  $\Lb\to p$ and $\Lb\to\Lambda$ form factors analyses have demonstrated that higher twist contributions can be numerically important
~\cite{Wang:2011uv,Wang:2015ndk,Han:2022srw,Huang:2022lfr,Feldmann:2023plv,Lu:2025gjt,Yang:2025yaw}. Formal power counting alone therefore does not necessarily imply a clear numerical hierarchy in beauty baryon decays. Moreover, even though some of subleading amplitudes do not play a dominant role in the branching fractions, they may still have significant impacts on the relative strong phases and thereby on the $\CP$ asymmetries. Meanwhile, the required nonperturbative inputs, such as light-cone distribution amplitudes (LCDAs), present further challenges. Baryon LCDAs involve more momentum variables and independent spin-Lorentz structures \cite{Braun:2000kw}, while their shapes and higher-twist components remain much less constrained. The nonzero spins of baryons give rise to a larger number of helicity amplitudes or partial wave components, thereby requiring the calculation of more hadronic amplitudes. Potential cancellations among these contributions to an integrated $\CP$ asymmetry can make the prediction more sensitive to the uncertainties in the individual amplitudes, requiring greater precision in their calculation \cite{Han:2024kgz}.

Model calculations emphasizing long-distance final-state interactions (FSI) likewise face several challenges. Strong rescattering couples the observed final states to all other accessible intermediate hadronic channels allowed by the symmetries, such that beauty baryon decays would involve more light and charmed baryons and mesons. Intermediate excited baryons and mesons are a particular source of uncertainty, since many of their masses, widths, strong couplings and transition form factors are poorly known. Reliable FSI estimates are therefore considerably hard to establish for beauty baryon decays. 

The smallness of $\CP$ asymmetries in two body $\Lambda^{0}_{b}$ decays have shifted both experimental and theoretical attentions toward multibody beauty baryon decays where potentially large $\CP$ violation may arise, however, the underlying dynamics become substantially more complicated, where different mechanisms could dominate different regions of phase space.
A quasi-two-body approximation is most reliable when an isolated narrow resonance dominates. In contrast, several broad baryon resonances can overlap with each other and with nonresonant contributions, requiring a coherent treatment of their magnitudes and phases. Although such problems also occur in meson decays \cite{Heuser:2025mnk,Bediaga:2020qxg}, the richer excited baryon spectra and limited knowledge of them make these problems particularly challenging in baryonic channels.

Flavor-$SU(3)$ symmetry offers a complementary way to relate decay amplitudes without calculating their strong dynamics directly. Its application to beauty baryon decays is limited by the small number of measured channels and the lack of polarization and angular information: helicity-summed measurements generally do not determine the complex amplitudes and relative strong phases for each helicity component separately. Symmetry-breaking corrections introduce further hadronic uncertainties. 

Because of these difficulties, the theoretical study of baryonic $\CP$ violation is significantly less developed than its mesonic counterpart. It therefore motivates the use of complementary methods to calculate or constrain the relevant amplitudes of beauty baryon decays. The following section reviews generalized and QCD factorization, perturbative QCD, and rescattering approaches, with particular attention to their phenomenologically applications, underlying assumptions and present limitations in baryon decays.

%% file: sec3-computation.tex
Predicting $\CP$-violating observables in hadron decays requires quantitative calculations of decay amplitudes, in particular of the hadronic matrix elements of the effective weak Hamiltonian. Short-distance electroweak and QCD corrections can be calculated perturbatively order by order and absorbed into the Wilson coefficients once the effective Hamiltonian has been established. By contrast, the hadronic matrix elements $\bra{f}\mathcal{O}_{i}\ket{i}$, defined by inserting effective operators between specific initial and final hadron states, encode dynamics associated simultaneously with the heavy-quark mass scale $m_{Q}$ and the nonperturbative scale $\Lambda_{\text{QCD}}$. In general, they cannot be determined fully by perturbative methods alone. A variety of theoretical approaches have been developed to calculate the hadronic matrix elements of effective operators in non-leptonic $B$-meson decays, including naive factorization~\cite{Bauer:1984zv,Bauer:1986bm}, generalized factorization~\cite{Ali:1997nh,Neubert:1997uc,Ali:1998eb}, QCD factorization (QCDF)~\cite{Beneke:1999br,Beneke:2000ry,Beneke:2001ev,Beneke:2003zv}, perturbative QCD (PQCD)~\cite{Lu:2000em,Keum:2000wi,Keum:2000ph}, soft-collinear effective theory (SCET)~\cite{Bauer:2000yr,Bauer:2001yt}, $SU(3)$ flavor symmetry~\cite{Zeppenfeld:1980ex,Savage:1989ub}, and hadronic rescattering methods~\cite{Cheng:2004ru,Cheng:2005wf}. However, the extension of these methods to exclusive non-leptonic heavy-baryon decays is still not sufficiently mature, and their applications remain comparatively limited. This section reviews the main frameworks used to calculate non-leptonic beauty-baryon decay amplitudes and to assess their implications for decay rates and $\CP$ asymmetries. 

\subsection{Topological Diagrams}
The weak decay amplitudes of exclusive non-leptonic heavy hadron decays can be described in terms of different topological diagrams, which are classified according to the topology of weak effective operator insertions and the associated quark flows \cite{Chau:1982da}. It should be emphasized that these topological diagrams are merely symbolic representations and are intended to include all strong-interaction effects, both perturbative and nonperturbative. It is a model independent way since all possible gluon exchanges, quark pair production and annihilation are implicitly included. This approach has been widely applied to heavy meson decays, particularly those of $B$ and $D$ mesons \cite{Chau:1986jb,Chau:1987tk,Chau:1986tn,Chau:1989tk,Gronau:1994rj,Chiang:2004nm}, and has been established on a rigorous theoretical footing. It has been established that, in the meson sector, the topological amplitudes are in one-to-one correspondence with the irreducible representations of the $\text{SU(3)}$ flavor symmetry group \cite{Wang:2020gmn,He:2018php}.

\begin{figure}
\begin{center}
    \includegraphics[width=0.6\textwidth]{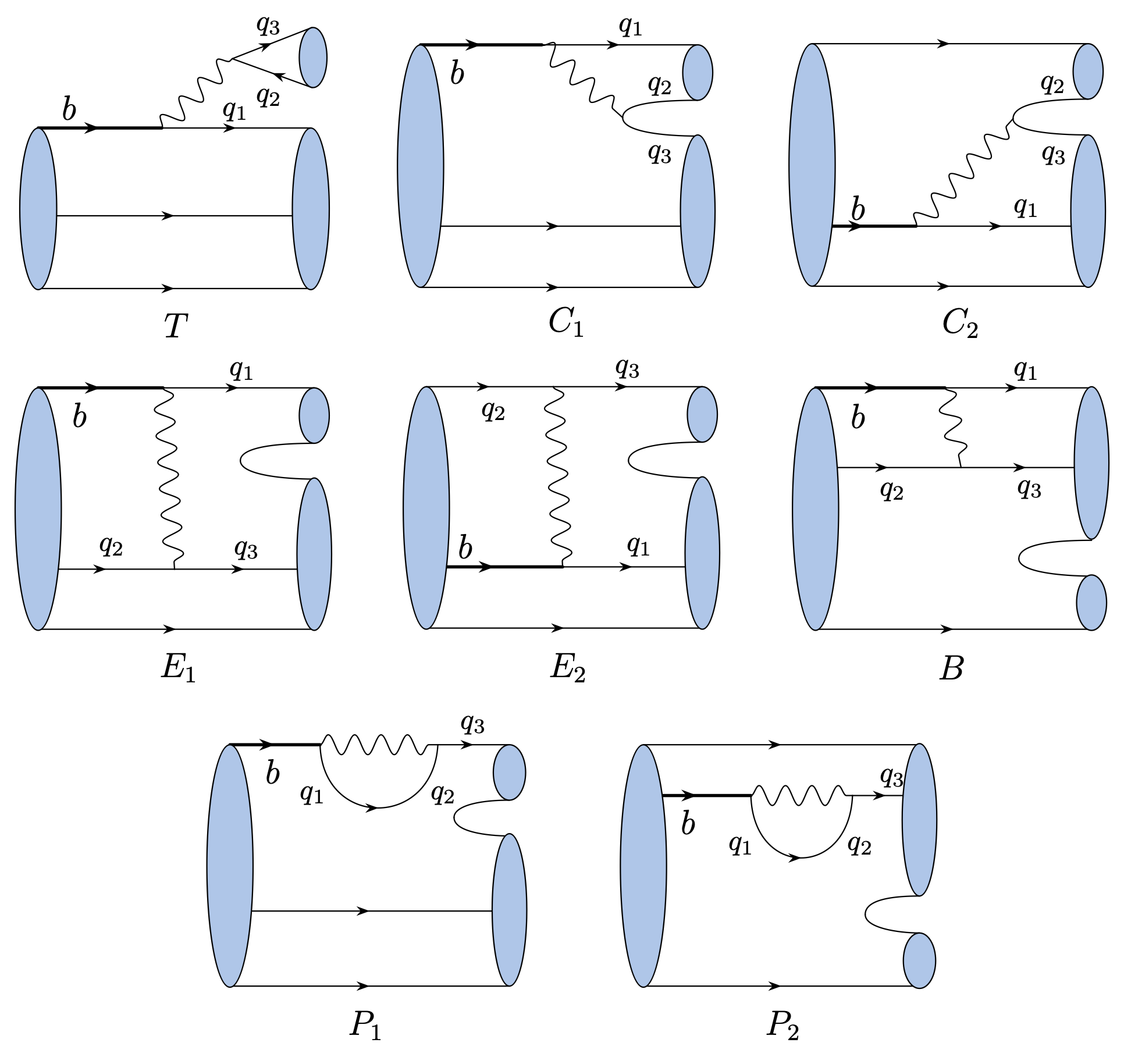}
    \captionsetup{type=figure}
    \captionof{figure}{Topological diagrams for two-body non-leptonic decays of the $\Lb$ baryon. The first two rows correspond to tree topologies, while the third row represents penguin topologies.}
    \label{fig:Topologicalbaryon}
\end{center}
\end{figure}

The extension of topological diagram description from mesons to baryon decays is natural. Indeed, this framework has been widely applied to charm and bottom baryon decays \cite{Chau:1995gk,Cheng:1996cs}. As depicted in Fig.~\ref{fig:Topologicalbaryon}, all possible topological diagrams for nonleptonic two-body $\Lb$ decays are shown as a representative example, without loss of generality. In each diagram, we denote the beauty-quark line by $b$ and the three quark lines associated with the local weak-operator insertion by $q_{1}$, $q_{2}$ and $q_{3}$, corresponding to the partonic transition $b\to q_{1}\bar q_{2}q_{3}$. For the current--current operators $\mathcal{O}_{1,2}^{p}$~\eqref{tree operators}, one has  $q_{1}=q_{2}=p$ and $q_{3}=q$, with $q=d,s$ and $p=u,c$. For the QCD and electroweak penguin operators in Eqs.~\eqref{QCD penguin operators} and \eqref{EW penguin operators}, $q_{1}=q_{2}=q'$ and $q_{3}=q$, where $q'$ runs over the quark flavors in the corresponding operator sums. The electromagnetic and chromomagnetic dipole operators $\mathcal{O}_{7\gamma}$ and $\mathcal{O}_{8g}$ are not four-quark operators themselves. The contributions of $\mathcal{O}{7\gamma}$ and $\mathcal{O}{8g}$ have the same flavor topologies with the penguin operators and are conventionally included in the amplitudes denoted by $P_1$ and $P_2$ in Fig.~\ref{fig:Topologicalbaryon}. Specifically, the topological diagrams are
\begin{itemize}
    \item $T$: color-allowed diagram with external $W$-emission.
    \item $C_1$ and $C_2$: color-suppressed internal $W$-emission diagrams, where the difference between them is that the quark generated from bottom quark weak decay flows into the final-state meson ($C_1$) or baryon ($C_2$).
    \item $E_{1}$, $E_{2}$ and $B$: three distinct types of $W$-exchange diagrams, distinguished by the flow of quarks produced from the weak vertex.
    \item $P_1$ and $P_2$: two types of penguin diagrams.
\end{itemize}
It is noted that the wavy lines in Fig.~\ref{fig:Topologicalbaryon} do not denote propagating $W$ bosons which has been integrated out in the effective Hamiltonian. They schematically denote local effective-operator insertions and are drawn only to make the quark-flow topology easier to visualize. The terms “$W$-emission” and “$W$-exchange” are conventional names for the corresponding quark-flow topologies.

The topological diagrams shown above provide an intuitive understanding of the underlying mechanisms of nonleptonic $\Lambda^{0}_b$ decays. The ultimate goal, however, is to develop dynamical frameworks within which these topological amplitudes can be calculated, thereby enabling quantitative predictions for decay rates and $\CP$ asymmetries in nonleptonic baryon decays.

\subsection{Naive Factorization}

Factorization is based on the separation of dynamics associated with widely different momentum scales. In suitable hard exclusive processes, a physical mechanism supporting this separation is color transparency~\cite{Bjorken:1988kk,Miller:2012ce}. A hard interaction can produce an energetic color-singlet configuration whose constituents have a small separation. Soft gluons, whose wavelengths are much larger than this separation, cannot resolve the individual color charges. This suppresses soft interactions between the compact color-singlet configuration and the rest of the process, providing a physical basis for factorization at leading power in the ratio of the soft scale $\Lambda$ to the hard scale $Q$. Corrections to this approximation are power suppressed in $\Lambda/Q$, although their precise scaling depends on the process under consideration.

The naive factorization approach was originally developed for exclusive non-leptonic decays of $D$ and $B$ mesons~\cite{Bauer:1984zv,Bauer:1986bm,Ellis:1975hr,Fakirov:1977ta,Cabibbo:1977zv}. Within this approach, the matrix elements of four-quark operators are approximated by products of two current matrix elements. For a non-leptonic two-body heavy-hadron decay admitting such a
factorized emission contribution, the decay amplitude is written as
\begin{equation}\label{eq:naivefactorization}
\begin{aligned}
\mathcal{A}_{\mathrm{NF}}(H\to h_{1}h_{2})
&\equiv
\bra{h_{1}h_{2}}\mathcal{H}_{\mathrm{eff}}\ket{H}_{\mathrm{NF}}
\\
&=
\frac{G_{F}}{\sqrt{2}}
\sum_{r}\lambda_{r}\sum_{i}C_{i}(\mu)\,
\bra{h_{1}}J^{(r)}_{1i}\ket{H}\,
\bra{h_{2}}J^{(r)}_{2i}\ket{0},
\end{aligned}
\end{equation}
where $\lambda_{r}$ denotes the relevant CKM factor and the sums include the operator contractions allowed by the flavor flow. The matrix element $\bra{h_{1}}J^{(r)}_{1i}\ket{H}$ is parametrized in terms of the $H\to h_{1}$ transition form factors, whereas $\bra{h_{2}}J^{(r)}_{2i}\ket{0}$ is expressed in terms of the decay constant of $h_{2}$. These nonperturbative quantities can be extracted from data or calculated using methods such as QCD sum rules and lattice QCD.

It can be found that the Wilson coefficients introduced in Eq.~\eqref{EFF HAM} enter the decay amplitudes \eqref{eq:naivefactorization} through the following color-combined quantities:
\begin{equation}\label{a-def}
\begin{aligned}
a_i &= C_i+{C_{i+1}}/{N_c}, \qquad i=1,3,5,7,9,\\
a_i &= C_i+{C_{i-1}}/{N_c}, \qquad i=2,4,6,8,10,
\end{aligned}
\end{equation}
where $N_c=3$ is the number of colors. The $1/N_c$ terms arise from color rearrangement of the four-quark operators in the factorized matrix elements. For the standard emission contractions considered here, the coefficients $a_{1,4,6,8,10}$ arise when the emitted meson $h_2$ is produced by a flavor-off-diagonal current, whereas $a_{2,3,5,7,9}$ arise for emission from a flavor-diagonal current. These two cases commonly correspond to charged- and flavor-neutral-meson emission, respectively.

Naive factorization provides a simple and widely used first approximation to non-leptonic decay amplitudes. It can give useful estimates of decay rates for many processes, particularly those dominated by the color-allowed $T$ topology. It nevertheless has two important shortcomings. At the theoretical level, the renormalization-scheme and scale dependence of the Wilson coefficients $C_{i}(\mu)$ should be canceled by that of the full matrix elements of the renormalized operators. The factorized products of current matrix elements do not reproduce the required dependence, so the resulting amplitude retains an unphysical scale and scheme dependence. At the phenomenological level, the factorized current matrix elements do not generate relative strong phases, and thus the direct $\CP$ asymmetry $A_{\CP}^{\mathrm{dir}}$ in Eq.~\eqref{eq:acp} vanishes within this approximation. Therefore, a quantitative description of $\CP$-violating observables requires frameworks that go beyond naive factorization.

\subsection{Generalized Factorization}

The generalized factorization approach was gradually established \cite{Ali:1997nh,Neubert:1997uc,Ali:1998eb,Deshpande:1997ar,Fleischer:1993gr,
Cheng:1998uy,Deshpande:1994pw} to overcome the shortcomings appearing in the naive factorization. This framework is based on two central ideas. First, selected perturbative corrections to the partonic matrix elements are combined with the Wilson coefficients $C_i(\mu)$ to define effective coefficients $C_i^{\mathrm{eff}}$. At the one-loop level, the effective coefficients can be written as
\begin{equation}
C_i^{\rm eff} = C_i(\mu)+V_i +\frac{\alpha_s(\mu)}{24\pi}A_i(C_t+C_p+C_g) +\frac{\alpha_{\rm em}}{8\pi}B_iC_e ,
\end{equation}
where $V_i$ denotes the vertex corrections, while the $C_t$, $C_p$, $C_g$ and $C_e$ terms collect the current--current penguin contractions, penguin-operator contractions, chromomagnetic contributions and electroweak-penguin corrections, respectively (see~\cite{Ali:1998eb} for details). The calculated corrections carry the renormalization-scheme and scale dependence required to compensate that of $C_i(\mu)$ at the same perturbative order. This rearrangement therefore reduces, and within the chosen prescription formally cancels to the calculated order, the corresponding explicit scheme and scale dependence. Absorptive parts of the loop corrections can also generate perturbative strong phases, providing necessary ingredients for generating nonzero $A^{\text{dir}}_{\CP}$. The decay amplitude is then written as
\begin{equation}\label{eq:generalized_factorization}
\mathcal{A}_{\mathrm{GF}}(H\to h_1h_2) = \frac{G_F}{\sqrt{2}} \sum_r\lambda_r\sum_i C_i^{\mathrm{eff}}\, \bra{h_1h_2}\mathcal{O}_i\ket{H}_{\mathrm{NF}},
\end{equation}
where the remaining operator matrix elements are reduced to products of current matrix elements as in naive factorization. The corresponding color-combined coefficients constructed from $C_i^{\mathrm{eff}}$ will be 
\begin{equation}\label{eq:aeff-def}
\begin{aligned}
a_i^{\mathrm{eff}} &= C_i^{\mathrm{eff}}+ {C_{i+1}^{\mathrm{eff}}}/{N_c}, \qquad i=1,3,5,7,9,\\
a_i^{\mathrm{eff}} &= C_i^{\mathrm{eff}}+ {C_{i-1}^{\mathrm{eff}}}/{N_c}, \qquad i=2,4,6,8,10.
\end{aligned}
\end{equation}
Second, residual nonfactorizable contributions not included in the partonic corrections are usually parametrized phenomenologically through the replacement $N_c=3\to N_c^{\mathrm{eff}}$ in the $1/N_c$ terms of $a_i^{\mathrm{eff}}$. The effective color parameter $N_c^{\mathrm{eff}}$ is introduced to absorb these residual contributions and may be constrained from data.

This treatment improves upon naive factorization but does not provide a systematic factorization theorem. The construction of the effective Wilson coefficients $C^{\text{eff}}_{i}$ comes with new problems of gauge and infrared-regulator dependence~\cite{Cheng:1999gs,Buras:1998us}. The perturbative loop calculation in the quark level generally requires an infrared regulator, off-shell external quarks for instance, to control potential soft and collinear divergences, thus the $C^{\text{eff}}_{i}$ naturally depends on regulator and gauge choice. The dependence on the gauge choice and on the infrared regulator should be canceled among all contributions in a complete physical amplitude. However, the remaining hadronic matrix element $\bra{h_1h_2}\mathcal{O}_i\ket{H}_{\mathrm{NF}}$ is replaced directly by its factorized expression in terms of decay constants and form factors and the expected cancellation is therefore not guaranteed. Moreover, the phenomenological modification of the color factors does not determine the underlying nonfactorizable dynamics and may need to be adjusted between different processes. Generalized factorization should therefore be regarded as a useful phenomenological extension of naive factorization rather than a fully self-consistent factorization framework.

Despite these theoretical limitations, generalized factorization has been widely used in phenomenological analyses of non-leptonic $B$-meson decays, including two-body $B\to PP$, $PV$ and $VV$ modes with $P(V)$ denoting a light pseudoscalar (vector) meson~\cite{Ali:1997nh,Ali:1998eb,Chen:1999nxa,Chen:1998dta,Du:1999im}, as well as three-body decays such as $B^\pm\to p\bar p h^\pm$~\cite{Geng:2006jt}. It has also been applied to beauty-baryon channels including $\Lb\to p\pi^-$, $pK^-$, $p\rho^-$ and $pK^{*-}$~\cite{Mohanta:2000nk,Hsiao:2014mua,Hsiao:2017tif,Geng:2021nkl}. The predicted branching fractions of $\Lb\to p\pi^-$ and $pK^-$ are broadly compatible with the measurements, whereas the corresponding direct $\CP$ asymmetries are not compatible with the available data~\cite{LHCb:2024iis}, as shown in Table~\ref{tab:directcp}. Generalized factorization predicts a direct $\CP$ asymmetry of nearly $20\%$ for $\Lb\to pK^{*-}$~\cite{Hsiao:2014mua,Geng:2021nkl}, whereas the LHCb analysis of $\Lb\to pK_S^0\pi^-$ measures ${A}_{\CP}=(-0.6\pm4.0\pm1.9)\%$ in the region dominated by the $K^{*-}(892)$ resonance~\cite{LHCb:2025ozp}. The experimental precision is therefore sufficient to show that the asymmetry is significantly smaller than the predicted effect.\footnote{This comparison is only indicative, since the measured resonance region may also contain nonresonant and interference contributions.} These comparisons illustrate that generalized factorization may provide useful estimates of decay rates for some channels, but has limited control over the nonfactorizable dynamics and strong phases required for predictions of $\CP$ asymmetries in beauty-baryon decays.

\subsection{QCD Factorization} 

The limitations of naive and generalized factorization motivated the development of the QCD factorization (QCDF) framework \cite{Beneke:1999br,Beneke:2000ry,Beneke:2001ev,Beneke:2003zv}. In the heavy-quark limit, $m_b\gg\Lambda_{\text{QCD}}$, QCDF organizes the hadronic matrix elements of a class of exclusive nonleptonic two-body $B$-meson decays, {\it e.g.} $B\to D\pi$ and $B\to\pi\pi$, as an expansion in $\Lambda_{\text{QCD}}/m_b$ and establishes a factorization formula for the leading-power contribution. The relevant hadronic matrix elements are factorized into perturbatively calculable hard scattering kernels and universal nonperturbative inputs, including transition form factors and light-cone distribution amplitudes (LCDAs). More explicitly, for $B\to M_{1}M_{2}$ decays where both final-state mesons are light, at leading power in the $\Lambda_{\text{QCD}}/m_b$ expansion, one has~\cite{Beneke:1999br,Beneke:2000ry}
\begin{equation}\label{eq:QCDF}
\begin{aligned}
\bra{M_{1}M_{2}}\mathcal{O}_{i}\ket{B}
={}&\sum_{j}F^{B\to M_{1}}_{j}(m^{2}_{2}) \int^{1}_{0}dx\,T^{I}_{ij}(x)\Phi_{M_{2}}(x) \\
&+\int^{1}_{0}d\xi\int^{1}_{0}dx\int^{1}_{0}dy\, T^{II}_{i}(\xi,x,y) \Phi_{B}(\xi)\Phi_{M_{1}}(y)\Phi_{M_{2}}(x).
\end{aligned}
\end{equation}
Here, $M_{1}$ denotes the final-state meson that contains the spectator quark from the $B$ meson, $M_{2}$ is the emitted meson, and $\Phi_{M_{1,2}}$ and $\Phi_{B}$ denote their LCDAs. The variables $\xi$, $x$ and $y$ are the corresponding longitudinal light-cone quark momentum fractions in the $B$, $M_2$ and $M_1$ LCDAs, respectively. The quantities $F^{B\to M_{1}}_{j}$ are transition form factors, and the quantities $T^{I}_{ij}$ and $T^{II}_{i}$ are hard-scattering kernels calculable in perturbation theory. The kernel $T^{I}_{ij}$ starts at $\mathcal{O}(\alpha_s^0)$ and receives higher-order contributions including vertex corrections, quark loops and $\mathcal{O}_{8g}$ insertions. The kernel $T^{II}_{i}$ describes hard spectator scattering and starts at $\mathcal{O}(\alpha_s)$. At $\mathcal{O}(\alpha_s^0)$, $T^{I}_{ij}$ is independent of the momentum fraction $x$, while $T^{II}_{i}$ is absent, so that the matrix element reduces to the product of a transition form factor and a decay constant, reproducing naive factorization. For heavy-light final states such as those in the $B\to D\pi$ decay, with the $D$ meson containing the spectator quark, the hard spectator scattering term is power suppressed in the heavy-quark limit because the spectator quark must remain soft to form the $D$ meson, and thus only the $T^{I}_{ij}$ term remains in the factorization formula~\eqref{eq:QCDF} at leading power.

The essential idea of QCDF is to separate perturbatively calculable short-distance interactions from nonperturbative long-distance dynamics. After the soft and collinear contributions have been absorbed into the form factors and LCDAs, the leading-power hard-scattering kernels are infrared finite and gauge invariant order by order. However, the factorization formula~\eqref{eq:QCDF} is established only for the kinematic configurations and decay classes to which the QCDF theorem applies. In particular, the same formula is not established when the emitted meson is itself heavy, as in $B\to\pi D$ with the $D$ meson emitted, or when both final-state mesons are heavy, as in $B\to D D_s$. Moreover, for the class of processes in which the leading-power QCDF theorem applies, power-suppressed contributions remain outside its controlled accuracy and may lead to endpoint-divergent convolutions. Some recent studies have explored the endpoint structure of selected power corrections in specific processes~\cite{Liu:2019oav,Lu:2022kos}, but these developments do not yet provide a general treatment of power corrections in nonleptonic two-body $B$ decays.

The QCDF approach has been successfully applied to non-leptonic $B$-meson decays. It has been systematically developed for two-body charmless modes, including $B\to PP,PV$ decays~\cite{Beneke:2001ev,Beneke:2003zv} and $B\to VV$ decays~\cite{Beneke:2006hg,Cheng:2008gxa}, as well as for channels containing a charm meson and a light meson $B\to DL$~\cite{Beneke:2000ry}. The absorptive parts of the perturbative vertex and penguin corrections provide calculable strong phases at leading power. Together with the nonperturbative inputs and power-suppressed contributions, these corrections support phenomenological predictions for branching fractions, $\CP$ asymmetries and polarization fractions. It is worth noting that several hard-scattering kernels and selected factorizable amplitudes have been calculated through next-to-next-to-leading order (NNLO) in $\alpha_s$~\cite{Huber:2016xod,Beneke:2009ek,Bell:2015koa,Bell:2020qus}. Nevertheless, residual power corrections and hadronic uncertainties remain important. For example, the branching fractions of $\bar{B}^{0}_{s}\to D^{+}_{s}\pi^{-}$ and $\bar{B}^{0}\to D^{+}K^{-}$ still require further theoretical investigation.

Applications of QCDF to beauty-baryon decays remain relatively limited. The heavy-light two-body decays $\Lb\to\Lambda_c^+ L$, with $L=\pi,\rho,K^{(*)},a_1$, are comparatively tractable. At leading power, their factorization closely parallels that of $\bar B\to D^{(*)}L$: the light meson is emitted from the weak current, while hard spectator scattering is power suppressed. Perturbative corrections to these amplitudes have been calculated through NNLO~\cite{Huber:2016xod}. Since these modes are dominated by a single CKM structure, no appreciable direct $\CP$ asymmetry is expected in the SM. By contrast, for charmless decays such as $\Lb\to p\pi^-$ and $\Lb\to pK^-$, a rigorous QCDF treatment incorporating the full three-quark structure of the baryons has not yet been established. These decays involve a richer set of topological amplitudes, including both tree and penguin contributions, and are expected to receive contributions from hard spectator scattering with two spectator quarks at leading power. The presence of baryonic light-cone distribution amplitudes and the more complicated momentum sharing among the three valence quarks furthermore render a systematic factorization analysis considerably more challenging.

A number of charmless two-body non-leptonic $\Lb$ decays have nevertheless been studied in QCDF-inspired calculations using a quark--diquark approximation, in which the two spectator quarks are represented by a single effective color-antitriplet degree of freedom ~\cite{Zhu:2018jet,Zhu:2016bra}. This approximation reduces the valence configuration to a meson-like two-body system and allows the mesonic QCDF formula to be used as a phenomenological model. It is not, however, a consequence of a baryonic factorization theorem. A hard gluon can resolve the internal two-quark structure of the diquark unless the latter is sufficiently compact on the scale set by the hard momentum transfer. Consequently, a point-like diquark cannot in general reproduce the full momentum sharing and hard spectator scattering among the three valence quarks, for which the complete baryonic LCDAs would be required. Predictions of strong phases and $\CP$ asymmetries obtained in this approximation therefore carry additional model dependence. Representative results for selected channels are collected in Table~\ref{tab:directcp}. For example, a positive direct $\CP$ asymmetry of order $10\%$ was predicted for $\Lb\to pK^-$~\cite{Zhu:2018jet}, whereas LHCb measures $A^{\rm dir}_{\CP}=(-1.1\pm0.7\pm0.4)\%$~\cite{LHCb:2024iis}. Similarly, in the $\Lb\to pK_S^0\pi^-$ analysis, the asymmetry $A^{\rm dir}_{\CP}=(-0.6\pm4.0\pm1.9)\%$ is measured in the $K^*(892)^-$-dominated region of $\Lb\to pK_S^0\pi^-$~\cite{LHCb:2025ozp}, which is substantially smaller than the $\mathcal{O}(20\%-30\%)$ asymmetry of the $\Lb\to pK^*(892)^-$ decay predicted in Ref.~\cite{Zhu:2018jet}.

Before concluding this subsection, it is worth emphasizing that relatively sizable nonfactorizable power corrections may arise whenever the formal leading-power factorizable contribution is suppressed, for example, by a small strong coupling $\alpha_{s}$, small Wilson coefficients, or small color factors. In mesonic QCDF, weak-annihilation amplitudes are power suppressed and are not controlled by the leading-power factorization theorem~\eqref{eq:QCDF}. They can nevertheless generate sizable strong phases and substantially affect predictions of $\CP$ asymmetries. Beauty-baryon decays contain a larger set of exchange-type contractions, making the corresponding subleading dynamics still more difficult to control. 
In practice, the leading-power factorizable contribution in charmless beauty-baryon decays requires the exchange of two hard or hard-collinear gluons, thereby starting at order  $\mathcal{O}(\alpha_s^2)$~\cite{Wang:2011uv}. 
Consequently, formally power-suppressed contributions are expected to play a more important role in baryon decays than in the corresponding meson decays.

\subsection{SCET Power Counting}

Some challenges in applying QCDF to beauty-baryon decays can be illuminated by soft-collinear effective theory (SCET) power counting. As a starting point, it is particularly instructive to examine the heavy-to-light baryon transition form factors, which provide useful insight into the scaling behavior of the relevant hadronic matrix elements. At large recoil, heavy-quark and large-energy symmetries reduce the QCD form factors for each transition $\Lb\to\mathcal{B}$, with, for example, $\mathcal{B}=\Lambda,p$, to a single leading-power SCET form factor $\xi_{\mathcal{B}}(E)$. In the SCET power counting~\cite{Wang:2011uv}, the leading-twist contribution arises when two hard-collinear gluon exchanges convert both soft spectator quarks in the $\Lb$ into energetic collinear quarks in the final-state baryon. The resulting convolution of the perturbative jet function $J(x_i,y_i)$ with the LCDAs of the $\Lb$ baryon and the final-state light baryon is endpoint finite at the order considered, for the assumed endpoint behavior of the LCDAs. Therefore, the leading-power contribution to $\xi_{\mathcal{B}}(E)$ is factorizable but starts at $\mathcal{O}(\alpha_s^2)$. Soft-overlap configurations in which the spectators do not both participate in the hard-collinear scattering are formally power suppressed in this counting. However, this does not imply that they are numerically small, since the leading-power hard-scattering term carries an $\alpha_s^2$ suppression. For example, the perturbative calculation gives $|\xi_\Lambda(q^2=0)|\sim10^{-2}$ for the hard-scattering contribution~\cite{Wang:2011uv}, whereas SCET sum rules give an estimate of order $0.3\text{--}0.4$ associated with soft dynamics~\cite{Feldmann:2011xf}. This striking difference suggests that formally power-suppressed nonfactorizable contributions may in practice numerically dominate heavy-to-light baryonic form factors. The power counting differs from that for heavy-to-light meson transitions such as $B\to\pi$, where soft-overlap and hard-scattering contributions both occur at leading power, while the latter starts at $\mathcal{O}(\alpha_s)$ rather than $\mathcal{O}(\alpha_s^2)$~\cite{Beneke:2000wa,Hill:2005ju}.

For non-leptonic bottom-baryon decays, an additional complication arises from the richer topology structure of baryonic matrix elements~\cite{Chau:1995gk}. SCET analyses of heavy-to-heavy decays organize the relevant Wick contractions into four topologies, denoted by $T$, $C$, $E$ and $B$~\cite{Leibovich:2003tw,Mantry:2003uz}, corresponding, respectively, to the colour-allowed tree topology, the colour-suppressed internal-emission topology, the exchange topology, and the baryon-specific bow-tie topology. In the notation of Fig.~\ref{fig:Topologicalbaryon}, these classes correspond to $T$, $C_2$, $E_2$ and $B$, respectively. For $\Lb\to\Lambda_c^+\pi^-$, the $T$ topology contributes at leading power and factorizes in analogy with $\bar B\to D^{(*)}\pi$~\cite{Zhang:2010jc,Huber:2016xod}. The $C$ and $E$ topologies require two insertions of the subleading soft-collinear SCET Lagrangian and are therefore suppressed by $\Lambda_{\rm QCD}/E_\pi$ at the amplitude level relative to $T$, where $E_\pi$ is the energy of the emitted pion. The bow-tie topology $B$, which has no analogue in mesonic decays, requires further subleading insertions and is suppressed by at least one additional power of $\Lambda_{\rm QCD}/E_\pi$. Schematically, for this class of decays~\cite{Leibovich:2003tw,Mantry:2003uz},
\begin{equation}\label{eq:scet_topology_power_counting}
T:C:E:B\sim 1:\frac{\Lambda_{\rm QCD}}{E_\pi}: \frac{\Lambda_{\rm QCD}}{E_\pi}: \left(\frac{\Lambda_{\rm QCD}}{E_\pi}\right)^2.
\end{equation}
This power counting also explains why some non-leptonic baryonic channels are naturally suppressed. For example, in $\Lb\to\Sigma_c^{(*)}\pi$, the leading $T$ contribution vanishes in the heavy-quark and isospin limits, so that the dominant nonzero amplitudes come from the $C,E$ topologies, and their decay rates are suppressed. A comparably systematic SCET power counting of the topological amplitudes in charmless beauty-baryon decays is not yet available and would be highly desirable, particularly for assessing contributions relevant to $\CP$ asymmetries.

It should be emphasized that a formal amplitude power counting for beauty-baryon decays does not necessarily provide a reliable numerical hierarchy. As the preceding discussion illustrates, nominally leading-power contributions may themselves be subject to perturbative or dynamical suppression, so that power-suppressed terms need not be numerically subleading; moreover, even when they are genuinely subleading at the amplitude level, their interference with leading amplitudes can have an important impact on certain observables. When an unsuppressed leading-power amplitude is present, power-suppressed topologies may provide only subleading corrections to a branching fraction and may therefore be neglected as an approximation. Such a simplification is, however, much less justified for $\CP$ asymmetries, since power-suppressed topologies can have a significant impact on the generation of strong phases. This is familiar from nonleptonic $B$-meson decays, where formally power-suppressed exchange and annihilation amplitudes can play an essential role in generating strong phases and shaping direct $\CP$ asymmetries~\cite{Keum:2000ph,Beneke:2003zv,Fleischer:2022rkm,Chay:2007ep,Chai:2022ptk}. The situation is expected to be even more pronounced in beauty baryon decays, where exchanging type contractions are considerably more numerous than in B meson decays. Reliable predictions for $\CP$ violating observables in beauty-baryon decays therefore require a much better understanding of subleading dynamics.

\subsection{Perturbative QCD Approach}

The perturbative QCD (PQCD) approach is an alternative $k_T$-factorization framework for exclusive heavy-hadron decays~\cite{Lu:2000em,Keum:2000wi,Keum:2000ph,Chai:2022ptk}. The central idea of PQCD is to retain the transverse momenta $k_T$ of the valence partons in the propagators appearing in the hard-scattering kernels. When a longitudinal momentum fraction approaches the endpoint region $x\to 0$, the transverse momentum modifies the propagators and regulates the endpoint singularities that arise in collinear factorization. Radiative corrections generate large transverse-momentum double logarithms $(\alpha_{s}\text{ln}^{2}k_{T})^{n}$, which has to be resummed to maintain a reliable perturbative expansion. Their resummation produces a Sudakov factor, conventionally written as $\exp[-S(Q,b)]$, which suppresses contributions from large transverse separations, where $Q$ denotes the dominant light-cone component of a meson momentum and $b$ is the impact parameter conjugate to $k_T$~\cite{Li:1992nu,Li:2010nn}. In addition, threshold logarithms of the form $(\alpha_s\ln^2 x_i)^n$ associated with the endpoint region are resummed into a threshold factor, often referred to as a jet function, $S_t(x_i)$~\cite{Li:1994cka,Li:1996gi,Li:1998is,Feng:2007qv}. For instance, when applying PQCD to the calculation of the $B\to \pi$ form factor, these resummation effects are intended to suppress soft-overlap configurations associated with large transverse separations and to make a perturbative hard-scattering description of heavy-to-light form factors possible.\footnote{This review emphasizes the phenomenological applications of theoretical frameworks, including PQCD, to $\CP$ violation in beauty-baryon decays, while leaving aside a detailed discussion of their formal foundations, theoretical consistency, and domains of applicability.}

The treatment of transverse momentum and the associated resummation also allows annihilation and other endpoint-sensitive topologies to be factorized within the PQCD framework.  In the standard implementation, the corresponding convolution integrals are rendered finite, while absorptive parts of the perturbative hard kernels can generate calculable strong phases. The PQCD framework has therefore been widely applied to non-leptonic $B$ meson decays, since it can treat different topological amplitudes in a unified framework. For charmless two-body decays, it has been used to study branching fractions and direct $\CP$ asymmetries~\cite{Ali:2007ff, Lu:2000hj,Li:2008ts,Qin:2014xta,Zou:2015iwa,Chai:2026rhh,Chai:2022ptk}, as well as polarization observables in $B\to VV$ modes~\cite{Ali:2007ff,Chai:2026rhh,Chai:2022ptk}. Selected next-to-leading-order corrections to decay amplitudes have also been included in specific analyses~\cite{Wang:2014mua,Zhang:2014bsa,Cheng:2014rka,Cheng:2018khi}. These applications provide a useful benchmark for extending the $k_T$-factorization framework to beauty-baryon decays, where the more complicated baryon wave functions and higher-twist contributions require separate consideration.

The extension of PQCD to beauty-baryon decays can first be examined through the heavy-to-light transition form factors $\Lb\to\mathcal{B}$, with $\mathcal{B}=p,\Lambda$~\cite{Han:2022srw,Yang:2025yaw}. In the $k_T$-factorization framework, a form factor can be written schematically as 
\begin{equation}\label{eq:pqcd_baryon_form_factor}
\begin{aligned}
F(\Lb\to\mathcal{B})
={}&
\int^{1}_{0} [dx]\,[dx']\,\int [d^2 b]\,[d^2b']\,
\Psi_{\Lambda_b}(x_i,b_i)\,
H(x_i,x_j',b_i,b_j',t)\,
\Psi_{\mathcal{B}}(x_j',b_j')S_{t}(x^{\prime}_{i})
\\
&\times
\exp\left[-S_{\Lambda_b}(t)-S_{\mathcal{B}}(t)\right],
\end{aligned}
\end{equation}
where $x_i$ and $x_i'$ denote the longitudinal momentum fractions of the valence quarks, while $b_i$ and $b_i'$ are the impact parameters conjugate to their transverse momenta. The functions $\Psi_{\Lambda_b}$ and $\Psi_{\mathcal{B}}$ describe the corresponding transverse-momentum-dependent baryon wave functions, and $H(x_i,x_j',b_i,b_j',t)$ is the perturbatively calculable hard-scattering kernel. The exponential factor contains the $k_T$ Sudakov suppression, and $t$ denotes the hard scale, which is usually chosen as the largest virtuality or energy scale appearing in the hard-scattering process, with the requirement $t\geq1\text{GeV}$ in order to avoid entering the non-perturbative region. A systematic threshold-resummation factor $S_t(x^{\prime}_i)$ has not yet been derived for these baryonic transitions and is therefore set to unity as a working approximation in the existing analyses~\cite{Han:2022srw}. 
Retaining the transverse momenta makes the framework applicable to the calculation of the contributions from higher twist baryon LCDAs, since the regulator $k_{T}$ in the hard kernel largely smears the sensitivity to the endpoint behavior of LCDAs.

Numerical studies of the $\Lb\to p$ and $\Lb\to\Lambda$ form factors in the PQCD approach indicate that higher-twist baryon LCDAs can be numerically important~\cite{Lu:2009cm,Han:2022srw,Yang:2025yaw}. Taking the $\Lb\to p$ transition for instance, retaining only the leading-twist $\Lb$ and proton LCDAs gives $f_1(0)$ of order $10^{-3}$, which is far below the values predicted by Lattice QCD~\cite{Detmold:2015aaa} and light-cone sum rule (LCSR)~\cite{Khodjamirian:2011jp}. It is natural to understand, since the SCET power counting indicates that the leading-power contribution suffers suppression from two hard collinear gluon exchanges. After higher-twist LCDAs are included, the form factors are enhanced by roughly two orders of magnitude. In the models considered in Ref.~\cite{Han:2022srw}, the dominant enhancement to $f_1(0)$ arises from the combination of the twist-$4$ $\Lb$ LCDA and the twist-$5$ proton LCDA. The resulting values are $f_1(0)\simeq 0.27$ and $g_1(0)\simeq 0.31$ for the exponential model of the $\Lambda^{0}_b$ LCDA, and $f_1(0)\simeq 0.24,~g_1(0)\simeq 0.27$ for the free-parton model of the $\Lb$ LCDA. These values are close to those obtained from lattice QCD~\cite{Detmold:2015aaa} and LCSR~\cite{Khodjamirian:2011jp} within the quoted theoretical uncertainties. The enhancement is associated with the endpoint behavior of the higher-twist LCDAs and the corresponding hard kernels. Thus, despite their nominal suppression in the twist expansion, higher-twist contributions can be numerically dominant and are essential for a realistic PQCD description of heavy-to-light beauty-baryon form factors.

Transition form factors, however, describe only the factorizable parts of nonleptonic amplitudes. A complete treatment of charmless decays such as $\Lb\to p\pi^-$ and $\Lb\to pK^-$ must also include the nonfactorizable emission, exchange and bow-tie contractions. In PQCD, these contributions are obtained by including the wave function and the Sudakov suppression factor of the emitted meson to Eq.~\eqref{eq:pqcd_baryon_form_factor} and evaluating the convolution with the corresponding hard-scattering kernels as in \cite{Han:2025tvc}. Early calculations~\cite{Lu:2009cm} based only on leading-twist baryon wave functions found the factorizable contribution to be strongly suppressed, as reflected by the small leading-twist $\Lb\to p$ form factor. More recent PQCD calculations of the $\Lambda^{0}_{b}\to ph$ decays, where $h$ runs over $\pi^{-},K^{-}$, vector mesons $\rho^{-},K^{*-}$ and axial-vector mesons $K^{-}_{1}(1270),K^{-}_{1}(1440)$, include higher-twist hadron LCDAs together with the full set of relevant topological amplitudes at the perturbation order of $\alpha^{2}_{s}$~\cite{Han:2024kgz,Han:2025tvc}. It is observed that the enhancement produced by higher-twist baryon LCDAs in the heavy-to-light baryonic form factors also appears in the amplitudes of charmless non-leptonic beauty baryon decays.

The resulting amplitudes exhibit a clear numerical hierarchy. For example, the tree amplitudes of $\Lb\to p\pi^-$ calculated in the PQCD approach satisfy $|T|\gg |E_2|>|C_2|>|B|$~\cite{Han:2024kgz,Han:2025tvc}. The factorizable contribution dominates the $T$ topology, while its nonfactorizable counterpart is suppressed. This differs from earlier leading-twist PQCD calculations and illustrates the numerical importance of higher-twist corrections. The suppression of the colour-suppressed, exchange and bow-tie amplitudes relative to $T$ is consistent with the topology hierarchy discussed in Sec.~3.5. Nevertheless, the SCET counting presented there was derived for heavy-to-heavy channels and does not constitute a formal power-counting theorem for these charmless decays. Moreover, as will be discussed below, the penguin counterparts of the subleading topologies provide important strong phases and play a crucial role in predictions of $\CP$ asymmetries.

Recent PQCD studies have investigated $\CP$ violation in two-body beauty-baryon decays, including the simplest charmless channels $\Lb\to p\pi^-$ and $\Lb\to pK^-$~\cite{Han:2024kgz,Han:2025tvc}. These modes are particularly interesting because their measured direct $\CP$ asymmetries are both at the percent level,
\begin{equation}
\begin{aligned}
A_{\CP}^{\rm dir}(\Lb\to p\pi^-)
&=(0.2\pm0.8\pm0.4)\%,\\
A_{\CP}^{\rm dir}(\Lb\to pK^-)
&=(-1.1\pm0.7\pm0.4)\%,
\end{aligned}
\end{equation}
in contrast to the sizable asymmetries observed in some corresponding charmless $B$-meson decays~\cite{LHCb:2024iis}. Earlier calculations based on generalized factorization and the quark--diquark approximation did not naturally reproduce this near-vanishing pattern~\cite{Mohanta:2000nk,Hsiao:2017tif,Geng:2021nkl,Zhu:2016bra,Zhu:2018jet}. Including higher-twist hadron LCDAs and subleading topological amplitudes, the recent PQCD analysis gives~\cite{Han:2024kgz,Han:2025tvc}
\begin{equation}
\begin{aligned}
A_{\CP}^{\rm dir}(\Lb\to p\pi^-) &=(5^{+2}_{-3})\%,\\
A_{\CP}^{\rm dir}(\Lb\to pK^-) &=(-6^{+3}_{-2})\%.
\end{aligned} \label{eq:pqcd_pipk_acp}
\end{equation}
These predictions are compared with results from the other approaches and with the available measurements in Table~\ref{tab:directcp}. Both are consistent with the measurements within the quoted uncertainties and provide a dynamical explanation of the unusually small integrated $\CP$ asymmetries in these baryonic decays.

\begin{figure}
    \centering
    \includegraphics[width=0.55\linewidth]{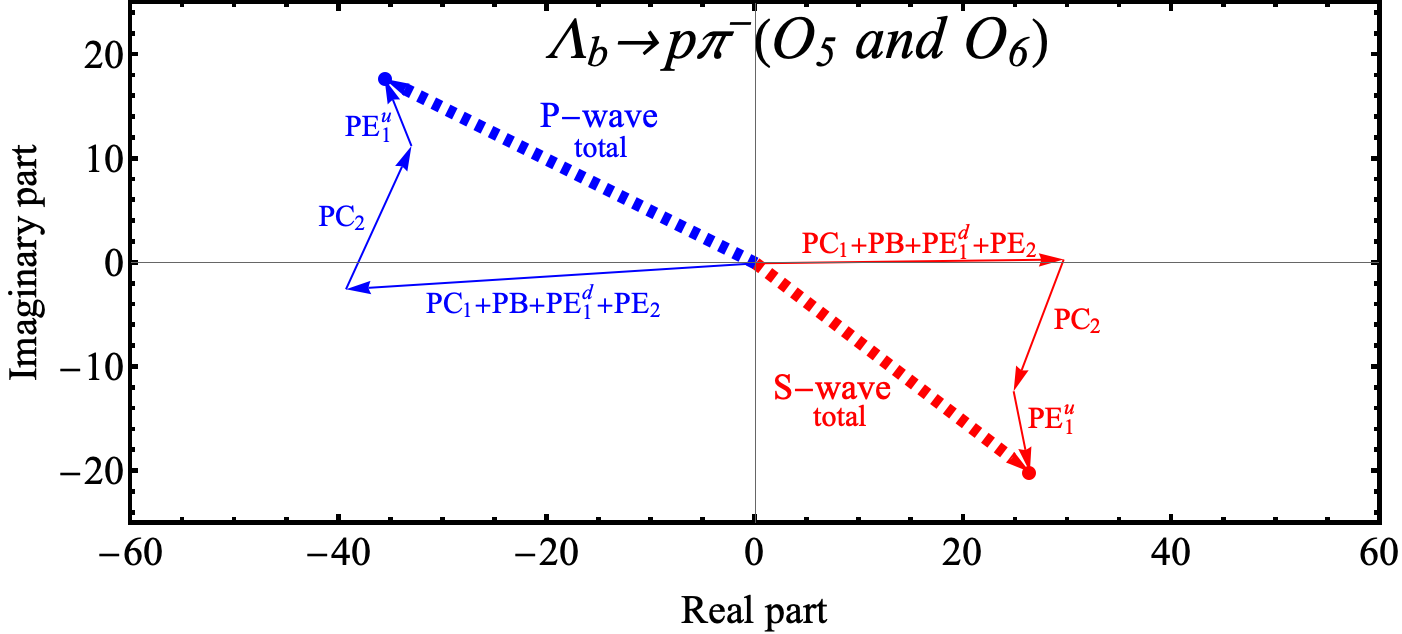}  
\vspace{5mm}
    \includegraphics[width=0.55\linewidth]{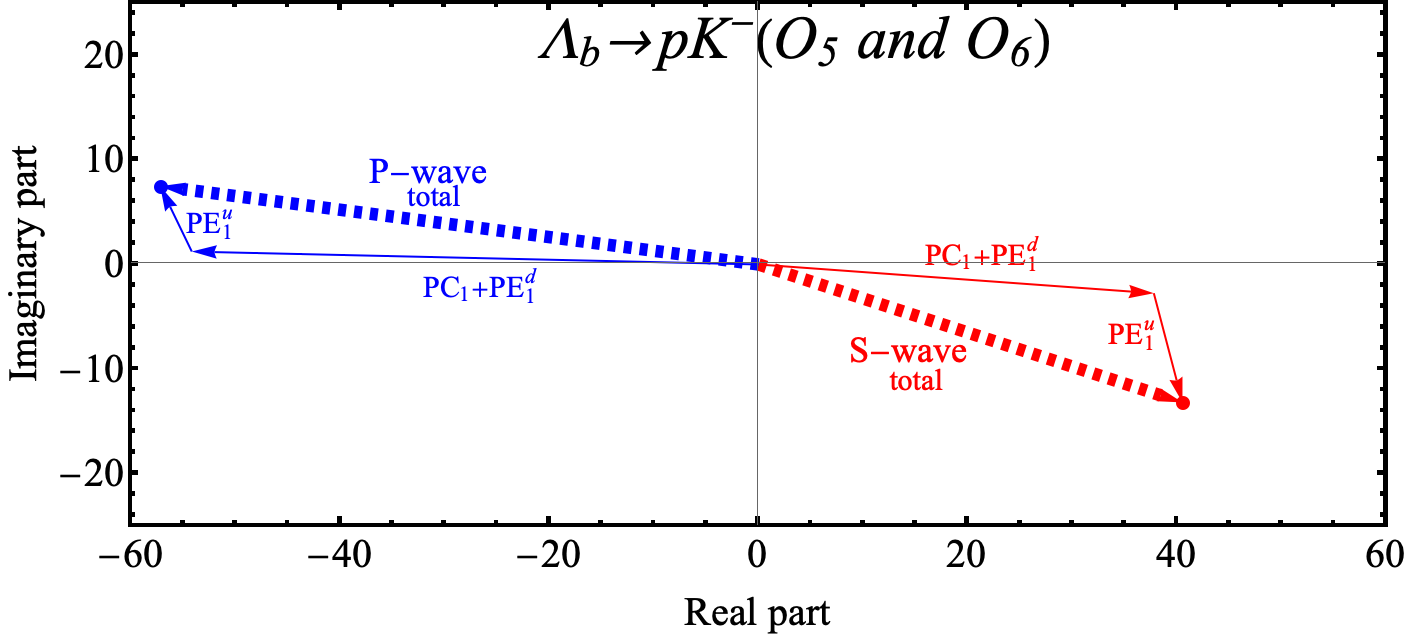}
    \caption{The QCD penguin operators $\mathcal{O}_{5}$ and $\mathcal{O}_{6}$ contributions to the $\Lb\to p\pi^-$ (upper panel) and $\Lb\to pK^-$ (lower panel) decay amplitudes, shown in the complex plane. The red (blue) vectors represent the $S$-wave ($P$-wave) penguin amplitudes. For $\Lb\to p\pi^-$, the combined $PC_2$ and $PE_1$ contributions dominate the strong phases and lead to opposite sign $S$- and $P$-wave $\CP$ asymmetries that partially cancel each other in the total $\CP$ asymmetry. For $\Lb\to pK^-$, the strong phases generated by $PC_1$ and $PE_1$ contributions from $\mathcal{O}_{5}$ and $\mathcal{O}_{6}$ are opposite in sign, while that is not true after combining with the other contributions, seeing Fig.~7 in \cite{Han:2025tvc}.}
    \label{fig:penguin-in-complex-plane}
\end{figure}

To identify the origin of these small integrated asymmetries, it is useful to resolve the decay amplitudes into their partial-wave components based on the PQCD calculation. For $\Lb\to p\pi^-$, the PQCD calculation gives representative partial-wave asymmetries $A_{\CP}^{S}\simeq17\%$ and $A_{\CP}^{P}\simeq-6\%$, which show that the individual partial-wave asymmetry can reach $\mathcal{O}(10\%)$ as in the charmless $B$ meson decays. Their opposite signs however lead to a substantial cancellation in the total asymmetry, resulting in $A^{\rm dir}_{\CP}(\Lambda^{0}_{b}\to p\pi^{-})\simeq5\%$.  This cancellation is illustrated in Fig.~\eqref{fig:penguin-in-complex-plane}, the nonfactorizable penguin amplitude $PC_2$ and the penguin exchange amplitude $PE_1$ (the penguin operator contributions to the $C_{2}$ and $E_1$ topologies) from operators $\mathcal{O}_{5}$ and $\mathcal{O}_{6}$ generate important strong phases in the two partial waves, but with opposite signs. Naively, one may understand this by convoluting a topological diagram with the effective weak operators and the most numerically dominant LCDA components of $\Lambda^{0}_{b}$, final proton and $\pi^{-},K^{-}$. The penguin operators $\mathcal{O}_{5}$ and $\mathcal{O}_{6}$, with chiral structure $(V-A)(V+A)$ or $(S-P)(S+P)$ after Fierz transformation, finally lead to $S_{PC_{2}}\approx-P_{PC_{2}}$ and $S_{PE_{1}}\approx -P_{PE_{1}}$, as shown in Fig.~\eqref{fig:penguin-in-complex-plane}. The tree operators, with chiral structure $(V-A)(V-A)$, hence give $S_{T}\approx P_{T}$.

For $\Lambda^{0}_{b}\to p\pi^{-}$, after combining with contributions from the other operators, the strong phases of $S$- and $P$-waves amplitudes remain opposite in sign, and therefore leading to final cancellation in total $\CP$ asymmetry. For $\Lb\to pK^-$, the strong phases between $S$- and $P$-wave amplitudes are no longer opposite in sign, and the corresponding partial wave $\CP$ asymmetries are approximately $A_{\CP}^{S}\simeq-5\%$ and $A_{\CP}^{P}\simeq-21\%$. Its $S$-wave $\CP$ asymmetry $A_{\CP}^{S}$ however is suppressed since the relevant strong-phase difference, $\delta_{\mathcal P}^{S}-\delta_{\mathcal T}^{S}$, is small of order $\mathcal{O}(0.1)$. Although both partial wave asymmetries are negative, the $S$ wave $\CP$ asymmetry is smaller, while it carries approximately $\kappa_S\simeq94\%$ in the rate. The smallness of total asymmetry in $\Lambda^{0}_{b}\to pK^{-}$ therefore results from the dominance of the relatively small $S$ wave $\CP$ asymmetry. In both channels, the strong phases originate primarily from nonfactorizable subleading penguin topologies, indicating their importance for predictions of $\CP$ asymmetries.

PQCD calculations have also provided predictions for $\CP$ violation in other two-body and quasi-two-body beauty-baryon decays, offering useful targets for future experimental searches. Although the cancellation between different partial-wave asymmetries is channel dependent, it is not unique to $\Lb\to p\pi^-$. It has also been found in $\Lb\to pK^{*-}$, $pa_1^-(1260)$ and $pK_1^-(1270)$, with the relatively small predicted total direct asymmetries being $(-5^{+10}_{-16})\%$, $(-1^{+4}_{-3})\%$ and $(9^{+8}_{-5})\%$, respectively~\cite{Han:2024kgz,Han:2025tvc}. In contrast, the individual partial-wave asymmetries are substantially larger, as shown in Table~II of Ref.~\cite{Han:2024kgz}. A measurement of the integrated direct $\CP$ asymmetry alone may therefore obscure sizable $\CP$-violating effects. Moreover, the individual partial waves are difficult to disentangle experimentally in a simple two-body final state because of its limited kinematic information. The subsequent decays of vector or axial-vector mesons provide additional angular information and can therefore give access to interference between different partial waves. For example, consider $\Lb\to pA$, with $A=a_1^-(1260)$ or $K_1^-(1270)$, followed by $a_1^-\to\pi^-\pi^+\pi^-$ or $K_1^-\to K^-\pi^+\pi^-$. The momenta of the three final-state mesons in the rest frame of $A$ define a decay plane. If $\theta$ denotes the angle between the normal vector $\boldsymbol n$ of this plane using a fixed ordering of the daughter momenta and the proton direction $\hat{\boldsymbol p}_p$ in the $\Lb$ rest frame, the angular distribution contains a term of the form $\operatorname{Re}\!\left(S^{T}P_{2}^{*}\right)\cos\theta$~\cite{Han:2025tvc,Wang:2024qff}, where $S^{T}$ and $P_2$ denote the relevant partial-wave amplitudes in the notation of Ref.~\cite{Han:2025tvc}. This motivates the up--down asymmetry
\begin{equation}\label{UP-Down asymmetry}
A_{\rm UD}
=
\frac{\Gamma(\cos\theta>0)-\Gamma(\cos\theta<0)}
{\Gamma(\cos\theta>0)+\Gamma(\cos\theta<0)},
\end{equation}
and the corresponding $\CP$ asymmetry
\begin{equation}\label{CPV in UP-Down asymmetry}
a_{\CP}^{\rm UD} = \frac{A_{\rm UD}+\bar A_{\rm UD}}{A_{\rm UD}-\bar A_{\rm UD}},
\end{equation}
where $\bar a_{\rm UD}$ denotes the observable in the charge-conjugate decay. The plus sign in the numerator accounts for the parity transformation of the up--down observable. The PQCD predictions are $a_{\CP}^{\rm UD}(\Lb\to pa_1^-(1260))\simeq-24\%$ and $a_{\CP}^{\rm UD}(\Lb\to pK_1^-(1270))\simeq+26\%$~\cite{Han:2025tvc}. These sizable $\CP$ asymmetries are promising for the future experimental measurements, and illustrate how the angular distribution of the subsequent decay of an excited meson can expose $\CP$-violating effects hidden by cancellations in the integrated asymmetry.

Beyond these channels, PQCD has been applied to the color-allowed charmed decays $\Lb\to\Lambda_c^+\pi^-,\Lambda_c^+ K^-$~\cite{Zhang:2022iun}, the color-suppressed charmonium modes $\Lb\to\Lambda J/\psi,\Lambda\psi(2S)$~\cite{Rui:2022sdc}, and the penguin-dominated decay $\Lb\to\Lambda\phi$~\cite{Rui:2022jff}. Different from $\Lb\to p\pi^-$, a recent PQCD study of $\Lb\to\Lambda D_{\pm}$ shows that the $\CP$ asymmetries in \(S\)- and \(P\)-wave have the same sign, leading to large total $\CP$ asymmetries in $\Lb\to \Lambda D$~\cite{Rui:2026ihu}. These results should nevertheless be viewed with some caution, since the energetic, large-recoil picture of PQCD is less directly applicable to channels containing heavy charm or charmonium states.

Despite the phenomenological progress described above, the PQCD predictions for both the branching fractions and $\CP$ asymmetries of $\Lb\to p\pi^-$ and $pK^-$ carry sizable theoretical uncertainties~\cite{Han:2024kgz,Han:2025tvc}. Reducing these uncertainties remains an important goal for more stringent comparisons with experimental data. Possible improvements include higher-order perturbative corrections, threshold resummation for baryonic processes, and an assessment of higher Fock-state contributions. Their numerical impact remains to be established, and their implementation is technically demanding because of the large number of contributing diagrams and the complicated momentum sharing among the constituent partons. Improved determinations of hadron LCDAs, including their shapes and higher-twist components, are also needed to improve the precision of these predictions.

The limited knowledge of nonperturbative inputs also restricts the range of channels accessible to present PQCD calculations. This is particularly relevant to quasi-two-body decays involving resonances such as $N^*(1520)$, $\Delta$, $\Lambda(1520)$ and $f_0(980)$, whose LCDAs are poorly constrained. Extending the factorization calculations to multibody decays further requires a treatment of overlapping resonances, nonresonant contributions, and their interference across phase space. Meanwhile, LHCb has extensively studied multibody channels such as $\Lb\to p\pi^+\pi^-\pi^-,~\Lb\to pK^-\pi^+\pi^- $ and $\Lb\to pK^-K^+\pi^- ,~\Lb\to pK^+K^-K^-$~\cite{LHCb:2017gjj, LHCb:2018fpt,LHCb:2019jyj,LHCb:2016yco,LHCb:2025ray}. Their rich resonant structures provide strong phases and interference mechanisms that can enhance $\CP$ asymmetries in selected regions of phase space~\cite{Zhang:2013oqa,Zhang:2022emj,Wang:2024oyi,Heuser:2025mnk}. Addressing this broader experimental landscape therefore calls for complementary descriptions of long-distance dynamics. The next subsection discusses final-state interactions and hadronic rescattering as approaches to these contributions.

\subsection{Final-State Interactions and Rescattering Mechanisms}

Final-state interactions are ubiquitous in hadronic decays and production processes~\cite{Sakurai:1964vpc,Wolfenstein:1990ks,Suzuki:1999uc}. After the weak decay has produced a set of hadrons, these particles can rescatter through strong interactions before reaching the observed final state. Such rescattering can modify both the magnitudes and the strong phases of decay amplitudes. It is therefore particularly relevant for direct $\CP$ violation, which requires the interference of at least two amplitudes with different weak phases and different strong phases. In principle, electromagnetic interactions may also induce final-state interactions. For example, in nuclear $\beta$ decay, the emitted electron interacts with the residual nucleus, leading to modifications of the electron energy spectrum~\cite{Brodine:1970cc}. In hadron decays, such effects are typically minor and can be neglected to a good approximation.

The relation between factorization and final-state rescattering is subtle. For decay classes in which leading-power factorization applies, long-distance soft rescattering is suppressed in the heavy-quark limit. At the hadronic level, this suppression can be understood as a cancellation among different intermediate channels~\cite{Beneke:2000ry}. At the physical heavy-quark mass, however, such cancellations need not be numerically complete, and power-suppressed contributions can remain relevant for decay amplitudes. The factorization frameworks discussed above may themselves contain selected nonfactorizable corrections to capture the power-suppressed contributions. Hadronic rescattering should therefore not be naively regarded as a contribution absent from factorization by definition; rather, it provides an alternative phenomenological description of the suppressed strong-interaction effects at the hadron level, and allows their possible impact at the physical heavy-quark mass to be investigated.

The effect of final-state rescattering in $\Lb$ decays can be represented as~\cite{Duan:2024zjv}
\begin{equation}
\mathcal{A}_{f}
= \sum_i \left(S^{1/2}\right)_{fi} \,\mathcal{A}^{(0)}_{i} ,
\label{eq:FSI}
\end{equation}
where $\mathcal{A}^{(0)}_{i} = \bra{i}\mathcal{H}\ket{\Lambda^{0}_{b}}$ denotes the weak-decay amplitude for producing a possible intermediate channel $i$ including short-distance QCD corrections, while $\left(S^{1/2}\right)_{fi}$ describes its subsequent strong-interaction evolution into the observed final state $f$ carrying the corresponding strong phases. This decomposition is schematic and its implementation is
model dependent. In practice, a complete sum over all allowed intermediate channels is not feasible. The following discussion therefore considers two phenomenological implementations: hadronic triangle-loop calculations with selected intermediate states, and the use of experimentally determined scattering partial waves for a specified two-body subsystem. 

\subsubsection{Hadronic loop mechanism}

Hadronic rescattering via triangle loops was developed as a means of studying long-distance effects in $B$-meson decays. These analyses showed that soft hadronic rescattering could have a non-negligible impact on decay amplitudes and $\CP$-violating observables~\cite{Cheng:2004ru}. This motivates its use as a phenomenological tool for investigating the possible impact of final-state rescattering in heavy baryon decays~\cite{Jia:2024pyb}, where the relevant long-distance dynamics are difficult to calculate from first principle. At the quark level, some of these hadronic rescattering amplitudes can be associated with nonfactorizable topologies through quark exchange or quark--antiquark annihilation. For example, light-quark exchange and annihilation may resemble internal-emission and exchange topologies, respectively, while the annihilation of an intermediate $c\bar c$ pair can be associated with the nonfactorizable charm-penguin contribution~\cite{Cheng:2004ru,Wang:2021rhd,Lai:2025ryi}. 

\begin{figure}
\begin{center}
    \includegraphics[width=0.4\textwidth]{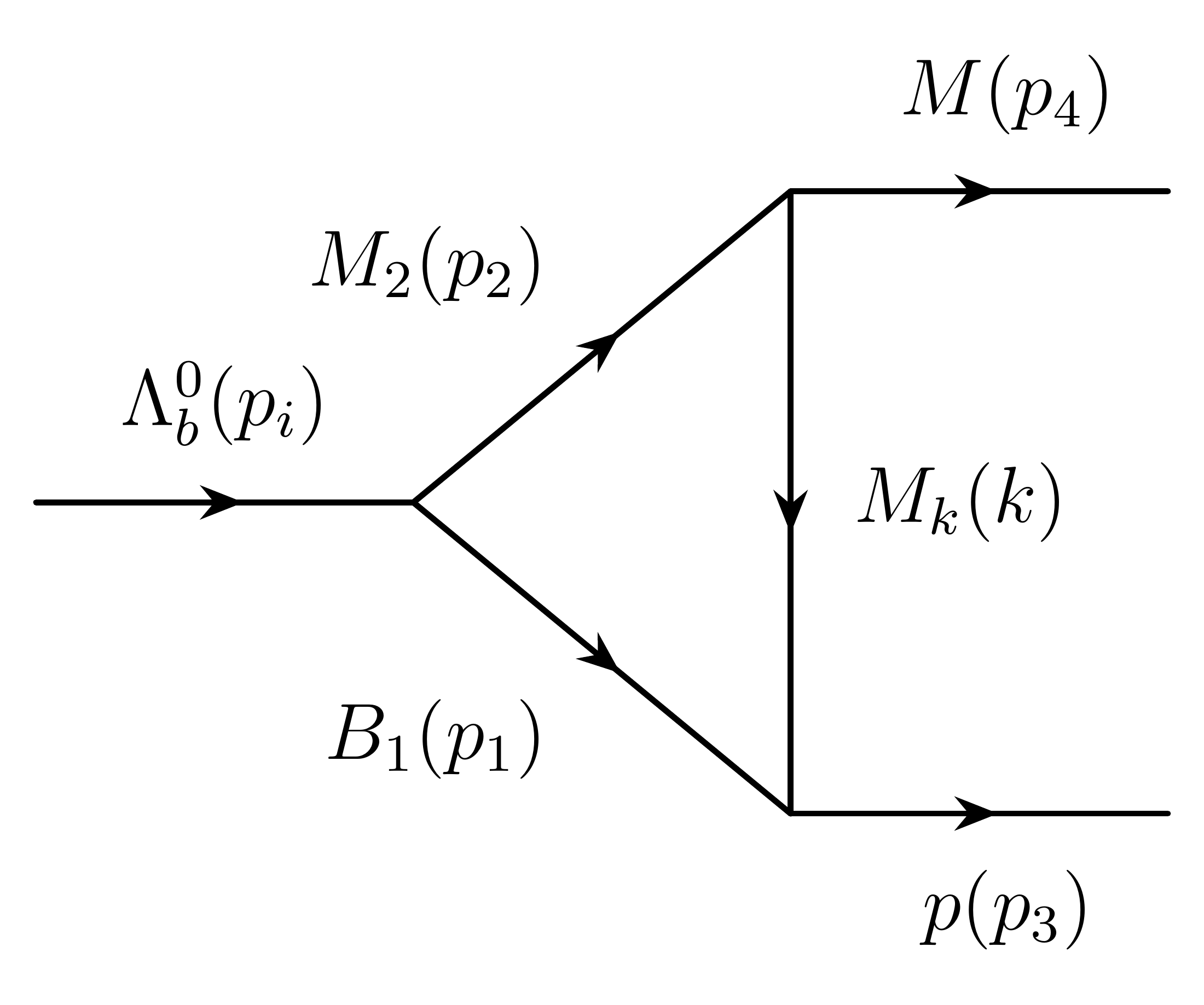}
    \captionsetup{type=figure}
    \captionof{figure}{Representative hadronic triangle-loop contribution to the decay $\Lb(p_i)\to p(p_3)M(p_4)$ via single-meson exchange. The arrows indicate the momentum-flow convention.}
    \label{fig:FSIprevious}
\end{center}
\end{figure}

In the hadronic rescattering description, the amplitude for a two-body beauty-baryon decay is written schematically as a sum of products of weak-production amplitudes for intermediate hadronic states and their rescattering amplitudes into the observed final state. A complete treatment consistent with unitarity would require summing over all allowed intermediate hadronic states, but this is generally impractical. A common working approximation is to retain only selected two-body intermediate states and describe their rescattering into the observed final state through single-particle exchange. This gives rise to a triangle-loop amplitude, as illustrated in Fig.~\ref{fig:FSIprevious}. Bubble contributions~\cite{Cheng:2004ru}, in which the two intermediate hadrons couple to a single $s$-channel hadronic state that subsequently produces the final state, are neglected in the calculations considered here, since no relevant resonance has a mass close to that of the initial beauty baryon. Then, the weak-production amplitudes are estimated using naive factorization, while the rescattering amplitudes are calculated from effective hadronic Lagrangians constructed to respect flavor and chiral symmetries, with the corresponding strong couplings as inputs.

In some phenomenological triangle-loop calculations, {\it e.g.}~\cite{Cheng:2004ru}, the contribution from on-shell intermediate states is assumed to dominate, and the rescattering amplitude is approximated by its absorptive part, obtained by cutting these states. The loop then reduces to a phase-space integral over the product of the weak-production and hadron-scattering amplitudes: $\operatorname{Abs}[\mathcal{A}_{f}] = \frac{1}{2}\sum_i \int d\Phi_i\, \mathcal{A}(\Lb\to i)\, \mathcal{T}^{*}(f\to i)$, where $i$ labels the intermediate states, $d\Phi_i$ is their Lorentz-invariant phase space, and $\mathcal{T}$ is the strong-scattering amplitude. However, it has been found that the dispersive parts of the hadronic loop amplitudes can also be important in the investigation of nonleptonic $\Lambda^{+}_{c}$ decays~\cite{Jia:2024pyb}, so the complete loop amplitudes need to be calculated. To illustrate the structure of a triangle-loop calculation, we consider $\Lb(p_i)\to p(p_3) M(p_4)$, with an intermediate octet baryon $B_1(p_1)$ and a meson $M_2(p_2)$, which rescatter through the exchange of a meson $M_k(k)$. The corresponding loop amplitude has the schematic form 
\begin{equation}\label{eq:triangle}
\mathcal{M}_{\triangle} = \int \frac{d^4k}{(2\pi)^4}\, \frac{\mathcal{N}(k)\,\mathcal{F}(\Lambda,k,m_k)} {[(p_2-k)^2 - m_1^2+i\epsilon] [(p_1+k)^2-m_2^2+i\epsilon] (k^2-m_k^2+i\epsilon)} ,
\end{equation}
where $\mathcal{N}(k)$ collects the weak-production and strong-interaction vertices, including their spinor, polarization, and propagator-numerator structures. The three denominators correspond to the internal baryon, meson, and exchanged vector meson, respectively. The form factor $\mathcal{F}$ parametrizes the off-shell behavior and possible substructure of the exchanged hadron; for some choices, it also regulates the ultraviolet behavior of the loop integral. Its functional form is model dependent. A typical form used in $\Lb$-decay calculations is~\cite{Duan:2024zjv}
\begin{equation}
\mathcal{F}(\Lambda,k,m_k) = \frac{\Lambda^4}{(k^2-m_k^2)^2+\Lambda^4} .
\end{equation}
In this parametrization, $\Lambda$ is a model scale associated with the form-factor prescription; it should not be interpreted as a universal cutoff or physical scale. Separate parameters $\Lambda_{\rm charm}$ and $\Lambda_{\rm charmless}$ are introduced to distinguish rescattering involving charmed intermediate states from that involving only light hadrons. Their values can be guided by the available branching-fraction and $\CP$-asymmetry information in $\Lb\to p\pi^-$ and $pK^-$, and be applied to other charmless beauty-baryon decay channels because the particles involved in the rescattering processes have comparable kinematic properties.

\begin{table*}[t]
    \footnotesize
	\centering
	\renewcommand{\arraystretch}{1.8}
    \caption{Selected direct $\CP$ asymmetries, in units of percentage, in two-body beauty-baryon decays predicted by different theoretical approaches, together with the available experimental results. The predictions for $\Lambda_b^0\to\Lambda K^0$ refer to the flavor-specific primary decay, which may be reconstructed experimentally through $K_S^0$. For the $\Lambda_b^0\to\Lambda D_\pm$ entries, the $D$ states can be reconstructed through $\CP$-eigenstate decays, such as $D\to\pi^+\pi^-,K^+K^-$ ($\CP$ even) and $D\to K_S^0\pi^0,K_S^0\omega$ ($\CP$ odd).}
	\label{tab:directcp}
	\begin{tabular*}{160mm}{c@{\extracolsep{\fill}}ccccc}
		\hline
        \hline
Decay mode
& GF  & QCDF (diquark)  & PQCD & FSIs   & Experiment  \\
		\hline
$\Lambda^{0}_b\to p\pi^-$ &   \makecell{$-3.9\pm0.2$~\cite{Hsiao:2014mua} \\ $-3.6\pm 0.3$~\cite{Geng:2021nkl}}      &   $-3.4^{+0.3}_{-0.4}$~\cite{Zhu:2018jet} & $5^{+2}_{-3}$~\cite{Han:2024kgz} & $-2^{+6}_{-5}$~\cite{Duan:2024zjv} & $0.2\pm 0.9$~\cite{LHCb:2024iis} \\
        \hline
$\Lambda^{0}_b \to  pK^-$ &   \makecell{$5.8\pm0.2$~\cite{Hsiao:2014mua} \\ $6.4\pm 0.3$~\cite{Geng:2021nkl} }  & $10.1^{+1.3}_{-2.0}$~\cite{Zhu:2018jet} &  $-6^{+3}_{-2}$~\cite{Han:2024kgz} &   $-9\pm2$~\cite{Duan:2024zjv}  & $-1.1\pm0.8$~\cite{LHCb:2024iis} \\
        \hline
$\Lambda^{0}_b\to p\rho^-$ &   \makecell{$-3.7\pm0.3$~\cite{Hsiao:2014mua} \\ $-3.3\pm 0.2$~\cite{Geng:2021nkl} }     &   $-3.2^{+0.3}_{-0.3}$~\cite{Zhu:2018jet} &  $3^{+3}_{-5}$~\cite{Han:2024kgz} & $-24^{+7}_{-3}$~\cite{Duan:2024zjv} & \\
        \hline
$\Lambda^{0}_b\to pK^{*-}$ &   \makecell{$19.6\pm1.5$~\cite{Hsiao:2014mua} \\ $19.3\pm0.8$~\cite{Geng:2021nkl}}  &  $31.1^{+2.8}_{-1.9}$~\cite{Zhu:2018jet}    &   $-5^{+10}_{-16}$~\cite{Han:2024kgz} & $2\pm4$~\cite{Duan:2024zjv} & $-0.6\pm4.5$~\cite{LHCb:2025ozp} \\
        \hline
$\Lambda_b^0\to\Lambda K^0$
& $0.2^{+0.1}_{-0.0}$~\cite{Hsiao:2017tif}
& $-20.6^{+1.7}_{-1.8}$~\cite{Zhu:2018jet}
& & & \\
\hline
$\Lambda_b^0\to\Lambda\eta$
& $0.4\pm0.2$~\cite{Hsiao:2017tif}
& $-3.4^{+0.6}_{-0.4}$~\cite{Zhu:2018jet}
& & & \\
\hline
$\Lambda^{0}_b\to \Lambda\rho^0 $ &   $2.3^{+0.2}_{-0.8}$~\cite{Hsiao:2017tif}  &  $25.3^{+7.5}_{-4.3}$~\cite{Zhu:2018jet}  &   &  & \\
        \hline
$\Lambda^{0}_b\to \Lambda K^{*0} $ &   $1.3\pm0.1$~\cite{Hsiao:2017tif}  &  $-25.1^{+4.1}_{-7.4}$~\cite{Zhu:2018jet}  &  &  & \\
\hline
$\Lambda^{0}_b\to \Lambda\phi $ &   $1.4^{+0.7}_{-0.2}$~\cite{Hsiao:2017tif}  &  $1.6^{+0.4}_{-0.3}$~\cite{Zhu:2018jet}  & $-1.0^{+1.0}_{-1.5}$~\cite{Rui:2022jff}  & $-0.5^{+2}_{-1}$~\cite{Duan:2024zjv} & \\
        \hline
$\Lambda_b^0\to\Lambda D_+$
& &
& $-44^{+10}_{-3}$~\cite{Rui:2026ihu}
& $28^{+7}_{-11}$~\cite{Li:2026zrs}
& \\
\hline
$\Lambda_b^0\to\Lambda D_-$
& &
& $72^{+4}_{-14}$~\cite{Rui:2026ihu}
& $-14^{+6}_{-4}$~\cite{Li:2026zrs}
& \\
\hline
$\Xi_b^-\to\Lambda K^-$
& $6.7^{+0.3}_{-0.2}\pm0.3$~\cite{Hsiao:2017tif}
& & & & \\
\hline
$\Xi_b^-\to\Lambda(1520)K^-$
& & & 
& $-3.4^{+7.9}_{-6.7}$~\cite{Feng:2026soj}
& $-5\pm12$~\cite{LHCb:2021enr} \\
        \hline\hline
        \end{tabular*}
\end{table*}

Hadronic-loop calculations have been applied to $\Lb$ decays~\cite{Duan:2024zjv}. Their predictions for direct $\CP$ asymmetries are listed in Table~\ref{tab:directcp}, alongside results from other theoretical approaches and experimental measurements. In this calculation, the long-distance topological amplitudes for the $p\pi^-$ and $pK^-$ modes are about an order of magnitude smaller than the factorizable $T$-topology amplitude, a hierarchy qualitatively consistent with power-counting expectations~\cite{Duan:2024zjv}. Nevertheless, the nonfactorizable charm-penguin contribution remains important for predicting the $\CP$ asymmetries because it provides sizable strong phases. The calculation also finds partial cancellation between the $S$- and $P$-wave $\CP$ asymmetries in both channels. Predictions have also been made for channels such as $\Lb\to p\rho^-$, $pK^{*-}$, and $\Lambda\rho^0$, including their $\CP$ asymmetries and polarization observables. For $\Lb\to pK^{*-}$, a direct $\CP$ asymmetry of about $2\%$ is obtained, close to the cited PQCD prediction~\cite{Han:2024kgz} but different from some generalized-factorization~\cite{Hsiao:2014mua,Geng:2021nkl} and diquark-model QCDF estimates~\cite{Zhu:2018jet}. In the $K^{*-}$-dominated region of the $\Lb\to pK_S^0\pi^-$ decay, LHCb measures $A^{\rm dir}_{\CP}=(-0.6\pm4.0\pm1.9)\%$~\cite{LHCb:2025ozp}. This result is consistent with zero and provides sensitivity to the larger asymmetries predicted in some approaches, although its precision is not yet sufficient to test predictions at the few-percent level. Since it is measured in a resonance region of a three-body decay, this asymmetry is not identical to that of the two-body $\Lb\to pK^{*-}$ decay.

The hadronic-loop approach involves unavoidable model dependence. An important issue is the truncation of the intermediate hadronic channels. Even when two-body rescattering is assumed to dominate, including all possible intermediate states is generally impractical. Existing calculations typically retain the lowest-lying light-baryon octet and decuplet, pseudoscalar and vector light mesons, together with selected charmed hadrons. Including additional excited states could make the treatment more complete, but their strong couplings and production amplitudes are often poorly known, introducing further model parameters. In addition, the results depend on the prescription for the form factor $\mathcal{F}(\Lambda,k,m_k)$, for which various models have been used. The triangle-loop framework should therefore be regarded as an effective phenomenological model. Nevertheless, it has been applied to selected resonant channels, including $\Lb\to N^*(1520,1535)M$ with $M=K_S,f_0,\rho,K^{*},\phi$, $\Lb\to\Lambda(1520)M$ with $M=\pi^0,f_0,\rho^0,K^{*0},\phi$, $\Xi_b^-\to\Lambda(1520)K^-$, and $\Lb\to D\Lambda$, as well as channels involving pentaquark states such as $\Lb\to P_c\pi^-$ and $P_cK^-$~\cite{Duan:2026oxg,Feng:2026soj,Shang:2026knt,Li:2026zrs}. These calculations provide inputs for resonant contributions to multibody decays through the subsequent decays of the produced resonances. A complete multibody description further requires a consistent treatment of finite-width effects, nonresonant contributions, and interference among the various amplitudes. The more general challenges of describing multibody amplitudes are discussed next.

\subsubsection{$N\pi$ re-scattering framework for $b$ baryon multi-body decays}\label{sec:npis}

Multibody decays introduce further challenges because different dynamical mechanisms can dominate different regions of phase space. For example, the invariant-mass distribution of $\Lb\to\Lambda K^+K^-$ can be displayed in a Dalitz plot using the two independent variables $m_{\Lambda K^+}$ and $m_{K^+K^-}$. The low-$m_{K^+K^-}$ region receives important contributions from light-meson resonances such as $\phi$ and $f_0(980)$~\cite{LHCb:2024yzj}, while the low-$m_{\Lambda K^+}$ region can involve several excited $N^*$ states~\cite{Wang:2024oyi}. Other regions may receive considerable contributions from nonresonant production, rescattering among different subsystems, and shorter-distance multibody mechanisms. Their relative importance depends on the decay channel and kinematic region. Understanding this variation across phase space is essential for describing the strong phases and identifying regions where $\CP$ asymmetries may be enhanced.

Even if all intermediate resonant contributions could be calculated reliably, a complete multibody description would still require nonresonant production and additional rescattering effects. Schematically, these ingredients may be organized as
\begin{equation}
\mathcal{A}_{\rm multi} = \sum_R\mathcal{A}_R +\mathcal{A}_{\rm nonres} +\mathcal{A}_{\rm resc},
\end{equation}
where the terms represent explicit resonance contributions, nonresonant production, and rescattering effects not already incorporated in the other terms, respectively. This separation is model dependent, and overlapping contributions must be treated consistently to avoid double counting. In particular, nonresonant contributions need not be small or slowly varying: they can carry strong phases and interfere coherently with resonant amplitudes. Their treatment is therefore also important for predictions of local and phase-space-integrated $\CP$ asymmetries.

A quasi-two-body description is most useful when a region is dominated by a few isolated, narrow resonances. The treatment becomes more challenging when many broad resonances overlap. For example, the low-$m_{p\pi^+\pi^-}$ region of the four-body decays $\Lb\to p\pi^+\pi^-\pi^-$ and $\Lb\to pK^-\pi^+\pi^-$ can receive contributions from numerous excited nucleon states. The relevant $p\pi\pi$ channels contain a rich spectrum extending from $N(1440)$ to masses around $2.5\,\mathrm{GeV}$, with many states having widths above $100\,\mathrm{MeV}$~\cite{Wang:2024oyi}. Their nearby masses and broad widths complicate the experimental separation of individual contributions and the theoretical identification of dominant amplitudes; uncertainties in their masses, widths, production amplitudes and strong couplings further limit predictive control. States with different spin and parity contribute to multiple partial waves and generate rich angular distributions. Interference among these amplitudes, which can have different strong phases, may produce sizable local $\CP$ asymmetries even when the phase-space-integrated asymmetry is small. Rescattering among particles associated with different resonance subsystems further complicates descriptions based on independently modeled resonances. These challenges affect both factorization-based and hadronic-loop calculations, although the nonperturbative inputs required by the two approaches differ. In such regions, experimentally constrained two-body scattering partial waves offer an alternative way to describe the strong interactions within a selected subsystem. Over the energy range covered by the scattering analysis, these amplitudes encode resonance and background effects, including their strong phases. Combined with a description of weak production, they provide another phenomenological approach to multibody beauty-baryon decays. The following discussion considers an implementation based on $N\pi$ scattering.

The use of experimentally determined scattering amplitudes to describe final-state interactions has precedents in meson decays. In charmless three-body $B^\pm$ decays, direct $\CP$ asymmetries can be enhanced in selected low-invariant-mass regions. Such patterns have been interpreted in terms of rescattering between the $\pi\pi$ and $K\bar K$ channels, using scattering amplitudes and inelasticities extracted from data rather than modeling each contribution through individual resonances~\cite{Bediaga:2020qxg}. Related coupled-channel effects have also been considered in $D^0\to\pi^+\pi^-$ and $D^0\to K^+K^-$, whose final states can rescatter through $S$-wave $\pi\pi\leftrightarrow K\bar K$ interactions~\cite{Bediaga:2022sxw}. A related data-driven approach has been applied to $N\pi$ rescattering in multibody $\Lb$ decays to estimate $\CP$ asymmetries~\cite{Wang:2024oyi}.

For multibody beauty baryon decays, the $N\pi$ system provides a natural application of this strategy. The $\Delta^{++}$ and many excited nucleon states have been identified and characterized through partial-wave analyses of $N\pi$ scattering, which constrain their masses and widths from data~\cite{Manley:1984jz}. The corresponding partial-wave amplitudes are available from analyses such as SAID~\cite{SAID} and Refs.~\cite{Ronchen:2012eg,Gasparyan:2003fp}. Using these amplitudes in place of separate resonance-by-resonance parametrizations reduces the corresponding model dependence and provides an empirical description of the energy-dependent strong phases in the selected $N\pi$ channel.

\begin{figure}
\begin{center}
    \includegraphics[width=0.5\textwidth]{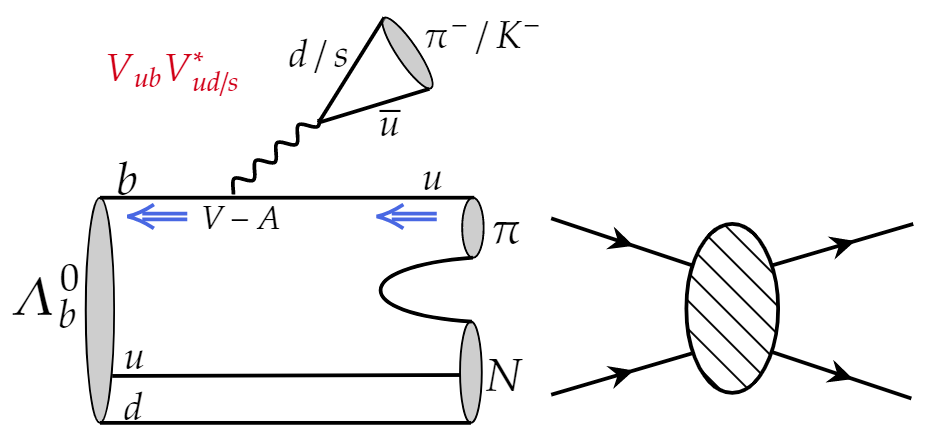}
    \vspace{5mm}
    
    \includegraphics[width=0.5\textwidth]{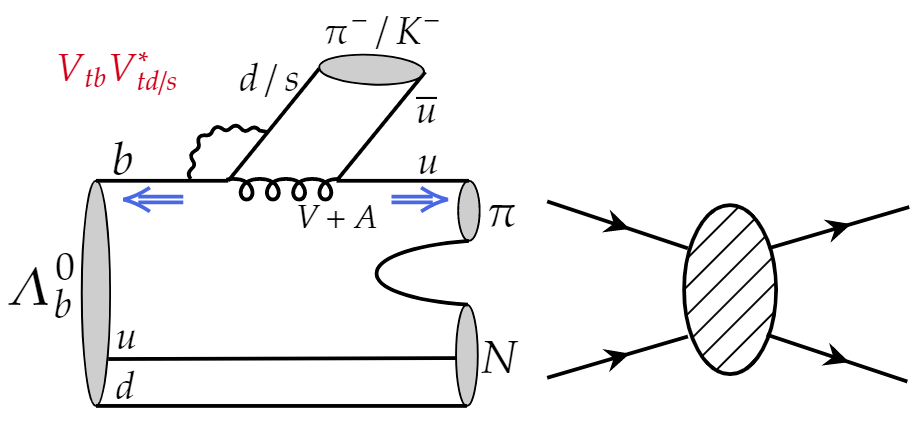}
    \captionof{figure}{Schematic illustration of the mechanism for $\CP$ violation induced by hadronic rescattering. The figures are taken from Ref.~\cite{Wang:2024oyi}. The decay is modeled as a two-step process: short-distance weak production of an $N\pi$ subsystem together with a bachelor meson $h^-$, followed by long-distance rescattering within the subsystem. Interfering tree and penguin contributions carry different CKM weak phases, while the relevant scattering partial waves provide strong phases; the $N\pi$ amplitudes are constrained by experimental data, including the SAID analysis. 
    }\label{fig:FSInpi}
\end{center}
\end{figure}

The mechanism considered for $\Lb$ decays is illustrated in Fig.~\ref{fig:FSInpi}~\cite{Wang:2024oyi}. In this picture, the weak decay first produces an $N\pi$ subsystem together with a bachelor meson $h^-=\pi^-$ or $K^-$, followed by rescattering within the subsystem into the observed final state.. At low $m_{N\pi}$, the $N$ and $\pi$ move nearly collinearly and recoil against the bachelor meson. In the color-transparency picture, an energetic bachelor meson produced in a compact color-singlet configuration has suppressed soft interactions with the recoiling system. This provides a kinematic motivation for neglecting soft rescattering between $h^-$ and the $N\pi$ subsystem in the approximation considered here, while retaining rescattering within the $N\pi$ subsystem. Schematically, the amplitude is written as $\mathcal{A}=S^{1/2}\mathcal{A}_0$, where $\mathcal{A}_0$ is the weak-production amplitude, including short-distance QCD corrections, and $S^{1/2}$ describes the subsequent rescattering in the selected subsystem. The production amplitude contains tree and penguin contributions proportional to $\lambda_u=V_{ub}V_{uq}^{*}$ and $\lambda_t=V_{tb}V_{tq}^{*}$, respectively, with $q=d$ or $s$. The current--current operators $\mathcal{O}_{1,2}$ and the QCD penguin operators $\mathcal{O}_{5,6}$ have different chiral structures and can therefore populate the $N\pi$ partial waves with different relative strengths. Since the scattering phases vary with the partial wave, the tree and penguin contributions would acquire different strong phases after rescattering.

Soft rescattering between the bachelor meson and the $N\pi$ subsystem is suppressed in this picture, but short-distance interactions between them are not excluded. In principle, such contributions can be included in $\mathcal{A}_0$ through quark-level corrections such as vertex corrections and hard-scattering contributions. A systematic factorization treatment would require nonperturbative inputs involving the $N\pi$ system, such as its light-cone distribution amplitudes, analogous to those used in analyses of $B\to(\pi\pi)\ell\nu_\ell$ form factors at low $m_{\pi\pi}$~\cite{Cheng:2019hpq}. Unfortunately, no corresponding factorization theorem has yet been established. The calculation of Ref.~\cite{Wang:2024oyi} instead represents these short-distance corrections phenomenologically through complex effective coefficients in generalized factorization, which provides another source of strong phases.
The long-distance factor $S^{1/2}$ is related to experimentally determined $N\pi$ scattering partial-wave amplitudes, which encode resonance and background effects in the analyzed channel and energy range without requiring an explicit model for each $N^*$ state.

In the factorization model used to estimate the production amplitude, the $\Lb\to N\pi$ transition is described by form factors, while the emitted meson contributes its decay constant. For a spin-$1/2$ $N\pi$ subsystem, the weak amplitude can be organized as
\begin{equation}\label{eq:A0}
\mathcal{A}_0 =
\bar{u}_{N\pi,\,1/2^+}(A+B\gamma_5)u_{\Lambda_b}
+\bar{u}_{N\pi,\,1/2^-}(\widetilde A+\widetilde B\gamma_5)u_{\Lambda_b}
+\cdots ,
\end{equation}
where only the $f_{1}\gamma_{\mu}$ and $g_{1}\gamma_{\mu}\gamma_{5}$ Lorentz structures are kept in the $\Lb\to N\pi$ transition form factors, and the ellipsis denotes higher-spin components. Up to a common normalization, the reduced amplitudes $A,B,\tilde{A},\tilde{B}$ (The amplitudes with the same spin but opposite parity are distinguished by a tilde in \(A,B\)) in the generalized-factorization treatment are
\begin{equation}
\begin{aligned}
A &= f_{h}(\lambda_u a^{\rm eff}_1-\lambda_t a^{\rm eff}_{46+})f_1^{1/2^+}m_-,
&\qquad
B &= f_{h}(\lambda_u a^{\rm eff}_1-\lambda_t a^{\rm eff}_{46-})g_1^{1/2^+}m_+,\\
\widetilde A &= f_{h}(-\lambda_u a^{\rm eff}_1+\lambda_t a^{\rm eff}_{46-})f_1^{1/2^-}m_-,
&\qquad
\widetilde B &= f_{h}(-\lambda_u a^{\rm eff}_1+\lambda_t a^{\rm eff}_{46+})g_1^{1/2^-}m_+,
\end{aligned}
\end{equation}
with $a^{\rm eff}_{46\pm}=a^{\rm eff}_4\pm R_h a^{\rm eff}_6$, $R_h=2m_h^2/[m_b(m_u+m_q)]$, and $m_\pm=m_{\Lambda_b}\pm m_{N\pi}$, $h^{-}$ decay constant $f_{h}$. Here $a^{\rm eff}_{1,4,6}$ are the effective coefficients defined in Eq.~\eqref{eq:aeff-def}, and $m_{N\pi}$ is the invariant mass of the subsystem. The $N\pi$ scattering partial wave amplitudes are provided by the SAID partial wave analysis~\cite{SAID} and can be found at \url{https://gwdac.phys.gwu.edu/}. The partial waves are conventionally labeled $L_{2I,2J}$, where $L$, $I$, and $J$ denote the orbital angular momentum, isospin, and total angular momentum. Taking $\Lambda^{0}_{b}\to (p\pi^{0})h^{-}$ as an example, the total decay amplitudes $\mathcal{A}=S^{1/2}\mathcal{A}_{0}$ is
\begin{equation}
\begin{aligned}
\mathcal{A}\bigl(\Lb\to(p\pi^0)h^-\bigr) = \bar{u}_{N\pi,\,1/2^+}(A+B\gamma_5)u_{\Lambda_b}\,P_{11}  +\bar{u}_{N\pi,\,1/2^-}(\widetilde A+\widetilde B\gamma_5)u_{\Lambda_b}\,S_{11} +\cdots ,
\end{aligned}
\end{equation} 
keeping the leading $S_{11}$ and $P_{11}$ waves. The partial wave components with isospin $I=3/2$ do not contribute here since the quark transition $b\to u$ changes the isospin by $\Delta I=1/2$ within the factorization hypothesis. The scattering amplitudes $S_{11}$ and $P_{11}$ supply energy-dependent strong phases to the corresponding production components. Combining the above derivations together, one has 
\begin{equation}
\begin{aligned}\label{eq:A}
\mathcal{A}(\Lb\to (p\pi^{0}) h^-) ={}&
\lambda_u f_h\bar u_{N\pi}  a^{\rm eff}_1 \bigg[
\left( P_{11} f_1^{1/2^+} -S_{11} f_1^{1/2^-} + \cdots \right)m_-+ \left( P_{11} g_1^{1/2^+} -S_{11} g_1^{1/2^-}+ \cdots \right)m_+\gamma_5
\bigg]
u_{\Lambda_b}\\
& -\lambda_t f_h \bar u_{N\pi}  \bigg[
\left( a^{\rm eff}_{46+} P_{11} f_1^{1/2^+} - a^{\rm eff}_{46-}S_{11} f_1^{1/2^-}  + \cdots  \right)m_-\\
&+\left(  a^{\rm eff}_{46-} P_{11} g_1^{1/2^+} - a^{\rm eff}_{46+}S_{11} g_1^{1/2^-} + \cdots \right)m_+\gamma_5
\bigg]u_{\Lambda_b} + \cdots ,
\end{aligned}
\end{equation}
where $f_{h}$ is the decay constant of $h=\pi,K$. 

For the four-body mode $\Lb\to(p\pi^+\pi^-)h^-$, the rescattering process $N\pi\to p\pi^{+}\pi^{-}$ is inelastic and involves a three-particle final subsystem. The calculation in Ref.~\cite{Wang:2024oyi} treats this contribution through a quasi-two-body $\Delta^{++}\pi^-$ channel, corresponding to $N\pi\to\Delta^{++}\pi^-$. The relevant transition partial waves are more involved than those for elastic $N\pi$ scattering and constitute additional input to the model. The different chiral structures of the tree and penguin operators also affect the relative partial-wave contributions in this channel.

Based on the amplitude in Eq.~\eqref{eq:A} for $\Lb\to(p\pi^0)h^-$, one can identify two structurally distinct contributions to the differential $\CP$ asymmetry. Retaining only the $S_{11}$ and $P_{11}$ partial waves, the angular distribution at fixed $m_{N\pi}$ can be written as
\begin{equation}\label{eq:dGammadcostheta}
\begin{aligned}
\left.\frac{d\Gamma}{d\cos\theta}\right|_{m_{N\pi}} \propto{}& |P_{11}|^2\left(|A|^2+\kappa^2|B|^2\right) +|S_{11}|^2\left(|\widetilde A|^2 +\kappa^2|\widetilde B|^2\right) +2\operatorname{Re}\left[ \left(A\widetilde A^* +\kappa^2B\widetilde B^*\right) P_{11}S_{11}^* \right]\cos\theta ,
\end{aligned}
\end{equation}
where $\theta$ is the helicity angle between the proton momentum in the $N\pi$ rest frame and the $N\pi$ momentum in the $\Lb$ rest frame, and the kinematic factor $\kappa=p_{N\pi}/(E_{N\pi}+m_{N\pi})$ with $p_{N\pi},E_{N\pi}$ as the momenta and energy of the $N\pi$ subsystem in the $\Lambda^{0}_{b}$ rest frame. The first two terms in Eq.~\eqref{eq:dGammadcostheta} describe the incoherent contributions of the individual partial waves. Their $\CP$ differences are determined by quantities such as $|A|^2-|\bar A|^2$, weighted by $|P_{11}|^2$, and do not involve interference between different $N\pi$ partial waves. For example, assuming real form factors, $|A|^{2}-|\bar{A}|^{2}\propto \operatorname{Im}[\lambda_{u}\lambda_{t}^*]\operatorname{Im}[a^{\rm eff}_{1}a^{\rm eff*}_{46+}]$, with the strong-phase difference only originating from the complex effective Wilson coefficients. The last term in Eq.~\eqref{eq:dGammadcostheta} provides a second source of differential $\CP$ violation through interference between the $S_{11}$ and $P_{11}$ waves. For example, the contribution
proportional to $A\widetilde A^*$ satisfies
\begin{equation}
	\begin{aligned}
\operatorname{Re}\left[A P_{11}\tilde A^{*}S^{*}_{11}\right]&-\operatorname{Re}\left[\bar{A}\bar{P}_{11}\bar{\tilde A}^{*}\bar{S}^{*}_{11}\right] \propto \operatorname{Re}\big[(\lambda_u^*\lambda_t-\lambda_u\lambda_t^*)(a^{\rm eff}_{46+} a^{\rm eff*}_1 - a^{\rm eff *}_{46-} a^{\rm eff}_{1})P_{11} S_{11}^* \big] .
	\end{aligned}
\end{equation}
Since $S_{11}$ and $P_{11}$ are complex, energy-dependent scattering amplitudes, this interference contribution varies with $m_{N\pi}$ and is sensitive to their relative strong phase. Being proportional to $\cos\theta$, it integrates to zero over the full angular range, but remains accessible in angular bins or through suitably weighted angular moments. More generally, the angular distribution can be expanded in Legendre polynomials,
\begin{equation}
\frac{d\Gamma}{d\cos\theta} = \sum_{n=0}^{\infty} \mathcal{L}_n P_n(\cos\theta).
\end{equation}
Suitably weighted angular moments can project the coefficients $\mathcal{L}_n$, thereby retaining interference information that may cancel in the angle-integrated rate. The $\CP$ differences of these moments provide complementary probes of $\CP$ violation. A systematic discussion of such observables is deferred to Sec.~\ref{sec:observable}.

In the numerical analysis, the $\Lb\to N\pi$ form factors required for the weak-production amplitude are poorly known. Fortunately, within the form-factor parametrization used here, their common normalization cancels in the $\CP$-asymmetry ratio, although the relative values can still affect the result. In the heavy-quark limit, one has $f_1=g_1$. Possible corrections of order $10$--$20\%$ are explored by comparing the benchmark choice $f_1=g_1=1$ with $(f_1,g_1)=(1.1,0.9)$ and $(0.9,1.1)$~\cite{Wang:2024oyi}. For the $\Lb\to(p\pi^0)h^-$ and $\Lb\to(\Delta^{++}\pi^-)h^-$ channels, the calculation considers $m_{p\pi^0}\in[1.1,2.5]\,\mathrm{GeV}$ and $m_{\Delta^{++}\pi^-}\in[1.2,1.9]\,\mathrm{GeV}$, respectively. The corresponding region-integrated asymmetries are
\begin{equation}
\begin{aligned}
A_{\rm CP}[\Lb \to (\Delta^{++}\pi^-) \pi^-] ={}& (-3.9\pm 0.3) \% , \\
A_{\rm CP}[\Lb \to (\Delta^{++}\pi^-) K^-] ={}& (5.8\pm 0.2) \% , \\
A_{\rm CP}[\Lb \to (p\pi^0) \pi^-] ={}& (-3.9\pm 0.1) \% , \\
A_{\rm CP}[\Lb \to (p\pi^0) K^-] ={}& (5.8\pm 0.1) \% . \\
\end{aligned}
\end{equation}
Here the quoted uncertainties represent only the spread induced by the variation of the form-factor inputs; other theoretical uncertainties are not included. The predicted asymmetry of about $5\%$ in the $\Delta^{++}\pi^-$-dominated region of $\Lb\to pK^-\pi^+\pi^-$ is compatible with the corresponding experimental result~\cite{Wang:2024oyi}. The region-integrated results show only a weak dependence on the form-factor scenario. Within this calculation, the $S$--$P$ interference term nearly vanishes after full angular integration, and the remaining integrated asymmetries are therefore controlled mainly by the strong phases in the effective Wilson coefficients. By contrast, the same analysis finds that local asymmetries can reach the $\mathcal{O}(10\%)$ level and, in some regions, approach $30\%$, providing targets for future measurements.

This approach uses scattering data to determine the strong-interaction phases in a chosen subsystem, but it does not remove the need to model weak production. In particular, interactions between the bachelor meson and the $N\pi$ system, as well as contributions from other subsystems, are not fixed by the $N\pi$ partial waves alone. A systematic factorization treatment involving $N\pi$ distribution amplitudes has not been established; the cited analysis instead represents additional short-distance phases phenomenologically through complex effective coefficients. The method is therefore complementary to triangle-loop calculations and does not by itself provide a complete description of multibody decays. Its predictions motivate differential measurements of invariant-mass and angular distributions, which can test the phase structure more directly than a single phase-space-integrated asymmetry.

The triangle-loop and scattering-data approaches discussed above, together with the factorization and PQCD frameworks reviewed in this section, all involve approximations and nonperturbative inputs that limit their quantitative precision. Their role should nevertheless be viewed in its historical context. Most of these theoretical works were carried out before the first observation of $\CP$ violation in a beauty-baryon decay and therefore served as exploratory rather than precision studies. Even when the numerical predictions were necessarily uncertain, these analyses provided useful benchmarks, identified promising decay channels, resonance regions and angular observables, and helped motivate and interpret subsequent experimental searches. With the growing experimental data set, these calculations can now be tested, refined and compared, providing a starting point for a more quantitative understanding of $\CP$ violation in beauty-baryon decays.



%% file: sec4-experiment.tex

\subsection{Machines and Experiments for Baryonic $C\!P$ Violation Searches}
This Section aims at discussing the machines and experiments that have contributed to the experimental knowledge in the $C\!P$ properties of baryon decays. What makes the LHCb experiment unique in this quest will be highlighted in the second part.     

\subsubsection{Pioneering experiments}

Early experimental searches for $C\!P$ violation in baryon decays covers several decades across fixed-target experiments, $e^+e^-$ colliders, and $p\bar{p}$ colliders. These historical efforts established already stringent bounds on direct $C\!P$ violation in charm and beauty baryon decays as well as in hyperon decays. This subsection distinguishes a non-comprehensive selection of the machines and experiments that have pioneered these searches over the last two decades. 

\paragraph{Searches for $C\!P$ violation in hyperon charmed decays at fixed target experiments}
\par\noindent

Hyperons are produced at the Low Energy Antiproton Ring (LEAR) at CERN (which operated from 1982 to 1996), via the threshold reaction on a fixed proton target $p\bar{p} \to \Lambda\bar{\Lambda}$. The PS185 experiment~\cite{PS185:1997oge} focuses on the measurement of the angular distributions of the final state particles in the self-analysing decay $\Lambda \to p\pi^-$ and derived with a percent precision the particle-antiparticle polarisation asymmetry, measured consistent with $C\!P$ symmetry. 

The HyperCP~\cite{HyperCP:2004zvh} and FOCUS~\cite{FOCUS:2005vxq} experiments make both use of primary high-energy proton beamline at Fermilab. HyperCP measures $C\!P$ asymmetries for hyperons produced on a fixed-target while FOCUS uses an intense neutral beam to study the charmed hadron photo-production, and in particular $\Lambda_c^+$ baryons. The Table~\ref{tab:earlymeasurements} gathers their findings. An agreement with $C\!P$ symmetry is observed.

\paragraph{Searches for $C\!P$ violation with $e^+e^-$ machines}
\par\noindent

The $e^+e^-$ colliders running at the $\Upsilon(4S)$ resonance are producing large samples of $e^+e^- \to c\bar{c}$ continuum events and permits $C\!P$ violation searches in $\Lambda_c^+$ decays. The $e^+e^-$ beam-energy-symmetric ring CESR, located at Cornell, was in operation in the decades 1980-2000 and operated above the charm threshold. The CLEO-c experiment searches for $C\!P$ breaking in the decay $\Lambda_c^+ \to \Lambda \pi^+$~\cite{CLEO:2000frb, CLEO:2003uko}. 
A recent measurement~\cite{Belle-II:2025xvc} by the Belle II collaboration at the asymmetric $B$-factory SUPERKEKB in Japan  reports an asymmetry in the decay $\Lambda_c^+ \to p K^+K^-$ as well consistent with $C\!P$ symmetry.    

\paragraph{Searches for $C\!P$ violation with $p \overline{p}$ colliders machines} 
\par\noindent

A search for a $C\!P$-violating asymmetry in two-body $\Lb$ decays is executed at the proton-antiproton collider Tevatron, prefiguring the searches that are conducted at the LHC. Large boost, capability to identify tracks displaced with respect to the collision point and particle identification measurements through time-of-flight and specific ionisation enables the CDF experiment to measure a $C\!P$ asymmetry in the $\Lb \to p K^-$ decay at $\mathcal{O}(10\%)$ precision, showing no significant charge asymmetry~\cite{CDF:2014pzb}.          

\begin{table}
\begin{center}
\captionsetup{type=table}
\caption{A selection of $C\!P$ measurements performed by experiments spanning over two decades and in several experimental environment (see text).}
\label{tab:earlymeasurements}
\begin{tabular} {lcccc}
Experiment & 	Machine  & Date   &  Channel & Asymmetry (${\cal  A}_{\Lambda}$ )  \\
\hline\hline

PS185	  &  LEAR  (CERN)       & 1994  &  $p \overline{p} \to \Lambda \overline{\Lambda}$ & $(0.6 \pm 1.5) \times10^{-2}$ \\
HyperCP	  &  E871 (Fermilab)    & 2001  &  $\Xi^- \to \Lambda (\to p \pi^-) \pi^-$  & $ (0.0 \pm 6.7)\times 10^{-4}$             \\ 
CLEO	 & CESR (Cornell)	    & 2001  &  $ \Lambda_c^+ \to \Lambda \pi^+$ & $-0.04 \pm 0.06$ \\ 
FOCUS	 &  E831 (Fermilab)     & 2005  &  $ \Lambda_c^+ \to p K^- K^+$   &  $0.07 \pm 0.08$ \\ 
CDF   	 &  Tevatron (Fermilab) & 2014  &  $\Lb \to pK^-$           &  $-0.10 \pm 0.09$  \\
Belle II &  SUPERKEKB (KEK)     & 2025  &  $\Lambda_c^+ \to p K^- K^+$    &  $(3.9 \pm 1.8)\times 10^{-2}$  \\
\hline
\end{tabular}
\end{center}
\end{table}

\subsubsection{Precision measurements of hyperon $\CP$ violation at BESIII}
The BESIII experiment~\cite{BESIII:2009fln}, operating at the BEPCII $e^+e^-$ collider, is dedicated to tau-charm physics. By exploiting quantum-entangled hyperon-antihyperon pairs produced in $J/\psi$ and $\psi'$ decays, BESIII has established a precision frontier for hyperon $\CP$ tests using the parity-violating decay parameter~\cite{Zhang:2026qjt}, even though the total hyperon yields may be lower than in previous experiments. With the world's largest samples of $10^{10}$ $J/\psi$ and $2.7 \times 10^{9}$ $\psi'$ events, full angular analyses enable direct comparisons of hyperon and antihyperon decay parameters, enabling precise $\CP$ violation measurements. As summarized in Table~\ref{tab:BESIII_CP_summary}, the $\CP$ asymmetry of $\alpha$ for the $\Lambda\to p\pi^-$ decay, $\alpha(\Lambda\to p\pi^-)$, is measured using the quantum-correlated $J/\psi \to \Lambda \bar{\Lambda}$ process, achieving a precision of about 0.5\%~\cite{BESIII:2018cnd,BESIII:2022qax}. Similar precision is obtained for the $\CP$ asymmetry of $\alpha(\Lambda\to n\pi^0)$~\cite{BESIII:2025wxe}. Recently, using the cascade decay \(J/\psi \to \Xi^-(\Lambda\pi^-)\bar{\Xi}^+(\bar{\Lambda}\pi^+)\), the most precise measurement of $\alpha(\Xi^-\to\Lambda\pi^-)$ is determined, with its $\CP$ asymmetry measured to an uncertainty of 0.5\%~\cite{BESIII:2026hgj}. The decay parameter and its $\CP$ asymmetry have also been measured for the $\Xi^0\to\Lambda\pi^0$ decay using the $J/\psi \to \Xi^0\bar{\Xi}^0$ process, although with lower precision~\cite{BESIII:2023drj}. The $\CP$ asymmetries for $\alpha(\Sigma^+\to p\pi^0)$ and $\alpha(\Sigma^0\to \gamma\Lambda)$ have also been measured with sub-percent uncertainties~\cite{BESIII:2023sgt,BESIII:2025jxt}. Beyond direct asymmetry measurements, BESIII has set the world's most stringent limit on the $\Lambda$ hyperon electric dipole moment (EDM) using the same entangled-pair technique, improving the sensitivity by about three orders of magnitude and reaching the $10^{-19}~e\cdot\mathrm{cm}$ level~\cite{BESIII:2025vgs}. Nevertheless, no $\CP$ violation has been established in hyperon decays to date.

\begin{table}[htbp]
  \centering
  \captionsetup{type=table}
  \caption{Summary of decay parameter $\alpha$ and CP-violating observables for a few hyperon decays, measured by the BESIII experiment using entangled hyperon-antihyperon pairs.}
  \label{tab:BESIII_CP_summary}
  \resizebox{\textwidth}{!}{
    \begin{tabular}{l c c c}
      \toprule
      \textbf{Decay mode} & \textbf{Production} & \(\left<\alpha\right>\) & \textbf{\(A_{CP}^\alpha\)} \\
      \midrule
      \(\Lambda \to p\pi^-\)~\cite{BESIII:2022qax} & \(J/\psi \to \Lambda\bar{\Lambda}\) & \(0.7542 \pm 0.0010 \pm 0.0024\) & \((-2.5 \pm 4.6 \pm 1.2)\times 10^{-3}\) \\
      
      \(\Lambda \to p\pi^-\)~\cite{BESIII:2026hgj} & \(J/\psi \to \Xi^-(\Lambda\pi^-)\bar{\Xi}^+(\bar{\Lambda}\pi^+)\) & \(0.7456 \pm 0.0022 \pm 0.0014\) & \((-2.9 \pm 4.3 \pm 0.7)\times 10^{-3}\) \\
     
      \(\Lambda \to n\pi^0\)~\cite{BESIII:2025wxe} & \(J/\psi \to \Lambda\bar{\Lambda}\) & $0.672\pm0.002\pm0.002$ & \((-6.0 \pm 7.0 \pm 2.0)\times 10^{-3}\) \\
      \midrule
      \(\Sigma^+ \to p\pi^0\)~\cite{BESIII:2025jxt} & \(J/\psi, \psi' \to \Sigma^+\bar{\Sigma}^-\) & \(-0.9869 \pm 0.0011 \pm 0.0016\) & \((-11.8 \pm 8.3 \pm 2.8)\times 10^{-3}\) \\

      \(\Sigma^0 \to \gamma\Lambda\)~\cite{BESIII:2023sgt} & \(J/\psi, \psi' \to \Sigma^0\bar{\Sigma}^0\) & \(-0.654 \pm 0.024 \pm 0.011\) & \((0.4 \pm 2.9 \pm 1.3)\times 10^{-3}\) \\
      \midrule
      \(\Xi^- \to \Lambda\pi^-\)~\cite{BESIII:2026hgj} & \(J/\psi \to \Xi^-\bar{\Xi}^+\) & \(-0.3770 \pm 0.0024 \pm 0.0014\) & \((-7.8 \pm 4.8 \pm 0.8)\times 10^{-3}\) \\
      
      \(\Xi^0 \to \Lambda\pi^0\)~\cite{BESIII:2023drj} & \(J/\psi \to \Xi^0\bar{\Xi}^0\) & \(0.514 \pm 0.006 \pm 0.015\) & \((-5.4 \pm 6.5 \pm 3.1)\times 10^{-2}\) \\
      \bottomrule
    \end{tabular}
  }
\end{table}

\subsubsection{The LHC machine and the LHCb experiment}


\paragraph{The Large Hadron Collider (LHC)}
\par\noindent
The LHC is the world highest-energy particle accelerator to date. 
It is installed in an underground tunnel of 26.7 km circumference (initially housing the Large $e^+e^-$ collider LEP) on the French-Swiss border near Geneva, Switzerland. 
The LHC features two beam-pipes which cross at four interaction regions where are placed the four main LHC experiments in order to record same charge protons or heavy-ions collisions. 
The LHC builds on the historical CERN accelerator complex, the last injector of protons for the LHC being the Super Proton Synchrotron (SPS), as sketched in Fig.~\ref{fig:CERN_acc_complex}.

The High-Energy Physics and the flavor Physics programs use mostly proton collisions with different luminosities at the different interaction points. The beauty production cross-section at LHC has been measured by the LHCb collaboration in the pseudorapidity range 2 $< \eta <$5 to be $\sigma(pp \to b \overline{b}X) = 144 \pm 1 \pm 21~\upmu$b ($72 \pm 0.3 \pm 6.8~\upmu$b), at a centre-of-mass energy $\sqrt{s} =13 (7)$ TeV. Few per mil of the inelastic $pp$ scatterings yield therefore the production of $b$ quarks. 

This abundant production of $b \overline{b}$ pairs proceeds mostly incoherently, by contrast with the $B$-factories. Each $b$-quark can therefore hadronize into any beautiful hadron species. In particular, about a fourth of the $b$-flavored hadrons produced at LHC are baryons. Among the four main experiments installed at LHC, the LHCb has specifically designed a trigger to collect in particular the hadronic decays $b$-baryons, forming a unique sample of  $b$-baryon decays in High Energy Physics.

The results presented in this section are therefore  coming from physics analyses of the events recorded by the LHCb experiment during the first two runs of the LHCb operation. The third and last run of the LHC operation has been completed in 2026 and the three-years long Shutdown will be used to install and commission a new accelerator: the High-Luminosity LHC. An upgrade of the LHCb spectrometer was undertaken for the last Run of the LHC;  the description of the LHCb detector will be limited to the historical detector.      

\begin{center}
    \centering
    \includegraphics[width=0.7\linewidth]{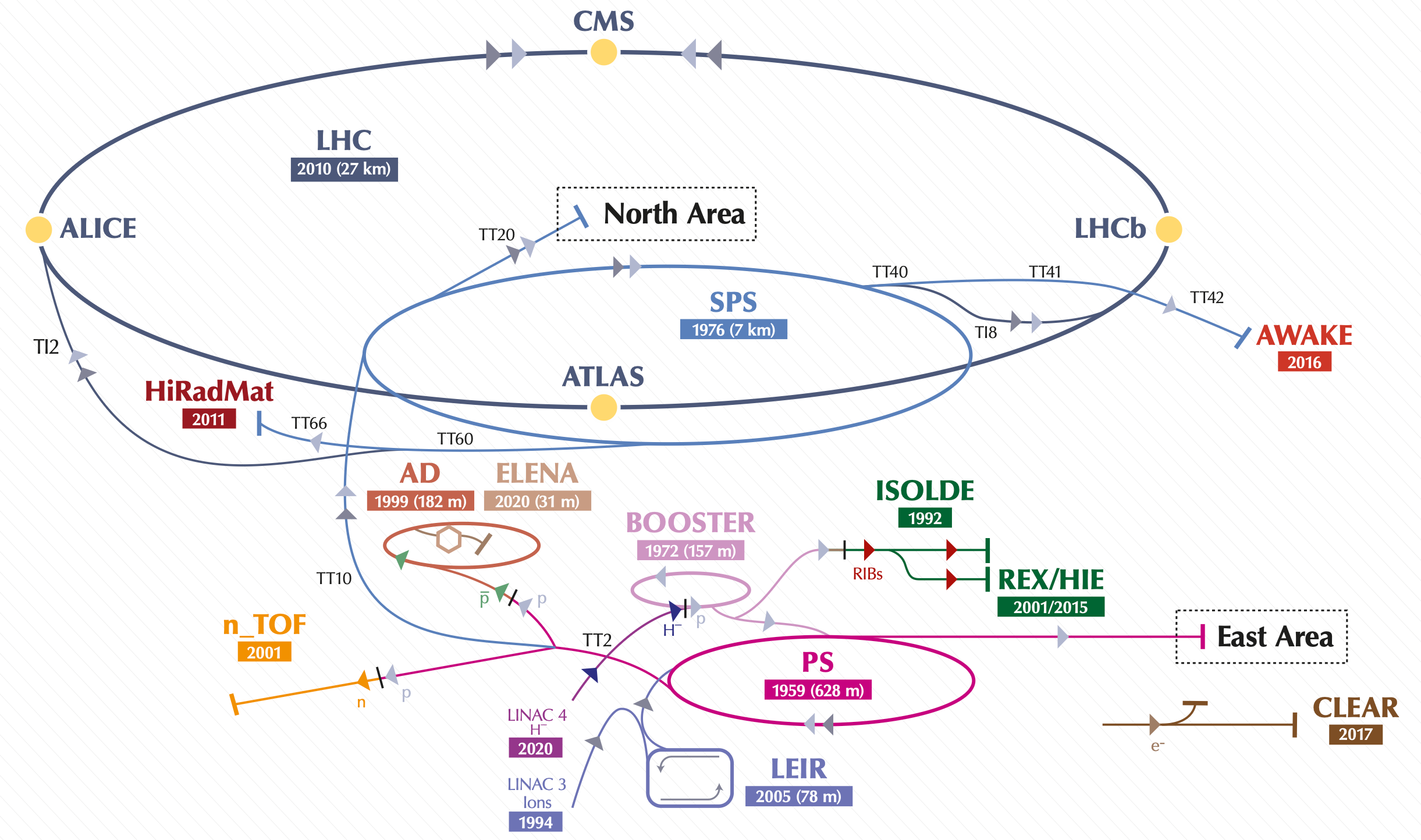}
    \captionsetup{type=figure}
    \captionof{figure}{The Large Hadron Collider CERN accelerator complex.}
    \label{fig:CERN_acc_complex}
\end{center}

\paragraph{The LHCb Experiment} 
\par\noindent
The LHCb experiment is one of the four main experiments at the LHC, located at the Interaction Point 8. The beauty production at LHC proceeds mainly through $b \overline{b}$ pair production, predominantly governed by the gluon fusion processes. The final state particles are therefore mostly distributed at large rapidities as illustrated in Figure \ref{fig:bbbar_production_LHCb}, justifying the geometrical acceptance of the experiment.  The LHCb detector is therefore a single-arm forward spectrometer, covering angles from 10-300 mrad in the horizontal plane and 10-250 mrad in the vertical plane (which corresponds to a pseudorapidity acceptance 2 $< \eta <$5).   

\begin{center}
    \centering
    \includegraphics[width=0.7\textwidth]{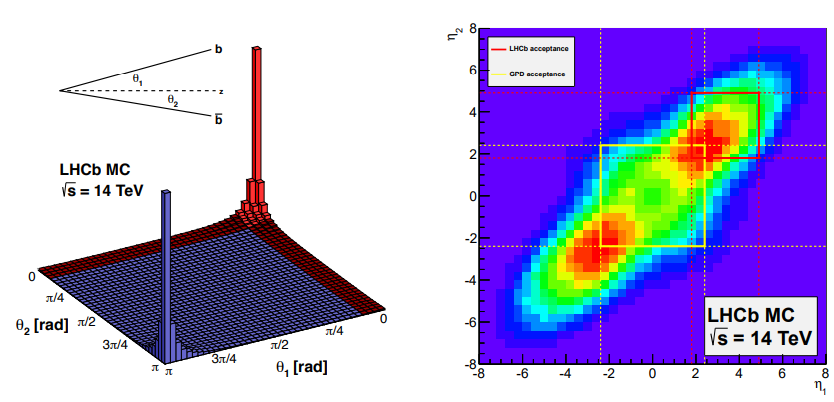}
    \captionsetup{type=figure}
\captionof{figure}{(Left)  Polar angle distribution of PYTHIA simulated $b\bar{b}$ pairs production at a center of mass energy $\sqrt{s} = 14$ TeV. (Right) Pseudorapidity distribution for the same events. The red region in both distributions represents  the LHCb detector acceptance while the blue and yellow parts displays ATLAS and CMS coverages.}
\label{fig:bbbar_production_LHCb}
\end{center}


The LHCb experiment and its triggering and operation during the Run I and II of the LHC were designed to cope with the overwhelming background accompanying the beauty production. First, the luminosity delivered by the LHC is levelled such that an average of about one interaction per bunch crossing at the interaction point is obtained. Second, a flexible trigger system, consisting of hardware and software stages, selects the events of interest for further analysis at a rate of 12.5 kHz (5 kHz) for Run II (Run I).  The first-level trigger is a hardware trigger implemented in the front-end electronics of the detectors. It selects each 25 ns events containing particles with high transverse energy or transverse momentum, reducing the event rate from 40 MHz to 1 MHz and enabling the full detector readout.  This first-level trigger decision is a combination of information recorded by the calorimeters and the muon chambers. The selected events are then passed to the software trigger, which makes use of all LHCb subsystems to perform heavy flavor selections based on the impact parameters of the tracks, secondary vertex displacements, and their kinematical and particle identification properties.   

\begin{figure}
\begin{center}
    \includegraphics[width=0.7\textwidth]{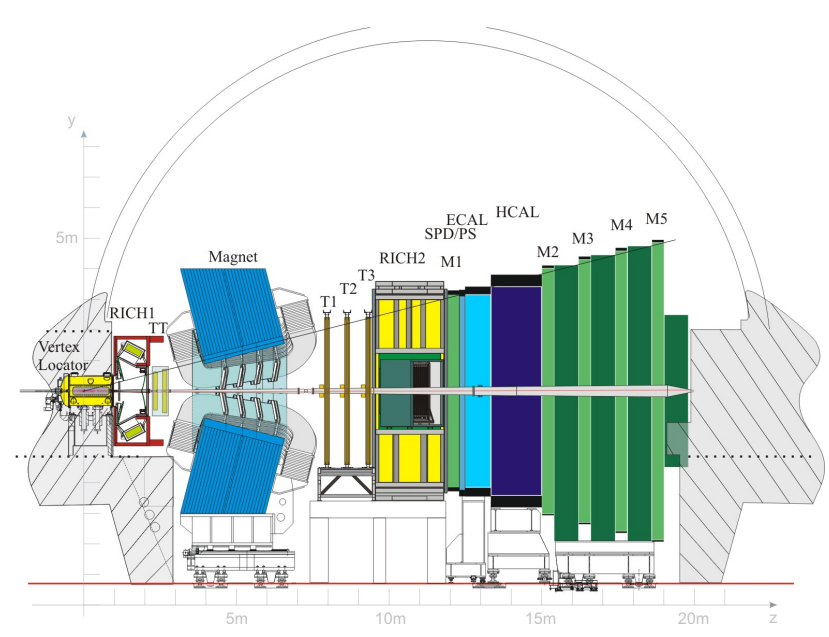}
    \captionsetup{type=figure}
\captionof{figure}{Side view of the LHCb detector design}
\label{fig:LHCb_detector_layout}
\end{center}
\end{figure}

The layout of the LHCb spectrometer during the LHC Run 2 operation is displayed  in Figure ~\ref{fig:LHCb_detector_layout}.  A detailed description is available in ~\cite{LHCb:2008vvz}. Let's summarise here its key features that allow for an accurate measurement of the $b$-flavored baryon decays to be performed.  The dipole magnet (4 Tm) deflects charged particles on the horizontal plane to accurately measure their momenta in a trajectography system. The tracking detectors comprise the VErtex LOcator (VELO) surrounding the interaction region and providing precise displacements of tracks from the primary vertex, a tracking station upstream of the magnet (TT), and tracking stations downstream of the magnet (T stations). The magnet field polarity is periodically switched to collect similar amount of data of each polarity to determine and correct for the experimental detection asymmetries of charged particles. Another fundamental detection system consists of a couple of Ring Imaging Cherenkov detectors, RICH1 and RICH2, providing charged hadron ($p, \, K, \,\pi$) discrimination over a wide range of momenta.  The calorimeter system consists of a Scintillating Pad Detector (SPD), a Preshower Detector (PS), a shashlik-type Electromagnetic Calorimeter (ECAL) followed by a Hadronic Calorimeter (HCAL). The ECAL measures accurately the photons and the electrons and the calorimeter system in its whole serves the selection of high transverse energy electromagnetic and hadron particles at the first-level of trigger. The muon detection system,  composed of five stations, equipped with multi-wire proportional chambers, completes the detector trigger at first-level and provides muon identification information.  

\begin{figure}
\begin{center}
    \includegraphics[width=0.5\textwidth]{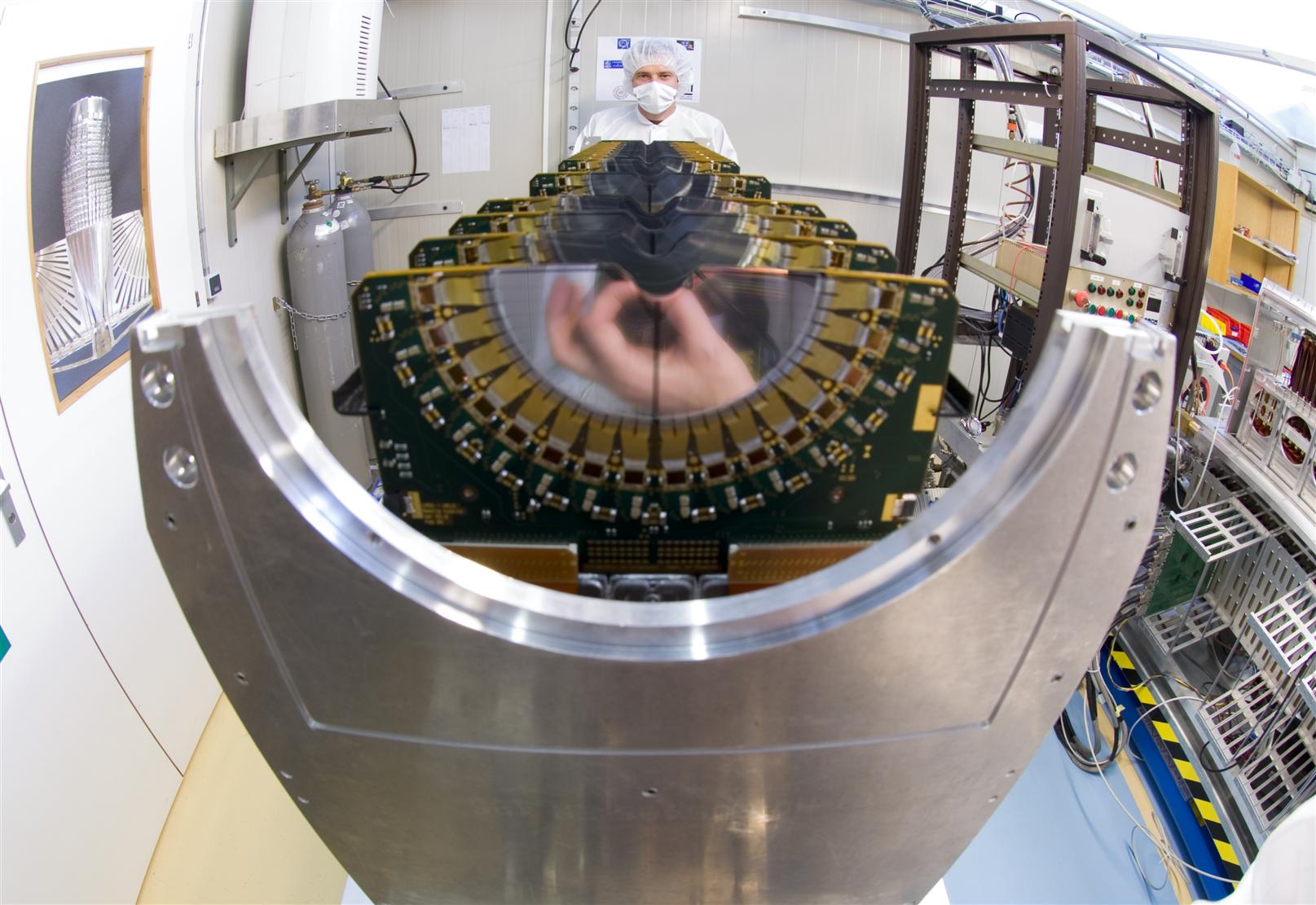}
    \captionsetup{type=figure}
\captionof{figure}{Photograph of one half of the VELO detector. Each half is composed of 21 modules placed perpendicular to the beam direction. Each module is equipped with two sensors mounted back-to-back: a sensor made of concentric silicon strip provides a measurement of the radial coordinate from the beam axis and the oppositely-mounted sensor measures the azimuthal angle of the hit. The last coordinate is given by the position of the station on the beamline.}
\label{fig:VELO}
\end{center}
\end{figure}

World-class measurements are and have been obtained with the LHCb spectrometer, while precision physics at a hadron machine was not taken for granted. The critical  explanation of this experimental success is the capability of the detector to measure the proton interaction vertices at each bunch crossing and single out particles that are displaced from the interaction point benefiting of the large boost experienced by the short-lived charmed and beautiful particles. It therefore selects the relevant heavy flavor physics among the immense background produced in proton-proton collisions. The VELO detector, displayed  in Figure \ref{fig:VELO}, is the key instrument in this endeavour and deserves a specific comment. It is constituted of a series of silicon modules, separated into two independent retractable halves surrounding the interaction point. The VELO is closed once stable beams are declared.  Its sensors reach a distance of 8 mm apart from the beam pipe, allowing for an accurate determination of all interaction vertices. 

\begin{figure}
\begin{center}
 \includegraphics[width=0.5\linewidth]{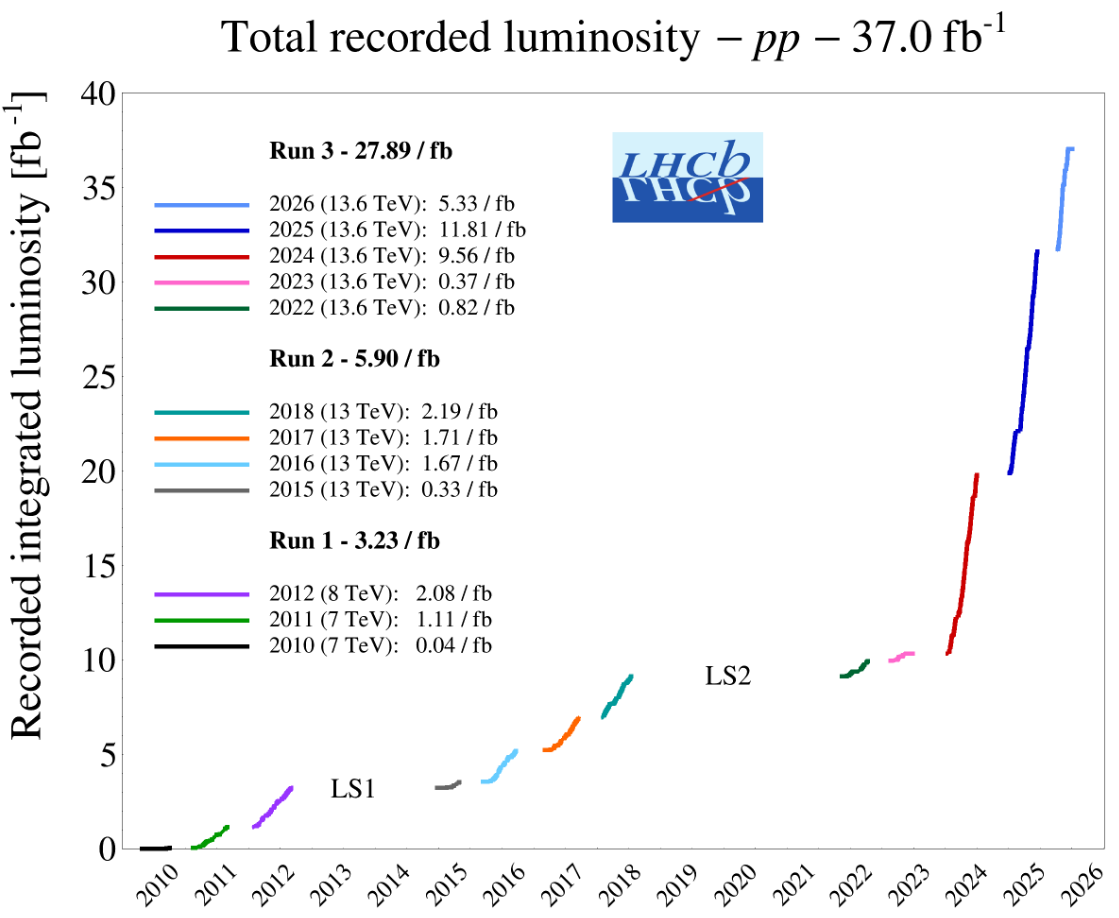}
 \captionsetup{type=figure}
    \captionof{figure}{Total integrated luminosity recorded by the LHCb experiment during proton-proton data taking periods~\cite{LHCbOperationsPlots}.}
    \label{fig:Lumi_LHCb}
\end{center}
\end{figure}

The amount of data collected by the LHCb experiment is reported in Fig.~\ref{fig:Lumi_LHCb}. This integrated luminosity figure of merit gathers the proton-proton collisions in Run I ($3.23~\mathrm{fb}^{-1}$), Run II ($5.90~\mathrm{fb}^{-1}$), and Run III ($27.89~\mathrm{fb}^{-1}$). The data recorded in the latter run were using the upgraded LHCb spectrometer featuring a new tracking system consisting of a pixel-based VELO and two new tracking detectors, the Upstream Tracker  and the Scintillating Fibre tracker, in place of TT and the three tracking Stations, respectively~\cite{LHCb:2012doh}. In addition, the LHCb spectrometer was operated with a nominal instantaneous luminosity five-times higher than during Run II and the detector was read-out at 40 MHz, with a flexible software-based trigger~\cite{LHCb:2014dlo}. The fully hadronic decays of $b$-flavored hadrons will particularly benefit from these improvements. The data samples discussed in this review are  multiplied by about a factor of 6 in the current data on tape!     

\subsection{Review of Baryonic $C\!P$ Violation Searches before First Observation}

Various experimental techniques and observables have been mobilised to search for $CP$-symmetry breaking in baryon decays: 
\begin{itemize} 
\item measurement of decay width asymmetries ($A_{C\!P}$), evaluating directly the decay rate difference between a baryon $X_b$ decay ($X_b \to f$) and its $C\!P$-conjugate decay ($\bar{X_b} \to \bar{f}$). 
\item the Triple-Product Asymmetry (TPA), utilizing in multi-body (3- or 4-body) decays the kinematical cross-products~\cite{Kayser:1989vw}; this observable is sensitive to $C\!P$-odd amplitudes \cite{Bensalem:2002ys, Durieux:2015zwa} and presents the property to factor out most of the experimental asymmetries.   
\item Energy-Test observable to highlight localised asymmetries in the phase-space of multibody decays~\cite{zech2003newtestmultivariatetwosample, Barter:2018xbc}. This does not require the a priori knowledge of the actual amplitude model of these decays.   
\end{itemize}

\subsubsection{Search for $C\!P$ violation in two-body decays}

\begin{figure}
\begin{center}
\includegraphics[width=0.4\textwidth]{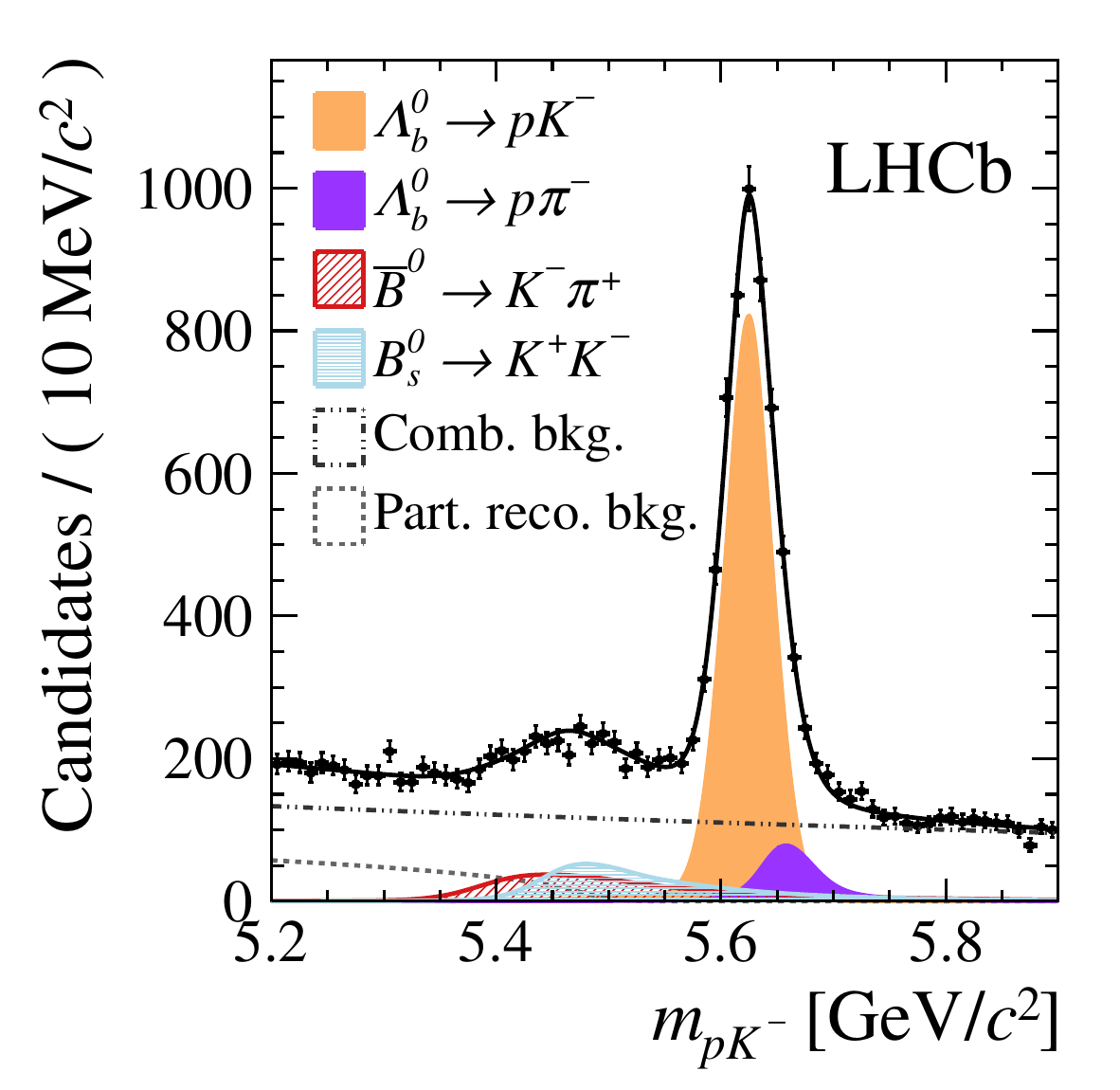}
\includegraphics[width=0.4\textwidth]{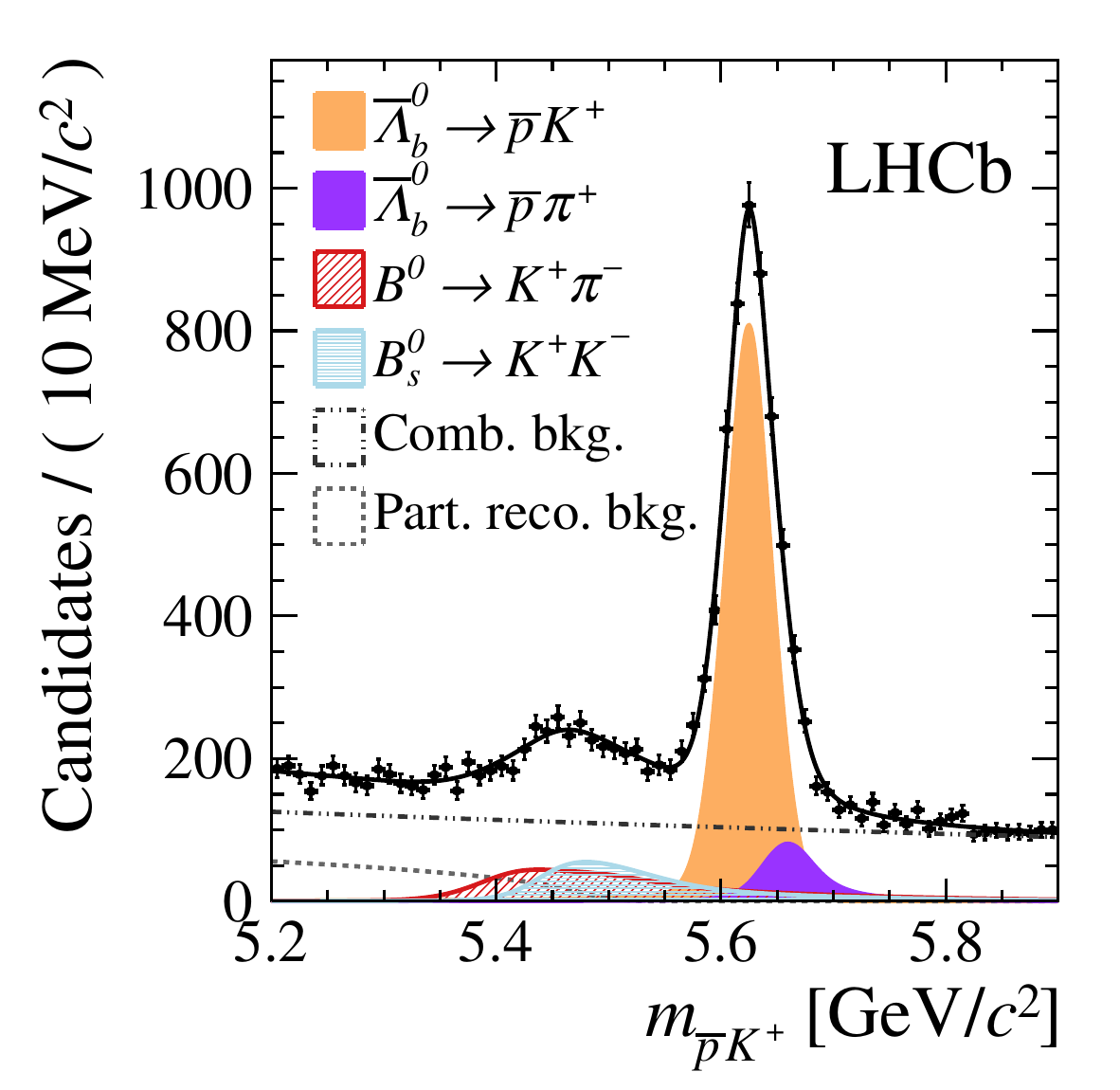}\\
\includegraphics[width=0.4\textwidth]{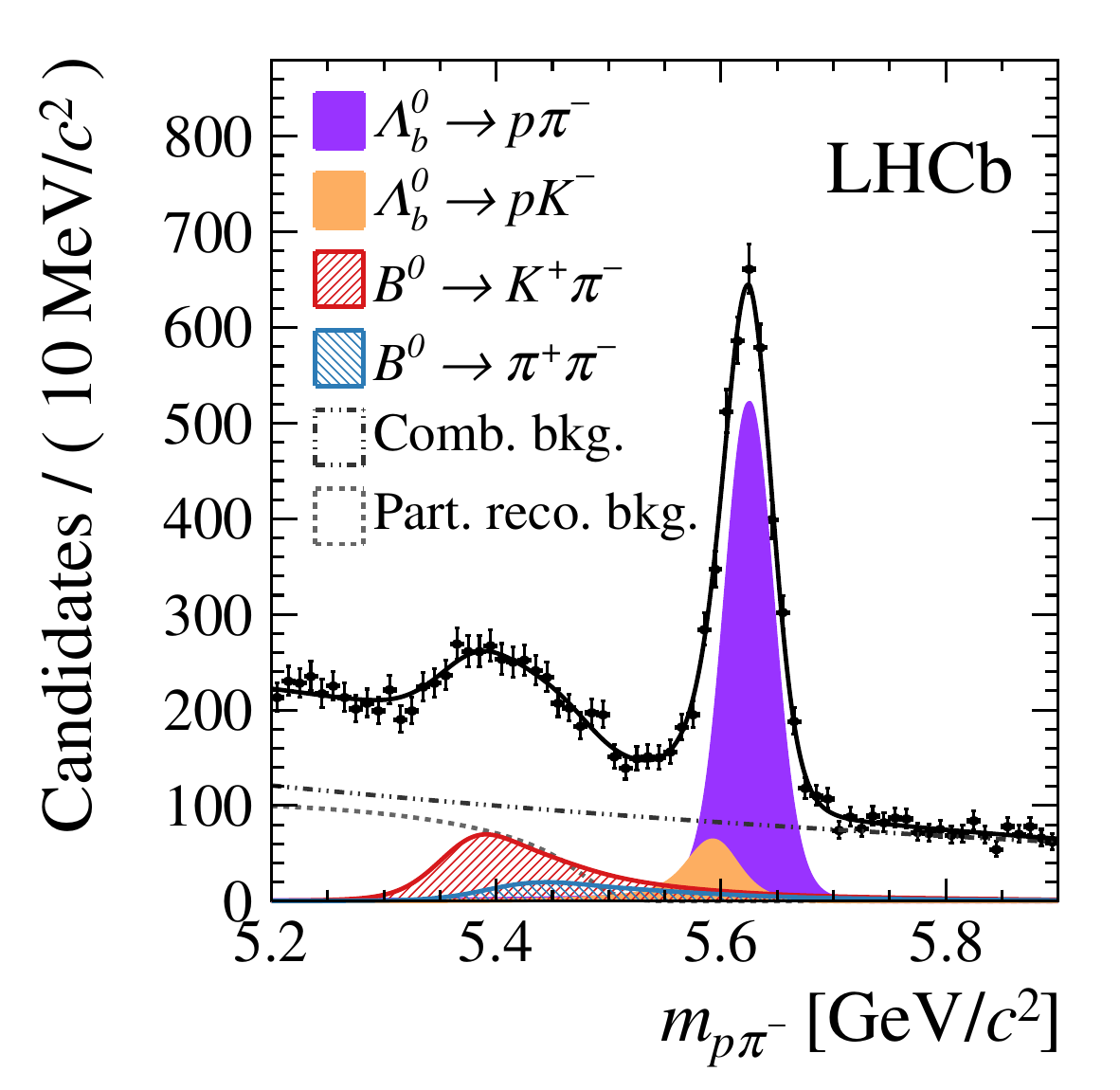}
\includegraphics[width=0.4\textwidth]{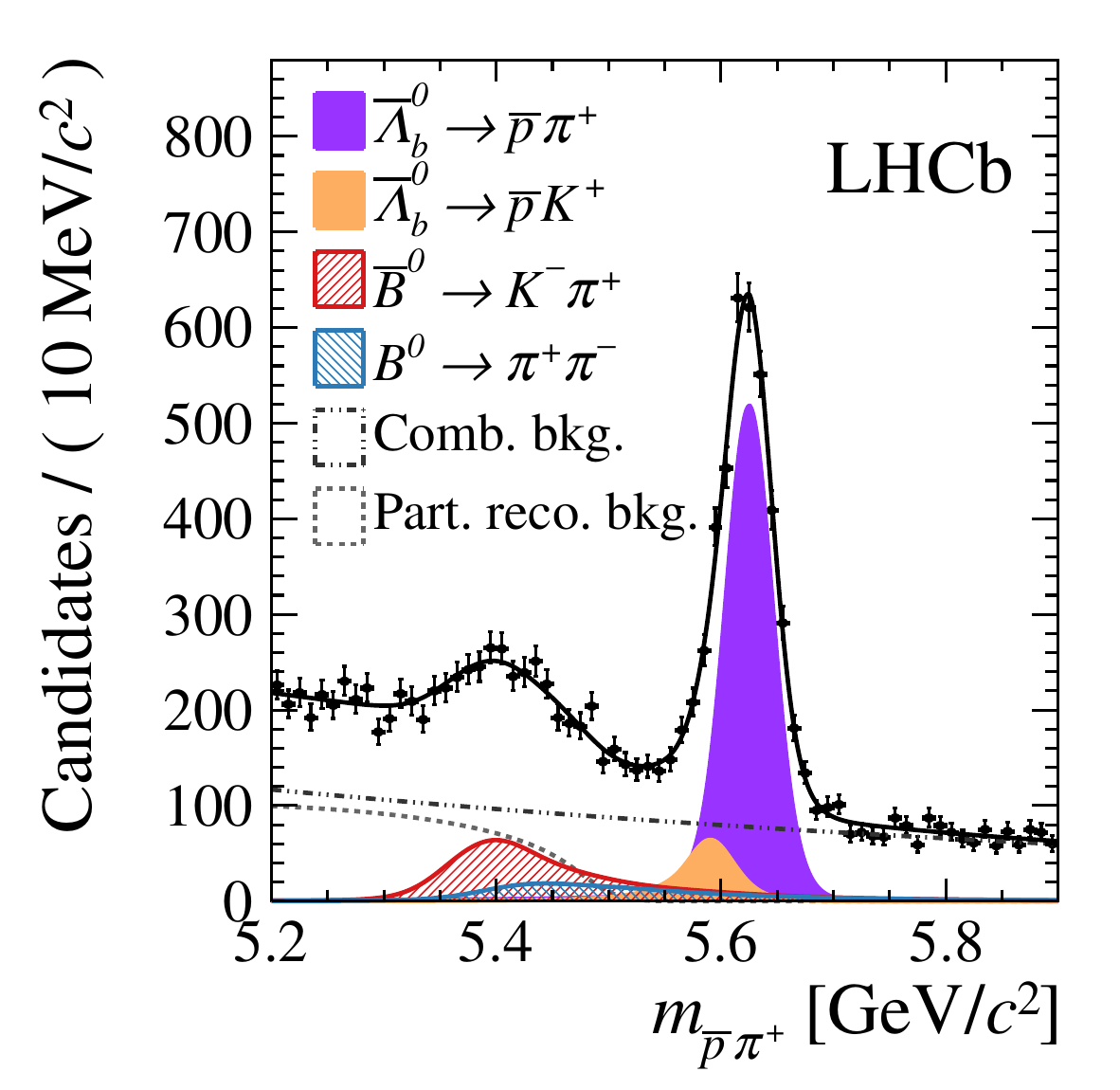}\\
\captionsetup{type=figure}
    \captionof{figure}{Invariant-mass distributions: (top left) $m_{pK^-}$, (top right) $m_{\overline{p}K^+}$, (bottom left) $m_{p \pi^-}$ and (bottom right) $m_{\overline{p}\pi^+}$ for candidates passing the selections. The results of the fits are superimposed.}
    \label{fig:Lb_2body}
\end{center}
\end{figure}

The charmless decays $\Lb \to p K^- (\pi^-)$ proceeds with tree-level transitions $b \to u \bar u d(s)$ and loop-level transitions $b \to s(d) q \bar q$ where the $q \bar q$ pair with $q={u,d,s}$, is produced by in the gluonic penguin. Weak and strong phases are present and can produce a $C\!P$ asymmetry. These modes are analogous to the meson decays $B \to K \pi$ where large asymmetries ${\cal O}(10\%)$ were measured at B-factories~\cite{BaBar:2004gyj, Belle:2004nch}. The LHCb experiment analysed the Run I data~\cite{LHCb:2018fly} and selected about 10 (6) thousands of $\Lb \to p K^-$ ($\Lb \to p \pi^-$) decays.  
Invariant mass distributions are displayed in Fig.~\ref{fig:Lb_2body}.

The $C\!P$ asymmetry as the rate of the corrected fitted yields are measured to be: 
\begin{equation}
A_{C\!P}(pK^-)= -0.020 \pm 0.013 \pm 0.019, \\
A_{C\!P}(p\pi^-)= -0.035 \pm 0.017 \pm 0.020.
\end{equation} 

It is remarkable that these two results are dominated by systematic uncertainties, originating mostly from the limited knowledge of the production and charge-difference detection asymmetries.  The difference in $C\!P$ asymmetries between the two decay modes is as well considered. This difference is to first order immune to systematic uncertainties, and reads $\Delta A_{C\!P} = A_{C\!P}(pK^-) - A_{C\!P}(p\pi^-)= 0.014 \pm 0.022 \pm 0.010$, showing no evidence of $C\!P$ violation.  

The data collected by the LHCb experiment during the LHC Run II are used to update these measurements. Significant improvements in the evaluation of systematic uncertainties are achieved and the combination of the two samples yielded the most precise determination of these $C\!P$ asymmetries to date, indicating no further evidence of $C\!P$ violation, 
\begin{equation}
A_{C\!P}(pK^-)= -0.011 \pm 0.007 \pm 0.004, \\
A_{C\!P}(p\pi^-)= -0.002 \pm 0.008 \pm 0.004.
\end{equation} 

\subsubsection{Search for $C\!P$ violation in three-body decays}

Charmless three-body $b$-baryon decays offer a rich phenomenology to search for $C\!P$ violation. Fully charged modes $\Xi_b^- \to pK^-K^-$ is explored and benefits from powerful particle identification selection. Decay modes with a $\Lambda$ or a $K^0_{\rm S}$ in the final state are as well studied, though subjected to a lower reconstruction efficiency of the neutral short-lived particles. $C\!P$-conjugated amplitudes are deciphered by means of a Dalitz plot analysis.   

\begin{figure}
\begin{center}
\includegraphics[width=0.4\textwidth]{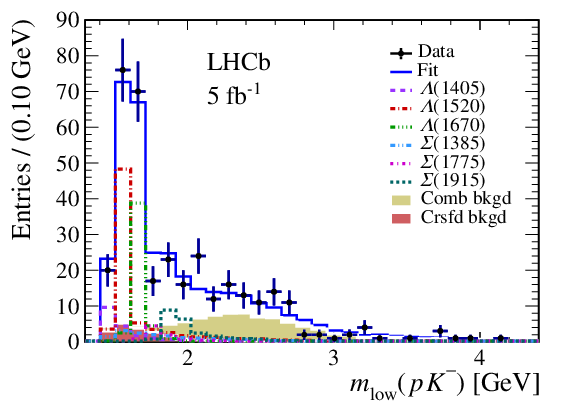} 
\includegraphics[width=0.4\textwidth]{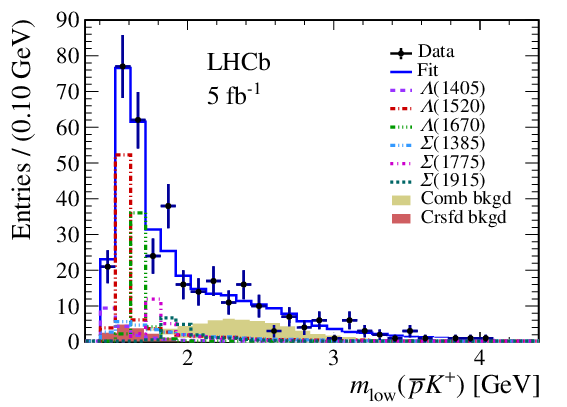}\\
\captionsetup{type=figure}
    \captionof{figure}{Invariant-mass distributions: (left) the smallest invariant mass $m_{pK^-}$ among the two possible combinations of negatively charged kaons, (right) $m_{\overline{p}K^+}$. A good description of the Dalitz plot is obtained. }
    \label{fig:Xib_3body}
\end{center}
\end{figure}

A good description of the phase-space distribution of the $\Xi_b^- \to pK^-K^-$ decay ~\cite{LHCb:2021enr} is obtained with an amplitude model containing contributions from a number of the hyperon-resonance series $\Sigma$ and $\Lambda$, namely $\Sigma(1385)^0$, $\Sigma(1775)$, $\Sigma(1950)$ , $\Lambda(1405)^0$, $\Lambda(1520)^0$, $\Lambda(1670)^0$. The Dalitz plot fit is shown on the Fig.~\ref{fig:Xib_3body}  with projections onto $(pK^-)$ and $(\overline{p}K^+)$ invariant masses axes. The fit fractions are determined and used in turn to derive the $C\!P$ asymmetries for each contributing resonance. No significant asymmetry is observed. The dominant amplitude $\Xi_b^- \to \Lambda(1520)^0 K^-$ yields $A_{C\!P}(\Lambda(1520)^0 K^-) = -0.05 \pm 0.09 \pm 0.08$.       

The LHCb data collected during the LHC Run I and II are analysed to search for the decay modes $\Lb \to \Lambda (\to p \pi^-) K^+K^-$, $\Lb \to \Lambda K^+\pi^-$ and $\Lb \to p K^0_{\rm S} \pi^-$. About 200 (100) signal decays are observed in the Run I data sample for $\Lb \to \Lambda K^+K^-$ ($\Lb \to \Lambda K^+\pi^-$). The decay mode $\Lb \to p K^0_{\rm S} \pi^-$ was initially studied on a fraction of the Run I data and a yield of about 100 events were observed as well. Though measurements of $C\!P$ asymmetries were performed, the statistics at hand was too scarce to be conclusive. Analyses including the data collected during the Run II are gathering larger yields~\cite{LHCb:2024yzj} and allows for a scrutiny a Dalitz plot of these decays to be performed.  

\begin{figure}
\begin{center}
\includegraphics[width=0.4\textwidth]{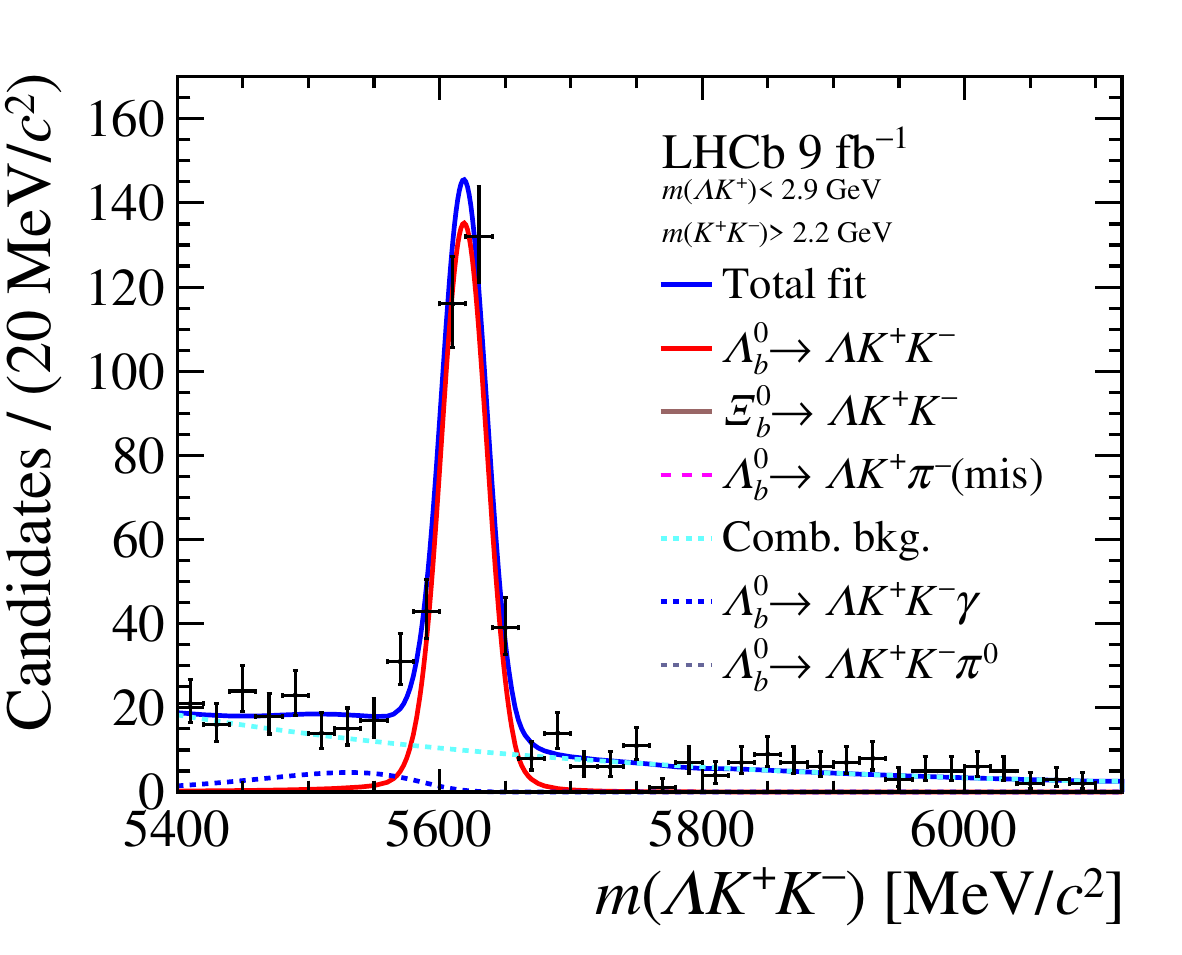} 
\includegraphics[width=0.4\textwidth]{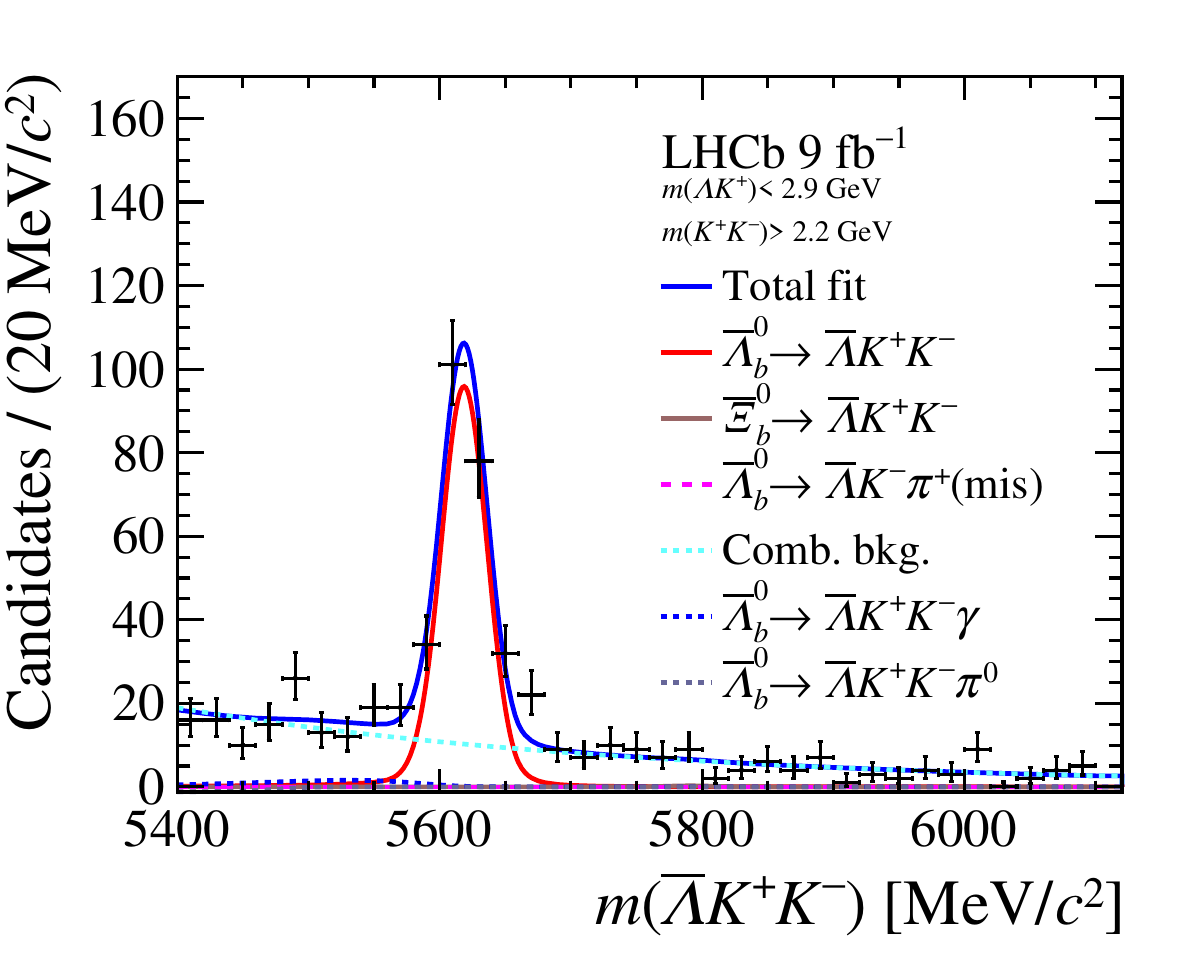}\\
\captionsetup{type=figure}
    \captionof{figure}{Invariant-mass distributions of (left) $\Lb \to \Lambda K^+K^-$ and (right) $\overline{\Lambda}_b^0 \to \overline{\Lambda} K^+K^-$ in $N^{*+}$ resonance-dominated regions. Fit results are shown as well and superimposed.}
    \label{fig:Lb_Lhh}
\end{center}
\end{figure}

The $C\!P$ asymmetries for the decay modes $\Lb \to \Lambda K^+K^-$ and $\Lb \to \Lambda K^+\pi^-$ are measured in the full and localised regions of the final-state phase space. Evidence is found for $C\!P$ violation in the $\Lb \to \Lambda K^+K^-$ channel and is  interpreted as originating primarily from an asymmetric $\Lb \to N^{*+}K^-$ decay amplitude. The distributions of invariant masses reconstructed per charge in the corresponding localised final-state phase space bin are displayed in Figure~\ref{fig:Lb_Lhh}. $C\!P$ asymmetries are measured relative to the CKM-favoured (tree-dominated and $C\!P$-conserving in the SM) decay $\Lb \to \Lambda_c^+ K^-$ as a control mode to cancel the production and most of the detection charge asymmetries. 

The difference of $C\!P$ asymmetries in the decay $\Lb \to \Lambda K^+K^-$ relative to the control channel $\Lb \to \Lambda_c^+ K^-$ is measured to be $\Delta A_{C\!P}(\Lambda K^+ K^-) = A_{C\!P}(\Lb \to \Lambda K^+ K^-) - A_{C\!P}(\Lb \to \Lambda_c^+ K^-) = 0.165  \pm 0.051$.  The $C\!P$ asymmetries obtained for the other decays are compatible with zero.

With a similar methodology, the decay $\Lb \to p K^0_{\rm S} \pi^-$ is studied with the sample of signal candidates collected by the LHCb experiment during the Run II. The $C\!P$ asymmetries are measured in different regions of the Dalitz plot but in contrast with the mode with $\Lambda$, no evidence of $C\!P$ violation is observed.     

\subsubsection{Search for $C\!P$ violation in multi-body decays of $b$-flavored baryons} 

\begin{figure}
\begin{center}
\includegraphics[width=0.4\textwidth]{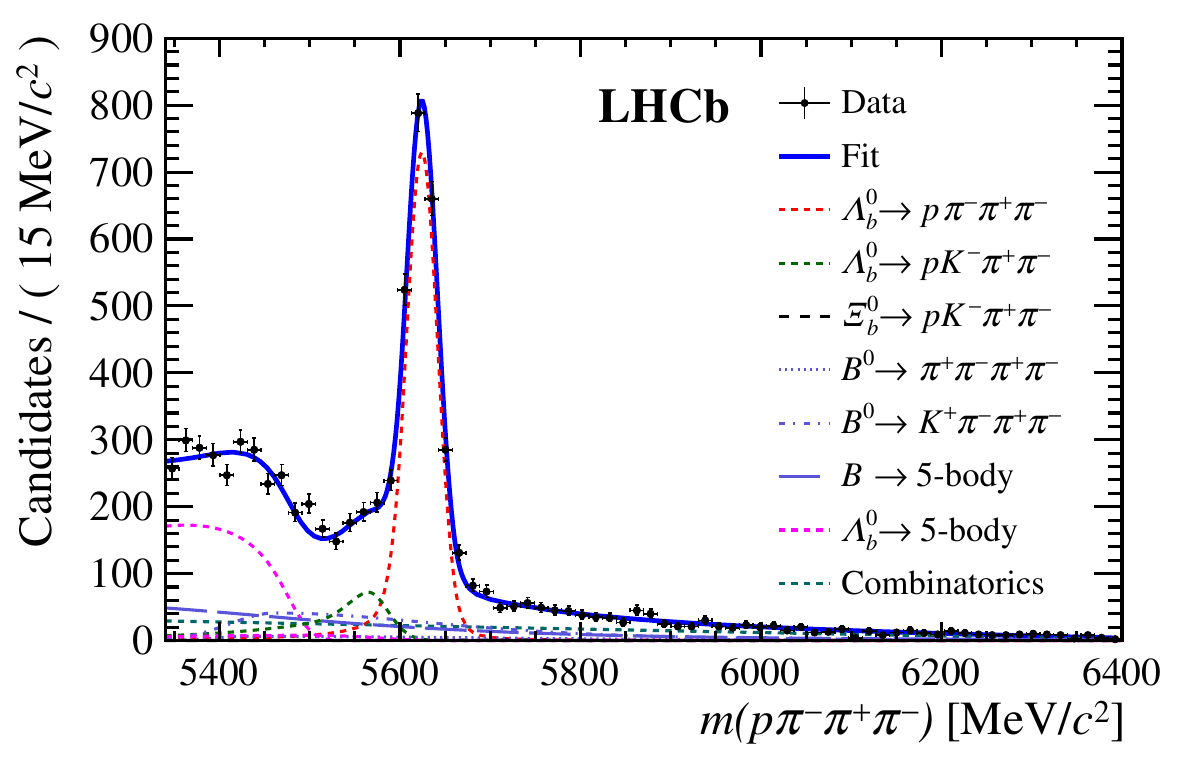}
\includegraphics[width=0.4\textwidth]{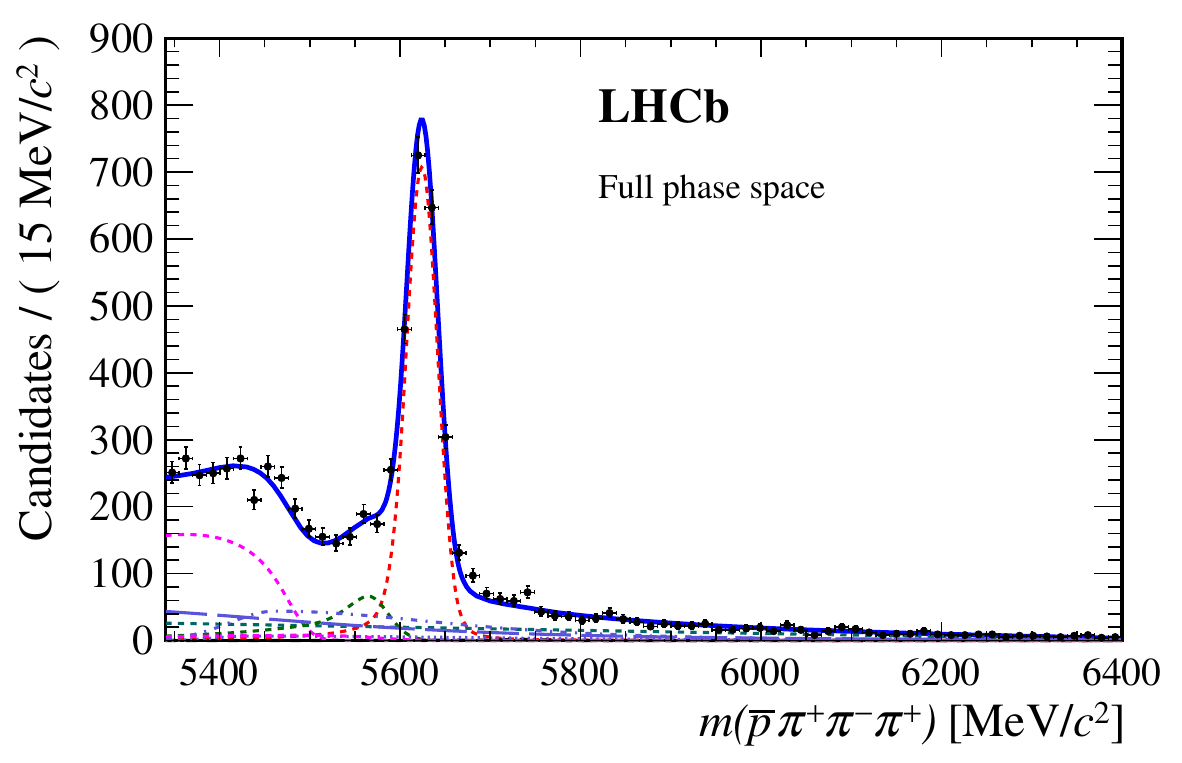}\\
\includegraphics[width=0.4\textwidth]{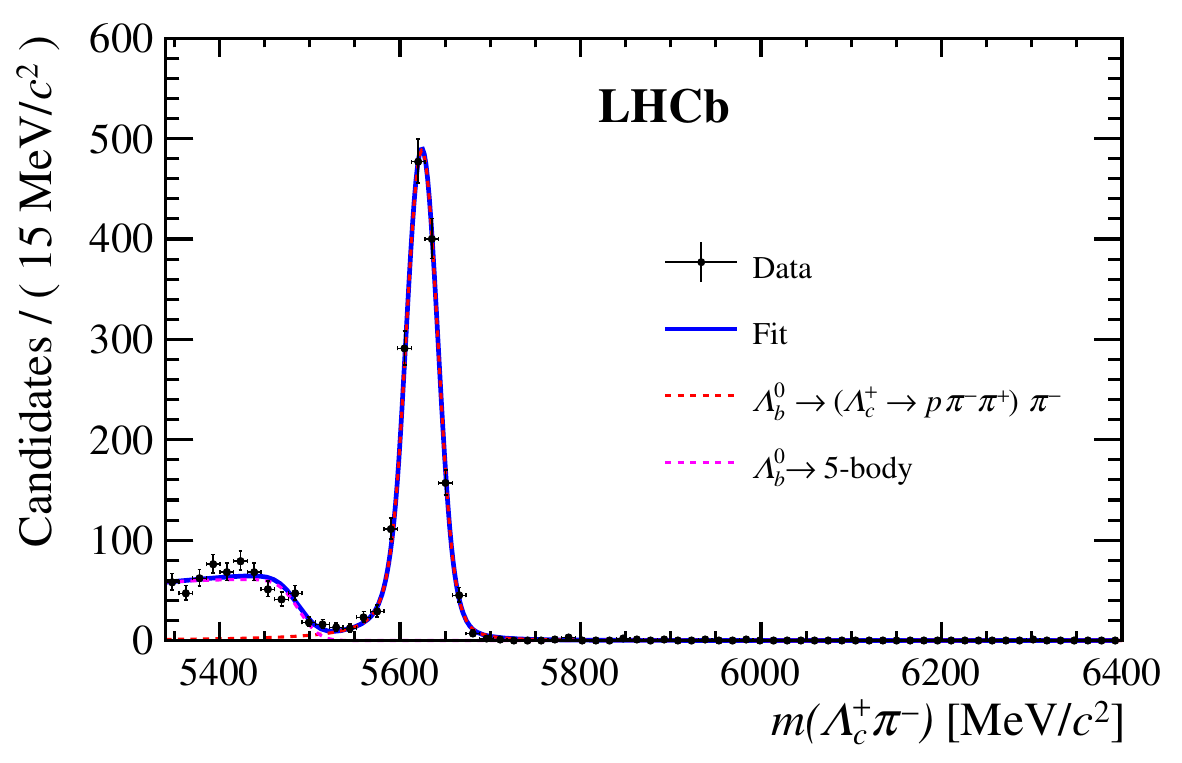}
\includegraphics[width=0.4\textwidth]{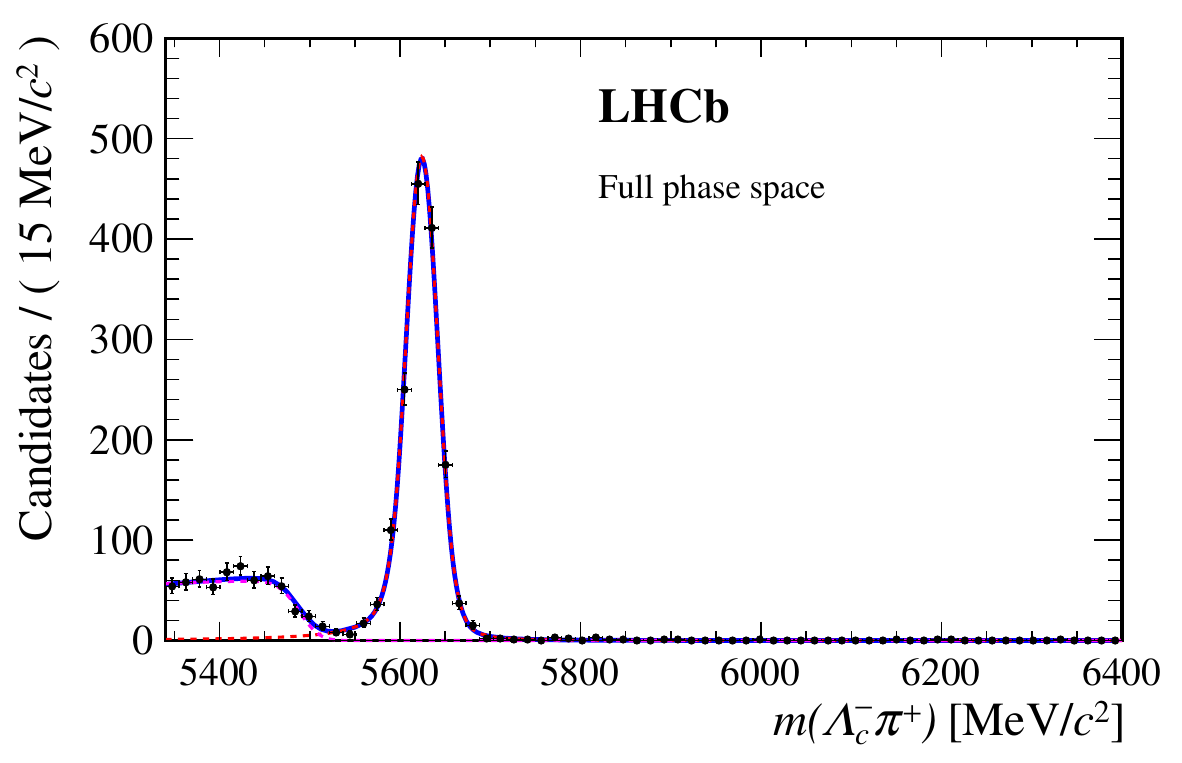}

\captionsetup{type=figure}
\captionof{figure}{ (top left) Invariant mass distribution of $\Lb\to p \pi^-\pi^+\pi^-$ and (top right) $\Lbr\to \overline{p}^-\pi^+\pi^+\pi^-$ decays, with fit projections superimposed. The control channels $\Lbr\to \Lambda_c^- \pi^+$ (bottom left) and (bottom right) $\Lb\to \Lambda_c^+ \pi^-$. Reproduced from~\cite{LHCb:2019jyj}.}
\label{fig:Lb2p3pi}
\end{center}
\end{figure}

The dataset collected during the LHC Run I for four-body fully charged decays of $\Lb$ and $\Xi_b^0$ baryons is comprehensively analysed by means of $\Delta A_{C\!P}$ technique featuring tree-level-dominated control modes with the same final state particles to cancel out the production and detection asymmetries. The analysis technique employed in this study~\cite{LHCb:2019jyj} will be fully developed in the chapter of this review  dedicated to the discovery. All possible charged hadron combinations are considered, resulting in six distinct experimental spectra. As an illustration, the invariant-mass distributions and fit results used to obtain raw yields of the $\Lb \to p \pi^- \pi^+ \pi^-$  and the corresponding charmed decay calibration channel $\Lb \to \Lambda_c^+ (\to p \pi^+\pi^-) \pi^-$  are shown in Figure \ref{fig:Lb2p3pi}. 

This specific mode proceeds via commensurate tree- and loop-diagram amplitudes in the SM.  In total, eighteen $C\!P$ asymmetries, both global and localised in the phase-space of the decays, are measured for $\Lb$ and $\Xi_b^0$ baryons. All measurements are compatible with $C\!P$ symmetry, ruling out $C\!P$ asymmetries as large as 10\%.  A hint for $C\!P$ violation at 2.7 standard deviation is however obtained for the decay mode $\Lb \to p K^- \pi^+ \pi^-$, justifying its specific study with the LHCb Run II data samples, detailed in the next section.  

The triple-Product Asymmetry (TPA) and the Energy Test are employed as alternative and complementary methods to study these fully-charged modes in general~\cite{LHCb:2018fpt} and singularly the specific decay $\Lb \to p \pi^-\pi^+\pi^-$~\cite{LHCb:2019oke}.  A total of about 27,000 signal decays is obtained and the inclusive $C\!P$ asymmetry measured using the TPA method was $a^{{\hat T}-\rm odd}_{C\!P}= (-0.7 \pm 0.7 \pm 0.2)\%$. A scrutiny of this observable is also performed in various regions of the phase-space. The method of the Energy Test is also considered, with the relevant distance observable built from final state two-body and three-body invariant masses. No indication of $C\!P$ violation is found in neither methods. Interestingly enough, parity violation in this weak hadronic decay is unraveled with both methods.    

\subsubsection{Other results and perspectives} 

The breadth of analyses of $C\!P$ violation in baryon decays with the datasets recorded by the LHCb experiment is wide and promising. This review of a selection of measurements preceding the discovery of $C\!P$ violation in baryon decays does not exhaust all searches undertaken with the first two runs of LHC. $C\!P$ asymmetries in flavor-Changing Neutral-Current (FCNC) $b \to s,d$ baryon decay transitions, mediated in the SM via electroweak loop diagrams, are studied with radiative ~\cite{LHCb:2021byf} or semileptonic final states~\cite{LHCb:2017slr,LHCb:2024ick}. Charmed baryon decays (see e.g. ~\cite{LHCb:2020zkk} or $b$-flavored baryon transitions with a charmed state in the final state (to determine the CKM $\gamma$ angle~\cite{LHCb:2021ohr}) are also be measured. These results are consistent with $C\!P$ symmetry but the analysis of the full sample of the dataset acquired during the LHC Run III opens a new era of precise explorations, benefiting from the efficient and flexible software trigger together with the upgraded tracking system.

%% file: sec5-discovery.tex
\subsection{Landmark Discovery of $\CP$ Violation in Baryons}

The first $\CP$ violation in baryon decays is observed by the LHCb experiment, corresponding to a difference between the rates of the four-body decay $\LbtopKpipi$ and its charge-conjugate process $\Lbr\to \overline{p}K^+\pi^-\pi^+$~\cite{LHCb:2025ray}. 
The measurement is an update of the previous analysis using the full proton–proton collision data collected during Run I (2011–2012) and Run II (2015–2018), corresponding to an integrated luminosity of about $9~\mathrm{fb}^{-1}$. 
The global $\CP$ asymmetry, averaged over the full phase space, is determined to be $A_{\CP}=(2.45\pm0.46\pm0.10)\%$, which differs from zero by $5.2$ standard deviations.
$\CP$ asymmetries are also measured in four localized phase-space regions defined according to various hadronic resonances $R$. The most significant local asymmetry is found in the region dominated by excited nucleons in the $p\pi^+\pi^-$ invariant-mass spectrum, corresponding to the $\Lb\to R(p\pi^+\pi^-)K^-$ decay topology, with $A_{\CP}=(5.4\pm0.9\pm0.1)\%$, differing from zero by $6.0$ standard deviations. These results provide the first experimental demonstration that baryons and antibaryons can behave differently in the interplay of weak and strong decays, similar to what is observed in mesons.

The decay $\LbtopKpipi$ has its own advantages in overcoming the experimental challenges in the search for $\CP$ violation. First, with quark content $bud$, the $\Lb$ baryon is the most frequently produced $b$-flavored baryon. In fact, its production cross section in $pp$ collisions at LHC energies is about half that of the $B^0$ meson~\cite{LHCb:2014ofc}. The branching fraction of $\LbtopKpipi$ is measured to be about $5\times10^{-5}$~\cite{LHCb:2017gjj}, the largest among all observed charmless $\Lb$ decay modes. It is even larger than that of $B^0\to K^+\pi^-$, in which the first direct $\CP$ violation in $b$-hadron decays was observed~\cite{BaBar:2004gyj,Belle:2004nch}. Moreover, since all final-state particles are charged, the decay is particularly well suited to reconstruction at LHCb. These features provide a relatively high-yield sample with good purity, enabling high-precision measurements.
This decay compares favorably with the $\Lb\to p K_\text{S}^0\pi^-$, $\Lb\to\Lambda h^+ h^-$ ($h=K$ or $\pi$), and $\Lb\to p K^-\pi^0$ decays, which have branching fractions of a similar order but much lower observed yields per unit luminosity owing to poorer experimental efficiency.

\begin{figure}
\begin{center}
\includegraphics[width=0.7\textwidth]{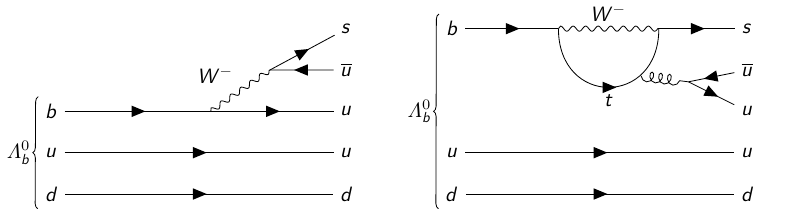}
\captionsetup{type=figure}
\captionof{figure}{Leading diagrams contributing to the $\LbtopKpipi$ decay. The light quarks from the $b$-quark decay and the spectator quarks hadronize into the final-state hadrons $pK^-\pi^+\pi^-$, with or without intermediate resonances. The left (right) diagram corresponds to the tree (loop) contribution. }\label{fig:FeynmanDigrams}
\end{center}
\end{figure}

At the quark level, the $\LbtopKpipi$ decay receives contributions from two main amplitudes: a tree-level $b\to su\bar{u}$ transition and a loop-level $b\to s$ transition. The leading diagrams for the decay are shown in Fig.~\ref{fig:FeynmanDigrams}.
The tree amplitude is proportional to the product of CKM matrix elements $V_{ub}V_{us}^*$, while the loop amplitude is proportional to $V_{tb}V_{ts}^*$. These two products of CKM matrix elements carry an irreducible relative weak phase. If the two amplitudes have comparable magnitudes and a nonzero relative strong-phase difference, a sizable $\CP$ asymmetry could arise. While the weak phases are completely determined by the CKM matrix, the strong phases are generated by final-state interactions.
The four-body final state of $\LbtopKpipi$ contains a large number of intermediate resonances, such as excited $\Lambda$ states, excited nucleon states $N^*$, excited $\Delta$ states, and $K^*$, $\rho$, and $f_0$ mesons. This rich resonance structure increases the probability of obtaining the strong-phase differences required for $\CP$ violation, in particular in localized phase-space regions of resonances. As a result, the $\CP$ asymmetry may vary significantly across the phase space of the decay owing to possible variations in the dynamics associated with these resonances. Local measurements in different resonant regions can provide information about the underlying dynamics that is not accessible from a single global average over the phase space.

Nevertheless, the measurement presents several serious challenges. The four-body dynamics are complicated, and a full amplitude analysis is technically demanding because of the many overlapping broad resonances whose properties were not precisely determined previously. The measured $\CP$ asymmetry is therefore an average over the selected phase-space region, or over a local region, and it cannot be directly interpreted in terms of a single underlying amplitude. In addition, the four-body decay is vulnerable to several peaking backgrounds. These sources must be subtracted to avoid bias in the physics interpretation of the results. They include the charmed background, in which the final-state particles are partially produced from charm hadrons. For example, the $\Lb\to p D^0\pi^-$ decay with $D^0\to K^-\pi^+$ has the same final state as the signal decay but is dominated by the $b\to c\overline{u}d$ process, and thus no $\CP$ violation is expected. Peaking backgrounds can also arise when at least one final-state particle is misidentified. For example, the $\overline{B}^0\to K^-\pi^+\pi^+\pi^-$ decay, with a $\pi^+$ misidentified as a $p$, can pollute the signal sample and may carry a large $\CP$ asymmetry.
Peaking backgrounds also arise from partially reconstructed decays, such as $\LbtopKpipi\pi^0$ with the $\pi^0$ not reconstructed. In addition, there is combinatorial background from random combinations of tracks; this is the dominant background given the huge hadronic cross section in $pp$ collisions relative to the signal decay.

\begin{figure}
\begin{center}
\includegraphics[width=0.5\textwidth]{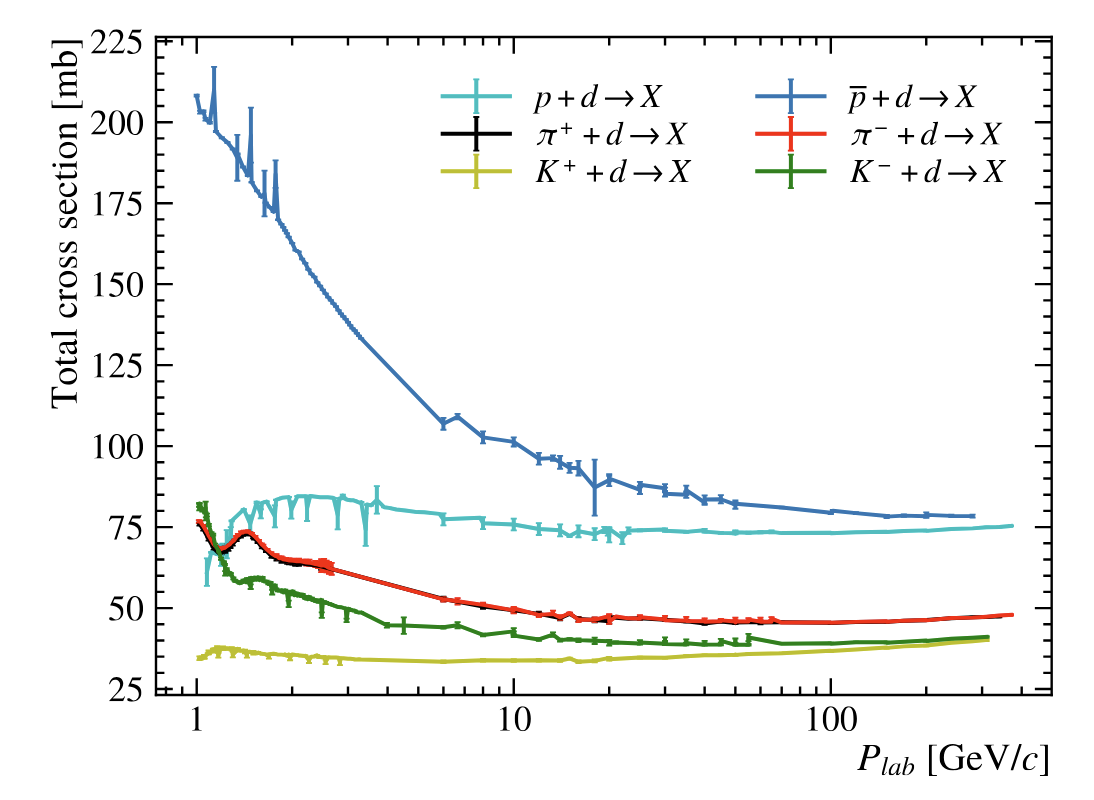}
\captionsetup{type=figure}
\captionof{figure}{ Total interaction cross section of charged hadrons impacting on a deuterium target. Reproduced from~\cite{ParticleDataGroup:2024cfk}.}
\label{fig:ParticleCrossSection}
\end{center}
\end{figure}

The detection asymmetry arises from different cross sections for particles and antiparticles interacting with the detector material. Fig.~\ref{fig:ParticleCrossSection} shows the total cross sections of protons, kaons, and pions, and their antiparticles, on a deuterium target. The valence $\overline{u}$ and $\overline{d}$ quarks inside the antiproton can annihilate with quarks in the detector material, which is not the case for the proton; as a result, the interaction cross section is higher for antiprotons. This is particular the case at low momentum, where the annihilation probability is higher. For kaons, the situation is similar but less pronounced. Since pions and the detector material contain almost equal numbers of valence $u$ and $d$ quarks, the difference between $\pi^+$ and $\pi^-$ interactions is negligible. A higher interaction rate implies a lower probability of traversing the entire LHCb tracking detector and hence a lower detection efficiency. The actual asymmetries between particles and antiparticles depend on the amount of material: more material leads to larger asymmetries. For LHCb Run I and Run II data, the proton asymmetry is measured to be about 2.5\% at momenta of a few GeV and decreases to 0.5\% above 100 GeV. For kaons, the asymmetry lies in the range 0.5\%--1\% from high to low momentum. For pions, the asymmetry is not significant: it is 0.3\% at low momentum and 0.1\% at high momentum. For the signal decay $\LbtopKpipi$, the proton detection asymmetry compensates that of the kaon, while those for $\pi^+$ and $\pi^-$ cancel each other.

Event selections used to reduce background or to perform the trigger also introduce artificial asymmetries between particles and antiparticles. The dominant one is due to the trigger selection. The hardware trigger decision consists of two categories: trigger on the signal (TOS) or trigger independent of the signal (TIS). The former is based on the energy deposited by final-state particles in the calorimeter, while the latter arises from other particles in the same event meeting the trigger requirement. Whereas protons lose only their kinetic energy in the calorimeter material, antiprotons can annihilate and deposit more energy, leading to a higher TOS efficiency. For kaons and pions, the trigger-efficiency asymmetry is less pronounced. The TIS trigger is usually attributed to decays of the other beauty hadron, which has the opposite $b$-quark flavor to the signal decay, in a $b\overline{b}$ pair-production event. Therefore, the TIS asymmetry has the opposite sign to the TOS asymmetry, but its magnitude is expected to be smaller because of the average over many possible decay modes. For Run I and Run II data, the total TOS asymmetry for the signal decay is measured to be 1.5\%, while the TIS asymmetry is not significant, within an uncertainty of 0.5\%.

The asymmetry due to particle-identification (PID) efficiencies arises from different acceptances and responses of the LHCb RICH detector to particles and antiparticles in the presence of the magnetic field. However, the overall PID asymmetry is measured to be only at the level of 0.1\%, making it subleading. The opposite charges of the final-state particles of $\Lb$ and $\Lbr$ also affect their acceptance in the magnetic field. The LHCb magnet polarity is reversed regularly during data taking, and almost equal amounts of data are collected for both polarities, allowing the acceptance effects to be approximately canceled.

The nuisance asymmetries due to production cross-section and final-state reconstruction\&detection are subtracted from the measured $\CP$ asymmetry using results from control samples. However, the corresponding uncertainties would be too large for a measurement aiming at $0.1\%$-level precision. Instead, a control channel is introduced to reduce the nuisance asymmetries and the associated uncertainties.
The decay $\Lb\to \Lambda_c^+(\to pK^-\pi^+)\pi^-$ provides an almost ideal control channel for the $\CP$ measurement of the $\LbtopKpipi$ decay. This decay is dominated by the CKM-favored $b\to c$ tree transition and is expected to have negligible $\CP$ violation at the current level of precision ($0.1\%$). Thus, its raw asymmetry, defined as the relative difference between the yields of $\Lb$ and $\Lbr$ decays, is dominated by asymmetries in production, detection, and selection.
Moreover, the control channel has the same initial state and the same final-state particles as the signal. As a result, by measuring the difference between the $\CP$ asymmetries of the signal and control channels, which is equivalent to the $\CP$ asymmetry of the signal channel, the nuisance asymmetries and the related systematic uncertainties largely cancel.
The subtraction of the control-channel asymmetry introduces a statistical uncertainty. Fortunately, the control channel has a large branching fraction, a high yield, and very high purity, which makes it possible to estimate the raw asymmetry with good statistical precision.

The data used for the first observation of $\CP$ violation in $\Lb$ decays were collected at $\sqrt{s}=7$ and $8$ TeV in 2011--2012 (Run I) and at $13$ TeV in 2015--2018 (Run II), corresponding to integrated luminosities of $3$ and $6~\mathrm{fb}^{-1}$, respectively. Simulated Monte Carlo (MC) samples are used to study the signal and background mass shapes and to train the multivariate classifiers. The simulation uses Pythia for the proton--proton collisions, EvtGen for the decays of unstable particles, Photos for final-state radiation, and Geant4 for the interactions of particles with the detector. Separate samples are produced for the signal channel $\LbtopKpipi$ and the control channel $\Lb\to \Lambda_c^+(\to pK^-\pi^+)\pi^-$. The signal sample is generated using a mixture of two models: a uniform phase-space model and a cocktail model containing several quasi-two-body decays that represent identified resonances. Background samples are generated for  decays  that could be misidentified as the signal decay, such as $\Lb\to p\pi^-\pi^+\pi^-$, $\Lb\to pK^-K^+\pi^-$, $\Lb\to \Lambda_c^+K^-$, and $\Xi_b^0\to pK^-\pi^+K^-$. Half of the simulated samples are generated with the magnetic field pointing upwards and half with it pointing downwards.

The event selection starts with the trigger. Only events selected by physics triggers are retained for further analysis. For the hardware trigger, positive decisions from the calorimeter or muon detector are required for either the signal decay or other tracks in the same event. At the software stage, a two-, three-, or four-track vertex formed from high-transverse-momentum tracks and displaced from any primary vertex is required.
The offline selection consists of two main steps. In the first step, loose preselections are applied, aimed at reducing obvious combinatorial and peaking backgrounds without significantly affecting the signal decay. In this preselection, the transverse momentum of the $\Lb$ candidate must be greater than $1500~\mathrm{MeV}/c$, and the sum of the transverse momenta of the daughter tracks must be greater than $3500~\mathrm{MeV}/c$. The pseudorapidity of each final-state particle is required to be within the acceptance of the RICH detectors so that the particle can be well identified.
Loose PID requirements are applied to all final-state particles; these help reduce the large cross-feed background from $b$-meson decays or from other four-body $\Lb$ decays. The fake background arising from exchanging the $p$ and $\pi^+$ mass hypotheses, or the $K^-$ and $\pi$ mass hypotheses, is also largely suppressed by the minimal PID requirements. The $\Lb$ vertex-fit quality, flight-distance significance, and impact-parameter significance are also required to satisfy specific criteria consistent with a true signal decay. The efficiencies of these selections are kept very high, usually above 99\%, as validated using simulated decays.

Next, a series of vetoes is applied to remove peaking backgrounds. Charmed resonances are identified in the invariant-mass spectra and removed by excluding candidates near the known mass peaks. For example, the $D^0\to K^-\pi^+$ peak is removed by requiring the $K^-\pi^+$ invariant mass to lie outside the range $[1.82,1.89]~\mathrm{GeV}/c^2$, and the $J/\psi\to\pi^+\pi^-$ peak is removed by excluding the range $[3.06,3.15]~\mathrm{GeV}/c^2$. The $\Lambda_c^+\to pK^-\pi^+$ peak is treated differently in the signal and control channels. In the signal channel, candidates with a $pK^-\pi^+$ mass below $2.32~\mathrm{GeV}/c^2$ are removed, which also removes partially reconstructed backgrounds on the low-mass side of the $\Lambda_c^+$ peak. In the control channel, candidates are retained only if the $pK^-\pi^+$ mass lies in the range $[2.255,2.30]~\mathrm{GeV}/c^2$.
Other charmed resonances, such as $\Xi_c^+$, $\chi_c$, $D^+$, $D_s^+$, and $D^0$, are also vetoed. Misidentification backgrounds are suppressed by vetoing peaks that appear under alternative mass hypotheses. For example, the $\phi\to K^+K^-$ decay is removed by excluding the $K^+K^-$ invariant-mass range $[1.01,1.03]~\mathrm{GeV}/c^2$, where the $K^+$ mass is assigned to the $p$ candidate in the final state. The $D^0$, $D^+$, $D_s^+$, $\Lambda_c^+$, $J/\psi$, $B^0$, and $B_s^0$ peaks are also vetoed under the corresponding alternative hypotheses. It is worth noting that $B$-meson decays such as $\overline{B}_{s}^0\to \pi^+ K^-\pi^+\pi^-$, with a $\pi^+$ misidentified as the $p$, are particularly dangerous, since they can produce a false signal given that large $\CP$ violation is possible in these meson decays.

The second step of the offline analysis is a multivariate analysis aimed at suppressing mainly combinatorial background. A boosted decision tree (BDT) classifier is trained with simulated signal events and with data events from the high-mass sideband, defined as $m(pK^-\pi^+\pi^-)>5840~\mathrm{MeV}/c^2$. This sideband is dominated by combinatorial background. The training variables include the transverse momentum, pseudorapidity, impact-parameter significance, vertex-separation significance, and vertex-fit quality of the $\Lb$ candidate, as well as isolation variables. Two isolation variables are used: the smallest change in the vertex $\chi^2$ when an additional track is included, and the asymmetry between the $p_\mathrm{T}$ of the $\Lb$ baryon and the scalar sum of the $p_\mathrm{T}$ of tracks within a cone around the $\Lb$ direction.
In the training, variables for proton, kaon, and pion tracks are also used, including the transverse momentum, impact-parameter significance, and the minimum and maximum values of these variables among the tracks. The training is performed separately for Run I and Run II data because the reconstruction and kinematic properties differ. No evidence of overtraining is found, as verified by Kolmogorov--Smirnov tests between the training and test samples.
After the BDT training, the selection criteria for the BDT response and the PID variables of the final-state particles are optimized simultaneously. Note that the variables for the $\pi^+$ and $\pi^-$ tracks in the BDT and PID selections are symmetrized by taking their sum or maximum/minimum values, so that the experimental effects for the two tracks largely cancel.
The optimization maximizes the figure of merit $\mathrm{FoM}=S'/\sqrt{S+B}$, where $S'$ is taken as the number of simulated signal events passing the selection, and $S+B$ is taken as the total number of events in data within $120~\mathrm{MeV}/c^2$ of the $\Lb$ mass. This mass range corresponds to about five times the experimental mass resolution, allowing more misidentified $\Lb$ decays to enter the optimization.
The signal efficiency of the final multivariate selection  is about 80\%, while more than 99\% of the combinatorial background events are removed.
After the BDT and PID selections, a supplementary veto is applied to remove smaller peaking backgrounds that become visible once the combinatorial background is reduced. These include additional charmed and misidentification peaks. For example, a small $\phi$ peak is removed with an additional kaon PID requirement. All possible $B$-meson decays with one or two misidentified tracks are checked, and no significant peaking structures are found.
For the control channel, the event selection is relatively simple: after applying the PID requirements used for the signal decay and requiring the $pK^-\pi^+$ invariant mass to be within $22.5~\mathrm{MeV}/c^2$ (about three times the experimental resolution) of the known $\Lambda_c^+$ mass, the sample is clean enough for the subsequent analysis.

The $\CP$ asymmetries for the signal and control channels are determined as
$A_{\CP}=A_\text{raw}-A_\text{prod}-A_\text{det}-A_\text{PID}-A_\text{trigger}$,
where $A_\text{raw}$ represents the relative difference between the $\Lb$ and $\Lbr$ yields in the final sample, and the other four asymmetries correspond to those of the production cross section, detection, PID, and trigger efficiencies, respectively.
The signal yield is extracted from an extended unbinned maximum-likelihood fit to the four-body invariant-mass spectrum $m(pK^-\pi^+\pi^-)$ in the range $[5400,6100]~\mathrm{MeV}/c^2$. The fit is performed simultaneously for $\Lb$ and $\Lbr$ decays, with shared shape parameters and independent yield parameters. Run I and Run II data are fitted separately. The signal and control channels are also treated separately because their experimental resolutions are found to differ slightly. The signal shape is modeled as a sum of a Gaussian function and two one-sided Crystal Ball functions~\cite{Root} with a common mean. The Crystal Ball function has a Gaussian core and a power-law tail on the left side to describe radiative losses. The parameters of the signal shape are determined from fits to simulated samples. In the fit to data, the peak position, the Gaussian width, and the average Crystal Ball width are allowed to float, while the other parameters are fixed to the values obtained from simulation. The simulated signal yield is about twenty times larger than that in data.

\begin{figure}
\begin{center}
\includegraphics[width=0.4\textwidth]{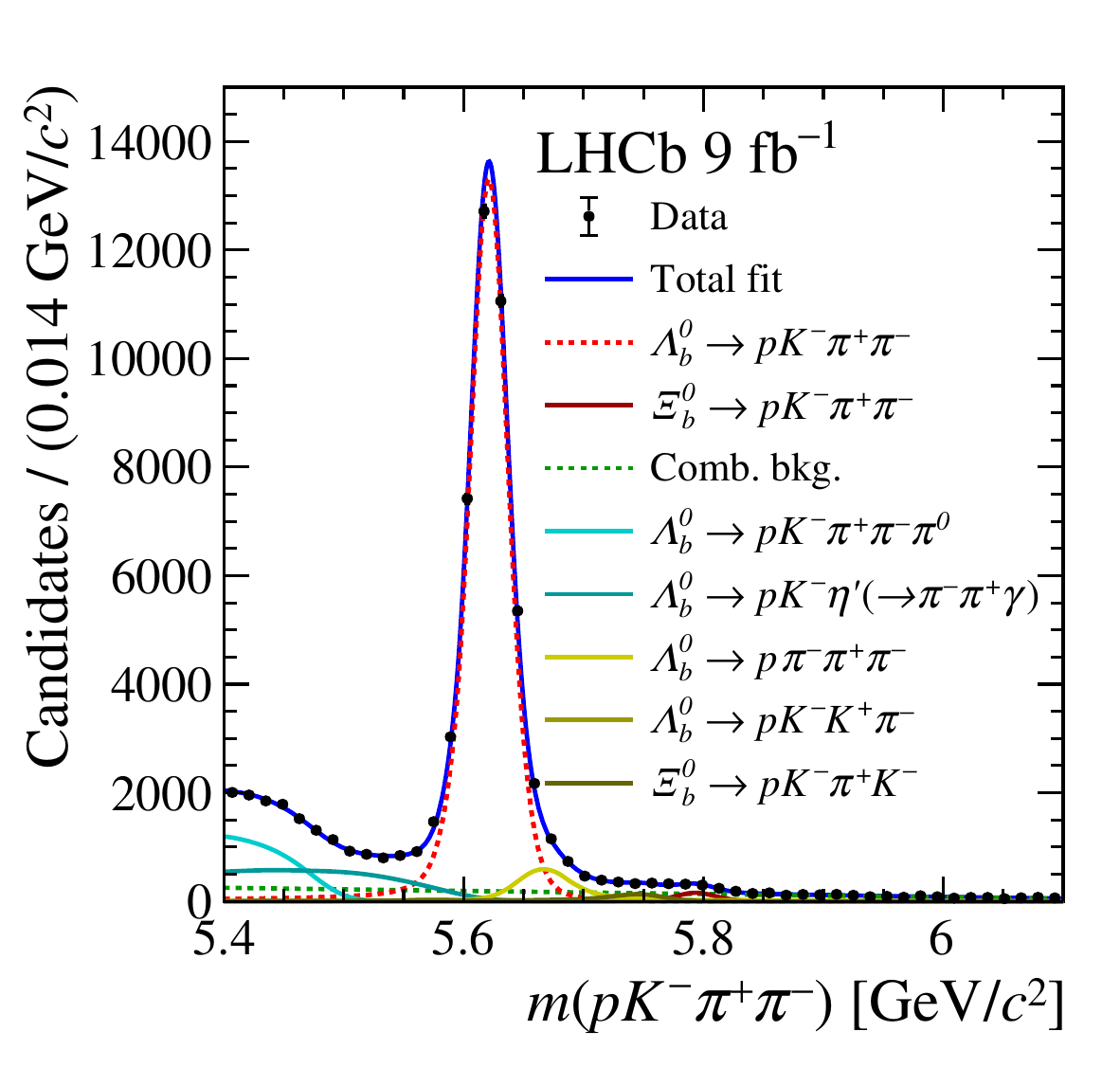}
\includegraphics[width=0.4\textwidth]{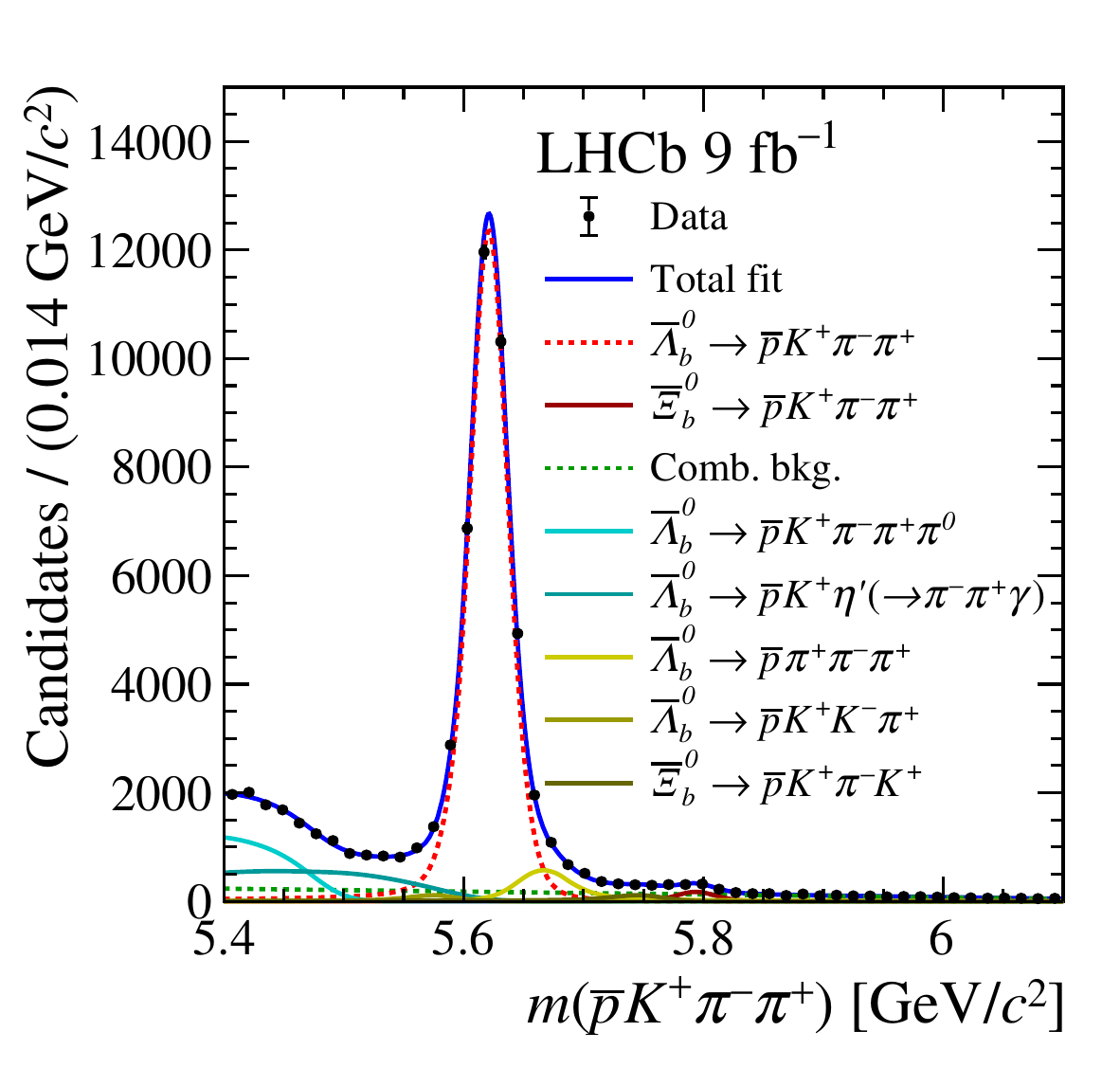}
\captionsetup{type=figure}
\captionof{figure}{ Invariant mass distribution of $\Lb\to p K^-\pi^+\pi^-$ (left) and $\Lbr\to \overline{p}K^+\pi^+\pi^-$ (right) decays, together with fit projections. Run I and Run II data samples are merged. Reproduced from~\cite{LHCb:2025ray}}\label{fig:LbTotFit}
\end{center}
\end{figure}

The background components are modeled as follows. The partially reconstructed background from $\LbtopKpipi\pi^0$ with a missing $\pi^0$ is described by an ARGUS function~\cite{ARGUS:1990hfq} convolved with a Gaussian resolution. The ARGUS threshold is fixed to the difference between the $\Lb$ and $\pi^0$ masses, $5485.023~\mathrm{MeV}/c^2$, while the other parameters are allowed to float. The partially reconstructed background from $\Lb\to pK^-\eta'(\to\pi^+\pi^-\gamma)$ with a missing photon  is modeled using the simulation sample~\cite{Cowan:2016tnm}. The misidentification backgrounds from $\Lb\to p\pi^-\pi^+\pi^-$, $\Lb\to pK^-K^+\pi^-$, and $\Xi_b^0\to pK^-\pi^+K^-$ are also modeled using nonparametric distributions obtained from simulation samples. The combinatorial background is modeled by a linear function. The $\Xi_b^0\to pK^-\pi^+\pi^-$ decay is similar to the $\Lb$ decay and is modeled with the same function as the $\Lb$ signal, but with the peak position shifted by the known mass difference between the $\Xi_b^0$ and $\Lb$ baryons, $172.5~\mathrm{MeV}/c^2$, and the width scaled by the ratio of their masses. The mass distributions of the $\Lb$ and $\Lbr$ candidates, together with the fit results, are shown in Fig.~\ref{fig:LbTotFit}. The fit quality is reasonably good.

\begin{figure}
\begin{center}
\includegraphics[width=0.4\textwidth]{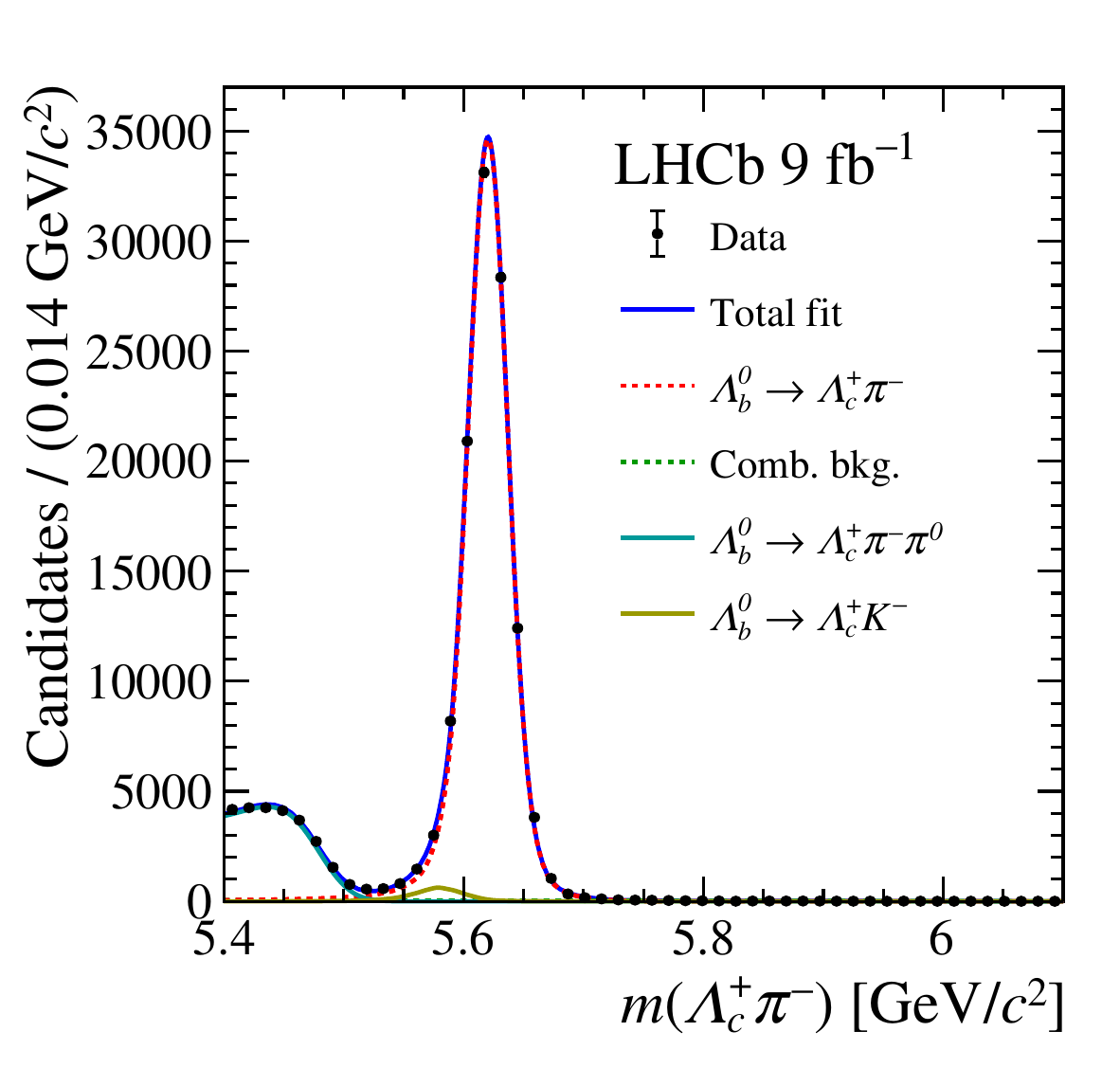}
\includegraphics[width=0.4\textwidth]{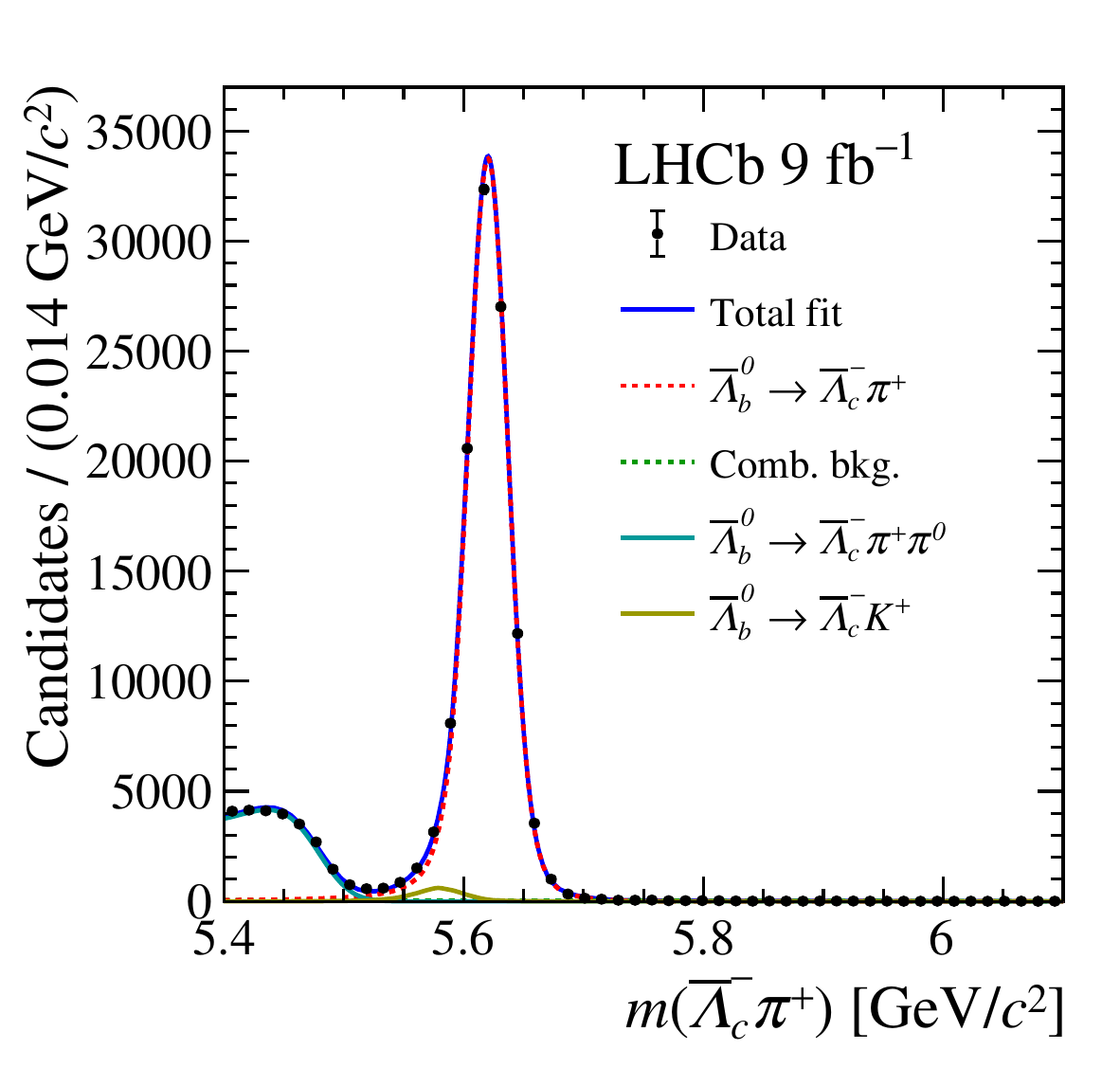}
\captionsetup{type=figure}
\captionof{figure}{Invariant mass distribution of $\Lb\to \Lambda_c^+(p K^-\pi^+)\pi^-$ (left) and $\Lbr\to \overline{\Lambda}_c^-(\overline{p}K^+\pi^-)\pi^+$ decays, together with fit projections. Run I and Run II data samples are merged. Reproduced from~\cite{LHCb:2025ray}.}
\label{fig:Lb2LcTotFit}
\end{center}
\end{figure}

For the control channel $\Lb\to \Lambda_c^+(\to pK^-\pi^+)\pi^-$, the fit uses the same signal model as for the $\Lb\to  pK^-\pi^+\pi^-$ decay, but with parameters determined independently from simulated samples. The background components include a partially reconstructed background from $\Lb\to \Lambda_c^+\pi^-\pi^0$, modeled by an ARGUS function; a misidentified background from $\Lb\to \Lambda_c^+K^-$, modeled using simulated decays; and a combinatorial background, modeled by a linear function. The ratio of the $\Lb\to \Lambda_c^+K^-$ and $\Lb\to \Lambda_c^+\pi^-$ yields is shared between the $\Lb$ and $\Lbr$ decays, since $\CP$ violation both decays are negligible and the relative efficiency is almost identical. The mass distributions and fit projections are shown in Fig.~\ref{fig:Lb2LcTotFit}, from which the high signal purity is evident.

The fit results for the signal channel give the total yields of $N(\LbtopKpipi)=(4.184\pm0.025)\times10^4$ and $N(\Lbr\to \overline{p}^-K^+\pi^-\pi^+)=(3.885\pm0.023)\times10^4$, yielding the raw asymmetry $A_\text{raw}=(3.71\pm0.39)\%$ averaged over Run I and Run II data. The separate measurements of raw asymemtries for Run I and Run II data are $(5.12\pm0.96)\%$ and $(3.42\pm0.43)\%$, respectively.
The total yield in the control channel is about $220\,000$, with the raw asymmetry determined to be $A_\text{raw}=(1.27\pm0.23)\%$, averaged over $(1.08\pm0.55)\%$ for Run I and $(1.32\pm0.25)\%$ for Run II. The raw asymmetries are consistent between the Run I and Run II data. The control channel shows that the total nuisance asymmetry is about 1\%.

The $\Lb$ production asymmetry $A_\text{prod}$ in $pp$ collisions is measured by LHCb to depend on its rapidity $y$ and transverse momentum $p_\text{T}$~\cite{LHCb:2021xyh}. Because of different selection efficiencies, the $p_\text{T}$--$y$ spectrum differs slightly between the signal and control channels; therefore, to properly cancel the production asymmetry, the $p_\text{T}$--$y$ distribution in the control channel is weighted to match that in the signal decay.
The reweighting is performed in two-dimensional bins of $p_\text{T}$ and $y$, in the range $2<p_\text{T}<30~\mathrm{GeV}/c$ and $2.1<y<4.5$. After reweighting, the raw asymmetry of the control channel changes only slightly, from $A_\text{raw}=(1.27\pm0.23)\%$ to $A'_\text{raw}=(1.25\pm0.23)\%$. The difference in raw asymmetry between the signal and control channels is thus determined to be $\Delta A_\text{raw}=(2.46\pm0.46)\%$, with the production asymmetry effectively canceled.

The final-state particles in the signal and control channels also have different kinematic distributions. In the control channel, the $pK^-\pi^+$ invariant mass is constrained to be near the known $\Lambda_c^+$ mass, which restricts the phase space available to the final-state particles. As a result, the momentum and pseudorapidity distributions of the final-state particles differ between the signal and control channels. This means that detection asymmetries do not cancel automatically and must be corrected using kinematic reweighting and per-particle calibrations of asymmetries.
The detection asymmetry is calculated separately for protons, kaons, and pions using the same strategy. First, the detection asymmetry for each particle species is determined in bins of momentum $p$ (and $\eta$ for pions). The detection asymmetries for protons and pions are obtained using a tag-and-probe method with the $\Lambda\to p\pi^-$ and $K_\text{S}^0\to\pi^+\pi^-$ decays, respectively~\cite{LHCb:2018fly,LHCb:2021ykz}. For kaons, the asymmetry is obtained by taking the difference between asymmetries of the $D^+\to K_\text{S}^0\pi^+$ and $D^+\to K^-\pi^+\pi^-$ decays, with the $K_\text{S}^0$ detection asymmetry estimated to be negligible~\cite{LHCb:2017bdt} (the remaining effect cancels in the difference between the control and signal channels). The kinematic-dependent detection asymmetries of the three types of final-state particle types are then weighted by the kinematic distributions of the signal and control channels to estimate the average detection asymmetries for the two channels.
The total detection asymmetry is $A_{\mathrm{det}}=A_{\mathrm{det}}^p-A_{\mathrm{det}}^K+(A_{\mathrm{det}}^{\pi^+}-A_{\mathrm{det}}^{\pi^-})$, where the minus sign arises because the detection asymmetry is defined as the relative difference in efficiency between positively and negatively charged particles. The signal--control differences are $\Delta A_{\mathrm{det}}=(0.055\pm0.128)\%$ for Run I and $\Delta A_{\mathrm{det}}=(0.081\pm0.050)\%$ for Run II. The uncertainties are estimated using pseudoexperiments, in which new detection-asymmetry tables are generated according to the central values and uncertainties of the original inputs. The distribution of the difference in an ensemble of pseudoexperiments is fitted with a Gaussian function, and its width is taken as the systematic uncertainty.

The PID asymmetry is evaluated similarly to the detection asymmetry. First, the asymmetries for protons, kaons, and pions are determined in bins of $p$ and $\eta$ using dedicated calibration samples~\cite{Anderlini:2016kco}. The proton asymmetry is calibrated using $\Lambda^0\to p\pi^-$ or $\Lambda_c^+\to p K^-\pi^+$ decays, while for kaons and pions, $D^{*+}$-tagged $D^0\to K^-\pi^+$ decays are used. The results are obtained separately for each magnet polarity and each data-taking period. The overall PID asymmetry is found to be at the $0.1\%$ level.
Averaging the kinematic-dependent PID asymmetry over the kinematic spectra of the signal and control channels allows the total PID asymmetry to be determined. The difference between the signal and control channels is measured to be $\Delta A_\text{PID}=(0.026\pm0.141)\%$ for Run I and $\Delta A_\text{PID}=(-0.028\pm0.002)\%$ for Run II, where the uncertainties are propagated using pseudoexperiments. The larger uncertainty for the Run I measurement is due to the limited statistics of the calibration samples.

The trigger asymmetry is dominated by the hardware trigger, while the software-trigger asymmetry is measured to be $0.01$--$0.02\%$ for Run I and $0.00$--$0.01\%$ for Run II and is therefore neglected. The hardware-trigger asymmetry is split into TOS and TIS contributions. First, the TOS efficiencies for protons, kaons, and pions as functions of the transverse energy are determined from dedicated calibration sample of $\Lambda_c^+\to p K^-\pi^+$ decays. The efficiency for a $\Lb$ candidate is calculated as
$\epsilon_{\Lb}=1-(1-\epsilon_p)(1-\epsilon_K)(1-\epsilon_{\pi^+})(1-\epsilon_{\pi^-})$,
taking into account overlaps between final-state particles. Using the corresponding calibration tables for the charge-conjugate final state, the efficiency for the $\Lbr$ baryon is obtained, and the asymmetry is then derived with an average value of 1.5\%.
Since the TIS asymmetry is by definition independent of the decay channel, it is determined using the control channel with the TISTOS method in bins of $\Lb$ transverse momentum. The total TIS asymmetry is then obtained by weighting the $p_\text{T}$-dependent asymmetry according to the $p_\text{T}$ spectrum of $\Lb/\Lbr$ baryons in data; it is, however, not significant, within an uncertainty of 0.5\%.
The total trigger asymmetry is calculated as a weighted average of the TIS and TOS asymmetries, where the weights represent the fractions of TIS and TOS decisions. The difference between the signal and control channel trigger asymmetry is $\Delta A_{\mathrm{trigger}}=(-0.039\pm0.029)\%$ for Run I and $\Delta A_{\mathrm{trigger}}=(-0.050\pm0.008)\%$ for Run II. It shows that asymmetries and the corresponding uncertainties in the signal and control channels cancel effectively. Summing over all sources of nuisance asymmetries, the difference between the signal and control channels is measured to be $(0.042\pm0.192)\%$ and $(0.003\pm0.051)\%$ for Run I and Run II data, respectively.

The choice of models used in the mass fit is a separate source of systematic uncertainty, which is estimated by repeating the fit with alternative models. The fixed parameters of the signal shape are varied within $\pm1$ standard deviation. The signal model is also replaced by the sum of a Gaussian function and a Hypatia function~\cite{Root}. The mass difference between the $\Xi_b^0$ and $\Lb$ baryons is varied by $\pm1$ standard deviation. The combinatorial background is changed from a linear function to an exponential function. The misidentification background from $\Lb\to p\pi^-\pi^+\pi^-$ is modeled with a uniform phase-space sample instead of the default cocktail sample. The partially reconstructed background from $\Lb\to pK^-\eta'(\to\pi^+\pi^-\gamma)$ is modeled by applying additional kinematic requirements to simulate data. The fit range is changed from $[5400,6100]~\mathrm{MeV}/c^2$ to $[5450,6100]~\mathrm{MeV}/c^2$, which reduces the contribution of the partially reconstructed $\LbtopKpipi\pi^0$ background by about a factor of two. For each alternative model, the difference in the raw asymmetry with respect to the baseline fit is taken as the corresponding systematic uncertainty, and the largest difference among the alternative models is assigned as the mass-fit systematic uncertainty. For Run I, the largest contribution comes from the combinatorial background model, $0.044\%$. For Run II, the largest contribution comes from the alternative signal model, $0.067\%$. The total systematic uncertainty is obtained by summing the nuisance-asymmetry uncertainty and the mass-fit uncertainty in quadrature. For Run I, the total is $0.198\%$; for Run II, the total is $0.084\%$; and for the combined Run I and Run II result, the total is $0.10\%$.

\begin{figure}
\begin{center}
\includegraphics[width=0.45\textwidth]{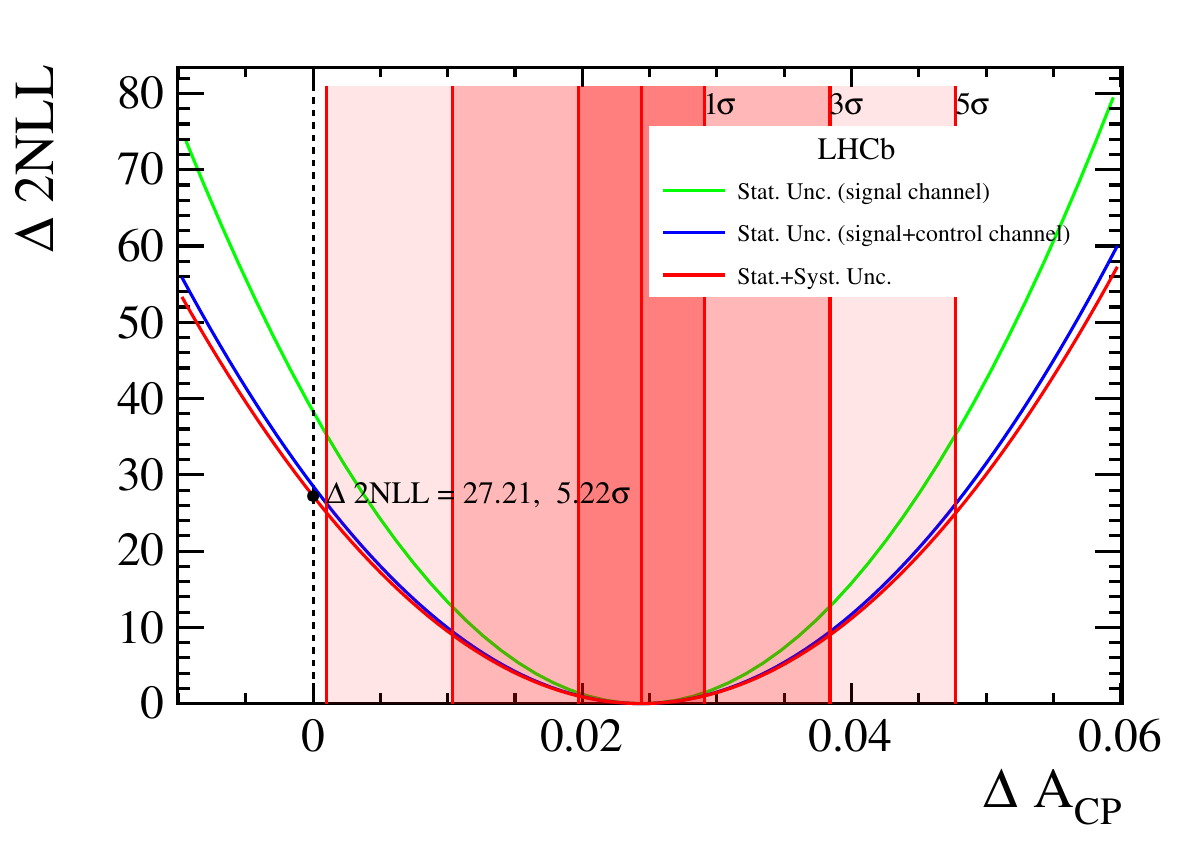}
\captionsetup{type=figure}
\captionof{figure}{The log-likelihood as a function of the difference in $\CP$ asymmetries between the signal and control channels, shifted by the log-likelihood of the nominal result. }
\label{fig:LLscan}
\end{center}
\end{figure}

Several cross-checks are performed to validate the measurement. The data are divided into different data-taking years, magnet polarities, hardware-trigger conditions, and intervals of $\Lb$ transverse momentum and rapidity. The raw asymmetries, as well as the difference between the signal and control raw asymmetries, are calculated for each subsample. The results are consistent with the average values within the statistical uncertainties. The consistency between the two magnet polarities is particularly important, as it tests the cancellation of detector-induced asymmetries. The consistency between different trigger conditions is also checked; the observed differences in the signal and control raw asymmetries are found to cancel in the signal--control difference. The analysis also verifies, using simulated samples, that the nuisance asymmetries cancel between the signal and control channels, and that the kinematic reweighting factors are relatively uniform across the phase space. The systematic uncertainty from the mass fit is cross-checked using pseudoexperiments, and the results are found to be consistent with those obtained from the direct fit variations.

Combining the Run I and Run II results gives the global $\CP$ asymmetry,
$A_{\CP}=(2.45\pm0.46\pm0.10)\%$, where the $\CP$ violation in the control channel is assumed to be negligible.
The significance of the deviation of the $\CP$ asymmetry from zero is calculated by dividing the central value by the total uncertainty, giving $5.2$ standard deviations. This is the first observation of $\CP$ violation in a baryon decay. The result is also validated using a profile-likelihood scan, which gives a consistent significance of $5.22$ standard deviations, as shown in Fig.~\ref{fig:LLscan}. According to the likelihood scan, $\CP$ asymmetries larger than $5\%$ are also excluded.

\begin{center}
\includegraphics[width=0.35\textwidth]{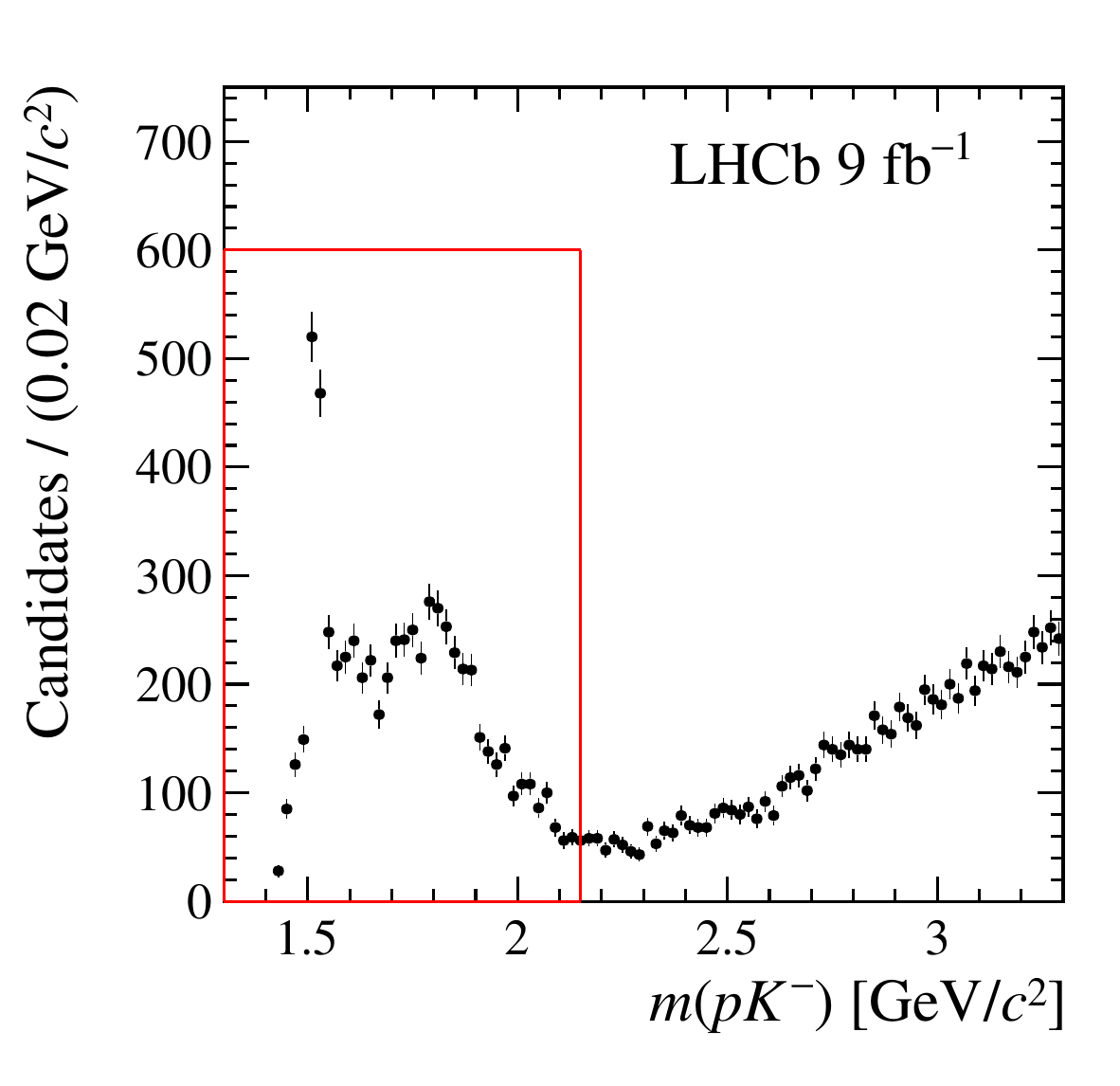}
\includegraphics[width=0.35\textwidth]{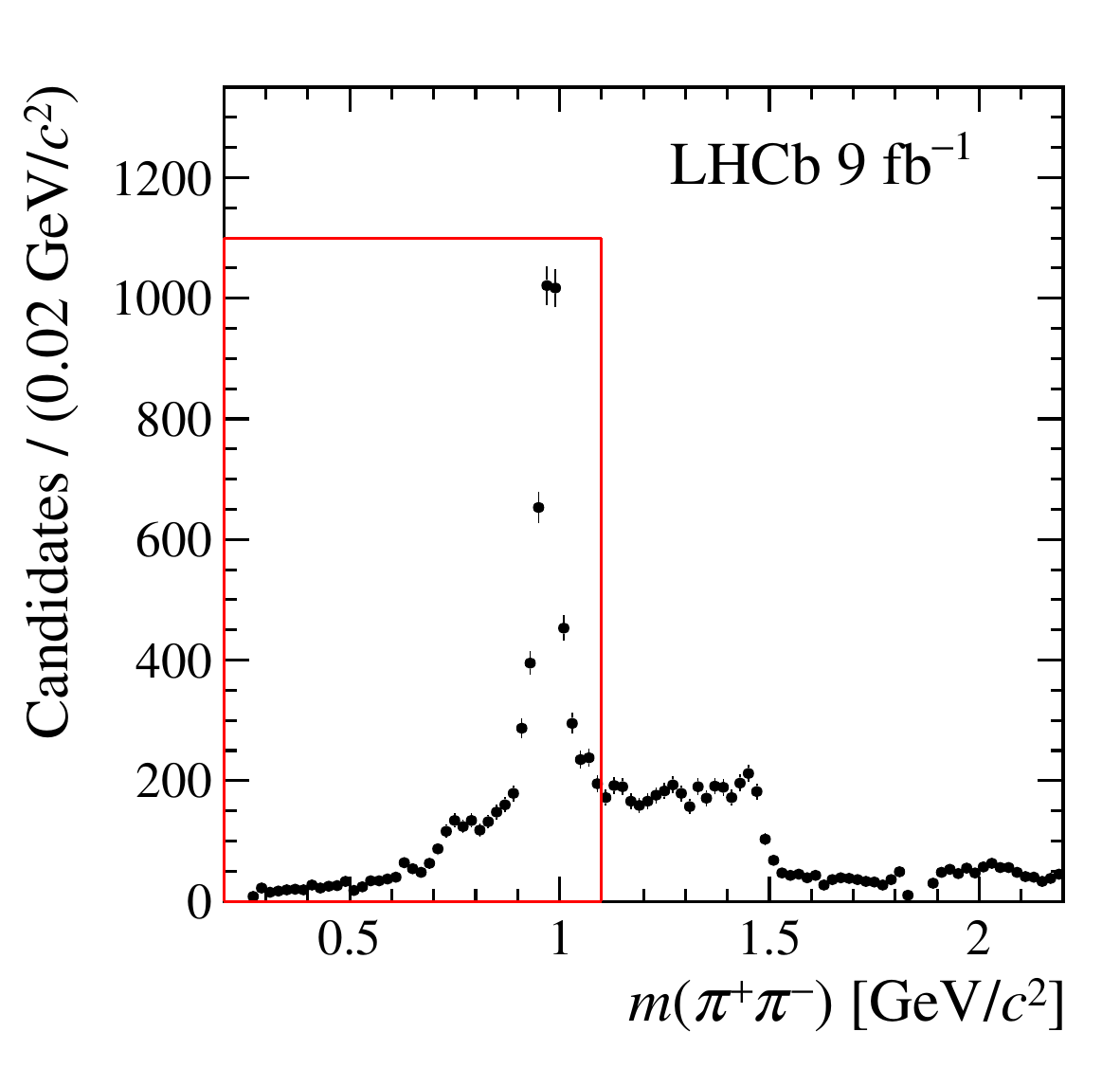}
\includegraphics[width=0.35\textwidth]{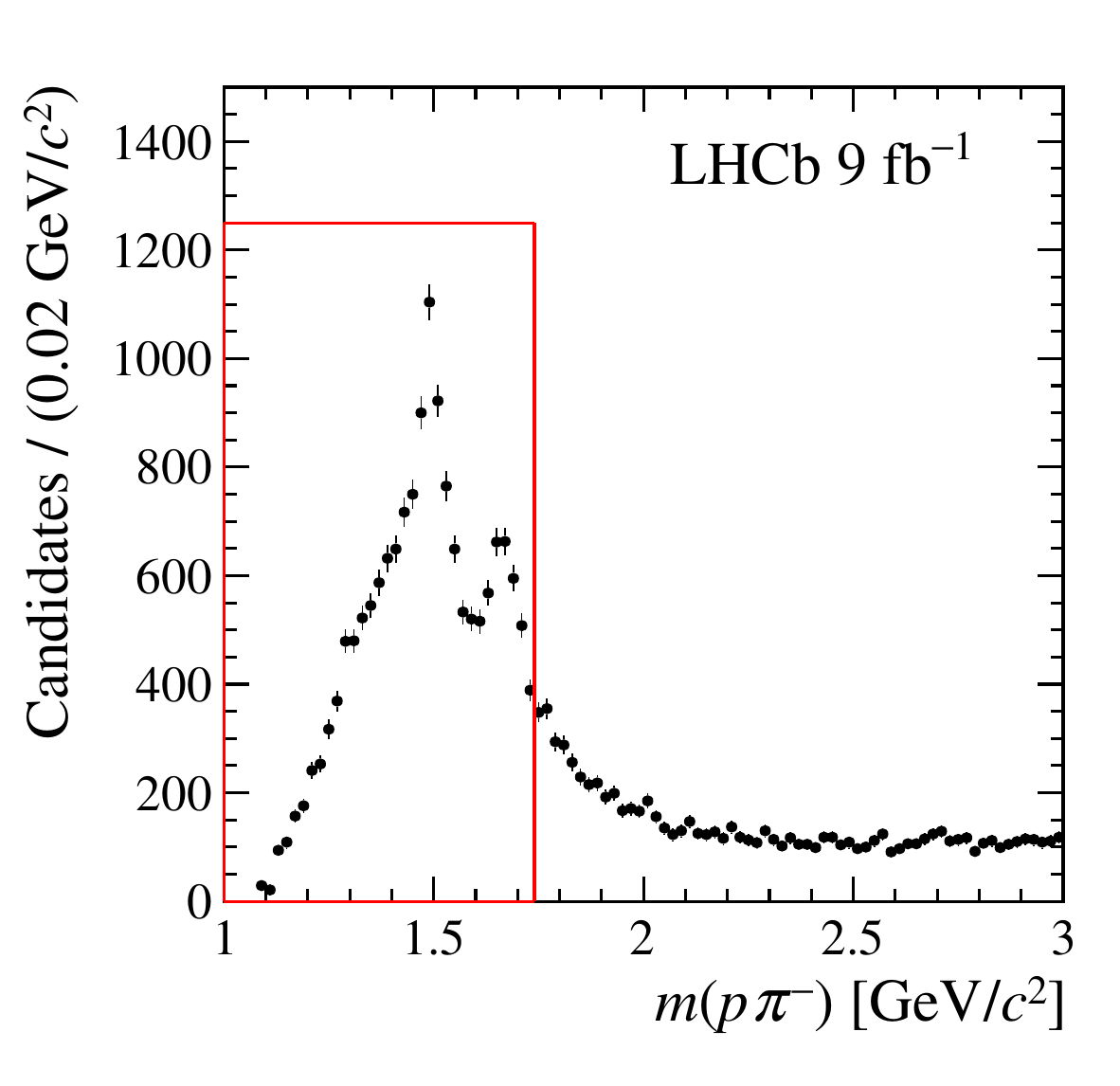}
\includegraphics[width=0.35\textwidth]{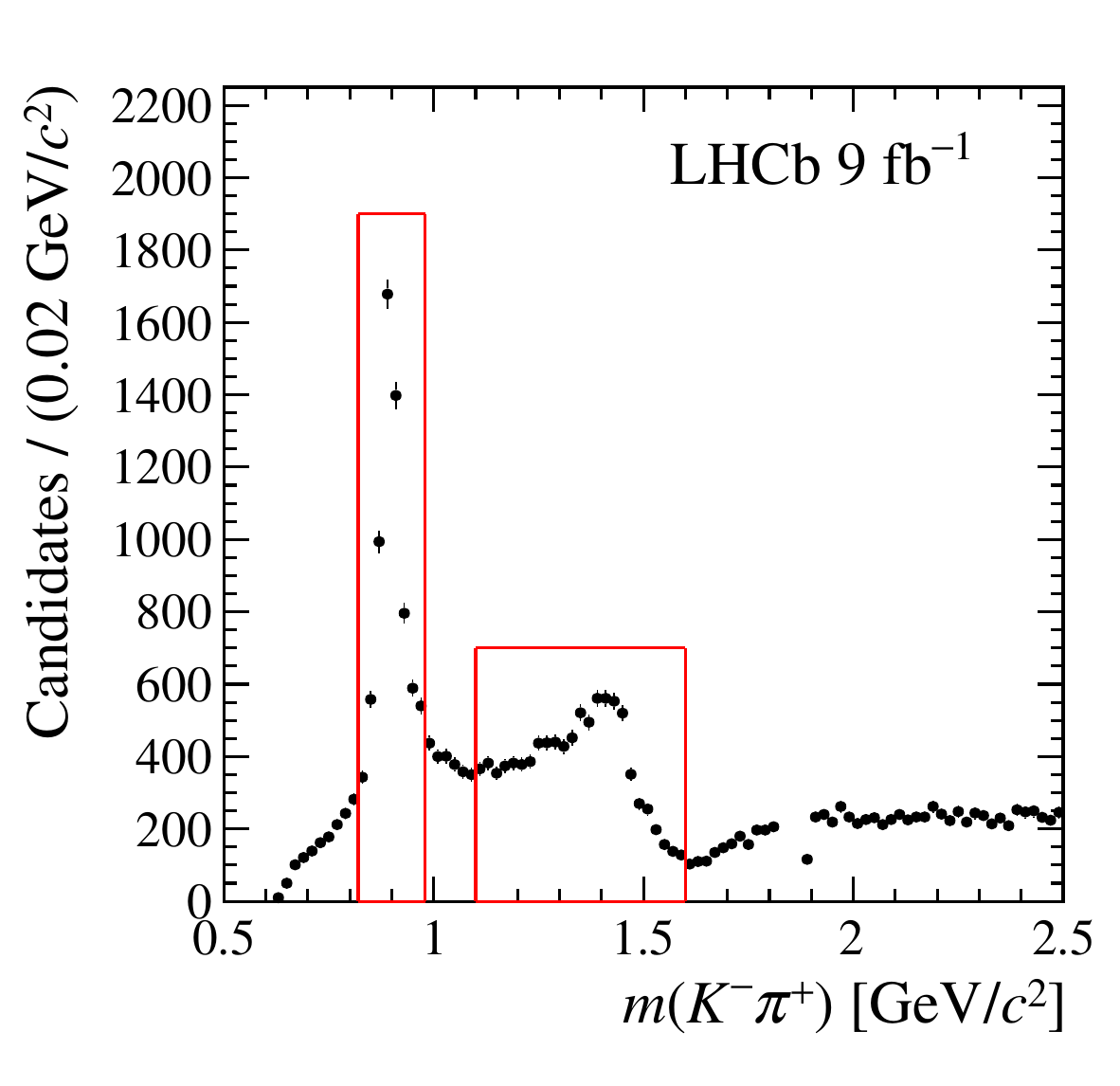}
\includegraphics[width=0.35\textwidth]{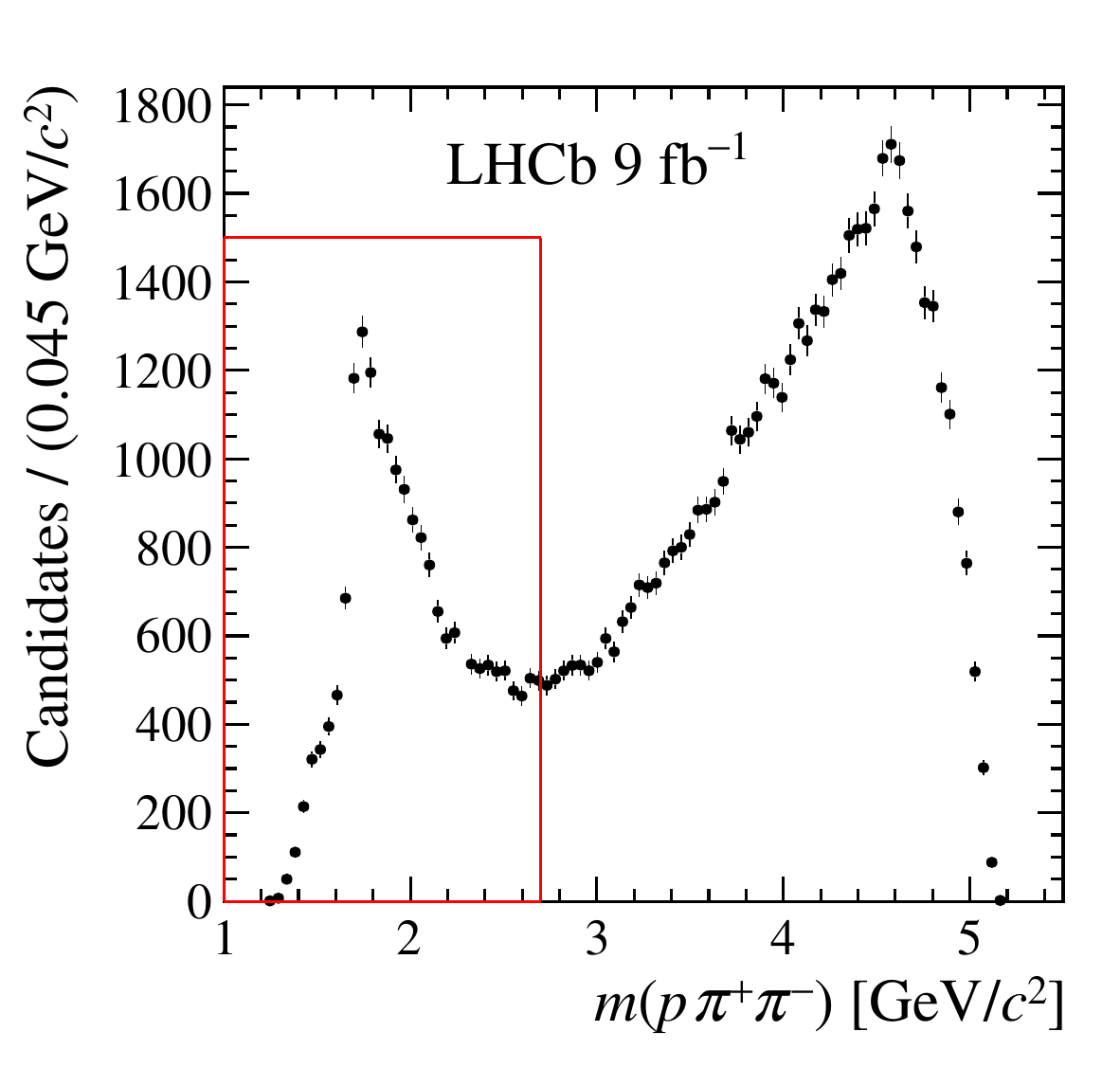}
\includegraphics[width=0.35\textwidth]{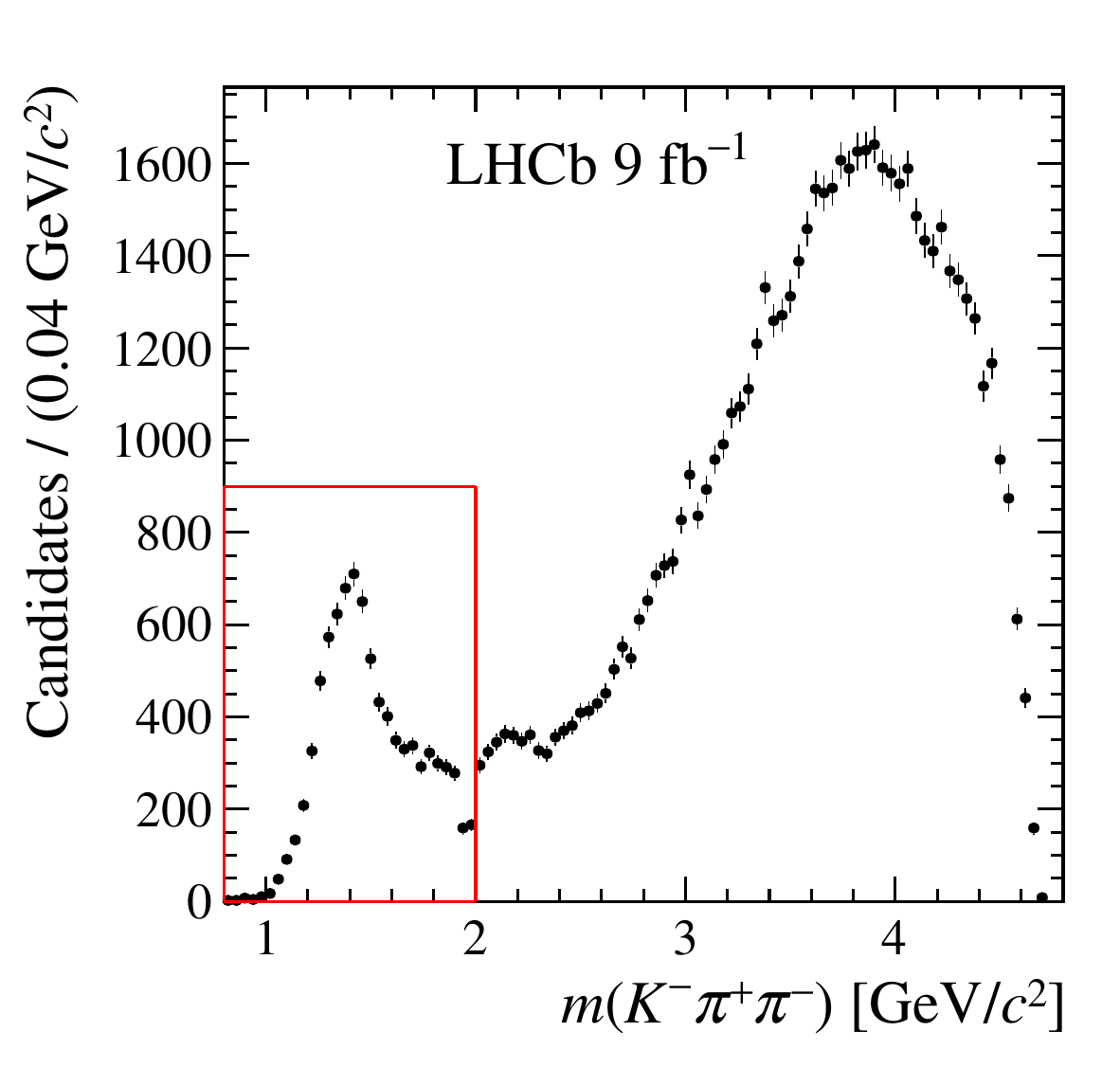}
\captionsetup{type=figure}
\captionof{figure}{
Invariant-mass distributions of two-body and three-body final-state particles: $pK^-$ (top left) and $\pi^+\pi^-$ (top right), corresponding to the $\Lb\to R(pK^-)R(\pi^+\pi^-)$ phase-space region; $p\pi^-$ (middle left) and $K^-\pi^+$ (middle right), corresponding to the $\Lb\to R(p\pi^-)R(K^-\pi^+)$ phase-space region; $p\pi^+\pi^-$ (bottom left), corresponding to the $\Lb\to R(p\pi^+\pi^-)K^-$ phase-space region; and $K^-\pi^+\pi^-$ (bottom right), corresponding to the $\Lb\to R(K^-\pi^+\pi^-)p$ phase-space region. In the plots, the $\Lb$ and $\Lbr$ samples are merged. The boxes in right color indicate the selected regions for the measurements. Reproduced from~\cite{LHCb:2025ray}.}\label{fig:LocalMassRegion}
\end{center}

The analysis also measures $\CP$ asymmetries in local regions of phase space, chosen according to the $\LbtopKpipi$ decay topologies. In total, there are seven topologies, corresponding to ($2+2$)-body or ($3+1$)-body configurations, and each topology receives contributions from several hadronic resonances $R$.
The $\Lb\to R(p\pi^+)R'(K^-\pi^-)$ decay topology is suppressed because no conventional hadrons are expected in the $K^-\pi^-$ final state. The $\Lb\to R(pK^-\pi^-)\pi^+$ and $\Lb\to R(pK^-\pi^+)\pi^-$ decays are mediated by excited $\Sigma^-$ and $\Sigma^+$ baryons, respectively. These two decays are less favored because the $ud$ diquark in the initial $\Lb$ baryon must break up into separate hadrons. Moreover, the $uu$ or $dd$ quarks in the $\Sigma$ baryons form an isospin triplet, whereas the $ud$ diquark in the $\Lb$ baryon is in an isospin-singlet state. These three regions are therefore not selected for localized $\CP$ measurements.  

The first localized region for the $\CP$ measurement, denoted by $R(pK^-)R(\pi^+\pi^-)$, requires $m_{pK^-}<2.2~\mathrm{GeV}/c^2$ and $m_{\pi^+\pi^-}<1.1~\mathrm{GeV}/c^2$. This region is dominated by excited $\Lambda$ resonances and light unflavored mesons such as $f_0(500)$, $\rho(770)$, and $f_0(980)$. The measured $\CP$ asymmetry is $(5.24\pm1.29\pm0.21)\%$ for the Run I and Run II average, with a  significance of $3.7$ standard deviations after accounting for the look-elsewhere effect.
The second region, denoted by $R(p\pi^-)R(K^-\pi^+)$, requires $m_{p\pi^-}<1.7~\mathrm{GeV}/c^2$ and either $0.8<m_{K^-\pi^+}<1.0~\mathrm{GeV}/c^2$ or $1.1<m_{K^-\pi^+}<1.6~\mathrm{GeV}/c^2$. This region is dominated by excited nucleon resonances and $K^{0}$ mesons. The measured $\CP$ asymmetry is $(2.73\pm0.82\pm0.14)\%$, with a significance of $2.9$ standard deviations.
The third region, denoted by $R(p\pi^+\pi^-)K^-$, requires $m_{p\pi^+\pi^-}<2.7~\mathrm{GeV}/c^2$. This region is dominated by excited nucleon or $\Delta^+$ resonances decaying to $p\pi^+\pi^-$. The measured $\CP$ asymmetry is $(5.39\pm0.86\pm0.10)\%$, with a significance of $6.0$ standard deviations after accounting for the look-elsewhere effect.
The fourth region, denoted $R(K^-\pi^+\pi^-)p$, requires $m_{K^-\pi^+\pi^-}<2.0~\mathrm{GeV}/c^2$. This region is dominated by excited $K^-$ resonances such as $K_1(1270)^-$, $K_1(1400)^-$, and $K^*(1410)^-$. The measured $\CP$ asymmetry is $(2.01\pm1.16\pm0.30)\%$.
The two-body and three-body invariant-mass distributions used to select the localized phase-space regions are shown in Fig.~\ref{fig:LocalMassRegion}. The mass-region selections and  $\CP$ asymmetries are summarized in Table~\ref{tab:PScut}.

\begin{table*}[tb]
\centering
\footnotesize
\caption{Measurements of $\CP$ asymmetries in the four selected phase-space regions, compared with the available theoretical estimates based on representative resonant channels. The listed channels do not provide a complete description of the corresponding experimental regions, and the theoretical results should therefore be regarded as phenomenological estimates rather than precision predictions.}
\label{tab:PScut}
\renewcommand{\arraystretch}{1.8}
\begin{tabular*}{\textwidth}{@{\extracolsep{\fill}}ccccc}
\hline
Decay topology
& Mass region (GeV$/c^2$)
& $A_{\CP}^{\rm exp}$
& $A_{\CP}^{\rm th}$
& Resonant channels \\
\hline

\multirow{2}{*}{$\Lb\to R(pK^-)R(\pi^+\pi^-)$}
& $m_{pK^-}<2.2$
& \multirow{2}{*}{$(5.3\pm1.3\pm0.2)\%$}
& \multirow{2}{*}{$\simeq 10.8\%$~\cite{Feng:2026soj}}
& \multirow{2}{*}{\shortstack{$\Lb\to\Lambda(1520)M$\\$M=f_0(500),\,\rho^0,\,f_0(980)$}} \\
& $m_{\pi^+\pi^-}<1.1$ & & & \\
\hline

\multirow{3}{*}{$\Lb\to R(p\pi^-)R(K^-\pi^+)$}
& $m_{p\pi^-}<1.7$
& \multirow{3}{*}{$(2.7\pm0.8\pm0.1)\%$}
& \multirow{3}{*}{$\simeq1.8\%$~\cite{Shang:2026knt}}
& \multirow{3}{*}{\shortstack{$\Lb\to N^*(1520)\bar K^{*0}$\\$\Lb\to N^*(1535)\bar K^{*0}$}} \\
& $0.8<m_{K^-\pi^+}<1.0$ & & & \\
& \text{or }$1.1<m_{K^-\pi^+}<1.6$ & & & \\
\hline

$\Lb\to R(p\pi^+\pi^-)K^-$
& $m_{p\pi^+\pi^-}<2.7$
& $(5.4\pm0.9\pm0.1)\%$
& $\simeq 5.8\%$~\cite{Wang:2024oyi}
& $\Lb\to(\Delta^{++}\pi^-)K^-$ \\
\hline

$\Lb\to R(K^-\pi^+\pi^-)p$
& $m_{K^-\pi^+\pi^-}<2.0$
& $(2.0\pm1.2\pm0.3)\%$
& $\simeq(7.1 - 8.4)\%$~\cite{Han:2025tvc}
& \shortstack{$\Lb\to pK_1^-(1270)$\\$\Lb\to pK_1^-(1400)$} \\
\hline
\end{tabular*}
\end{table*}

The local results show that the $\CP$ asymmetry is not uniform across the phase space. The most significant local asymmetry is found in the $R(p\pi^+\pi^-)K^-$ region, indicating that specific intermediate resonances play a key role in generating the observed $\CP$ violation. The $R(pK^-)R(\pi^+\pi^-)$ region also shows a significant asymmetry, whereas the $R(p\pi^-)R(K^-\pi^+)$ and $R(K^-\pi^+\pi^-)p$ regions are less significant. The pattern of local asymmetries is consistent with the expectation that the strong phases and the relative magnitudes of the tree and loop amplitudes vary among different resonances. The results are also consistent with previous Run I measurements, and the Run I and Run II results agree with each other within about two standard deviations.

In conclusion, the LHCb collaboration has reported the first observation of $\CP$ violation in a baryon decay, using the decay $\LbtopKpipi$ and its charge-conjugate process. The measurement uses the full Run I and Run II data sets, corresponding to an integrated luminosity of about $9~\mathrm{fb}^{-1}$, and employs the control channel $\Lb\to \Lambda_c^+(\to pK^-\pi^+)\pi^-$ to cancel production, detection, PID, and trigger asymmetries. The global $\CP$ asymmetry is $A_{\CP}=(2.45\pm0.46\pm0.10)\%$, which deviates from zero by $5.2$ standard deviations. Local measurements in four phase-space regions reveal a resonance-dependent pattern, with the most significant asymmetry in the $R(p\pi^+\pi^-)K^-$ region corresponding to excited nucleons, where $A_{\CP}=(5.4\pm0.9\pm0.1)\%$ and the significance is $6.0$ standard deviations. The results are consistent across data-taking periods, magnet polarities, trigger conditions, and kinematic intervals, and they agree with previous Run I measurements. This observation demonstrates that baryons and antibaryons can behave differently in weak decays and provides new constraints on the dynamics of baryon decays.

\subsection{Theoretical Interpretation}

This subsection examines possible theoretical interpretations of the $\CP$ asymmetries measured in the four phase-space regions of $\Lb\to pK^-\pi^+\pi^-$ considered in the LHCb analysis. For comparison, the available theoretical estimates of the $\CP$ asymmetry in each region, together with the corresponding resonant decay channels, are listed in Table~\ref{tab:PScut}. Although these resonances do not necessarily provide a complete description of the respective regions, they offer useful clues to the possible dynamical origins of the observed asymmetries.

The first region, denoted by $R(pK^-)R(\pi^+\pi^-)$ and defined by $m_{pK^-}<2.2~\mathrm{GeV}/c^2$ and $m_{\pi^+\pi^-}<1.1~\mathrm{GeV}/c^2$, is expected to receive substantial contributions from excited $\Lambda$ resonances and light-meson resonances such as $f_0(500)$, $f_0(980)$ and $\rho(770)$; its $\CP$ asymmetry is measured to be $(5.3\pm1.3)\%$. Meanwhile, the $\pi^+\pi^-$ invariant-mass distribution exhibits a pronounced enhancement near $1.0~\mathrm{GeV}$~\cite{LHCb:2025ray}, indicating that the $f_0(980)$ contribution may be particularly important in this region; see Fig.~\ref{fig:LocalMassRegion}. FSI calculations may therefore provide a useful framework for interpreting the $\CP$ asymmetry in this apparently resonance-dominated region. The decays $\Lb\to\Lambda(1520)M$, with $M=f_0(500)$, $\rho^0$ or $f_0(980)$, have been studied within the FSI framework~\cite{Feng:2026soj}. The predicted primary branching fractions are approximately $9.3\times10^{-7}$, $8.5\times10^{-7}$ and $6.0\times10^{-6}$ for these three channels, respectively. After weighting them by the corresponding effective partial widths for $M\to\pi^+\pi^-$, their relative contributions to $\Lb\to\Lambda(1520)\pi^+\pi^-$ are broadly consistent with the experimental spectrum shown in Fig.~\ref{fig:LocalMassRegion}. The corresponding primary-decay $\CP$ asymmetries are predicted to be approximately $1\%$, $26\%$ and $3\%$, respectively. Weighting these asymmetries by the primary-decay rates and the effective secondary-decay rates gives an estimated $\CP$ asymmetry of approximately $10.8\%$ for the full decay chain $\Lb\to\Lambda(1520)\pi^+\pi^-$~\cite{Feng:2026soj}. This value is not fully consistent with the experimentally measured asymmetry, but has the same sign and is of the same order of magnitude. The prediction is particularly sensitive to the poorly known $f_0(980)\to\pi\pi$ coupling, which enters the effective secondary-decay fractions, and should therefore be regarded as a model-dependent estimate. Moreover, while the combined contributions of $f_0(500)$, $f_0(980)$ and $\rho(770)$ broadly capture the measured $\pi^+\pi^-$ spectrum, the $m_{pK^-}$ distribution clearly requires contributions beyond $\Lambda(1520)$, with additional resonant structures visible around $m_{pK^-}\simeq1.7~\mathrm{GeV}$. A more complete theoretical description therefore requires the inclusion of additional resonances to determine whether the observed distributions can be fully understood. 

The second region, denoted by $R(p\pi^{-})R(K^{-}\pi^{+})$, is defined by $m_{p\pi^{-}}<1.7~\mathrm{GeV}$ and either $m_{K^{-}\pi^{+}}\in[0.8,1.0]~\mathrm{GeV}$ or $m_{K^{-}\pi^{+}}\in[1.1,1.6]~\mathrm{GeV}$. It is expected to receive substantial contributions from excited nucleon resonances and $K^{*}$ mesons, and the corresponding $\CP$ asymmetry is measured to be approximately $(2.7\pm0.8)\%$. The quasi-two-body decays $\Lambda^{0}_{b}\to N^{*}(1520)\bar K^{*0}$ and $\Lambda^{0}_{b}\to N^{*}(1535)\bar K^{*0}$ have been studied within the FSI framework~\cite{Shang:2026knt}. Their predicted primary branching fractions are $(1.58^{+1.69}_{-0.86})\times10^{-4}$ and $(1.66^{+1.90}_{-0.95})\times10^{-4}$, respectively, while the corresponding $\CP$ asymmetries are $(2.7^{+3.1}_{-1.9})\%$ and $(0.8^{+0.3}_{-0.2})\%$. Since both the primary branching fractions and the effective secondary-decay rates into the selected final state are comparable, the two contributions may be assigned approximately equal weights, giving the central-value average $1.8\%$, basically consistent with the experimental measurement. Nevertheless, the predicted branching fractions for both channels are considerably larger than the experimental constraints, so this interpretation remains tentative. Moreover, the measured invariant-mass spectra exhibit additional resonant structures beyond these two quasi-two-body channels, indicating that they cannot provide a complete description of the region.

Among the four Dalitz regions reported in the LHCb analysis, the third region, denoted by $R(p\pi^{+}\pi^{-})K^{-}$ and defined by $m_{p\pi^{+}\pi^{-}}<2.7~\mathrm{GeV}$, is the only one for which the statistical significance exceeds $5\sigma$. An estimate within the SM based on the $N\pi$ rescattering mechanism was obtained for this region~\cite{Wang:2024oyi}, as discussed in Sec.~\ref{sec:npis}. The predicted asymmetry is $A_{\rm CP}[\Lb \to (\Delta^{++}\pi^-) K^-]=(5.8\pm0.2)\%$ for $m_{\Delta^{++}\pi^-}<1.9~\mathrm{GeV}$, in good agreement with the experimental result $(5.4\pm0.9\pm0.1)\%$. In the SM, the relevant weak phase is supplied by the CKM ratio $(V^{*}_{tb}V_{ts})/(V^{*}_{ub}V_{us})$. The $N\pi$ rescattering mechanism involves two sources of strong phases: (\textbf{i}) the partial-wave amplitudes for $N\pi\to\Delta^{++}\pi^{-}$ scattering and (\textbf{ii}) the effective Wilson coefficients $a^{\rm eff}_{4,6}$. Under the factorization hypothesis, the first source can be regarded as part of the $N\pi$ wave functions, whereas the latter is obtained from perturbative quark-level calculations and is therefore insensitive to long-distance $N\pi$ rescattering. The total asymmetry $A_{\rm CP}[\Lb \to (\Delta^{++}\pi^-) K^-]$ is obtained by integrating over the Dalitz distribution in the region $m_{\Delta^{++}\pi^-}<1.9~\mathrm{GeV}$, so that the detailed partial-wave structure of the $N\pi$ wave functions is not directly resolved. Although this rescattering-based estimate is intrinsically model dependent, its agreement with the experimental result suggests that the observed asymmetry in this region can be accommodated within the SM. The theoretical uncertainties may nevertheless be underestimated, leaving more room for possible NP contributions than indicated by the quoted numerical result.

The fourth region, denoted by $pR(K^{-}\pi^{+}\pi^{-})$, is defined by $m_{K^{-}\pi^{+}\pi^{-}}<2.0~\mathrm{GeV}$ and is expected to receive substantial contributions from excited strange-meson resonances, including $K_{1}(1270)$, $K_{1}(1400)$ and $K^{*}(1410)$. The $\CP$ asymmetry measured in this region is $(2.0\pm1.2)\%$. Within the PQCD framework, the primary two-body modes $\Lb\to pK_1^-(1270)$ and $\Lb\to pK_1^-(1400)$, which contribute to this region through the subsequent $K_1^-\to K^-\pi^+\pi^-$ decays, have been calculated~\cite{Han:2025tvc}. Since the $K_{1}(1270)$-$K_{1}(1400)$ mix angle $\theta_{K}$ is not yet well determined, two commonly used benchmark values, $\theta_K=30^\circ$ and $60^\circ$, are adopted. For $\theta_K=30^\circ$, the predicted $\CP$ asymmetries for $\Lb\to pK_1^-(1270)$ and $\Lb\to pK_1^-(1400)$ are $(9^{+8}_{-5})\%$ and $(6^{+7}_{-9})\%$, respectively, while for $\theta_K=60^\circ$ they are $(7^{+5}_{-6})\%$ and $(8^{+18}_{-14})\%$. The corresponding central branching fractions are $5.48\times10^{-6}$ and $1.25\times10^{-6}$ for $\theta_K=30^\circ$, and $6.28\times10^{-6}$ and $0.53\times10^{-6}$ for $\theta_K=60^\circ$, indicating that the $K_1(1270)$ contribution is expected to be dominant. If these primary branching fractions are used as weights, while differences between the effective $K_1\to K^-\pi^+\pi^-$ decay fractions and interference between the two resonances are neglected, the rate-weighted central asymmetry is approximately $8.4\%$ for $\theta_K=30^\circ$ and $7.1\%$ for $\theta_K=60^\circ$. The predicted asymmetry of the dominant $K_1(1270)$ channel therefore depends only mildly on $\theta_K$. The $K^{*}(1410)$ contribution has not been included because of the lack of a suitable LCDA. Although the central weighted estimates are larger than the experimental central value, their sizable theoretical uncertainties overlap with the measurement. The calculated $K_1$ channels may therefore provide a possible clue to the observed asymmetry, but do not yet constitute a complete description of this region.

It should be emphasized again that the theoretical interpretations of the measured asymmetries in the four Dalitz regions discussed above remain largely phenomenological. They should therefore not yet be regarded as precision tests of the SM in baryon decays. Further theoretical progress will require a deeper and more quantitative understanding of the underlying decay dynamics.

\subsection{Flavor Symmetry Analysis}

Before systematically controlled high-precision studies of beauty-baryon decays become feasible, flavor-symmetry analyses provide a useful framework for organizing decay amplitudes, relating observables across different channels, and guiding experimental investigations. The smallness of up-, down- and strange-quark masses compared to QCD scale $\Lambda_{\text{QCD}}$ makes $SU(3)$ flavor an approximate symmetry of strong interaction. It obviously provides a potential simplification for the theoretical studies of heavy hadron weak transitions. On the one hand, $SU(3)$ flavor symmetry analysis reduces the number of independent hadronic matrix elements, thereby simplifying the theoretical treatment. On the other hand, it allows us to establish some relations among branching fractions and $\CP$ violations in different decays, and to get information about the other decays by employing existed experimental measurements. It should be emphasized, however, that the flavor symmetry approach by itself cannot determine the decay matrix elements. It is usually applied in conjunction with other dynamical approaches, such as factorization frameworks and hadronic rescattering, discussed in Sec.~\ref{sec:compute}, and can also be combined with experimental inputs to do global fits under appropriate theoretical approximations. With the first observation of $\CP$ violation in $\Lb\to pK^{-}\pi^{+}\pi^{-}$, flavor symmetry analyses are expected to become increasingly important. It may provide valuable insight for future experimental investigations, and help uncover new patterns of $\CP$ violation in beauty baryon decays.

The basic idea of flavor symmetry approach is rather straightforward. The light quarks $(u,d,s)$ and antiquarks $(\bar{u},\bar{d},\bar{s})$ can be assigned to the fundamental representations $\textbf{3}$ and $\bar{\textbf{3}}$ of $SU(3)$ group respectively. 
The hadrons appearing in decays are naturally classified according to the irreducible representations of $SU(3)$ group. For example, the heavy mesons $(B^+,B^0,B_s^0)=(\bar{b}u,\bar{b}d,\bar{b}s)$ transforms as a triplet $\textbf{3}$ in flavor space. The same is true for the $D$ mesons $(\bar{D}^{0},D^{-},D^{-}_{s})$. The light pseudoscalar mesons $\pi^{\pm},K^{\pm},\pi^{0},K^{0},\bar{K}^{0}$ and $\eta_{8}$ form an octet representation $\mathbf{8}$, usually denoted as $M^{i}_{j}$, according to the decomposition rule $\mathbf{3}\otimes\bar{\mathbf{3}}=\mathbf{8}\oplus 1$. The heavy beauty and charm quark are  $SU(3)$ flavor singlet, and the beauty and charm baryons are classified by $1\otimes\mathbf{3}\otimes\mathbf{3}=\bar{\mathbf{3}}\oplus \mathbf{6}$, where $\bar{\mathbf{3}}$ and $\mathbf{6}$ are antitriplet and sextet representations.

In order to proceed with the analysis, one must identify the most general $SU(3)$ transformation structure of the effective Hamiltonian. Since decay amplitudes are determined by matrix elements $\bra{F}\mathcal{H}_{\textbf{eff}}\ket{I}$, the effective Hamiltonian $\mathcal{H}_{\textbf{eff}}$ must transform under those irreducible representations that appear in the direct products of the representation spaces to which the initial state $\ket{I}$ and $\ket{F}$ belong.
Otherwise, the associated matrix elements are vanishing according to the Wigner-Eckart theorem, which imply the decays are forbidden in the exact $SU(3)$ flavor symmetry limit. 
The effective operators in Eqs.~\eqref{tree operators}, \eqref{QCD penguin operators}, \eqref{EW penguin operators}, and \eqref{b to c operator} can be regarded as $SU(3)$ tensors. Their decomposition into irreducible representations can therefore be carried out straightforwardly using the Young tableau technique  \cite{Georgi:2000vve}. 

As the simplest example, for the quark level transition $b\to qc\bar{c}$, the tree operator such as $(\bar{c}b)_{V-A}(\bar{q}c)_{V-A}$ can be directly identified as transforming as an $\bar{\textbf{3}}$ representation. At the hadron level, all two body decays of the type $B\to D\bar{D}$ arising from above operators with $q=s(d)$ for Cabibbo allowed (suppressed) decays can be determined through two parameters. This can be understood in a straightforward way. Firstly, consider couplings of the irreducible representation of initial $B$ meson, $B^{i}$, with those of the final mesons, $D^{j}$ and $\bar{D}^{k}$. The tensor product of the external states can only contain two forms, i.e., $B^{i}(D_{j}\bar{D}^{j})$ and $(B^{j}\bar{D}_{j})D^{i}$, in order to form a $SU(3)$ flavor singlet with Hamiltonian. 
The decay amplitudes can therefore be parameterized as
$ A(B\to D\bar{D})=a (B^{i}\mathcal{H}_{i})(D_{j}\bar{D}^{j})+b(B^{j}\bar{D}_{j})(D^{i}\mathcal{H}_{i})$, where $a,b$ define two independent reduced matrix elements. 
The amplitudes of different channels then differ only by the Clebsch-Gordan coefficients appearing in the contraction of representation vectors. To be precise, for the decay $\bar{B}^{0}_{s}\to D^{+}D^{-}$, one can identify $B^{i}=B^{3},D^{i}=D^{2}$ and $ \bar{D}^{j}=\bar{D}^{2}$, yielding $ A(\bar{B}^{0}_{s}\to D^{+}D^{-})=a$. Similarly, one obtains $ A(\bar{B}^{0}_{s}\to D^{+}_{s}D^{-}_{s})=a+b$ and $ A({B}^{-}\to D^{+}_{s}\bar{D}^{0})=b$.
It is clear that experimental measurements of their branching fractions can be used to constrain the magnitudes and relative phase of $a$ and $b$, up to discrete ambiguities. Once these parameters are fixed, the amplitudes of other channels related by flavor symmetry, such as $\bar{B}^{0}\to D^{+}_{s}{D}^{-}$, can be predicted. This simplest example already illustrates the basic idea underlying flavor symmetry analyses. The treatment of more complicated decays follows the same general principles. 

Flavor-symmetry methods have been widely applied to $B$- and $D$-meson decays \cite{Zeppenfeld:1980ex,
Savage:1989ub,
Grinstein:1996us,
Wang:2020gmn,
Gronau:1994rj,Deshpande:2000jp,
Grossman:2003qp,
Gronau:2013mda,Bertholet:2018tmx,
Gronau:1994bn,
Hsiao:2015iiu,
Grossman:2012ry,
He:1998rq,
Cheng:2016ejf,
Cheng:2014rfa,
Cheng:2012xb,
Bhattacharya:2008ss}, and even stable tetraquark decays \cite{Li:2018bkh,
Xing:2019hjg,
Xing:2019wil}. Extensive theoretical analyses have also been carried out for charm baryon decays based on rich experimental data of branching ratio and asymmetry parameters \cite{Lu:2016ogy,
Geng:2017mxn,
Geng:2018plk,
Geng:2018bow,
Geng:2018upx,
Zhao:2018mov,
Geng:2018rse,
Wang:2017azm,
Wang:2019dls,
Geng:2019xbo,
Cen:2019ims,
Jia:2019zxi,Xing:2024nvg,Sun:2024mmk,Xing:2023dni}. For beauty baryons, however, analogous investigations are more challenging due to the lack of necessary experimental inputs, particularly for asymmetry parameters. For charmless decays of beauty baryons into a light baryon and a pseudoscalar meson, such as $\Lambda^{0}_{b}\to \Delta^{+}\pi^{-}$ and $p\pi^{-}$, many sum rule relations among decaying amplitudes and $\CP$ asymmetries have been derived \cite{Dery:2020lbc,
He:2015fsa,
Roy:2019cky,
Roy:2020nyx,He:2015fwa,Wang:2025bdl}. However, owing to the lack of dynamical input, such analyses generally do not give quantitative predictions to $\CP$ asymmetries. As an alternative and simpler approach, one may consider the isospin symmetry between the $u$ and $d$ quarks, or the $U$-spin symmetry relating the $d$ and $s$ quarks. These $SU(2)$ symmetries analysis also give rise to many sum rule relations among decay amplitudes and $\CP$ asymmetries \cite{Fu:2025wwx,
Wen:2025ibn,
Gavrilova:2022hbx,
Zhang:2025jnw,
Roy:2025nao}. Some quantitative predictions for $\CP$ asymmetries could even be given by ignoring subleading contributions. For example, based on the $U$-spin symmetry, the $\CP$ asymmetries for the four body decay \cite{He:2025msg,
Chen:2025puj}
\begin{equation} 
\begin{aligned}
A_{\CP}^{\rm dir} \bigl(\Lb \to p\pi^-\pi^+\pi^-\bigr) |_{\rm U\text{-}spin} &= (-12\pm6)\% ,\\
A_{\CP}^{\rm dir} \bigl(\Lb \to pK^+ K^-\pi^-\bigr) |_{\rm U\text{-}spin} &= (-0.3\pm3.0)\% ,
\end{aligned}
\end{equation}
and for three body decays \cite{Wang:2024rwf}
\begin{equation} 
\begin{aligned}
A_{\CP}^{\rm dir}(\Lb\to \Sigma^0 K^+K^-)|_{\text{U-spin}} &= (8.3 \pm 2.8)\%, \\ 
A_{\CP}^{\rm dir}(\Xi_b^0\to \Lambda \pi^+\pi^-)|_{\text{U-spin}} &=(-9.0 \pm 5.1)\%, \\
A_{\CP}^{\rm dir}(\Lb\to \Sigma^0 K^+\pi^-)|_{\text{U-spin}} &= (-11.8 \pm 5.8)\%, \\ 
A_{\CP}^{\rm dir}(\Xi_b^0\to \Sigma^0 K^-\pi^+)|_{\text{U-spin}} &=(27 \pm 13)\%
\end{aligned}
\end{equation}
are predicted and therefore provide good probes for future experimental tests.

%% file: sec6-observable.tex
The nonzero spins of baryons give rise to rich polarization phenomena in both their production and decay. Consequently, $\CP$-violating observables beyond the conventional direct $\CP$ asymmetry $A_{\CP}^{\rm dir}$ can be constructed from correlations among spins and momenta and their transformation properties under $\CP$. A classic example is the $\CP$ asymmetry associated with the longitudinal polarization of the final-state baryon, $\langle \vec S\cdot\hat p\rangle$, commonly described by a decay asymmetry parameter in spin-$1/2$ baryon decays~\cite{Lee:1957qs}. Such observables probe interference among helicity or partial-wave amplitudes. Because different observables select different real or imaginary interference terms, they provide complementary sensitivity to weak- and strong-phase differences and can help discriminate among theoretical descriptions of beauty-baryon decays.

In recent years, polarization observables, including decay asymmetry parameters in spin-$1/2$ baryon decays, have been extensively studied in hyperon and heavy-baryon decays. In particular, with the development of modern partial-wave analysis technique, helicity or partial-wave amplitudes can be determined in weak decay chains. Numerous hyperon and charm-baryon decays have been studied by BESIII~\cite{BESIII:2026baz,BESIII:2026dkj,BESIII:2026mye,BESIII:2026hgj,BESIII:2025wxe,BESIII:2025jxt,BESIII:2023drj,BESIII:2023sgt,BESIII:2022qax} and Belle~\cite{Belle:2022uod}, including searches for $\CP$ violation based on decay asymmetry parameters. At BESIII, baryon--antibaryon pair production provides access to spin correlations and, in suitable channels, transverse polarization, enabling precision measurements of decay asymmetry parameters in hyperon and charm-baryon decays~\cite{BESIII:2018cnd}. To date, BESIII has reported the most precise determination of the decay asymmetry parameter $\alpha$ in $\Lambda\to p\pi^-$~\cite{BESIII:2022qax}. Likewise, hyperons produced in charm-baryon decays can be polarized, thereby facilitating measurements of the asymmetry parameters in their subsequent parity-violating decays. Using the decay chains $\Lambda_c^+\to\Lambda\pi^+\to(p\pi^-)\pi^+$ and $\Lambda_c^+\to\Sigma^0\pi^+\to(\Lambda\gamma)\pi^+\to(p\pi^-\gamma)\pi^+$, Belle searched for $\CP$ violation in the decay asymmetry parameter of $\Lambda\to p\pi^-$~\cite{Belle:2022uod}. For directly produced beauty baryons at LHCb, a sizable initial-state polarization is more difficult to exploit. Nevertheless, multibody decay distributions allow new $\CP$-violating observables to be constructed from final-state momenta, {\it e.g.}, the momentum triple-product asymmetries measured in $\Lb$ decays~\cite{LHCb:2016yco,LHCb:2018fpt,LHCb:2019oke,LHCb:2025mnp}.

In this section, we review phenomenological studies of $\CP$-violating observables constructed from baryon spin and multidimensional decay kinematics, with particular emphasis on beauty baryons. We first discuss why phase-space-integrated rate asymmetries may obscure $\CP$-violating information. We then introduce a broader class of $\CP$ violating observables associated with baryon polarizations, including those arising from decay asymmetry parameters, partial waves and multi-dimensional angular analyses.

\subsection{Beyond Integrated Rate Asymmetries}

In general, $\CP$ violation is signaled if, for at least one pair of $\CP$-conjugate phase-space configurations, the corresponding differential decay rates satisfy
\begin{equation}
\begin{aligned}
d\Gamma\!\left[ i(p_i,\lambda_i)   \to  f_1(p_1,\lambda_1)+\cdots+f_m(p_m,\lambda_m) \right] \neq
d\Gamma\!\left[ \bar i(-p_i,-\lambda_i) \to \bar f_1(-p_1,-\lambda_1)+\cdots+\bar f_m(-p_m,-\lambda_m) \right].
\end{aligned}
\end{equation}
Here, $p_k$ and $\lambda_k$ denote the three-momenta and helicities of the particles, respectively, which both reverse sign under the $\CP$ transformation. The conventional direct $\CP$ asymmetry $A_{\CP}^{\rm dir}$ compares decay rates integrated over phase space and therefore discards information contained in the momentum distributions~\cite{Kayser:1989vw,Nowakowski:1988dx,Nowakowski:1989ju}. These distributions encode interference among different partial-wave or helicity amplitudes and therefore contain information about spin and polarization in the decay. The integrated asymmetry can consequently be diluted by cancellations between different phase-space regions or partial-wave components, as illustrated by the PQCD and FSI calculations of $\Lb$ decays discussed in Sec.~3. Moreover, $A_{\CP}^{\rm dir}$ is proportional to $\sin\Delta\delta\sin\Delta\phi$, as shown in Eq.~\eqref{eq:acpn}. When the strong-phase difference $\Delta\delta$ is small, as occurs in several hyperon decays into a nucleon and a pion, the resulting rate asymmetry is strongly suppressed~\cite{Donoghue:1986hh}. This motivates searches by BESIII and Belle for additional $\CP$-violating effects encoded in angular distributions rather than relying only on integrated rate asymmetries.

Angular information is also important in specific meson decays. Prominent examples include $B\to V_1V_2$ and $B\to K^*\ell^+\ell^-$, where nontrivial angular distributions provide complementary tests of theoretical predictions and probes of possible physics beyond the SM~\cite{Valencia:1988it,Datta:2003mj,Matias:2012xw,Descotes-Genon:2012isb,LHCb:2013ghj}. Baryon decays contain additional spin degrees of freedom and often involve more helicity or partial-wave amplitudes. Their multidimensional distributions can therefore provide more opportunities to identify $\CP$-violating effects than the mesonic ones.

From the theoretical perspective, reliable calculations of strong phases remain challenging because of the complexity of nonperturbative strong-interaction dynamics. In selected channels, strong-phase information can be constrained using data-driven analyses of processes such as $\pi\pi$ and $N\pi$ scattering, or estimated within QCD-based factorization frameworks and hadronic-rescattering models. However, these theoretical approaches have their respective limitations, making complementary experimental information important both for testing their descriptions and for constraining relative phases. In particular, observables sensitive to interference between different partial-wave or helicity amplitudes provide such an experimental handle: their characteristic angular dependence allows the interference terms to be measured, thereby constraining combinations of relative strong and weak phases. For example, $\CP$-odd observables involving the Lee--Yang parameter $\beta$ in hyperon decays can depend on $\cos\Delta\delta$ rather than $\sin\Delta\delta$. Such observables turn the complexity of baryon phase space distributions into an advantage, complementing theoretical calculations and experimental information in studies of $\CP$ violation~\cite{BESIII:2021ypr}.

The advantages of multidimensional analyses can be summarized in three aspects. First, they provide access to interference between different partial-wave or helicity amplitudes. Even if the $\CP$ asymmetries associated with the $S$- and $P$-wave contributions cancel in $A_{\CP}^{\rm dir}$, $\CP$-violating differences involving $\operatorname{Re}(SP^*)$ or $\operatorname{Im}(SP^*)$ can remain nonzero. These terms multiply different angular structures and can therefore be separated experimentally. Second, angular distributions can reflect interference between different resonant contributions. For example, both the $f_0$ and $\rho^0$ resonances can contribute in the low-$m_{\pi^+\pi^-}$ region of $\Lb\to pK^-\pi^+\pi^-$. Their interference generates terms proportional to $\operatorname{Re}(\mathcal{A}_{f_0}\mathcal{A}_{\rho^0}^*)$ with characteristic angular dependence~\cite{Zhang:2013oqa}. Such contributions may vanish after complete angular integration but remain accessible in angular bins or through suitably weighted observables. Third, they can provide information about imaginary parts of amplitude interferences, such as $\operatorname{Im}(SP^*)$, which are often named $T$-odd correlations with $T$ being the time reversal~\cite{Wang:2022fih}. The corresponding $\CP$-odd combinations can be proportional to $\cos\Delta\delta\sin\Delta\phi$, rather than $\sin\Delta\delta\sin\Delta\phi$, and thus remain sensitive when the strong-phase difference is small. These examples demonstrate that angular analyses can reveal $\CP$-violating information that is inaccessible through the integrated asymmetry $A_{\CP}^{\rm dir}$ alone. The spin and angular formalism needed to construct these observables is introduced in the following subsection.

\subsection{Spin-Relevant Observables and Angular Distributions}\label{Angular Distributions}

The angular dependence of decay and scattering amplitudes is constrained by rotational symmetry. For a $2\to2$ scattering process involving spinless particles, such as $ab\to cd$, the amplitude can be expanded in Legendre polynomials,
\begin{equation}
\mathcal{M}(\theta)
=\sum_{l}(2l+1)P_l(\cos\theta)\,f_l,
\end{equation}
where $\theta$ is the scattering angle and $f_l$ is the partial-wave amplitude. For processes involving particles with nonzero spin, however, this expansion is less convenient because the angular dependence also involves spin couplings. The helicity-amplitude formalism~\cite{Chou:1958polarized,Jacob:1959at} provides a useful framework for describing such processes. The helicity $\lambda$ is the projection of a particle's spin along its direction of motion, $\lambda=\vec{S}\cdot\vec{p}/|\vec{p}|$. For a massive particle with spin $s$, the plane-wave helicity states $\ket{\vec p,\lambda}$, with $\lambda=-s,\ldots,s$, form a complete basis. Under a spatial rotation, the momentum and spin are rotated simultaneously, so that the helicity eigenvalue $\lambda$ remains unchanged. These states can therefore be projected onto states of definite total angular momentum $j$, denoted by $\ket{p,j,m;\lambda}$, where $p=|\vec p|$ and $m$ is the projection of $j$ along a chosen quantization axis, which need not coincide with the direction of motion. The helicity formalism is equally applicable to massless particles, such as photons, and is particularly useful in high energy processes where the masses of light fermions are negligible~\cite{Chou:1958polarized}.

Within this formalism, the angular distributions of particle decay and scattering processes can be formulated in a particularly convenient way. A helicity state $\ket{p,\theta,\phi,\lambda}$ with momentum along the direction $(\theta,\phi)$ is obtained by rotating a reference state $\ket{p,0,0,\lambda}$ whose momentum is aligned with the $z$ axis:
\begin{equation}
\ket{p,\theta,\phi,\lambda} =R(\phi,\theta,0)\ket{p,0,0,\lambda}.
\end{equation}
Following the Jacob--Wick convention~\cite{Jacob:1959at}, the state can be expanded in states of definite total angular momentum,
\begin{equation}\label{plane wave state}
\ket{p,\theta,\phi,\lambda}
=\sum_{j,m}
\sqrt{\frac{2j+1}{4\pi}}\,
D^{j}_{m,\lambda}(\phi,\theta,0)
\ket{p,j,m;\lambda},
\end{equation}
where $D^{j}_{m,\lambda}(\phi,\theta,0)$ is a Wigner $D$-function. The helicity state of a two-particle system is the direct product of the states of its constituents, $\ket{\vec p_1,\lambda_1;\vec p_2,\lambda_2} = \ket{\vec p_1,\lambda_1}\otimes \ket{\vec p_2,\lambda_2}$. In the rest frame of the decaying particle, momentum conservation requires the two final-state particles to move back to back. Denoting their common momentum magnitude by $p$ and the direction of particle 1 by $(\theta,\phi)$, we define $\ket{p,\theta,\phi;\lambda_1,\lambda_2} \equiv \ket{p\hat n(\theta,\phi),\lambda_1;  -p\hat n(\theta,\phi),\lambda_2}$, where the momentum direction of particle 2 is $(\pi-\theta,\phi+\pi)$. Since its momentum points opposite to that of particle 1, the projection of the two-particle helicity state along $\hat n(\theta,\phi)$ is $\lambda_1-\lambda_2$. For a two-body decay of an initial state with spin $j$ and spin projection $m$, the decay amplitude is therefore
\begin{equation}
\langle p,\theta,\phi;\lambda_1,\lambda_2|\mathcal{H}|jm\rangle = \sqrt{\frac{2j+1}{4\pi}}\, D^{j*}_{m,\lambda_1-\lambda_2}(\phi,\theta,0) \mathcal{H}_{\lambda_1,\lambda_2},
\end{equation}
where $\mathcal{H}_{\lambda_1,\lambda_2}$ is the angle-independent helicity amplitude. This expression follows from rotational invariance, or equivalently from the Wigner--Eckart theorem: the initial and final states must have the same total angular momentum, and the dependence on the angular coordinates is encoded in the Wigner $D$-functions. The helicity amplitudes contain the remaining dynamical information. Squaring the amplitude and summing over the final-state helicities gives
\begin{equation}\label{eq:differential distribution}
\begin{aligned}
f(\theta,\phi)_{m',m} = \sum_{\lambda_1,\lambda_2} \frac{2j+1}{4\pi} D^{j*}_{m,\lambda_1-\lambda_2}(\phi,\theta,0) D^{j}_{m',\lambda_1-\lambda_2}(\phi,\theta,0) \times \mathcal{H}_{\lambda_1,\lambda_2} \mathcal{H}^{*}_{\lambda_1,\lambda_2}.
\end{aligned}
\end{equation}
For an initial state with total angular momentum $j$ described by the spin-density matrix $\rho_{m,m'}$, the angular distribution is obtained by contracting this function with the density matrix,
\begin{equation}\label{eq:angular distribution}
\frac{d\Gamma}{d\Omega} \propto \sum_{m,m'} f(\theta,\phi)_{m',m}\,\rho_{m,m'} = \operatorname{Tr}\!\left[\rho\cdot f(\theta,\phi)\right].
\end{equation}

The formalism can be extended to cascade decays by retaining the helicity indices of an intermediate particle rather than summing over them. For example, in the same process described above, the spin-density matrix of the final-state baryon with helicity indices $\lambda_1,\lambda_1'$ can be written as
\begin{equation}\label{eq:spin-density-matrix}
\begin{aligned}
\rho^{f}_{\lambda_1,\lambda_1'}(\theta,\phi) \propto \left(\frac{d\Gamma}{d\Omega}\right)_{\lambda_{1},\lambda^{\prime}_{1}} \propto  \sum_{m,m',\lambda_2} \frac{2j+1}{4\pi} D^{j*}_{m,\lambda_1-\lambda_2}(\phi,\theta,0) D^{j}_{m',\lambda_1'-\lambda_2}(\phi,\theta,0) \times \mathcal{H}_{\lambda_1,\lambda_2} \mathcal{H}^{*}_{\lambda_1',\lambda_2} \rho_{m,m'}.
\end{aligned}
\end{equation}
This density matrix can then serve as the initial spin-density matrix for the subsequent decay of that baryon. For instance, in $\Lb\to\Lambda\pi^0\to(p\pi^-)\pi^0$, one first obtains the density matrix of the intermediate $\Lambda$ and then uses it to construct the angular distribution of the full decay chain. The polarization of a particle is defined by its expectation value of the spin operator $\hat{\vec S}$. For a normalized spin-density matrix $\rho^f$, the $i$th component is $\langle S_i\rangle=\operatorname{Tr}(S_i\rho^f)$. Taking a spin-$1/2$ particle as an example, the polarization components are conventionally written as $\mathcal{P}_i=\operatorname{Tr}(\sigma_i\rho^f)$, where $\sigma_i$ are the Pauli matrices. Equations~\eqref{eq:differential distribution} and~\eqref{eq:spin-density-matrix} therefore establish the relation between the polarization observables and the helicity amplitudes of decay processes.

\subsection{One-dimensional angular analysis and asymmetry parameter}

The full angular distributions of many cascade and multibody decays depend on several independent angular variables and are therefore intrinsically multidimensional. A complete angular analysis can consequently be experimentally challenging. Integrating over all but one angular variable reduces the distribution to a one-dimensional form and substantially simplifies the analysis, although at the cost of losing part of the multidimensional information. Various asymmetry parameters can be defined from such one-dimensional distributions. In the following, we discuss three representative examples and the corresponding $\CP$-violating observables. Complete multidimensional angular analyses will be considered in the next subsection.

The Lee--Yang asymmetry parameters were originally introduced to characterize parity violation in two-body decays of spin-$1/2$ baryons~\cite{Lee:1957qs}. As an illustrative example, consider the decay $\Lb\to p\pi^-$. In terms of the helicity amplitudes introduced in the preceding subsection, we define $h_\pm \equiv \mathcal H_{\pm 1/2,0}$, where the first helicity index is associated with the final proton and the second helicity index vanishes because the pion has spin zero. For an initial pure spin state $\ket{\psi_i} = c_+\ket{+\frac12} + c_-\ket{-\frac12}$ and with the spin-quantization $z$-axis chosen along the proton momentum, the conditional final state is
\begin{equation}
\ket{\psi_f} \propto c_+h_+\ket{+\frac12} + c_-h_-\ket{-\frac12}.
\end{equation}
Thus, the spin state of the daughter baryon depends on both the initial spin state and the helicity amplitudes of the decay.

For a generally polarized initial baryon, the spin state is described by a density matrix. Taking the proton momentum as the $z$ axis, let the initial polarization vector be $\vec{\mathcal P} = \mathcal P ( \sin\theta\cos\phi,\, \sin\theta\sin\phi,\, \cos\theta )$, such that the angle between the initial baryon polarization and the final baryon memontum is $\theta$, as illustrated by panel (a) of Fig.~\ref{fig-LambdaPolarization}. The initial spin-density matrix is then $\rho_i =  ( I+\vec{\mathcal P}\cdot\vec{\sigma} ) /2$. A pure spin state corresponds to $\mathcal P=1$, whereas $0<\mathcal P<1$ describes a partially polarized ensemble and $\mathcal P=0$ an unpolarized one. The unnormalized spin-density matrix of the final proton is given by 
\begin{equation}\label{eq:rhof}
\widetilde\rho_f = T\rho_iT^\dagger = \frac12
\begin{pmatrix}
|h_+|^2(1+\mathcal P\cos\theta) & \mathcal P\sin\theta\,e^{-i\phi}h_+h_-^* \\
\mathcal P\sin\theta\,e^{i\phi}h_-h_+^* & |h_-|^2(1-\mathcal P\cos\theta)
\end{pmatrix}, \qquad 
T= \begin{pmatrix} h_+&0\\ 0&h_- \end{pmatrix}.
\end{equation}
Taking its trace, which corresponds to summing over the proton spins, and subsequently normalizing and integrating over the azimuthal angle, we can obtain the angular distribution 
\begin{equation}\label{eqangular distribution}
\frac{1}{\Gamma}\frac{d\Gamma}{d\cos\theta} = \frac12 \left(1+\mathcal P\alpha\cos\theta\right),
\end{equation}
where the Lee--Yang decay asymmetry parameter is given by
$\alpha = (|h_+|^2-|h_-|^2)/(|h_+|^2+|h_-|^2)$, as defined in Eq.~(\ref{eq:alpha-helicity}),
or equivalently by $\alpha=2\operatorname{Re}(S^*P)/(|S|^2+|P|^2)$ in terms of partial-wave amplitudes as shown in Eq.~(\ref{eq:alpha-SP}). 
Thus, $\alpha$ characterizes the correlation between the initial polarization and the proton momentum. Moreover, the longitudinal polarization of
the final proton can also be obtained from the final proton density matrix \eqref{eq:rhof} by 
\begin{equation}
\mathcal P_L(\theta) = \frac{\operatorname{Tr} \left(\sigma_z\widetilde\rho_f\right)} {\operatorname{Tr}\widetilde\rho_f}
= \frac{\alpha+\mathcal P\cos\theta}{1+\alpha\mathcal P\cos\theta}.
\end{equation}
For an unpolarized initial baryon, $\mathcal P=0$, the angular distribution is isotropic and $\mathcal P_L=\alpha$. The parameter $\alpha$ therefore has two related interpretations: it governs the angular correlation with the initial polarization and, for an unpolarized parent, equals the longitudinal polarization of the daughter baryon. 

Throughout this review, we use $\alpha$ to denote the decay asymmetry parameter in $\mathcal B_i\to\mathcal B_fP$, where $\mathcal B_i$ and $\mathcal B_f$ are spin-$1/2$ baryons and $P$ is a pseudoscalar meson. It should not be confused with the CKM unitarity angle $\alpha$ in Eq.~\eqref{eq:ut_angles}. Because parity reverses the helicities, it interchanges the two helicity rates entering Eq.~\eqref{eq:alpha-helicity}. The parameter $\alpha$ is therefore parity odd and vanishes if the decay conserves parity. For a polarized parent baryon, $\alpha$ can be extracted from the angular distribution in Eq.~\eqref{eqangular distribution} through the forward--backward asymmetry
\begin{equation}
A_{\rm FB}
\equiv
\frac{1}{\Gamma}
\int_{-1}^{1}
\frac{d\Gamma}{d\cos\theta}\,
\operatorname{sgn}(\cos\theta)\,d\cos\theta
=
\frac{\mathcal P\alpha}{2}.
\end{equation}
Thus, $\mathcal P\alpha$ measures the difference between the $\cos\theta>0$ and $\cos\theta<0$ regions. The term proportional to $\alpha$ is odd in $\cos\theta$ and therefore vanishes upon integration over the full angular range. It should be emphasized that the initial polarization $\mathcal P$ is crucial for the extraction of $\alpha$. In the unpolarized limit, $\mathcal P=0$, the distribution becomes isotropic even if $\alpha$ is nonzero. This does not imply that $\alpha$ vanishes: as shown above, it remains equal to the longitudinal polarization of the daughter baryon and may be accessed if an appropriate spin analyzer is available. More generally, nonzero initial-state polarization enriches the angular distributions of decay and scattering processes.

\begin{table*}[t]
    \footnotesize
    \centering
    \renewcommand{\arraystretch}{1.5}
    \caption{Predictions for the decay asymmetry parameter $\alpha$, its $\CP$-even average  $\langle\alpha\rangle=(\alpha-\bar\alpha)/2$, and the $\CP$-odd combination $A_{\CP}^{\alpha}=(\alpha+\bar\alpha)/2$ in $\Lb\to ph^-$ decays. A dash indicates that the corresponding charge-conjugate combination was not reported.} \label{tab:observables}
    \begin{tabular*}{150mm}{c@{\extracolsep{\fill}}ccccc}
        \hline\hline
Decay mode
& Observable & GF~\cite{Mohanta:2000nk} & QCDF (diquark)~\cite{Zhu:2018jet} & PQCD~\cite{Han:2024kgz,Han:2025tvc} & FSIs~\cite{Duan:2024zjv} \\
        \hline
$\Lb\to p\pi^-$ & $\alpha$ & $-0.77$ & $-0.98^{+0.01}_{-0.01}$ & $-0.94^{+0.03}_{-0.02}$ & $-0.75^{+0.18}_{-0.13}$ \\
& $\langle\alpha\rangle$ & -- & -- & $-0.96^{+0.02}_{-0.02}$ & $-0.08^{+0.02}_{-0.01}$ \\
& $A_{\CP}^{\alpha}$ & -- & -- & $0.02^{+0.01}_{-0.02}$ & $-0.67^{+0.20}_{-0.15}$ \\
        \hline
$\Lb\to pK^-$ & $\alpha$ & $-0.77$ & $0.27^{+0.19}_{-0.13}$ & $0.23^{+0.15}_{-0.12}$ & $0.97^{+0.02}_{-0.05}$ \\
& $\langle\alpha\rangle$ & -- & -- & $0.20^{+0.18}_{-0.14}$ & $0.20^{+0.02}_{-0.05}$ \\
& $A_{\CP}^{\alpha}$ & -- & -- & $0.04^{+0.03}_{-0.04}$ & $0.77^{+0.05}_{-0.07}$ \\
        \hline      \hline
    \end{tabular*}
\end{table*}

For the charge-conjugate decay, it is useful to introduce the $\CP$-even average and the $\CP$-odd combination
\begin{equation}
\langle\alpha\rangle
=
\frac{\alpha-\bar\alpha}{2},
\qquad
A_{\CP}^{\alpha}
=
\frac{\alpha+\bar\alpha}{2},
\end{equation}
so that $\alpha=\langle\alpha\rangle+A_{\CP}^{\alpha}$. More generally, let $\mathcal O$ be an observable with a definite parity signature $\eta_{\mathcal O}=\pm1$, and let $\bar{\mathcal O}$ denote the corresponding observable in the charge-conjugate decay using the same angular convention. Under $\CP$ conservation, $\mathcal O=\eta_{\mathcal O}\bar{\mathcal O}$, and a normalized $\CP$ asymmetry can be defined as
\begin{equation}\label{CPV def}
a_{\CP}^{\mathcal O}
=
\frac{\mathcal O-\eta_{\mathcal O}\bar{\mathcal O}}
{\mathcal O+\eta_{\mathcal O}\bar{\mathcal O}}.
\end{equation}
Since $\alpha$ is parity odd, its normalized $\CP$ asymmetry is $a_{\CP}^{\alpha} = {(\alpha+\bar\alpha)}/{(\alpha-\bar\alpha)} ={A_{\CP}^{\alpha}}/{\langle\alpha\rangle}$\footnote{The notation for this observable is not uniform in the literature. Recent PQCD and hadronic-rescattering calculations use $A_{\CP}^{\alpha}$ or $a_{\CP}^{\alpha}$ for the unnormalized combination $(\alpha+\bar\alpha)/2$. Rate-weighted definitions, $(\alpha\Gamma+\bar\alpha\bar\Gamma)/(\Gamma+\bar\Gamma)$ and $(\alpha\Gamma+\bar\alpha\bar\Gamma)/(\alpha\Gamma-\bar\alpha\bar\Gamma)$, were adopted in Refs.~\cite{Geng:2023nia} and~\cite{Donoghue:1986hh}, respectively. An alternative normalization rule to loyally reflect statistical uncertainties was proposed in Ref.~\cite{Qi:2025zna}. Here, we distinguish the unnormalized combination $A_{\CP}^{\alpha}$ from the normalized asymmetry $a_{\CP}^{\alpha}$ commonly used in the BESIII and Belle measurements.}.
By decomposing the $S$- and $P$-wave amplitudes into their tree and penguin components, as in Eqs.~\eqref{eq:SPWwave} and~\eqref{eq:SPbarWwave}, these $\CP$-odd quantities can be expressed in terms of the corresponding weak- and strong-phase differences~\cite{Donoghue:1986hh}. Theoretical predictions for the $\alpha$, $\langle\alpha\rangle$, and $A_{\CP}^{\alpha}$ in $\Lb\to ph^-$ decays using various approaches are shown in Table~\ref{tab:observables}. It is observed that the predictions differ substantially among theoretical approaches, particularly for $\Lb\to pK^-$. Measurements of both the decay parameters and their charge-conjugate counterparts would therefore provide useful tests of the underlying theoretical descriptions. Analogous decay parameters can be defined for modes containing a final-state vector meson, such as $\Lb\to pK^{*-}$ and $p\rho^-$~\cite{Hong:2022prk}, with predictions given in Refs.~\cite{Han:2024kgz,Duan:2024zjv,Mohanta:2000nk}.

Experimentally, extracting $\alpha$ for the modes listed in Table~\ref{tab:observables} through Eq.~\eqref{eqangular distribution} requires a known and nonzero initial $\Lb$ polarization. Exploiting a sufficiently large and well-controlled beauty-baryon polarization at LHCb remains challenging. An alternative is provided by cascade decays: a baryon produced in a parity-violating weak decay can be polarized, thereby allowing the decay parameter $\alpha$ of its subsequent parity-violating decay to be measured. A possible beauty-baryon realization at LHCb is the observed decay $\Xi_b^-\to\Lambda_b^0\pi^-$, which can provide a polarized $\Lambda_b^0$ sample~\cite{LHCb:2015une}. This polarization could be exploited to probe $\alpha$ and its $\CP$ asymmetry in modes such as $\Lambda_b^0\to p\pi^-$ and $pK^-$. This strategy has been applied to charm baryons and hyperons, which are produced as daughter particles in primary weak decays, since they can acquire polarization that can be analyzed through their subsequent parity-violating decays. For example, Belle studied the decay chain $\Lambda_c^+\to\Lambda\pi^+\to(p\pi^-)\pi^+$~\cite{Belle:2022uod}. The decay of an unpolarized $\Lambda_c^+$ produces a longitudinally polarized $\Lambda$, with its polarization characterized by $\alpha_{\Lambda_c}$, and the one-dimensional angular distribution takes the form
\begin{equation}
\frac{d\Gamma}{d\cos\theta}
\propto
1+\alpha_{\Lambda_c}\alpha_\Lambda\cos\theta,
\end{equation}
where $\theta$ is the helicity angle between the proton momentum in the $\Lambda$ rest frame and the flight direction of the $\Lambda$ in the $\Lambda_c^+$ rest frame. This distribution is analogous to Eq.~\eqref{eqangular distribution}, with $\mathcal P$ replaced by $\alpha_{\Lambda_c}$. If $\CP$ violation in the Cabibbo-favored primary decay vanishes as predicted by the SM, such that $\bar\alpha_{\Lambda_c}=-\alpha_{\Lambda_c}$, the primary decay parameter cancels in the $\CP$ asymmetry constructed from the product $\alpha_{\Lambda_c}\alpha_\Lambda$. The normalized asymmetry $a_{\CP}^{\alpha}$ associated with the secondary decay $\Lambda\to p\pi^-$ can then be extracted~\cite{Wang:2022tcm}. Conversely, for decay chains where $\CP$ violation in the secondary decay is negligible, the same construction can be used to extract $a_{\CP}^{\alpha}$ for the primary decay. The method can also be extended to inclusive decays such as $\Lambda_c^+\to\Lambda X$. In this case, $\alpha_{\Lambda_c}$ is replaced by an average polarization parameter $\mathcal P=\langle P(q^2)\rangle$, and the angular distribution becomes $1+\mathcal P\alpha_\Lambda\cos\theta$. Because the inclusive sample combines many primary decay modes, some of which may themselves exhibit $\CP$ violation, the particle and antiparticle polarizations cannot in general be assumed to satisfy $\bar{\mathcal P}=-\mathcal P$. The resulting measurement therefore probes a compound $\CP$ asymmetry involving both the inclusive production of the polarized $\Lambda$ and its subsequent decay~\cite{BESIII:2026mye}.

\begin{figure}
\begin{center}
\includegraphics[width=0.45\textwidth]{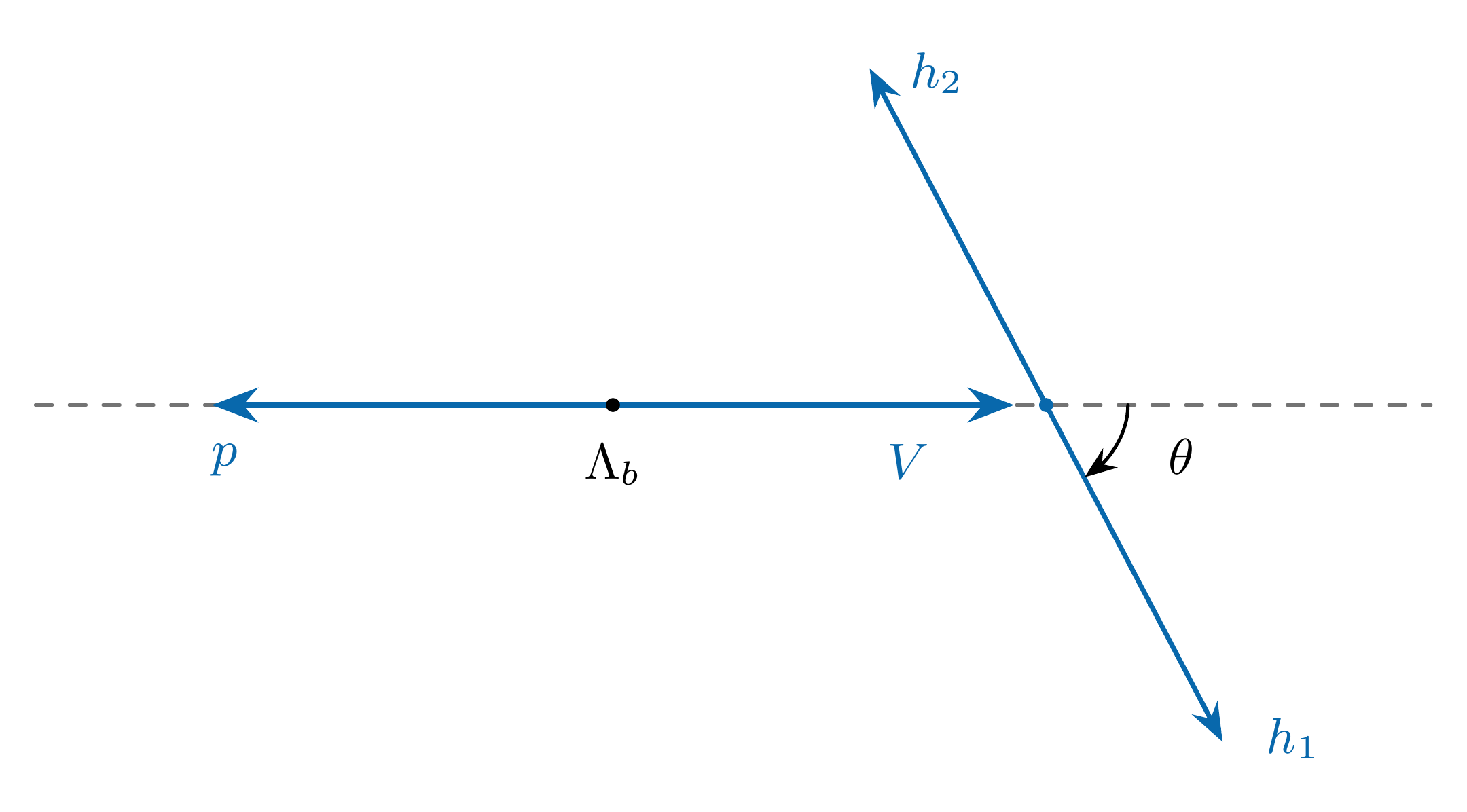}
\includegraphics[width=0.45\textwidth]{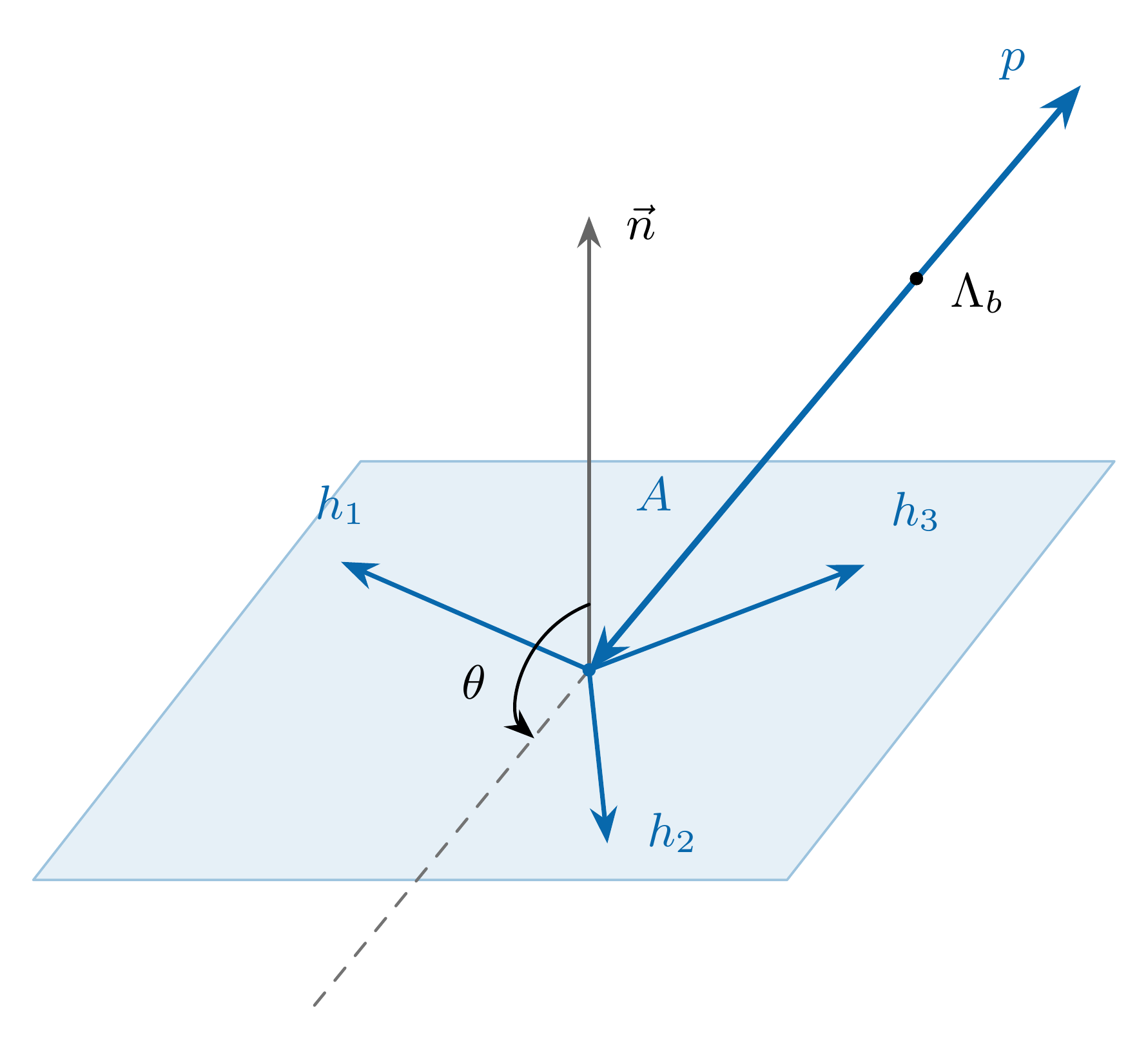}
\captionsetup{type=figure}
\captionof{figure}{The left panel shows the kinematic configuration of the three-body decay $\Lb\to pV\to ph_1h_2$, where $\theta$ is the angle between the momentum of $h_1$ in the $V$ rest frame and the direction of motion of $V$ in the $\Lb$ rest frame. The right panel shows the corresponding configuration for the four-body decay $\Lb\to pA\to ph_1h_2h_3$, where $\theta$ is the angle between the normal to the $A\to h_1h_2h_3$ decay plane and the direction of motion of $A$ in the $\Lb$ rest frame. A fixed ordering of the daughter momenta is used to define the orientation of the normal vector.}
\label{fig:LbtoPV and PA}
\end{center}
\end{figure}

Although extracting $\alpha$ and its associated $\CP$ asymmetry in the two-body modes $\Lb\to ph^-$ from polarization-dependent one-dimensional angular distributions is experimentally challenging because it requires a polarized initial $\Lb$, multibody $\Lb$ decays provide a richer laboratory for constructing one-dimensional angular observables without this requirement. For example, both the three-body decay $\Lb\to pV\to ph_1h_2$ and the four-body decay $\Lb\to pA\to p(h_1h_2h_3)$ allow one-dimensional angular observables and their associated $\CP$-violating quantities to be constructed. For $\Lb\to pV\to ph_1h_2$, shown in the left panel of Fig.~\ref{fig:LbtoPV and PA}, the angular distribution can be written as
\begin{equation}\label{LbtoPV}
\frac{d\Gamma}{d\cos\theta}\propto1+\mathcal J P_2(\cos\theta),
\end{equation}
where $P_2(\cos\theta)=(3\cos^2\theta-1)/2$ and $\theta$ is the helicity angle between the momentum of $h_1$ in the $V$ rest frame and the $V$ momentum in the $\Lb$ rest frame. The coefficient $\mathcal J$ is given by
\begin{equation}
\mathcal J=\frac{2|\mathcal H_{\frac12,0}|^2+2|\mathcal H_{-\frac12,0}|^2-|\mathcal H_{-\frac12,-1}|^2-|\mathcal H_{\frac12,1}|^2}{|\mathcal H_{\frac12,0}|^2+|\mathcal H_{-\frac12,0}|^2+|\mathcal H_{-\frac12,-1}|^2+|\mathcal H_{\frac12,1}|^2}.
\end{equation}
Unlike $\alpha$, the parameter $\mathcal J$ is parity even, consistent with parity conservation in the strong decay $V\to h_1h_2$. Its normalized $\CP$ asymmetry can therefore be defined using Eq.~\eqref{CPV def} with $\eta_{\mathcal J}=+1$. The PQCD calculation of Ref.~\cite{Han:2025tvc} instead quotes the $\CP$-averaged quantity $\langle\mathcal J\rangle=(\mathcal J+\bar{\mathcal J})/2$ and the non-normalized $\CP$-odd combination $A_{\CP}^{\mathcal J}=(\mathcal J-\bar{\mathcal J})/2$:
\begin{equation}
\begin{aligned}
\langle\mathcal J\rangle(\Lb\to p\rho^-)&=1.67^{+0.06}_{-0.08}, \qquad A_{\CP}^{\mathcal J}(\Lb\to p\rho^-)=-0.01\pm0.02,\\
\langle\mathcal J\rangle(\Lb\to pK^{*-})&=1.63^{+0.05}_{-0.21}, \qquad A_{\CP}^{\mathcal J}(\Lb\to pK^{*-})=0.04^{+0.06}_{-0.08}.
\end{aligned}
\end{equation}
Although the predicted values of $A_{\CP}^{\mathcal J}$ are small, $\mathcal J$ itself remains sensitive to the relative magnitudes of the helicity amplitudes and can therefore provide useful tests of theoretical descriptions of these decays.

For the four-body decay $\Lb\to pA\to p(h_1h_2h_3)$, shown in the right panel of Fig.~\ref{fig:LbtoPV and PA}, the angular distribution takes the form
\begin{equation}\label{Lb to pA}
\begin{aligned}
\frac{d\Gamma}{d\cos\theta}\propto{}&\left(|\mathcal H_{-\frac12,-1}|^2+|\mathcal H_{\frac12,1}|^2\right)\frac{1+\cos^2\theta}{2}+\left(|\mathcal H_{-\frac12,0}|^2+|\mathcal H_{\frac12,0}|^2\right)\sin^2\theta\\
&+R\left(|\mathcal H_{\frac12,1}|^2-|\mathcal H_{-\frac12,-1}|^2\right)\cos\theta,
\end{aligned}
\end{equation}
where $\theta$ is the angle between the normal to the $A\to h_1h_2h_3$ decay plane and the direction of motion of $A$ in the $\Lb$ rest frame, while $R$ is a nonperturbative analyzing parameter associated with the strong decay $A\to h_1h_2h_3$. The orientation of the decay-plane normal must be defined using a fixed ordering of the three daughter momenta. The up--down asymmetry $A_{\rm UD}$, defined by the normalized difference between the $\cos\theta>0$ and $\cos\theta<0$ regions, and its associated $\CP$ asymmetry $a_{\CP}^{\rm UD}$ are given in Eqs.~\eqref{UP-Down asymmetry} and~\eqref{CPV in UP-Down asymmetry}. Since $A_{\rm UD}$ is parity odd, the normalized $\CP$ asymmetry involves $A_{\rm UD}+\bar A_{\rm UD}$ in the numerator. With the nonperturbative parameter $R$ canceling in the asymmetry $a_{\CP}^{\rm UD}$, PQCD calculations predict sizable values of this observable:
\begin{equation}
\begin{aligned}
a_{\CP}^{\rm UD}(\Lb\to pa_1^-(1260))&=(-24^{+8}_{-13})\%,\\
a_{\CP}^{\rm UD}(\Lb\to pK_1^-(1270))\big|_{\theta_K=30^\circ}&=(26^{+4}_{-10})\%,\\
a_{\CP}^{\rm UD}(\Lb\to pK_1^-(1270))\big|_{\theta_K=60^\circ}&=(40^{+4}_{-9})\%,
\end{aligned}
\end{equation}
which may be accessible in future measurements~\cite{Han:2025tvc}. Here $\theta_K$ denotes the mixing angle between the strange axial-vector states $K_{1A}$ and $K_{1B}$, $K_1^-(1270) = \sin\theta\ket{K_{1A}} + \cos\theta\ket{K_{1B}}$, with quantum numbers ${}^3P_1$ and ${}^1P_1$, respectively. Since the value of $\theta_K$ is not yet well determined, the benchmark choices $\theta_K=30^\circ$ and $60^\circ$ are used to illustrate the corresponding dependence.

Asymmetry parameters extracted from one-dimensional angular analyses can further arise from interference between different resonant contributions. Taking the three-body decay $B^\pm\to\pi^+\pi^-\pi^\pm$ as an example, large local $\CP$ asymmetries have been observed in the region of the $\rho^0(770)$ resonance~\cite{LHCb:2014mir,LHCb:2019jta}. The observed phase-space dependence cannot be described solely by $\CP$ violation in the quasi-two-body decay $B^\pm\to\pi^\pm\rho^0$. It has been shown theoretically that the inclusion of the scalar resonance $f_0(500)$ and its interference with the $\rho^0(770)$ contribution can generate sizable local $\CP$-violating effects~\cite{Zhang:2013oqa}. Since the strong decays $\rho^0\to\pi^+\pi^-$ and $f_0\to\pi^+\pi^-$ proceed through $P$- and $S$-wave configurations, respectively, the amplitude can be written schematically as $M=M_\rho\cos\theta+M_f$, where $M_\rho$ and $M_f$ denote the amplitudes for $B^\pm\to\rho^0(\to\pi^+\pi^-)h^\pm$ and $B^\pm\to f_0(\to\pi^+\pi^-)h^\pm$, respectively. The angle $\theta$ is the helicity angle, defined in the $\pi^+\pi^-$ rest frame as the angle between the momenta of the parent $B$ meson and the $\pi^+$ meson. The squared amplitude is
\begin{equation}
|M|^2 = |M_f|^2+ |M_\rho|^2\cos^2\theta+ 2\,\operatorname{Re}(M_fM_\rho^*)\cos\theta.
\end{equation}
The first two terms are the individual $f_0$ and $\rho^0$ contributions, while the last term arises from their interference. At fixed $m_{\pi^+\pi^-}$, $\cos\theta$ is related to the invariant-mass variable $m_{h\pi^+}^2$, so that the interference term contributes with opposite signs in different regions of the Dalitz plot. It can be isolated through the forward--backward asymmetry
\begin{equation}
\begin{aligned}
A^{\rm FB} &= \frac{ \displaystyle \int_{-1}^1|M|^2\operatorname{sgn}(\cos\theta)\,d\cos\theta}{\displaystyle \int_{-1}^1|M|^2\,d\cos\theta } 
= \frac{ \operatorname{Re}(M_fM_\rho^*)}{|M_f|^2+\frac13|M_\rho|^2}.
\end{aligned}
\end{equation}
Because the helicity angle is constructed from the angle between two momenta and is parity even, the corresponding $\CP$-odd combination $a_{\CP}^{\rm FB}$ involves the difference between the particle and antiparticle forward--backward asymmetries. A resonance-based description alone cannot precisely determine this observable because the relative strong phase between $M_f$ and $M_\rho$ is not known from first principles, even if their tree and penguin components are estimated using factorization approaches. Incorporating experimentally constrained $\pi\pi$ scattering amplitudes can provide additional information on these strong phases~\cite{Bediaga:2020qxg}. Resonance interference nevertheless offers a useful phenomenological strategy for constructing $\CP$-violating observables in multibody beauty-baryon decays.

In a set of beauty-baryon decays such as $\Lb\to ph^-\rho^0(f_0)\to ph^-\pi^+\pi^-$ and $\Lb\to\Lambda\rho^0(f_0)\to\Lambda\pi^+\pi^-$, $A^{\rm FB}$ and $a_{\CP}^{\rm FB}$ can be defined in the low-$m_{\pi^+\pi^-}$ region. Related channels, including $\Lb\to\Lambda(1520)\pi^+\pi^-$, $\Lambda(1520)K^+K^-$ and $\Lambda(1520)K^+\pi^-$, have been studied using hadronic-loop calculations~\cite{Feng:2026soj}. Predictions for their forward--backward $\CP$ asymmetries have also been obtained. In particular, a model-dependent value of approximately $-50\%$ is predicted for $\Lb\to\Lambda(1520)\pi^+\pi^-$, making this channel an interesting target for future measurements~\cite{Feng:2026soj}.

More generally, beauty-baryon decay amplitudes can contain $S$-, $P$- and higher-partial-wave contributions. A natural generalization is to expand a one-dimensional angular distribution in Legendre polynomials~\cite{Zhang:2021fdd},
\begin{equation}
\frac{d\Gamma}{d\cos\theta}
=
\sum_j
\left\langle w^{(j)}\right\rangle
P_j(\cos\theta),
\end{equation}
where $\langle\cdots\rangle$ denotes that the distribution has been integrated over the remaining phase-space variables. For a cascade decay $\Lb\to Rh_3\cdots h_n$, followed by $R\to h_1h_2$, the angle $\theta$ may be defined in the rest frame of $R$ as the angle between the momentum of $h_1$ and the direction of the parent $\Lb$. Here, $h_i$ can be either a meson or a baryon. By the orthogonality of the Legendre polynomials, the coefficients can be extracted as
\begin{equation}
\left\langle w^{(j)}\right\rangle = \frac{2j+1}{2} \int_{-1}^1 P_j(\cos\theta) \frac{d\Gamma}{d\cos\theta} \,d\cos\theta.
\end{equation}
A corresponding set of moment asymmetries, often termed partial-wave $\CP$ asymmetries, can be defined by
\begin{equation}
a_{\CP}^{(j)} = \frac{ \left\langle w^{(j)}\right\rangle- \left\langle\bar w^{(j)}\right\rangle }{ \left\langle w^{(j)}\right\rangle+ \left\langle\bar w^{(j)}\right\rangle }.
\end{equation}
With this convention, the parameter $\mathcal J$ in Eq.~\eqref{LbtoPV} corresponds to the normalized second-order Legendre coefficient $\langle w^{(2)}\rangle$. These moment asymmetries are also proportional to sines of strong phases, analogous to the direct asymmetry $A_{\CP}^{\rm dir}$. Their advantage is that they isolate particular angular structures and interference contributions that are not separately accessible through the phase-space-integrated rate asymmetry. 

Such observables have been studied phenomenologically in a variety of multibody beauty-baryon decays~\cite{Zhang:2021fdd,Zhang:2022emj}. In the analysis of $\Lb\to DN\to(K^+\pi^-)(p\pi^-)$ in Ref.~\cite{Shen:2023eln}, including intermediate contributions from $D=D^0,\bar D^0$ and $N=N(1440),N(1520)$, the moment asymmetry $a_{\CP}^{(1)}$ was found to remain sizable even for strong-phase configurations in which the conventional direct asymmetry nearly vanishes, illustrating the complementary sensitivity of the two observables. Although precise predictions remain difficult because the relevant strong phases are poorly constrained, several moment asymmetries have also been estimated using the $N\pi$-rescattering framework discussed in Sec.~\ref{sec:compute}. A particularly striking example is the normalized $\CP$ asymmetry associated with the first Legendre moment, $P_1(\cos\theta)=\cos\theta$, in $\Lb\to(\Delta^{++}\pi^-)K^-$. After integration over $m_{\Delta^{++}\pi^-}\in[1.2,1.9]\,\mathrm{GeV}$, its magnitude is estimated to reach
\begin{equation}
\left|a_{\CP}^{(1)}\!\left[\Lb\to(\Delta^{++}\pi^-)K^-\right]\right|\sim40\%\text{--}70\%.
\end{equation}
The predicted sign and magnitude, however, vary substantially among the form-factor scenarios considered, indicating considerable sensitivity to the poorly known weak-production form factors. This moment asymmetry should therefore be regarded as a promising experimental target rather than a precision prediction~\cite{Wang:2024oyi}.

\subsection{Multi-dimensional Angular Analysis}

One-dimensional angular analyses often provide transparent physical interpretations by integrating over the remaining angular variables and defining suitable asymmetry parameters. This simplification, however, inevitably discards part of the information contained in the full decay distribution. In particular, interference terms among different helicity or partial-wave amplitudes may vanish after angular integration and therefore remain inaccessible in a one-dimensional analysis. Multidimensional angular analyses retain these correlations and allow additional $\CP$-violating observables to be constructed. Several representative examples are discussed below.

\begin{figure}
\begin{center}
\includegraphics[width=0.72\textwidth]{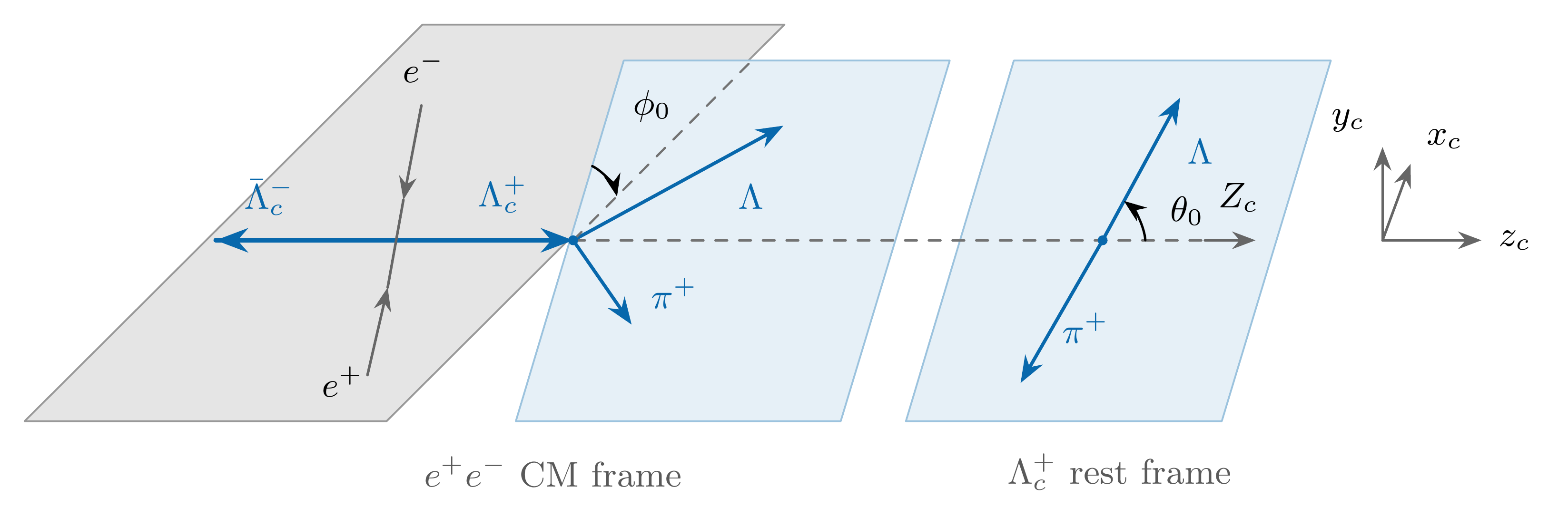}
\includegraphics[width=0.72\textwidth]{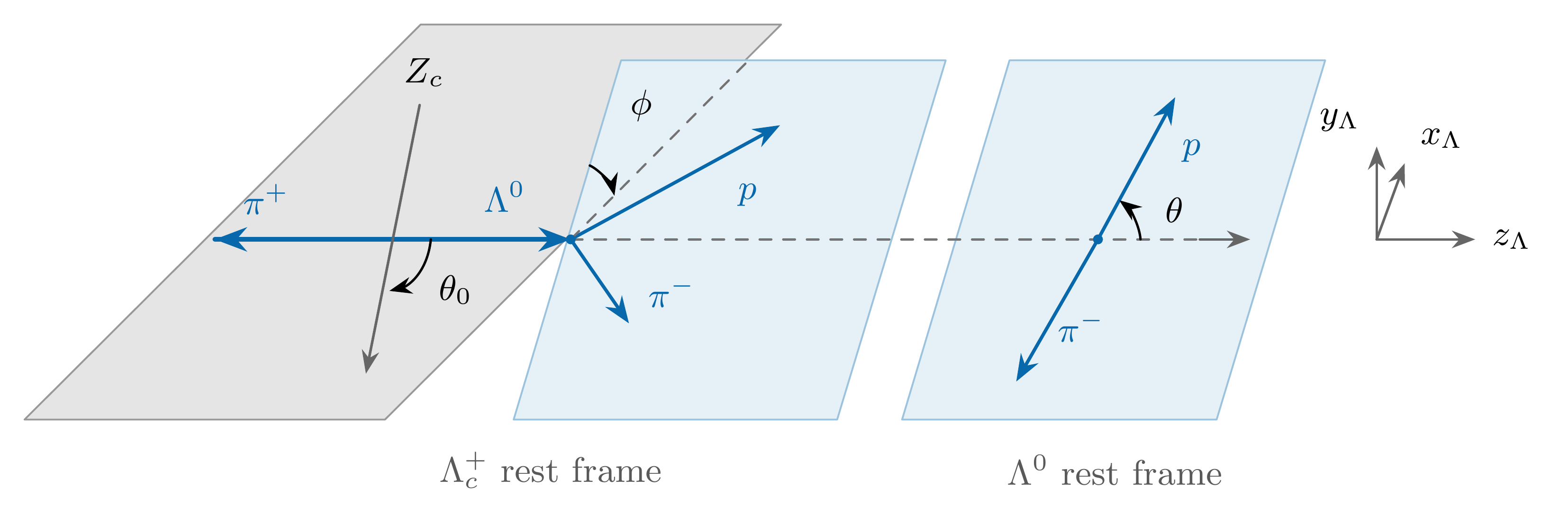}
\captionsetup{type=figure}
    \captionof{figure}{Coordinate systems and helicity angles for the production of a $\Lambda_c^+\bar\Lambda_c^-$ pair and the cascade decay $\Lambda_c^+\to\Lambda\pi^+\to(p\pi^-)\pi^+$. In the $e^+e^-$ center-of-mass frame, $z$ denotes the beam direction and $z_c$ is chosen along the $\Lambda_c^+$ momentum. The remaining axes are defined by $y_c\propto z\times z_c$ and $x_c=y_c\times z_c$. The direction of the $\Lambda$ momentum in the $\Lambda_c^+$ rest frame is specified by $(\theta_0,\phi_0)$. In the $\Lambda$ rest frame, $z_\Lambda$ is chosen along the $\Lambda$ momentum in the $\Lambda_c^+$ frame, while $y_\Lambda\propto z_c\times z_\Lambda$ and $x_\Lambda=y_\Lambda\times z_\Lambda$. The proton direction in this frame is specified by $(\theta,\phi)$. The same construction applies to other cascade decays, such as $\Xi^0\to\Lambda\pi^0\to(p\pi^-)\pi^0$, after replacing $\Lambda_c^+$ and $\pi^+$ by $\Xi^0$ and $\pi^0$, respectively.}
    \label{fig:Lc to LambdaPi}
\end{center}
\end{figure}

As discussed in the previous subsection, the decay parameter $\alpha$ defined in Eq.~\eqref{eq:alpha-helicity} characterizes the longitudinal polarization of the daughter baryon when the parent is unpolarized. More generally, the spin of the daughter baryon can also have components transverse to its momentum. Accessing these components requires sufficient independent kinematic directions to define the transverse axes. Such directions may be provided by the production kinematics---for example, by the beam direction and the baryon momentum in $e^+e^-\to\mathcal B\bar{\mathcal B}$---as well as by the additional momenta available in cascade decays. As an example, consider the decay chain $\Lambda_c^+\to\Lambda\pi^+\to(p\pi^-)\pi^+$ with $\Lambda_c^+$ produced in $e^+e^-$ collisions. The coordinate systems are illustrated in Fig.~\ref{fig:Lc to LambdaPi}. The $z_c$ axis is chosen along the $\Lambda_c^+$ momentum in the $e^+e^-$ center-of-mass frame, while $y_c$ is normal to the production plane and $x_c=y_c\times z_c$. The angles $(\theta_0,\phi_0)$ specify the direction of the $\Lambda$ momentum in the $\Lambda_c^+$ rest frame. The $z_\Lambda$ axis is chosen along the $\Lambda$ momentum in the $\Lambda_c^+$ rest frame, with $y_\Lambda$ normal to the plane spanned by $z_c$ and $z_\Lambda$ and $x_\Lambda=y_\Lambda\times z_\Lambda$. The angles $(\theta,\phi)$ then specify the proton direction in the $\Lambda$ rest frame in this coordinate system. Let the polarization vector of the initial $\Lambda_c^+$ be $\vec{\mathcal P}=(\mathcal P_x,\mathcal P_y,\mathcal P_z)$ in the frame $(x_c,y_c,z_c)$. Using the angular formalism introduced in Sec.~\ref{Angular Distributions}, the full decay distribution can be written as
\begin{equation}\label{Full analysis example 1}
\begin{aligned}
\frac{d\Gamma}{d\cos\theta_0\,d\phi_0\,d\cos\theta\,d\phi}\propto{}&1+\alpha_{\Lambda_c}\alpha_\Lambda\cos\theta+K_1\left(\alpha_{\Lambda_c}+\alpha_\Lambda\cos\theta\right)\\
&+K_2\alpha_\Lambda\sin\theta\left(\gamma_{\Lambda_c}\cos\phi-\beta_{\Lambda_c}\sin\phi\right)+K_3\alpha_\Lambda\sin\theta\left(\gamma_{\Lambda_c}\sin\phi+\beta_{\Lambda_c}\cos\phi\right),
\end{aligned}
\end{equation}
where $K_1=\sin\theta_0(\mathcal P_x\cos\phi_0+\mathcal P_y\sin\phi_0)+\mathcal P_z\cos\theta_0$, $K_2=\cos\theta_0(\mathcal P_x\cos\phi_0+\mathcal P_y\sin\phi_0)-\mathcal P_z\sin\theta_0$ and $K_3=\mathcal P_y\cos\phi_0-\mathcal P_x\sin\phi_0$ are the projections of the initial polarization onto the helicity axes defined by the $\Lambda$ momentum. After integration over $\theta_0$, $\phi_0$, and $\phi$, Eq.~\eqref{Full analysis example 1} reduces to the one-dimensional distribution $1+\alpha_{\Lambda_c}\alpha_\Lambda\cos\theta$ discussed above. The two additional Lee--Yang parameters in the primary decay are
\begin{equation}\label{eq:betagamma}
\beta_{\Lambda_c}=\frac{2\,\operatorname{Im}(h_+h_-^*)}{|h_+|^2+|h_-|^2},\qquad \gamma_{\Lambda_c}=\frac{2\,\operatorname{Re}(h_+h_-^*)}{|h_+|^2+|h_-|^2},
\end{equation}
where $h_\pm$ are the helicity amplitudes for $\Lambda_c^+\to\Lambda\pi^+$. Together with $\alpha_{\Lambda_c}$, these parameters determine how the polarization of the parent $\Lambda_c^+$ is transferred to the longitudinal and transverse polarization components of the daughter $\Lambda$. Equivalently, using $h_\pm=(S\pm P)/\sqrt{2}$, they can be written as
\begin{equation}
\beta_{\Lambda_c}=\frac{2\,\operatorname{Im}(S^*P)}{|S|^2+|P|^2},\qquad \gamma_{\Lambda_c}=\frac{|S|^2-|P|^2}{|S|^2+|P|^2}.
\end{equation}
An analogous set $(\alpha_\Lambda,\beta_\Lambda,\gamma_\Lambda)$ can be defined for the secondary decay $\Lambda\to p\pi^-$. However, only $\alpha_\Lambda$ appears in Eq.~\eqref{Full analysis example 1}, because the proton spin is not fully analyzed through a subsequent decay or an independent spin measurement. The normalized $\CP$ asymmetries associated with $\beta$ and $\gamma$ can be defined, in analogy with $a_{\CP}^{\alpha}$, as
\begin{equation}
a_{\CP}^{\beta}=\frac{\beta+\bar\beta}{\beta-\bar\beta},\qquad a_{\CP}^{\gamma}=\frac{\gamma-\bar\gamma}{\gamma+\bar\gamma}.
\end{equation}

On the experimental side, BESIII has collected large samples of hyperon--antihyperon and $\Lambda_c^+\bar\Lambda_c^-$ pairs, providing a favorable environment for measuring the Lee--Yang parameters. For production processes such as $e^+e^-\to J/\psi\to\Xi\bar\Xi,\Lambda\bar\Lambda,\Sigma\bar\Sigma$ and $e^+e^-\to\gamma^*\to\Lambda_c^+\bar\Lambda_c^-$ with unpolarized beams, parity conservation restricts the single-baryon polarization to the direction normal to the production plane. A nonzero transverse component $\mathcal{P}_y$ can be generated by a relative phase between the production helicity amplitudes, whereas $\mathcal{P}_x$ and $\mathcal{P}_z$ vanish in this production configuration. This transverse polarization enriches the angular distribution in Eq.~\eqref{Full analysis example 1}. In addition, the quantum entanglement of the baryon--antibaryon pair provides access to spin correlations even when the individual polarizations are small. An important example is the BESIII analysis of entangled $\Xi\bar\Xi$ pairs. In constructing the joint angular distribution, the angular analysis illustrated in Fig.~\ref{fig:Lc to LambdaPi} is applied not only to the baryon decay chain $\Xi\to\Lambda\pi\to(p\pi)\pi$, but also to the charge-conjugate chain $\bar\Xi\to\bar\Lambda\pi\to(\bar p\pi)\pi$, using the corresponding helicity angles on each side. The spin correlations between the entangled $\Xi$ and $\bar\Xi$ connect the two sets of angular variables, allowing the decay parameters $\alpha$, $\beta$, and $\gamma$ and their charge-conjugate counterparts to be determined simultaneously~\cite{BESIII:2021ypr}. The parameter $\beta$ is particularly interesting because $\CP$-odd observables constructed from it probe the imaginary part of the interference between the $S$- and $P$-wave amplitudes. Decomposing both partial-wave amplitudes into contributions with different weak phases, as in Eqs.~\eqref{eq:SPWwave} and~\eqref{eq:SPbarWwave}, one finds that the numerator $\beta+\bar\beta$ can contain interference terms proportional to $\cos\Delta\delta\,\sin\Delta\phi$. This differs from the $\sin\Delta\delta\,\sin\Delta\phi$ dependence of the direct rate asymmetry in Eq.~\eqref{eq:acp}. Observables involving $\beta$ may therefore retain sensitivity to $\CP$ violation when the relevant strong-phase differences are small~\cite{Wang:2022fih}. This feature makes them particularly promising for hyperon decays such as $\Xi\to\Lambda\pi$. The BESIII measurements show an enhanced central value of $(\beta+\bar{\beta})/(\alpha-\bar{\alpha})$ at the order of $10^{-3}$ although it suffers large uncertainties~\cite{BESIII:2021ypr}.

For multibody beauty-baryon decays, the situation is generally more complicated. Unlike the cascade decays discussed above, whose intermediate states can be cleanly identified, a multibody beauty-baryon final state often receives coherent contributions from several overlapping resonances with different spin-parity quantum numbers. Their amplitudes can interfere even within the same region of the Dalitz plot. Consequently, it is far from straightforward to establish the full angular distribution for most multibody beauty baryon decays. For example, contributions schematically written as $\Lb\to N^*K^*\to(N\pi)(K\pi)$ may involve several overlapping $N^*$ states and cannot in general be separated experimentally without a dedicated amplitude analysis. The angular distribution can therefore be constructed only after the relevant intermediate configurations have been specified, as illustrated by the decay $\Lb\to pA\to p(h_1h_2h_3)$ discussed above.

Taking $\Lb$ decays as examples, some commonly encountered kinematic topologies in three- and four-body modes can be organized as follows:
\begin{itemize}
\item \textbf{Three-body decays:}
\begin{enumerate}
\item[(i)] an excited baryon accompanied by a light meson, such as $\Lb\to N^{*0}K_S^0\to(p\pi^-)K_S^0$;
\item[(ii)] a light baryon accompanied by a meson resonance, such as $\Lb\to p\rho^-\to p(\pi^-\pi^0)$.
\end{enumerate}
\item \textbf{Four-body decays:}
\begin{enumerate}
\item[(i)] an intermediate baryon accompanied by a meson resonance, such as $\Lb\to\Lambda\phi\to(p\pi^-)(K^+K^-)$;
\item[(ii)] a light baryon accompanied by a resonance decaying into three mesons, such as $\Lb\to pK_1^-\to p(K^-\pi^+\pi^-)$.
\end{enumerate}
\end{itemize}
This classification refers only to the kinematic topology of the decay chain. Within each category, it allows various intermediate states carrying different spin-parity quantum numbers, which may contribute coherently to the same final state.

\begin{figure}
\begin{center}
    \includegraphics[width=0.72\textwidth]{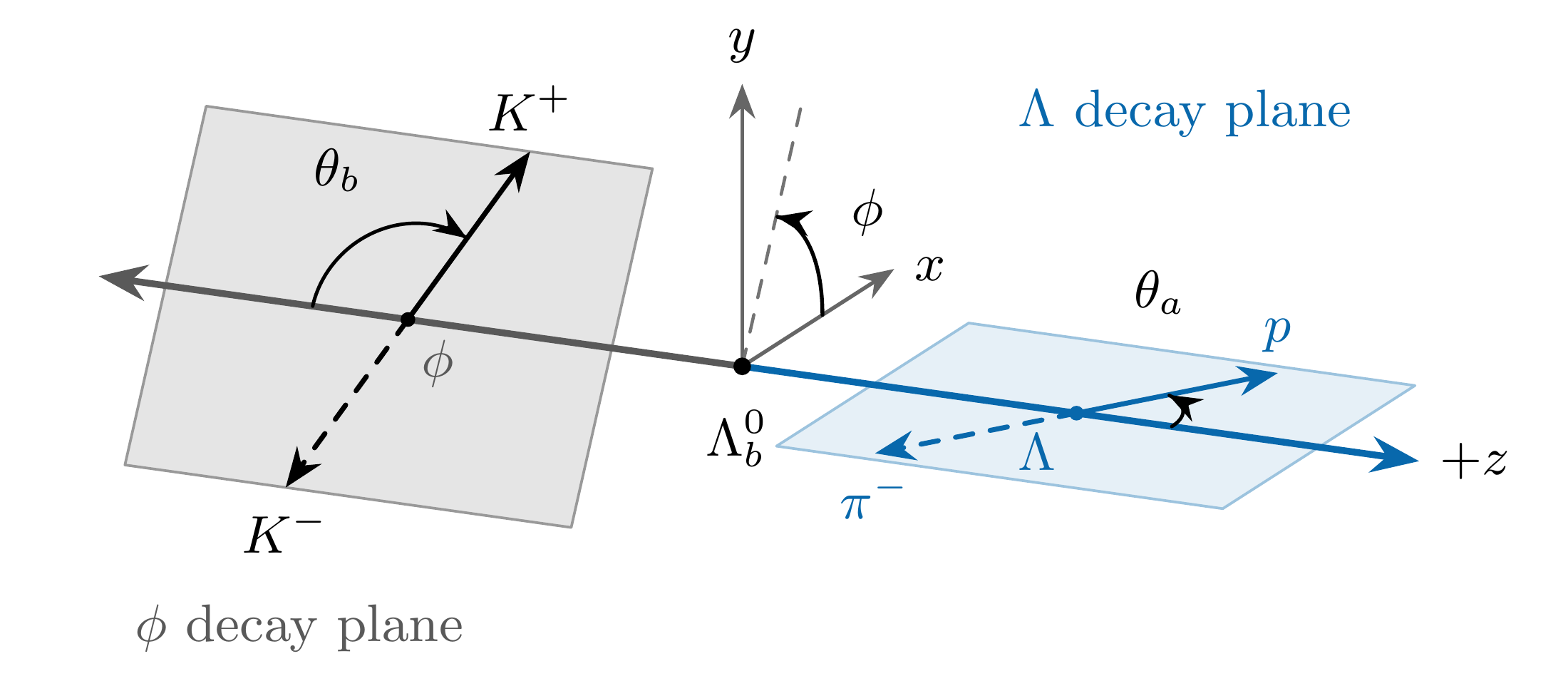}
    \captionsetup{type=figure}
    \captionof{figure}{Angles for the cascade decay $\Lb\to\Lambda\phi\to(p\pi^-)(K^+K^-)$. The angles $\theta_a$ and $\theta_b$ are the helicity angles of the proton and $K^+$ in the $\Lambda$ and $\phi$ rest frames, respectively, while $\phi$ is the relative azimuthal angle between the two decay planes.}
    \label{fig:angle4body}
\end{center}
\end{figure}

\begin{table*}[htbp]
\centering
\footnotesize
\renewcommand{\arraystretch}{2.4}
\caption{Contributions to the angular distribution of $\Lb\to\Lambda\phi\to(p\pi^-)(K^+K^-)$ for an unpolarized initial $\Lb$.  Using the helicity amplitudes introduced in Sec.~\ref{Angular Distributions}, we define $A_\pm\equiv\mathcal H_{\pm\frac12,0}$ and $B_\pm\equiv\mathcal H_{\pm\frac12,\pm1}$, where the two indices denote the helicities of the $\Lambda$ and $\phi$, respectively. Each row represents a term with a distinct angular dependence. The final column gives the normalized weight function $W_i(\Omega)$ that projects the corresponding angular coefficient through $\int d\Omega\,W_i(\Omega)\,d\Gamma/d\Omega$, up to a common normalization, where $d\Omega=d\cos\theta_a\,d\cos\theta_b\,d\phi/(8\pi)$. The weights assume uniform angular acceptance; acceptance corrections are required in an experimental application. The extension to polarized $\Lb$ decays can be found in Ref.~\cite{Durieux:2016nqr}.}
\label{angular distribution of LambdaPhi}
\begin{tabular}{|ccccccc|c|}
\hline
\multicolumn{7}{|c|}{Angular-distribution term}
& \multicolumn{1}{c|}{Projection weight $W_i(\Omega)$} \\
\hline
$+3$
& $|A_+|^2+|A_-|^2$
& & & & $\cos^2\theta_b$ &
& $\dfrac{5\cos^2\theta_b-1}{2}$ \\

$+3/2$
& $|B_+|^2+|B_-|^2$
& & & & $\sin^2\theta_b$ &
& $\dfrac{3-5\cos^2\theta_b}{2}$ \\

$+3$
& $|A_+|^2-|A_-|^2$
& $\alpha_\Lambda$ & & $\cos\theta_a$ & $\cos^2\theta_b$ &
& $\dfrac{3}{2}\cos\theta_a\left(5\cos^2\theta_b-1\right)$ \\

$+3/2$
& $|B_+|^2-|B_-|^2$
& $\alpha_\Lambda$ & & $\cos\theta_a$ & $\sin^2\theta_b$ &
& $\dfrac{3}{2}\cos\theta_a\left(3-5\cos^2\theta_b\right)$ \\

$3/\sqrt{2}$
& $\operatorname{Re}(A_+^*B_- - A_-B_+^*)$
& $\alpha_\Lambda$ & & $\sin\theta_a$ & $\sin2\theta_b$ & $\cos\phi$
& $\dfrac{15\sqrt{2}}{8}\sin\theta_a\sin2\theta_b\cos\phi$ \\

$3/\sqrt{2}$
& $\operatorname{Im}(A_+^*B_- - A_-B_+^*)$
& $\alpha_\Lambda$ & & $\sin\theta_a$ & $\sin2\theta_b$ & $\sin\phi$
& $\dfrac{15\sqrt{2}}{8}\sin\theta_a\sin2\theta_b\sin\phi$ \\
\hline
\end{tabular}
\end{table*}

For $\Lb\to pA\to p(h_1h_2h_3)$, the angular distribution for an unpolarized $\Lb$ was given in Eq.~\eqref{Lb to pA}; its extension to a polarized initial state was studied in Ref.~\cite{Wang:2024qff}. Angular distributions for the first classes of three- and four-body decays defined above, including the effects of the initial beauty-baryon polarization, were systematically derived in Ref.~\cite{Durieux:2016nqr}. As an example, the contributions to the unpolarized $\Lb\to\Lambda\phi$ distribution are summarized in Table~\ref{angular distribution of LambdaPhi}, where the projection weight functions to obtain the corresponding angular coefficients, the combinations of helicity-amplitude bilinears, are also listed. The angle definition is illustrated in Fig.~\ref{fig:angle4body}. In addition to the helicity-amplitude moduli, the distribution provides access to the interference combinations
\begin{equation}
\operatorname{Re}(A_+^*B_- - A_-B_+^*),\qquad \operatorname{Im}(A_+^*B_- - A_-B_+^*),
\end{equation}
which multiply different angular functions. The real and imaginary combinations play roles analogous to those of $\gamma$ and $\beta$ as given in Eq.~\eqref{eq:betagamma}, respectively, in spin-$1/2$ baryon decays.

The time-reversal transformation properties of helicity-amplitude interference terms provide a general basis for classifying the corresponding angular observables and their $\CP$ asymmetries~\cite{Wang:2022fih}. For the interference structures considered here, observables associated with the real and imaginary parts of the amplitude products are generally $T$-even and $T$-odd, respectively, where $T$-even ($T$-odd) denotes a term that remains unchanged (changes sign) under time reversal. This property follows from the antiunitarity of the time-reversal operator: the complex conjugation involved in the transformation reverses the sign of the imaginary part of an interference term while leaving its real part unchanged. By comparing these interference coefficients with their charge-conjugate counterparts, additional $\CP$-violating observables can be constructed in analogy with those associated with the decay parameters $\alpha$, $\beta$, and $\gamma$. The $\CP$ asymmetries constructed from the $T$-odd and $T$-even interference terms depend on the strong-phase difference as $\cos\Delta\delta$\footnote{Analogous strong-phase dependence can arise in $\CP$ asymmetries constructed from the imaginary parts of interference terms between different finite-width intermediate resonances in heavy-hadron cascade decays~\cite{Qi:2024zau}.} and $\sin\Delta\delta$, respectively, while both are proportional to $\sin\Delta\phi$, where $\Delta\phi$ is the weak-phase difference. They therefore provide complementary sensitivity to $\CP$ violation.

Multidimensional angular analyses of multibody beauty-baryon decays offer considerable opportunities for future searches for $\CP$ violation. A complete angular analysis should, in principle, include the coherent superposition of all relevant resonant and nonresonant contributions. Most angular-distribution formulae currently available for beauty-baryon decays instead assume the dominance of one or a few intermediate states, an approximation that is most reliable when the corresponding contributions can be experimentally isolated. Full amplitude analyses have already been successfully applied to several three-body charm-baryon decays, such as $\Lambda_c^+\to pK^+K^-$ and $\Lambda\pi^+\pi^0$, for which the number of relevant amplitudes remains manageable~\cite{BESIII:2022udq,BESIII:2026dkj}. The treatment becomes considerably more challenging for four-body decays because the number of amplitudes and fit parameters grows rapidly. It is therefore desirable to develop model-independent or minimally model-dependent angular-moment analyses organized according to the spin-parity quantum numbers of the contributing intermediate states, without specifying individual resonance line shapes, masses, or widths. Such analyses yield projection weight functions that allow different amplitude-interference structures to be separated and extracted, thereby providing access to a comprehensive set of corresponding $\CP$-violating observables. A step in this direction has recently been taken in the analysis of $B^0\to p\bar pK^+\pi^-$~\cite{Zhang:2025mne}, while analogous analyses for the numerous kinematic topologies available in beauty-baryon decays remain to be developed.

Before closing this section, it is useful to emphasize that beauty-baryon decays may also provide complementary determinations of the CKM unitarity-triangle angles. With the continuing progress of LHCb in heavy-baryon physics, several baryon-based methods have been proposed. One approach exploits the interference between even- and odd-partial-wave amplitudes constrained by Bose symmetry in three-body $\Xi_b$ decays, whose angular distributions are sensitive to $\gamma$~\cite{Sinha:2021mmx}. A second strategy exploits the interference between the $b\to c\bar us$ and $b\to u\bar cs$ tree amplitudes in $\Lb\to\Lambda D$ decays~\cite{Giri:2001ju}. For an unpolarized $\Lb$ sample, the experimentally accessible decay parameter $\alpha$ probes the $S$--$P$ interference, and including it in addition to the decay rates can improve the sensitivity to $\gamma$ by up to $60\%$~\cite{Zhang:2021sit}, depending on the hadronic parameter values~\cite{Rui:2026ihu,Li:2026zrs}. A third method exploits the joint spin--flavor quantum entanglement of the $\Lambda$ baryon and the neutral $D$ meson. Measurements of a set of Lee--Yang asymmetry parameters can reconstruct this final-state quantum configuration and enable a model-independent determination of $\gamma$~\cite{Du:2026qsy}. These proposals exploit distinct interference structures and experimental observables. As the available LHCb data samples continue to grow, they may develop into complementary routes for determining $\gamma$. More broadly, they illustrate how the spin and multidimensional information discussed in this section can be used not only to search for $\CP$ violation, but also to determine the underlying CKM weak phases.

%% file: sec7-summary.tex
Violation of the $\CP$ symmetry is important both for testing the SM and for the dynamical generation of the cosmic baryon asymmetry~\cite{Sakharov:1967dj}. However, the $\CP$ violation generated by quark mixing in the SM is far too weak to account for the observed asymmetry~\cite{Gavela:1993ts,Huet:1994jb}, motivating searches for additional $\CP$-violating sources in hadron decays, which can arise in new physics frameworks such as two-Higgs-doublet, supersymmetric, and neutrino-extension models~\cite{Aliev:1999ap,Mohanta:2000za,Chen:2002rg,Das:2021prm}. In 2025, more than six decades after the first observation of $\CP$ violation in the meson sector, LHCb reported the first observation of $\CP$ violation in a baryon decay~\cite{LHCb:2025ray}. This milestone opens a new arena for testing the origin and structure of $\CP$ violation and for searching for additional $\CP$-violating interactions.

The long delay between the discoveries of $\CP$ violation in mesons and baryons reflects both experimental challenges and dynamical features specific to baryon decays. Beauty baryons are heavier than the dominant beauty mesons, placing their pair-production thresholds beyond the reach of conventional $B$ factories; at LHCb, they are also generally produced with smaller fragmentation fractions than beauty mesons. Dynamically, their nonzero spins allow several helicity or partial-wave amplitudes to contribute. Theoretical studies have shown that $\CP$-violating effects associated with different partial waves can partially cancel, suppressing the phase-space-integrated rate asymmetry even when some individual contributions are sizable. These complications also make baryons a distinctive laboratory. Their spins and rich multibody kinematics allow polarization information and interference among decay amplitudes to be accessed through angular distributions, providing observables with no direct analogue in two-body decays of spinless particles.

This final section first summarizes how experimental searches, theoretical calculations and new analysis methods converged on the first observation, and then outlines a staged roadmap from broader exploration toward precision tests of the SM and searches for additional $\CP$-violating interactions.

\subsection{Summary}

The first observation of baryonic $\CP$ violation was not the outcome of a single theoretical model or a single experimental technique. It emerged from a long program in which null measurements established experimental control, increasingly diverse observables exposed different forms of $\CP$-odd information, and phenomenological calculations identified mechanisms that could suppress integrated asymmetries while enhancing selected partial waves or regions of phase space. The discovery should therefore be understood as the culmination of mutually reinforcing developments rather than as an isolated measurement.

Before the observation, experiments had already explored a broad range of beauty-baryon channels and analysis strategies. Searches in the two-body modes $\Lb\to pK^-$ and $p\pi^-$ established how production and detection asymmetries could be controlled in fully hadronic final states~\cite{CDF:2014pzb,LHCb:2018fly}. Three-body studies included amplitude analyses of $\Xi_b^-\to pK^-K^-$ and measurements of $\Lb$ decays containing $\Lambda$, $K_S^0$ and light hadrons, thereby beginning to resolve resonance-dependent structures~\cite{LHCb:2021enr,LHCb:2024yzj}. Four-body measurements extended the program to inclusive and regional rate asymmetries, triple-product observables and model-independent energy tests in channels such as $\Lb\to p\pi^+\pi^-\pi^-$ and several charmless $\Lb$ and $\Xi_b^0$ modes~\cite{LHCb:2018fpt,LHCb:2019oke,LHCb:2019jyj}. Although these studies did not establish $\CP$ violation, they characterized signal yields and backgrounds, developed and validated methods for controlling production and detection asymmetries, and showed that multibody beauty-baryon samples could be used to compare baryon and antibaryon decay distributions across phase space, rather than only their integrated decay rates.

This experimental accumulation was accompanied by substantial progress in reconstruction and analysis. Precise vertexing and charged-hadron identification made it possible to isolate fully hadronic beauty-baryon decays in a hadron-collider environment. Reversal of the magnetic-field polarity, carefully selected control modes, and kinematic reweighting reduced the sensitivity to production, trigger and detection asymmetries. At the same time, Dalitz-plot studies, phase-space binning, triple-product asymmetries and unbinned comparison methods allowed the search to move beyond a single integrated rate. The first observation ultimately depended on this combination of statistical power and control of percent-level instrumental effects~\cite{LHCb:2025ray}.

Theory developed along several complementary lines. Generalized factorization and quark--diquark implementations of QCD factorization provided early estimates for two-body amplitudes; PQCD provided a framework for calculating short-distance QCD contributions at the partonic level, including hard spectator scattering and nonfactorizable topologies; hadronic triangle loops and experimentally constrained scattering amplitudes modeled selected long-distance rescattering effects; and flavor symmetries related channels without attempting to determine all reduced amplitudes dynamically~\cite{Mohanta:2000nk,Zhu:2018jet,Han:2024kgz,Duan:2024zjv,Wang:2024oyi,Dery:2020lbc}. None of these methods presently supplies a universal model-independent description. 

Generalized factorization generates only the perturbative strong phases encoded in effective quark-level coefficients and does not systematically describe hadronic rescattering or other nonperturbative sources of strong phases; existing QCDF-inspired calculations for beauty-baryon decays rely on the quark--diquark approximation, which is not derived from a baryonic factorization theorem and cannot fully reproduce the three-quark structure of the baryon; a rigorous factorization proof for PQCD calculations of beauty-baryon decays is still lacking, and the framework has not yet been developed for multibody modes; hadronic-loop results depend on the truncation of intermediate states and phenomenological off-shell form-factor prescriptions; scattering-based descriptions still require a model for weak production; flavor symmetry analyses alone cannot determine the dynamical matrix elements and suffer from the corresponding symmetry breaking effects. Nevertheless, these calculations played an important exploratory role before the discovery: they estimated plausible scales, proposed sensitive channels and observables, and clarified how strong phases and partial-wave interference could shape the measured asymmetries.

One particularly useful lesson came from the PQCD analysis of two-body $\Lb$ decays. For $\Lb\to p\pi^-$, representative partial-wave asymmetries of about $17\%$ in the $S$ wave and $-6\%$ in the $P$ wave were found to cancel partially, leaving an integrated asymmetry of only about $5\%$. In $\Lb\to pK^-$, the rate is dominated by an $S$ wave with a comparatively small predicted asymmetry, while the more asymmetric $P$ wave carries little weight~\cite{Han:2024kgz,Han:2025tvc}. These results supplied a dynamical explanation for why null or small two-body rate asymmetries do not imply an absence of appreciable $\CP$-violating information. They also motivated observables that separate partial waves or project their interference rather than integrating over it.

A second notable development was the use of experimentally determined $N\pi$ scattering partial waves to describe rescattering within a selected low-invariant-mass subsystem. Before the experimental observation, this approach predicted $A_{\rm CP}[\Lb\to(\Delta^{++}\pi^-)K^-]=(5.8\pm0.2)\%$ after integration over $m_{\Delta^{++}\pi^-}\in[1.2,1.9]\,\mathrm{GeV}$~\cite{Wang:2024oyi}. The quoted small uncertainty represents only the variation among the form-factor choices considered and is not a complete estimate of the model uncertainty. The predicted value is numerically close to the asymmetry subsequently observed in the low-$m_{p\pi^+\pi^-}$ region, although the theoretical and experimental regions, and hence their amplitude content, are not identical.

The decisive step was the LHCb measurement of $\Lb\to pK^-\pi^+\pi^-$. The phase-space-integrated asymmetry was measured to be $A_{\CP}=(2.45\pm0.46\pm0.10)\%$, differing from zero by $5.2$ standard deviations and establishing $\CP$ violation in a baryon decay for the first time~\cite{LHCb:2025ray}. The most significant regional result, $A_{\CP}=(5.4\pm0.9\pm0.1)\%$ with a significance of $6.0$ standard deviations, was found for $m_{p\pi^+\pi^-}<2.7\,\mathrm{GeV}$ in a topology receiving contributions from excited nucleon and $\Delta$ states. Several of the regional asymmetries are also broadly compatible in scale with phenomenological calculations of selected resonance contributions relevant to the corresponding topologies~\cite{Wang:2024oyi,Feng:2026soj,Shang:2026knt,Han:2025tvc}. Such comparisons are necessarily qualitative, because these calculations describe particular quasi-two-body or resonant contributions rather than the complete coherent amplitude within each experimental region. The comparisons therefore indicate that effects of the observed size can arise from SM weak phases combined with hadronic dynamics, but should not be regarded as precision predictions of the measured regions or as unique identifications of the responsible resonances.

The experimental resources now available are substantially stronger than before the observation. There is an established signal, a set of benchmark two-, three- and four-body measurements, and several null results that constrain large effects in the observables and phase-space regions tested. The Run~III sample and future LHCb upgrades should greatly enlarge the yields of fully hadronic beauty-baryon decays. The LHCb Run~III period has recently concluded, delivering a total integrated luminosity of about 28~fb$^{-1}$ with enhanced trigger efficiency. This yields roughly an order of magnitude more beauty-hadron data than Run~I/II, enabling a broader set of $\Lambda_b^0$ decay measurements to reach $\mathcal{O}(0.1\%)$ precision. 
Equally important, the methods needed to control nuisance asymmetries and multidimensional efficiencies have been demonstrated on data. The immediate opportunity is therefore not limited to improving the significance of the discovery channel, but extends to determining whether related effects can be revealed by other observables, occur in other regions of phase space, persist across different hadronic final states generated by the same quark-level transition, or appear in different quark-level transitions and beauty-baryon species.

The computational toolkit is more heterogeneous but also broader than it was a decade ago. Short-distance factorization frameworks, partonic PQCD calculations, hadronic rescattering models, scattering-data inputs and flavor-symmetry relations probe different components of the amplitudes and carry different systematic limitations. Their disagreements are themselves informative, because measurements of branching fractions, polarization observables and $\CP$ asymmetries can help diagnose which dynamical ingredients may be missing or overestimated. At the present stage, it is therefore more informative to test whether a calculation consistently describes the signs, relative magnitudes and kinematic dependence of several related observables and decay channels than to emphasize its agreement with a single measured central value.

Important phenomenological developments include the construction of a broad range of observables designed to probe $\CP$ violation by exploiting the spin degrees of freedom and multidimensional kinematics of baryon decays. In this review, we have organized these observables within a common angular-distribution language. The phase-space-integrated rate asymmetry remains the simplest benchmark, while local rate asymmetries retain variations across invariant masses and other kinematic variables. One-dimensional angular distributions support forward--backward or up--down asymmetries and Legendre moments that project selected angular structures. Under suitable polarization or cascade-decay conditions, the Lee--Yang parameter $\alpha$ can also be obtained from a one-dimensional distribution. The parameters $\beta$ and $\gamma$, however, generally require sensitivity to transverse polarization, azimuthal dependence or joint multidimensional angular distributions. More generally, multidimensional angular coefficients and their projected moments measure polarization and interference among helicity or partial-wave amplitudes. Together, these observables can recover $\CP$-violating information that is hidden by integrations and can test the phase structure of theoretical amplitudes more sharply than a rate alone.

Taken together, these advances have positioned both experiment and theory to move beyond the first observation and develop a broader program of beauty-baryon $\CP$-violation studies. We now turn to a staged roadmap, beginning with the exploration of additional channels and observables and progressing toward precision tests of the SM and searches for additional $\CP$-violating interactions.

\subsection{Prospects}

The first observation changes the status of the subject, but it does not immediately turn beauty-baryon $\CP$ violation into a precision program. Precision tests require several well-measured channels and theoretical descriptions under sufficient control; neither condition has yet been achieved across the field. A realistic roadmap should therefore distinguish a near-term phase of empirical accumulation from a longer-term program aimed at precision. In the next stage, the practical priority remains to map out the empirical landscape: which classes of decays exhibit measurable effects, how these effects vary across phase space, and which angular and polarization-sensitive observables provide complementary information.

Experimentally, the near-term program should broaden along several axes. Measurements should cover $\Lb$, $\Xi_b$ and, where feasible, $\Omega_b$ decays; nonleptonic modes with charmless and charm-containing final states, as well as semileptonic and radiative decays; and two-, three- and four-body final states. These classes involve different trade-offs among signal yields, reconstruction efficiencies, background conditions, and the kinematic and spin information available for constructing $\CP$-odd observables. The observable program should expand from integrated and local rate asymmetries to one-dimensional angular moments and multidimensional angular coefficients. The latter are especially valuable because they retain polarization information and interference between amplitudes that may cancel in the total rate or disappear upon projection onto a single angular variable. Repeating the discovery measurement with higher statistics and establishing consistent patterns in related channels will be essential for confirming the robustness of the observed effects, enabling credible dynamical interpretations, and constraining and guiding the development of theoretical frameworks.

This expansion must be accompanied by common experimental standards. Charge-dependent efficiencies, production asymmetries, trigger effects and nonuniform multidimensional acceptance should be propagated with correlations and validated using control modes. Local phase-space asymmetries do not require an amplitude model for their definition. By contrast, attributing an observed pattern to particular resonances, strong phases or short-distance topologies does require a coherent amplitude analysis and an assessment of model dependence. Keeping these two tasks distinct will make both the measurements and their interpretation more transparent.

In the near term, existing theoretical approaches should continue to extend their calculations to more channels and observables despite their limitations. At this stage, their main value is not yet to provide percent-precision SM benchmarks, but to estimate the likely magnitudes of $\CP$-violating effects, identify promising channels and resonance or angular regions in which interference is enhanced, and expose cancellations that could hide $\CP$ violation in an integrated measurement. Whenever the required amplitudes are available, predictions for branching fractions should be accompanied by predictions for rate asymmetries, polarization observables and angular-moment asymmetries. Systematic comparisons with data across related channels and observables can then reveal which dynamical ingredients are adequately described or remain missing, thereby guiding improvements to the respective frameworks.

The longer-term objective is to develop systematically improvable QCD descriptions for selected classes of decays. This requires factorization formulae with explicit domains of validity, a complete power counting for baryonic topologies, and controlled formulations or parameterizations of power-suppressed contributions. The necessary nonperturbative inputs include LCDAs for the relevant baryons and mesons, multihadron distribution amplitudes for quasi-two-body regions, and transition form factors, all with quantified uncertainties. Lattice-QCD calculations and QCD sum rules can provide complementary constraints on these quantities, while experimentally determined scattering amplitudes can separately constrain long-distance final-state interactions in hybrid or data-driven descriptions. A staged theoretical program should first determine and validate the baryonic transition form factors, using semileptonic branching fractions and angular distributions. The next objective should be to establish the topology power counting and leading-power factorization formulae for nonleptonic decays and to test them in channels whose branching fractions are dominated by a single leading amplitude. Once the leading-power description is under control, subleading topologies and power-suppressed contributions, including their strong phases, should be systematically calculated or parameterized. Achieving this final level of control is essential for turning predictions of $\CP$ asymmetries and amplitude-interference observables from exploratory estimates into precision tests.

Light-cone sum rules may provide one complementary route~\cite{Khodjamirian:2000mi,Khodjamirian:2023wol}. Besides their familiar use for form factors, one may construct correlation functions intended to access a nonleptonic two-body amplitude directly. The external light-cone state can be chosen as the initial beauty baryon, a final light baryon, or a final light meson. The other hadrons are represented by interpolating currents with the appropriate quantum numbers. After complete sets of hadronic states are inserted in the corresponding channels, the desired nonleptonic matrix element appears in the residue of the relevant hadron poles and can be extracted by matching a double dispersion representation to the light-cone expansion, supplemented by Borel transformations and continuum subtraction~\cite{Shi:2022kfa,Shi:2024plf}. Such constructions could connect hadronic distribution amplitudes to specific nonfactorizable or power-suppressed contributions. They do not replace a factorization theorem, however, and their practical reliability will depend on the choice of correlator, continuum treatment and available distribution amplitudes.

A parallel priority is to generalize angular analyses beyond a single isolated resonance. Recent work has demonstrated the value of retaining the full angular correlations in specific four-body cascade topologies~\cite{Zhang:2025mne}, but many existing formulae still assume one quasi-two-body chain. Realistic multi-body beauty-baryon decays may instead contain several overlapping resonant contributions with different spin-parity quantum numbers, together with nonresonant amplitudes, all contributing coherently to the same final state. A useful formalism should therefore construct complete angular bases accommodating multiple $J^P$ contributions and derive projection weights that isolate experimentally accessible combinations of helicity or partial-wave amplitudes and their interferences. Minimally model-dependent moment analyses could be organized according to the quantum numbers that may contribute within a chosen mass region. They would provide experiments with finite sets of angular coefficients whose $\CP$-odd combinations can be measured and subsequently used to constrain or validate more detailed amplitude models.

Multidimensional distributions can also furnish null tests for additional $\CP$-violating sources beyond the SM. A recent study considered the cascade $\Lb\to\Lambda_c^+(\to\Lambda\pi^+)\ell^-\bar\nu_\ell$ in the presence of general vector, axial-vector, scalar, pseudoscalar and tensor interactions carrying new weak phases~\cite{Li:2026szn}. In the absence of absorptive phases, the particle--antiparticle rate asymmetry remains zero because the decay rates are invariant under complex conjugation of the Wilson coefficients. By contrast, the genuine $\CP$-odd angular coefficient $K_{3s}$, which multiplies the sines of three distribution angles, vanishes in the SM but is sensitive to imaginary interference terms generated by additional weak phases. In addition, $K_{3s}$ can probe interference terms that are absent from the corresponding $\CP$-odd angular coefficients in $\bar B\to D^*(\to D\pi)\ell^-\bar\nu_\ell$. This illustrates the potential of spin-dependent angular coefficients in beauty-baryon decays to uncover new $\CP$-violating effects hidden in mesonic systems and probe their Lorentz structures.

LHCb will remain the central facility for this program because of its large and diverse beauty-baryon samples, with Run~III data and the planned Upgrade~II offering substantial gains for fully hadronic modes. Future high-luminosity $e^+e^-$ colliders operating near the $Z$ pole, such as CEPC, FCC-ee or an ILC $Z$ program, could provide a complementary environment with cleaner event geometry, well-defined initial conditions and longitudinally polarized $b$ quarks produced through the parity-violating $Zb\bar b$ coupling and from polarized initial states~\cite{CEPCStudyGroup:2018ghi,Ai:2024nmn,FCC:2018byv,ILC:2013jhg}. The polarization of the resulting beauty baryons would open additional one- and multidimensional angular observables, which are absent in unpolarized beauty-baryon decays.

Experience from hyperon and charm-baryon studies should continue to inform this development (see the review~\cite{Cheng:2025kpp} for example). Entangled baryon--antibaryon samples, self-analyzing weak decays and joint angular fits have already enabled simultaneous determinations of Lee--Yang parameters and their charge-conjugate counterparts~\cite{BESIII:2021ypr}. These measurements do not directly determine beauty-baryon amplitudes, but they test the angular formalism and statistical tools needed for more complicated beauty-baryon systems. Conversely, beauty-baryon measurements can extend these methods to amplitudes with potentially sizable weak-phase differences in $b$-quark transitions and a wider range of short-distance operators.

Beyond the direct task of the baryon program---namely, testing the Standard Model realization of $\CP$ violation and systematically searching for new weak phases in this newly accessible sector---baryonic $\CP$ observables can, within specified effective theories or new physics models, be combined with constraints from meson decays, electric-dipole-moment searches and collider measurements~\cite{Hiller:2001zj,Cirigliano:2019vfc,Fuchs:2020uoc,Hou:2023ran,Liu:inprep}. Within a given framework, the same $\CP$-violating couplings or overlapping effective operators can contribute to several classes of observables, allowing their complementary sensitivities to resolve flavor and interaction structures that no single sector can determine. When these interactions also participate in baryogenesis, such a multi-front program provides one of the most comprehensive means of testing the relevant parameter space, even if the associated new particles remain beyond the direct reach of present experiments.

The first observation of baryonic $\CP$ violation therefore marks the beginning rather than the culmination of this program. During the coming period, more channels, more differential information and more theoretically motivated observables must be accumulated. In parallel, the most promising modes should be developed toward controlled QCD predictions and coherent multidimensional analyses. The combined advances in experimental measurements, theoretical calculations and phenomenological analyses that enabled the first discovery will remain essential both for building a systematic understanding of $\CP$ violation in the baryon sector and for probing possible sources of $\CP$ violation beyond the SM.